\documentclass[longauth]{aa} 

\usepackage{graphicx}
\usepackage{txfonts}
\usepackage[colorlinks=true,allcolors=blue]{hyperref}
\usepackage{xspace}
\xspaceaddexceptions{] \ }
\usepackage{threeparttable}
\usepackage{capt-of}
\usepackage{bm}
\usepackage{adjustbox}

\usepackage{caption}
\DeclareCaptionFormat{cont}{#1 (Continued)#2#3\par}

\newcommand{\oiii}{[O\,\textsc{iii}]\xspace}

\newcommand{\nii}{[N\,\textsc{ii}]\xspace}
\newcommand{\sii}{[S\,\textsc{ii}]\xspace}

\newcommand{\ha}{H$\alpha$\xspace}
\newcommand{\hb}{H$\beta$\xspace}

\newcommand{\cii}{[C\,\textsc{ii}]\xspace}

\newcommand{\hei}{He\,\textsc{i}\xspace}
\newcommand{\heii}{He\,\textsc{ii}\xspace}

\newcommand{\lbol}{$L_\textrm{bol}$\xspace}
\newcommand{\vout}{$v_\textrm{out}$\xspace}

\newcommand{\mdot}{$\dot{M}_\textrm{out}$\xspace}
\newcommand{\edot}{$\dot{E}_\textrm{out}$\xspace}

\newcommand{\vmax}{$v_\mathrm{max}$\xspace}
\newcommand{\rmax}{$R_\mathrm{max}$\xspace}
\newcommand{\rweig}{$R_\mathrm{fw}$\xspace}

\newcommand{\kms}{km~s$^{-1}$\xspace}
\newcommand{\ergs}{erg~s$^{-1}$\xspace}
\newcommand{\ergscm}{erg~s$^{-1}$~cm$^{-2}$\xspace}
\newcommand{\Msun}{$M_\sun$\xspace}
\newcommand{\Msunyr}{$M_\sun$~yr$^{-1}$\xspace}

\begin{document}

    \title{GA-NIFS: Powerful and frequent outflows in moderate-luminosity active galactic nuclei at $z\sim3-6$
   }

   \author{Giacomo Venturi
        \inst{\ref{inst:SNS}}\fnmsep\inst{\ref{inst:OAA}}\fnmsep\thanks{\href{mailto:giacomo.venturi@inaf.it}{giacomo.venturi@inaf.it}.}
        \and
        Stefano Carniani
        \inst{\ref{inst:SNS}}
        \and
        Elena Bertola
        \inst{\ref{inst:OAA}}
        \and
        Chiara Circosta
        \inst{\ref{inst:IRAM}}\fnmsep\inst{\ref{inst:ESAC}}
        \and
        Eleonora Parlanti
        \inst{\ref{inst:SNS}}
        \and
        Michele Perna
        \inst{\ref{inst:CAB}}
        \and
        Santiago Arribas
        \inst{\ref{inst:CAB}}
        \and
        Torsten Böker
        \inst{\ref{inst:ESA-STScI}}
        \and
        Andrew Bunker
        \inst{\ref{inst:Oxf}}
        \and
        Stéphane Charlot
        \inst{\ref{inst:Sorb}}
        \and
        Francesco D'Eugenio
        \inst{\ref{inst:KICC}}\fnmsep\inst{\ref{inst:Cavendish}}
        \and
        Roberto Maiolino
        \inst{\ref{inst:KICC}}\fnmsep\inst{\ref{inst:Cavendish}}\fnmsep\inst{\ref{inst:UCL}}
        \and
        Bruno Rodríguez del Pino
        \inst{\ref{inst:CAB}}
        \and
        Hannah Übler
        \inst{\ref{inst:MPE}}
        \and
        Giovanni Cresci
        \inst{\ref{inst:OAA}}
        \and
        Gareth C. Jones
        \inst{\ref{inst:KICC}}\fnmsep\inst{\ref{inst:Cavendish}}
        \and
        Nimisha Kumari
        \inst{\ref{inst:AURA}}
        \and
        Isabella Lamperti
        \inst{\ref{inst:UniFi}}\fnmsep\inst{\ref{inst:OAA}}
        \and
        Madeline A. Marshall
        \inst{\ref{inst:LANL}}
        \and
        Jan Scholtz
        \inst{\ref{inst:KICC}}\fnmsep\inst{\ref{inst:Cavendish}}
        \and
        Sandra Zamora
        \inst{\ref{inst:SNS}}
          }

   \institute{Scuola Normale Superiore, Piazza dei Cavalieri 7, I-56126 Pisa, Italy \label{inst:SNS}
            \and
            INAF -- Osservatorio Astrofisico di Arcetri, Largo E. Fermi 5, I-50125 Firenze, Italy \label{inst:OAA}
            \and
            Institut de Radioastronomie Millimétrique (IRAM), 300 Rue de la Piscine, 38400 Saint-Martin-d’Hères, France \label{inst:IRAM}
            \and
            European Space Agency, ESAC, Villanueva de la Cañada, E-28692 Madrid, Spain \label{inst:ESAC}
            \and
            Centro de Astrobiolog\'{\i}a (CAB), CSIC-INTA, Ctra. de Ajalvir km 4, Torrej\'on de Ardoz, E-28850, Madrid, Spain \label{inst:CAB}
            \and
            European Space Agency, c/o Space Telescope Science Institute, 3700 San Martin Drive, Baltimore MD 21218, USA \label{inst:ESA-STScI}
            \and
            Department of Physics, University of Oxford, Denys Wilkinson Building, Keble Road, Oxford OX1 3RH, UK \label{inst:Oxf}
            \and
            Sorbonne Universit\'e, CNRS, UMR 7095, Institut d'Astrophysique de Paris, 98 bis bd Arago, 75014 Paris, France \label{inst:Sorb}
            \and
            Kavli Institute for Cosmology, University of Cambridge, Madingley Road, Cambridge CB3 0HA, UK
            \label{inst:KICC}
            \and
            Cavendish Laboratory, University of Cambridge, 19 JJ Thomson Avenue, Cambridge CB3 0HE, UK
            \label{inst:Cavendish}
            \and
            Department of Physics and Astronomy, University College London, Gower Street, London WC1E 6BT, UK
            \label{inst:UCL}
            \and
            Max-Planck-Institut f\"ur extraterrestrische Physik, Gie{\ss}enbachstra{\ss}e 1, 85748 Garching, Germany \label{inst:MPE}
            \and
            AURA for ESA, Space Telescope Science Institute, 3700 San Martin Drive, Baltimore, MD 21218, USA
            \label{inst:AURA}
            \and
            Dipartimento di Fisica e Astronomia, Universit\`a di Firenze, Via G. Sansone 1, 50019, Sesto F.no (Firenze), Italy \label{inst:UniFi}
            \and
            Los Alamos National Laboratory, Los Alamos, NM 87545, USA \label{inst:LANL}
             }

   \date{Received 10 December 2025; accepted 4 June 2026}
 
  \abstract
  {
    The period between $z \sim 3-6$ (`cosmic morning'), a key transformational phase in galaxy evolution preceding `cosmic noon' ($z \sim 1-3$), is very poorly explored in terms of feedback from active galactic nuclei (AGN) acting through gas outflows driven by energetic radiation.
    
    In this work, we study the properties of outflows in AGN (mostly X-ray-selected) from the GOODS-S field, exploiting  JWST NIRSpec IFU observations as part of the Galaxy Assembly with NIRSpec IFS (GA-NIFS) GTO survey.
    Together with its twin subsample from the COSMOS field reported in a previous GA-NIFS work, this constitutes the largest spatially resolved sample of AGN outflows at these redshifts to date, comprising 16 targets with outflows (out of a total of 19 AGN), and probes the unexplored regime of AGN at $z \sim 3-6$ with bolometric luminosities of \lbol $\sim 10^{45-46}$~\ergs. 
    We mapped the rest-frame optical ionised gas emission lines (e.g. \oiii, \ha) at sub-kiloparsec scales and spectrally isolated the broad wings that trace fast outflows from the gas at rest in the host galaxies.
    
    The incidence of ionised outflows in the GOODS-S + COSMOS GA-NIFS sample is high (>75\%), among the highest at any redshift; 
    however, a more homogeneous outflow detection method and sample selection would be required to draw firm conclusions on the redshift evolution of the outflow incidence.   
    We inferred outflow velocities between $\sim$600--2000~\kms, maximum radii of $\lesssim$1 up to $\sim$4~kpc, and ionised gas mass outflow rates of $\sim$0.1 up to $\gtrsim$100~\Msunyr, which in some cases can exceed the host galaxy star formation rate (SFR).
    The assumed electron density of 1000~cm$^{-3}$ is conservative; lower values, as measured in some outflow studies and suggested by theory, would imply inversely proportionally higher mass outflow rates.
    We find that the outflow properties inferred for the GOODS-S + COSMOS GA-NIFS AGN sample and their relations to \lbol and SFR generally align with those observed for other spatially resolved literature samples of AGN outflows across different redshifts and AGN luminosities.
    Nonetheless, after accounting for any luminosity bias, our analysis suggests a cosmic evolution of the outflow properties, with higher median mass outflow rates (and possibly also mass loading factors) at higher redshifts, especially at $z>3$, which indicates that AGN outflows were stronger in the early Universe than at later times, and thus potentially more capable of affecting their host galaxy.
  }


   \maketitle

\nolinenumbers

\section{Introduction}\label{sec:intro}
The \textit{James Webb} Space Telescope \citep[JWST;][]{Gardner2006, Gardner2023} is revolutionising our view of the early Universe through its high-sensitivity instruments observing in the infrared (IR) band. In particular, the Near Infrared Spectrograph \citep[NIRSpec;][]{Jakobsen2022} integral field unit \citep[IFU;][]{Boker2022, Rigby2023} enables us to spatially resolve high-$z$ systems in unprecedented detail. NIRSpec IFU studies have investigated bright unobscured and extremely red quasars (QSOs) with powerful gas outflows and their environment ($z \sim 1-3$, `cosmic noon', \citealt{Wylezalek2022, Cresci2023, Veilleux2023, Vayner2024a}; $z \sim 4.5$, \citealt{Vayner2025}; $z \sim 6$, \citealt{Loiacono2024, Decarli2024, Marshall2025b}; and $z \sim 7.5$, \citealt{Liu2024}), highly obscured active galactic nuclei (AGN; $z \sim 4-5$; \citealt{Suh2025, Solimano2025}), multi-black-hole systems \citep[$z \sim 5$;][]{Ubler2025}, and radio galaxies ($z \sim 3-4$; \citealt{Saxena2024, Wang2024, Roy2024, Roy2026}), as well as the metal content, gas physical properties, and stellar populations of single galaxies and proto-clusters from $z \sim 1-3$ \citep{Welch2024} up to $z \sim 6-8$ \citep{Hashimoto2023, Venturi2024, Fujimoto2025, Messa2025} and beyond ($z \sim 11$; \citealt{Maiolino2024, Scholtz2024a, Xu2024}).

The largest effort so far in exploiting the IFU capabilities of NIRSpec to study the high-$z$ Universe is the Galaxy Assembly with NIRSpec Integral Field Spectroscopy (GA-NIFS)\footnote{\url{https://ga-nifs.github.io/}.} Guaranteed Time Observations (GTO) survey (PIs R. Maiolino and S. Arribas; see \citealt{Perna2023a} for an early overview).
The goal of the survey is to characterise the internal structure and the close environment of a sample of 55 AGN and star-forming galaxies at $z \sim 2-11$, to investigate the physical processes that drive galaxy evolution across cosmic time.
The NIRSpec IFU observations comprise high-resolution ($R$ = 2700) grating data, to study the rest-optical emission lines, and also low-resolution ($R$ = 100) prism data for all of the star-forming galaxies and a subset of the AGN (mostly type-2), to study the stellar continuum emission.
Briefly, in AGN and QSOs at $z \sim 3-7$ the GA-NIFS survey has found over-massive black holes (relative to the local black hole–host mass relation), powerful outflows, and several galaxy companions, including dual AGN, as well as hidden black holes in star-forming galaxies and sub-millimetre galaxies \citep{Ubler2023, Ubler2024a, Ubler2024b, Perna2023b, Perna2025a, Perna2025b, Marshall2023, Marshall2025a, Parlanti2024, DEugenio2024, Zamora2025, Perez-Gonzalez2025, Bertola2025}. It has revealed crowded environments and intense merging activity around massive star-forming galaxies up to $z \sim 9,$ as well as revealing their internal structure, chemistry, and outflows \citep{Arribas2024, Jones2024a, Jones2025, Rodriguez-DelPino2024, Ji2024, Lamperti2024, Parlanti2025, Marconcini2024a,  Marconcini2025, Scholtz2025}.

\begin{figure}
    \centering
    \hspace{1cm}
    \includegraphics[width=0.8\linewidth]{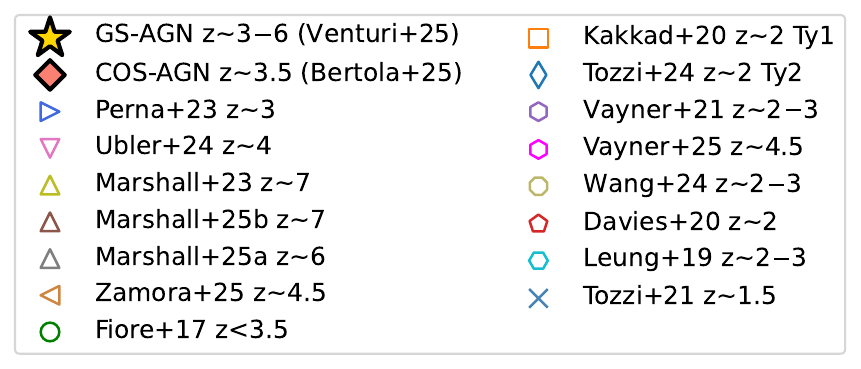}\\
    \includegraphics[width=0.9\linewidth]{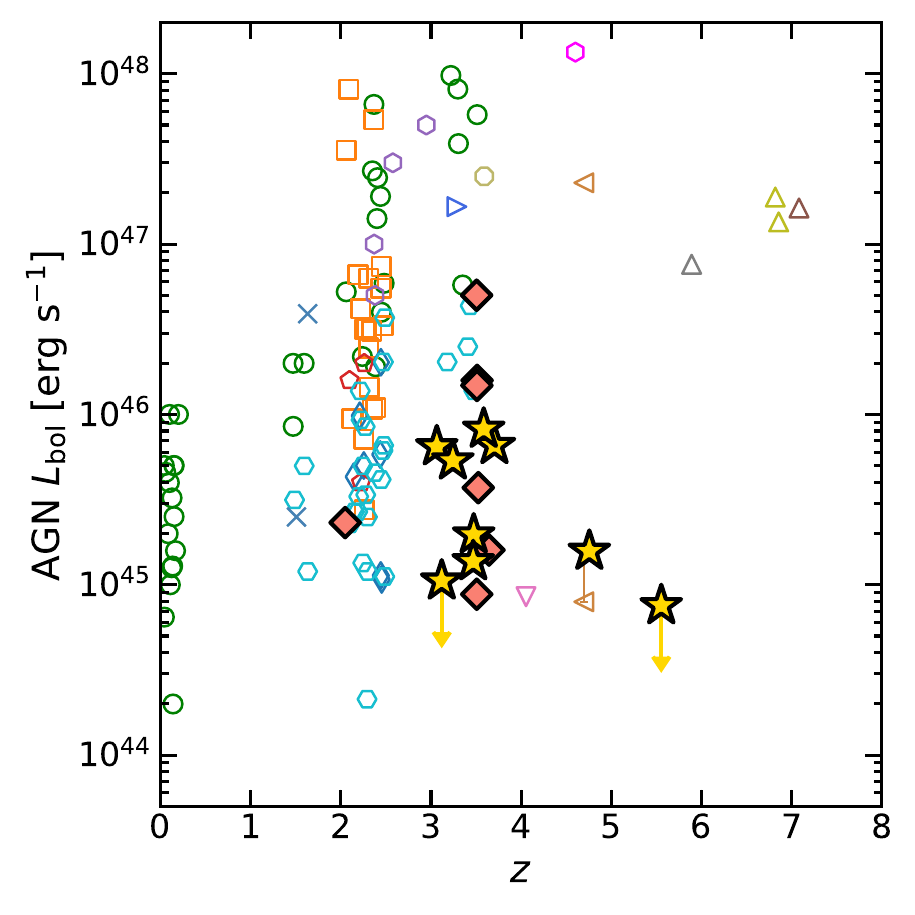}
    \caption{AGN bolometric luminosity (\lbol) versus redshift ($z$). The sources observed as part of the GA-NIFS survey are reported with golden stars (GS-AGN; this work) and salmon diamonds (COS-AGN; \citealt{Bertola2025}); out of the sample of 19 GS- and COS-AGN, only the 16 sources with detected outflows are shown. We also plot the targets included in other spatially resolved studies of AGN outflows at different $z$ and \lbol, comprising other GA-NIFS sources not part of the two samples above (the $z \sim 3$ QSO from \citealt{Perna2023b}, the $z \sim 7$ QSOs from \citealt{Marshall2023, Marshall2025a}, the $z \sim 4$ source in GOODS-N from \citealt{Ubler2024b}, and the QSO + AGN-host SMG from \citealt{Zamora2025}). The plot shows that the parameter space at $z \gtrsim 3$ and \lbol $\lesssim$ $10^{46}$~\ergs had remained uncharted until the advent of GA-NIFS.
    }
    \label{fig:Lbolvz}
\end{figure}

The properties of gas outflows accelerated by AGN and their impact on the host galaxies have been mainly studied through spatially resolved (mostly IFU) observations either in the local and low-$z$ Universe ($z \lesssim 1$; see compilation in \citealt{Fiore2017} and e.g. \citealt{Speranza2024}) and at cosmic noon ($z \sim 1-3$; \citealt{Fiore2017} compilation and e.g. \citealt{Genzel2014, ForsterSchreiber2019,  Leung2019, Kakkad2020, Tozzi2024}) or, at higher redshifts ($z \gtrsim 3$), only in powerful QSOs with high bolometric luminosities ($L_{\rm bol} \gtrsim 10^{47}$~\ergs; e.g. \citealt{Perna2023b, Marshall2023, Marshall2025a} from GA-NIFS and \citealt{Vayner2025}; see Fig.~\ref{fig:Lbolvz} and also Fig.~1 in \citealt{Bertola2025} for non-spatially resolved observations).
The parameter space that comprises redshifts between $z \sim 3-6$ (`cosmic morning') and bolometric luminosities of $L_{\rm bol} \lesssim 10^{46}$~\ergs, typical of less extreme AGN (i.e. not QSOs), has not been systematically probed with IFU observations so far. This is an important transitional phase in galaxy evolution during which galaxies undergo deep transformations that lead to the peak of star formation and AGN activity at cosmic noon. Moreover, the above bolometric luminosity range is particularly important since it comprises the bulk of the AGN population at these redshifts \citep[e.g.][]{Pouliasis2024}.

The GA-NIFS survey covers this gap by targeting a sample of AGN that match the above parameter space with NIRSpec IFU. In this study, we expand on the work of \cite{Bertola2025}, also part of GA-NIFS, who analysed the AGN in the COSMOS field (hereafter, COS-AGN), comprising six systems (two of which are dual AGN, which implies a total of eight AGN).
Specifically, we analyse nine other targets (two of which are dual AGN, which results in a total of 11 AGN) drawn from the GOODS-S field (hereafter, GS-AGN).
The full GS-AGN (this work) + COS-AGN \citep{Bertola2025} GA-NIFS sample thus comprises a total of 19 AGN, 16 of which have detected outflows (see Sect.~\ref{sec:outf_incid}).
We aim to obtain the spatially resolved properties of the outflows, compare them to those from other spatially resolved studies of ionised outflows in the literature (see Fig.~\ref{fig:Lbolvz}), and study their evolution with redshift, by using a consistent method to calculate the outflow properties for all of them.

Most of the AGN in both the GS and the COS sample were selected from X-rays (absorption-corrected X-ray luminosity $L_\mathrm{X,\,corr} > 10^{44}$~\ergs from \citealt{Luo2017} for GS and \citealt{Marchesi2016} for COS). These were complemented with additional AGN identified from the GA-NIFS NIRSpec IFU data, either secondary AGN companions revealed through emission-line diagnostic ratios (two for the COS-AGN, i.e. COS~1638-B and COS~1656-B, and two for the GS-AGN, i.e. GS~551-B and GS~10578-B; \citealt{Perna2025b}, \citealt{Bertola2025}, and Appendix~\ref{sec:app_emline_maps}), or single AGN in targets originally selected as star-forming or distant red galaxies (three in the GS sample). These are GS~3073, identified from its broad permitted lines \citep{Ubler2023}, and GS~19293 and GS~20936, assessed through line ratios \citep{Perna2025b}; GS~3073 and GS~19293 are X-ray undetected while GS~20936 is below the above $L_\mathrm{X,\,corr}$ threshold.

This work is structured as follows. In Sect.~\ref{sec:data_anal} we describe the observations and their reduction and analysis. Section~\ref{sec:outf_incid} discusses the incidence of ionised outflows in the GS- + COS-AGN GA-NIFS sample as compared to other studies at different redshifts.  Section~\ref{sec:outf_props} illustrates the derivation of the outflow properties. Section~\ref{sec:outf_vs_agn+gal_props} focuses on the comparison between the outflow properties and the AGN bolometric luminosity and host-galaxy properties for both the targets from this work and those from the existing spatially resolved AGN outflow studies in the literature. Finally, Sect.~\ref{sec:zevol} investigates the evolution with redshift of the outflow properties and Sect.~\ref{sec:concl} summarises our conclusions.
Throughout this work, the reported wavelengths are in vacuum and we adopt a flat $\Lambda$CDM cosmology with $H_0$ $\simeq$ 67.7~km~s$^{-1}$~Mpc$^{-1}$, $\Omega_\mathrm{M}$ $\simeq$ 0.31, and $\Omega_\mathrm{\Lambda}$ $\simeq$ 0.69 \citep{Planck2020}.

\begin{table}
    \centering
    \caption{Details of the observations used for this work.}
    \resizebox{\columnwidth}{!}{
    \begin{tabular}{lcccccc}
    \hline\hline
    Name & RA [deg] & Dec [deg] & Grating/filter & $t_\mathrm{exp}$ [s] \\
    \hline
        GS~133   & 53.02062 & --27.74223 & G235H/F170LP & 3560  \\
        GS~539   & 53.12208 & --27.93878 & G235H/F170LP & 3501  \\
                 &          &            & G395H/F290LP & 3501  \\
        GS~551\tablefootmark{a}   & 53.12435 & --27.85163 & G235H/F170LP & 3560  \\
        GS~774   & 53.17440 & --27.86740 & G235H/F170LP & 3560  \\
        GS~811   & 53.18466 & --27.88097 & G235H/F170LP & 3560  \\
        GS~3073  & 53.07886 & --27.88416 & G395H/F290LP & 18207 \\
        GS~10578\tablefootmark{a} & 53.16531 & --27.81413 & G235H/F170LP & 14706 \\
        GS~19293 & 53.05393 & --27.74771 & G235H/F170LP & 14706 \\
        GS~20936 & 53.11790 & --27.73438 & G235H/F170LP & 14706 \\
    \hline
    \end{tabular}}
    \label{tab:obs}
    \tablefoot{
    From left to right: target name, RA, Dec, NIRSpec grating/filter configuration, and on-source exposure time.\\
    \tablefoottext{a}{Dual AGN. The coordinates refer to the primary AGN.}
    }
\end{table}

\section{Data description, reduction, and analysis}\label{sec:data_anal}
The observations used in this work were acquired with JWST NIRSpec IFU under programme \#1216 (PI N. Luetzgendorf) as part of the GA-NIFS survey.
The spectra cover the rest-frame optical band (down to 2885 \AA\ and up to 8040 \AA, depending on the target redshift).
We obtained high-resolution ($R$ $\simeq$ 2700) grating data, which allow for an accurate modelling of the complex emission line profiles arising in the presence of outflowing gas, characterised by broad wings \citep[e.g.][]{Cresci2023, Veilleux2023, Perna2023b, Bertola2025}.
The observations were acquired in either the G235H/F170P (for $z$ $\lesssim$ 3.7) or G395H/F290P (for $z$ $\gtrsim$ 5.0) grating/filter configurations, or in both (for 3.7 $\lesssim$ $z$ $\lesssim$ 5.0), in order to cover both the \oiii$\lambda\lambda$4960,5008 + \hb and the \nii$\lambda\lambda$6550,85 + \ha emission line complexes. 
Information on the observations is given in Table~\ref{tab:obs}.
The different exposure times, ranging from slightly less than 1 h for some sources to about 4--5 h for others, are due to the fact that the former had been originally selected as confirmed X-ray AGN while the latter as (fainter) star-forming or distant red galaxies, and were either X-ray undetected (GS~3073 and GS~19293) or below the adopted X-ray luminosity threshold (GS~20936), but turned out to host an AGN only through the analysis of the NIRSpec data (see Sect.~\ref{sec:intro}); the only exception is GS~10578, originally already a confirmed X-ray AGN, but included as a distant red galaxy and thus observed with longer exposure times.
For more details on the observations employed for this work, we refer to \cite{Perna2025b}.

The raw data were downloaded from the MAST archive and processed with a custom version of the JWST Science Calibration Pipeline version 1.8.2 with CRDS context jwst\_1068.pmap, as described in detail in \cite{Perna2023b}. The modifications improved the data quality, in particular the masking of cosmic rays, open shutter leakage, and outliers, and the subtraction of the 1/$f$ correlated noise. To reject strong outliers, a modified version of LACOSMIC \citep{vanDokkum2001} was used before producing the final data cubes \citep{DEugenio2024}.
The final cubes were combined with the drizzle method with a pixel scale of 0.05$''$.

In the following, we describe the analysis of the science-ready data cubes performed in this work. 
The goal of the data analysis is two-fold: i) extracting the emission line fluxes on a spaxel-by-spaxel basis, to produce emission-line maps, and ii) obtaining the properties of the AGN outflows, by combining integrated spectra and the spatially resolved information.
In general, our analysis aims to spectrally isolate broad emission-line wings due to outflowing gas from the narrow line core tracing gas at rest.
For two of the objects, namely GS~539 and GS~811, the signal-to-noise (S/N) per spectral channel of the emission lines was too poor to allow for the proposed analysis, even when integrating multiple spaxels.
Therefore, for these two objects, we spectrally re-binned the data cube by a factor of 2 (GS~539) and 3 (GS~811); we tested different re-binning factors, and found that these two provided the best compromise between S/N and spectral resolution (needed to isolate broad outflow, narrow systemic, and BLR components).

\subsection{Spaxel-by-spaxel fitting}\label{sec:outf_spaxel_anal}

\begin{figure*}
    \centering
    \includegraphics[width=0.245\linewidth,trim={1cm 0 1cm 0},clip]{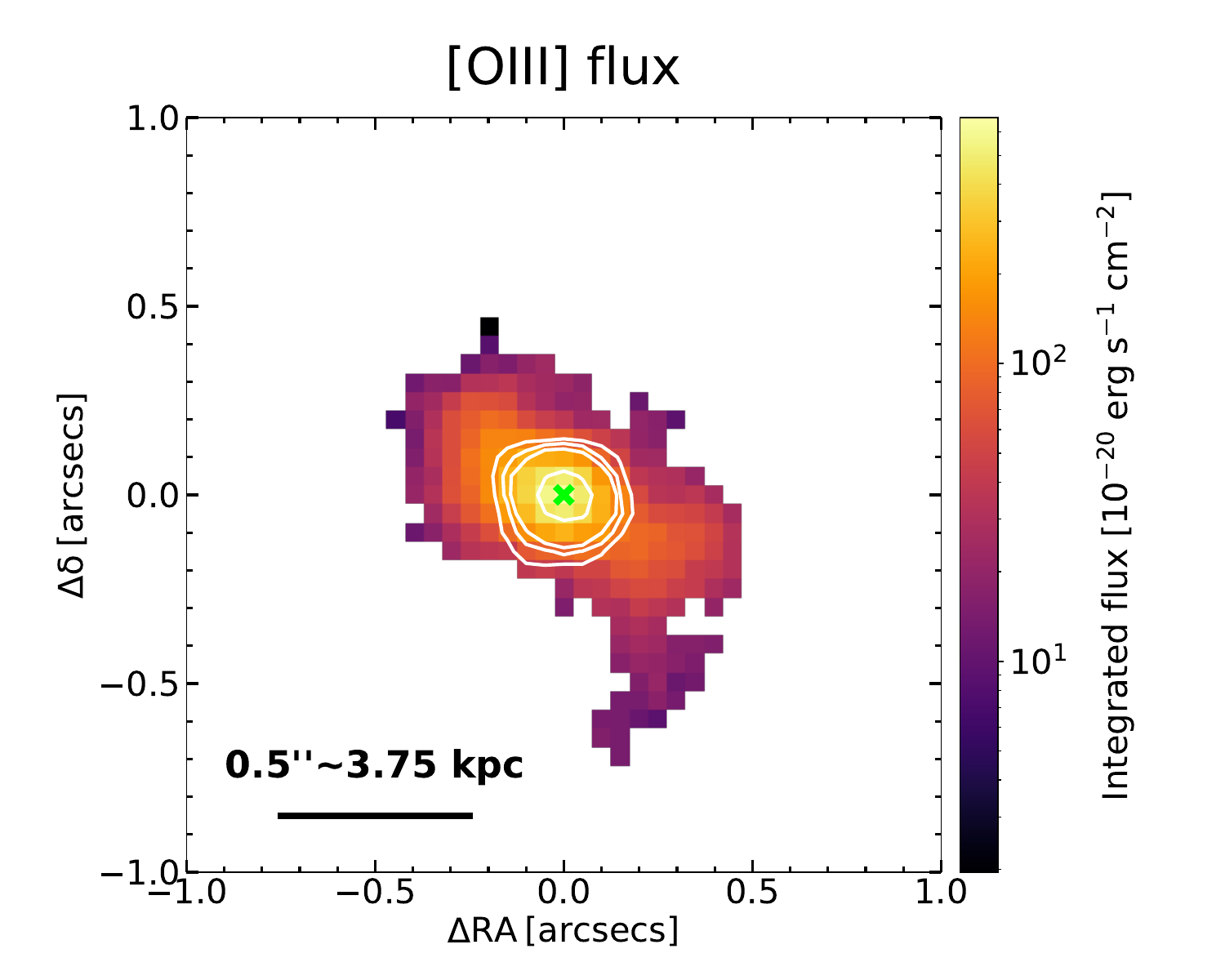}
    \includegraphics[width=0.245\linewidth,trim={1cm 0 1cm 0},clip]{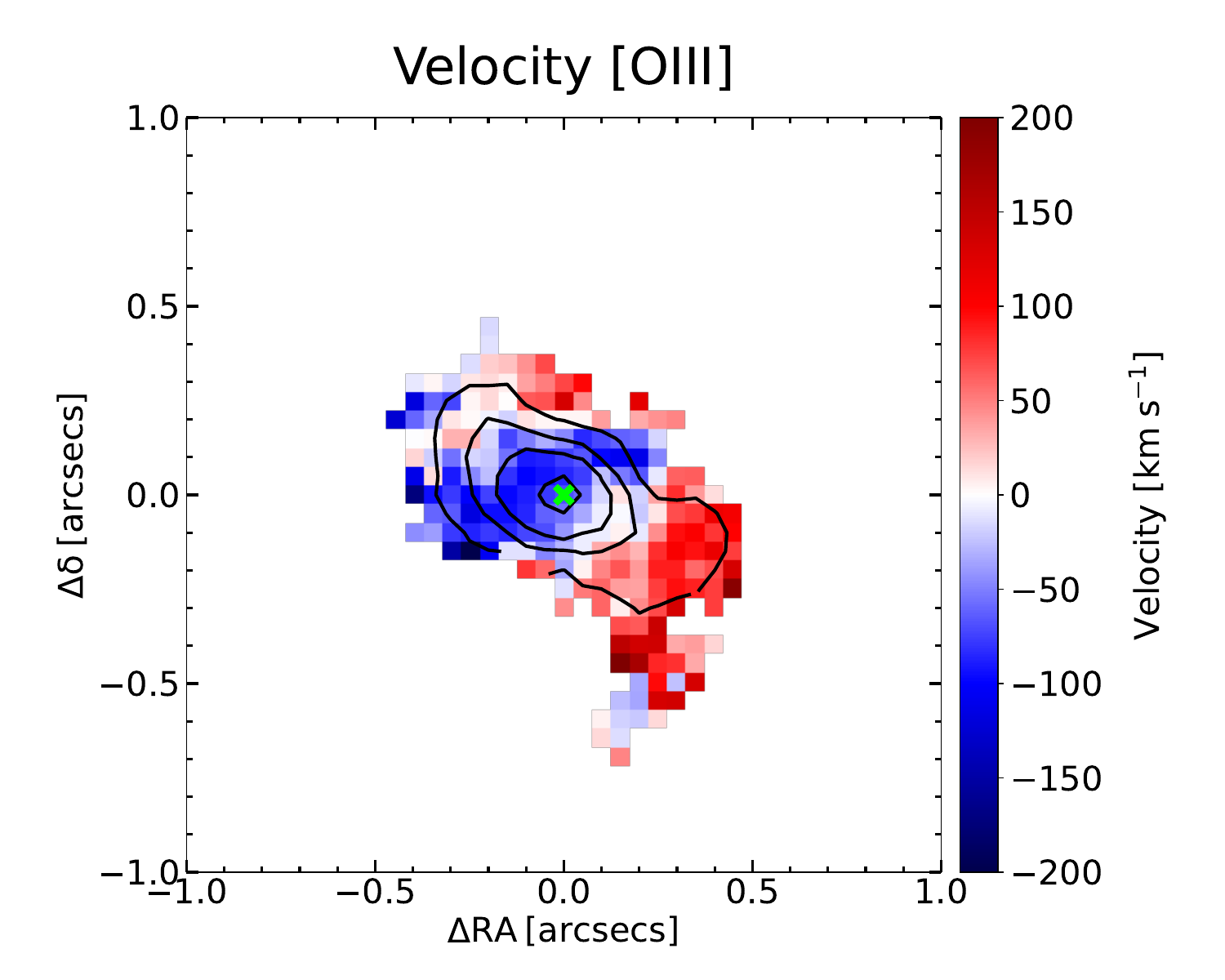}
    \includegraphics[width=0.245\linewidth,trim={1cm 0 1cm 0},clip]{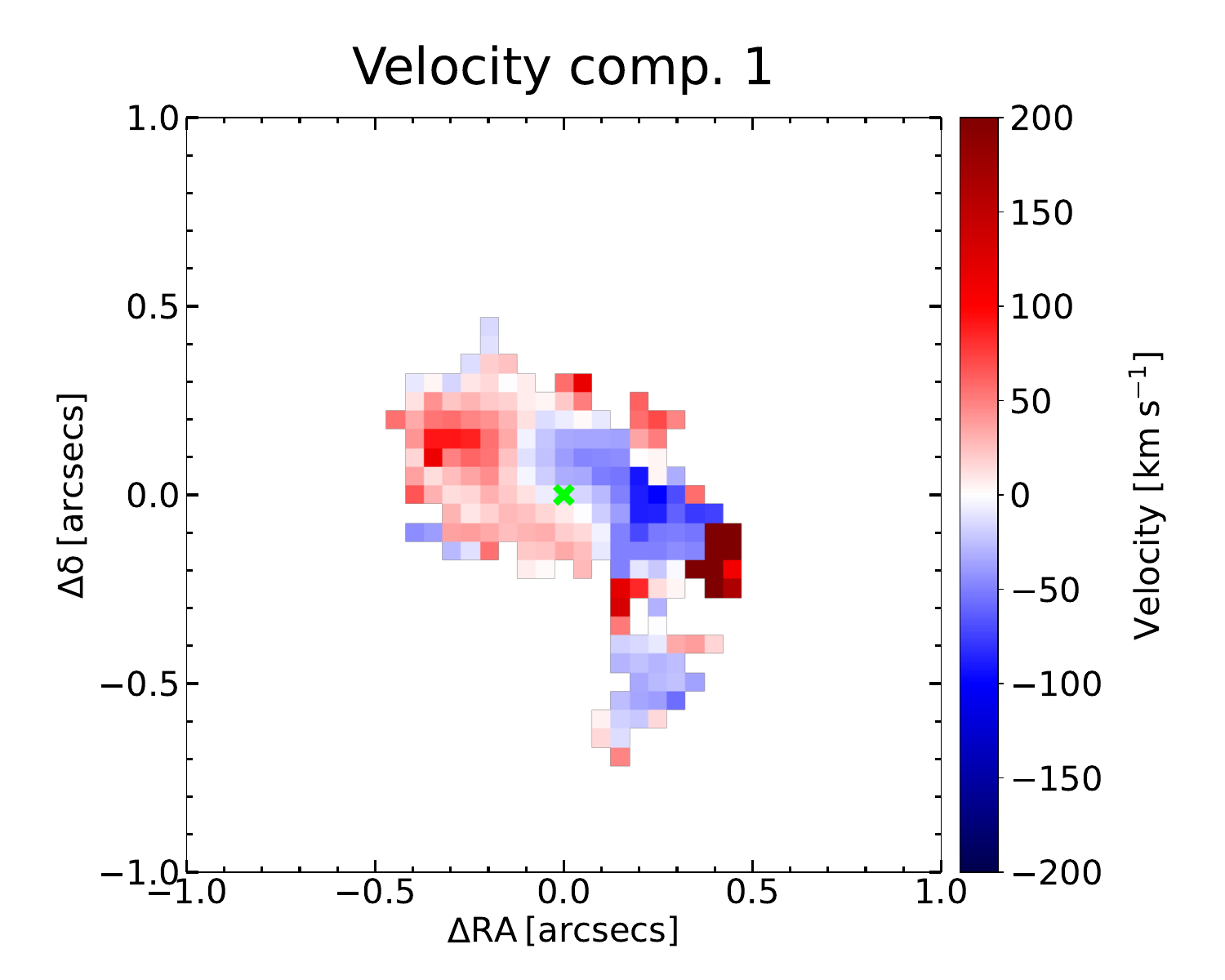}
    \includegraphics[width=0.245\linewidth,trim={1cm 0 1cm 0},clip]{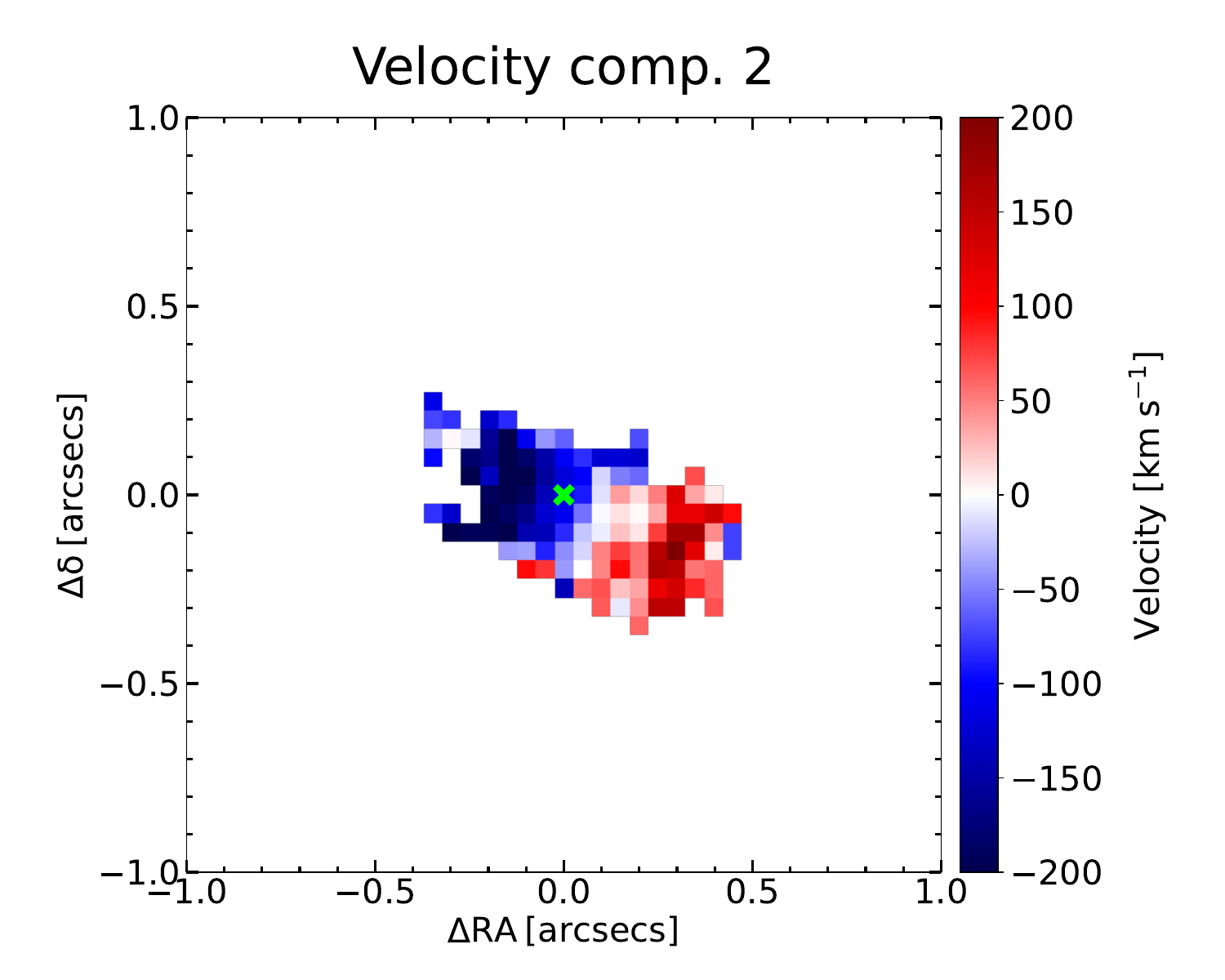}\\
    \includegraphics[width=0.245\linewidth,trim={1cm 0 1cm 0},clip]{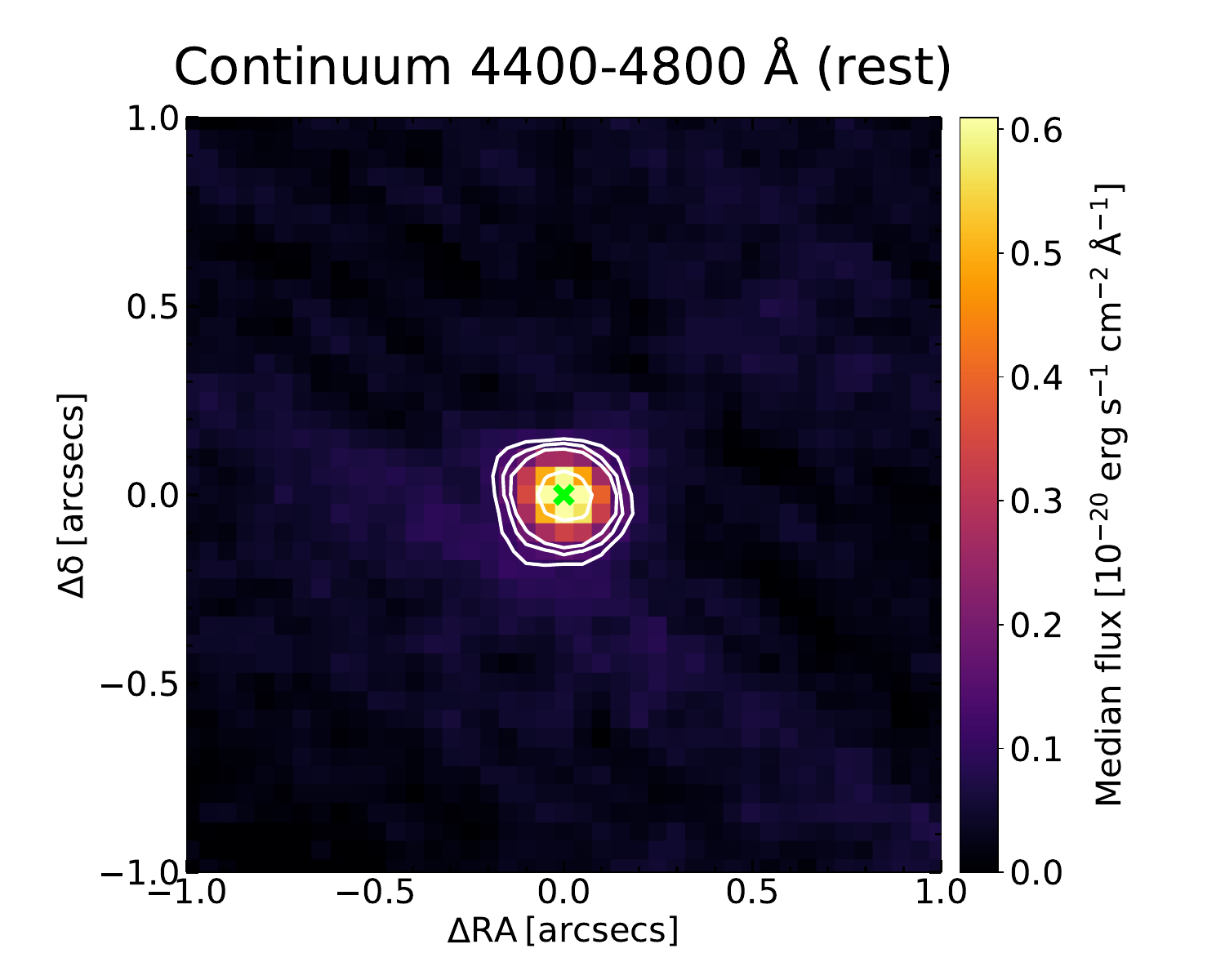}
    \includegraphics[width=0.245\linewidth,trim={1cm 0 1cm 0},clip]{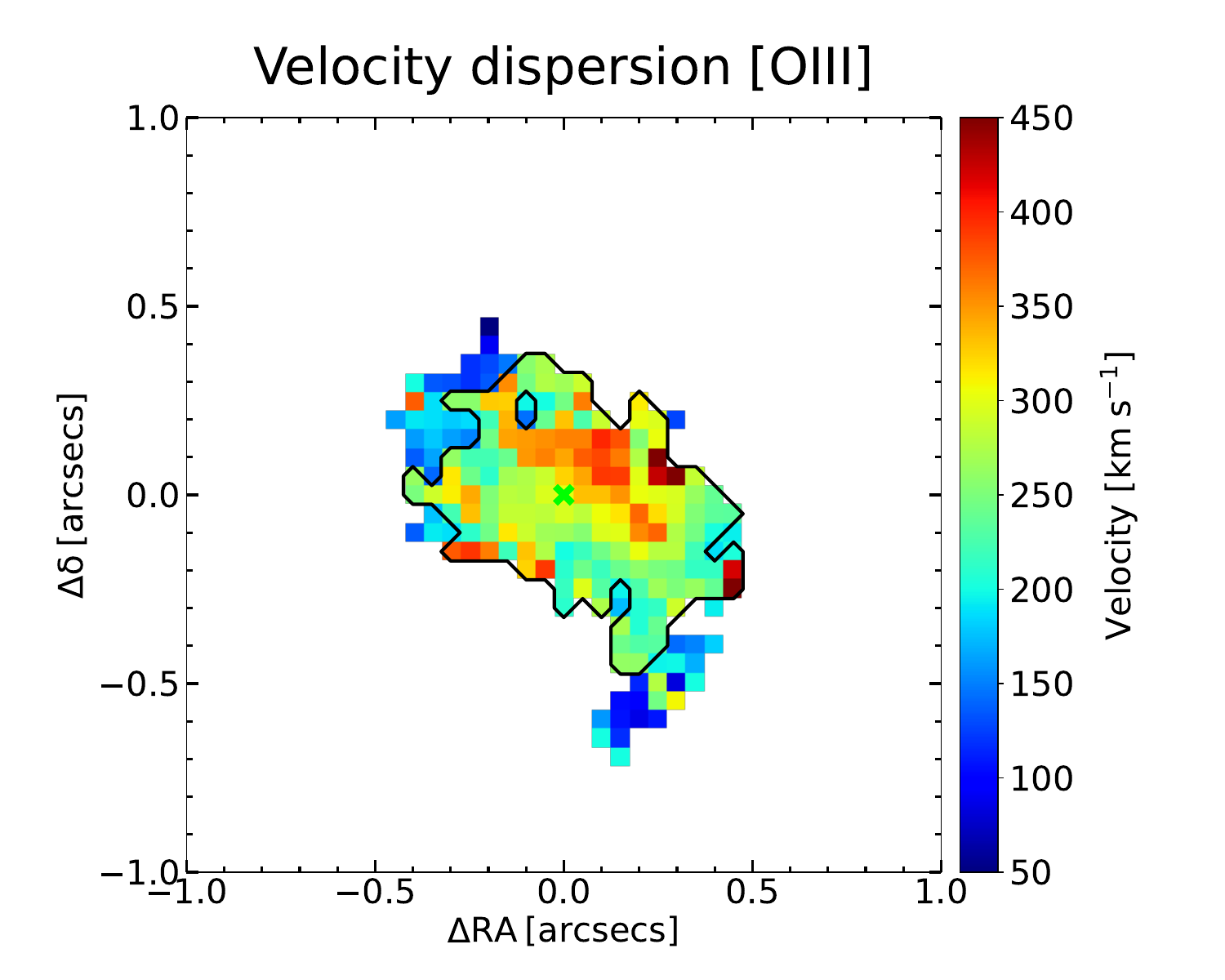}
    \includegraphics[width=0.245\linewidth,trim={1cm 0 1cm 0},clip]{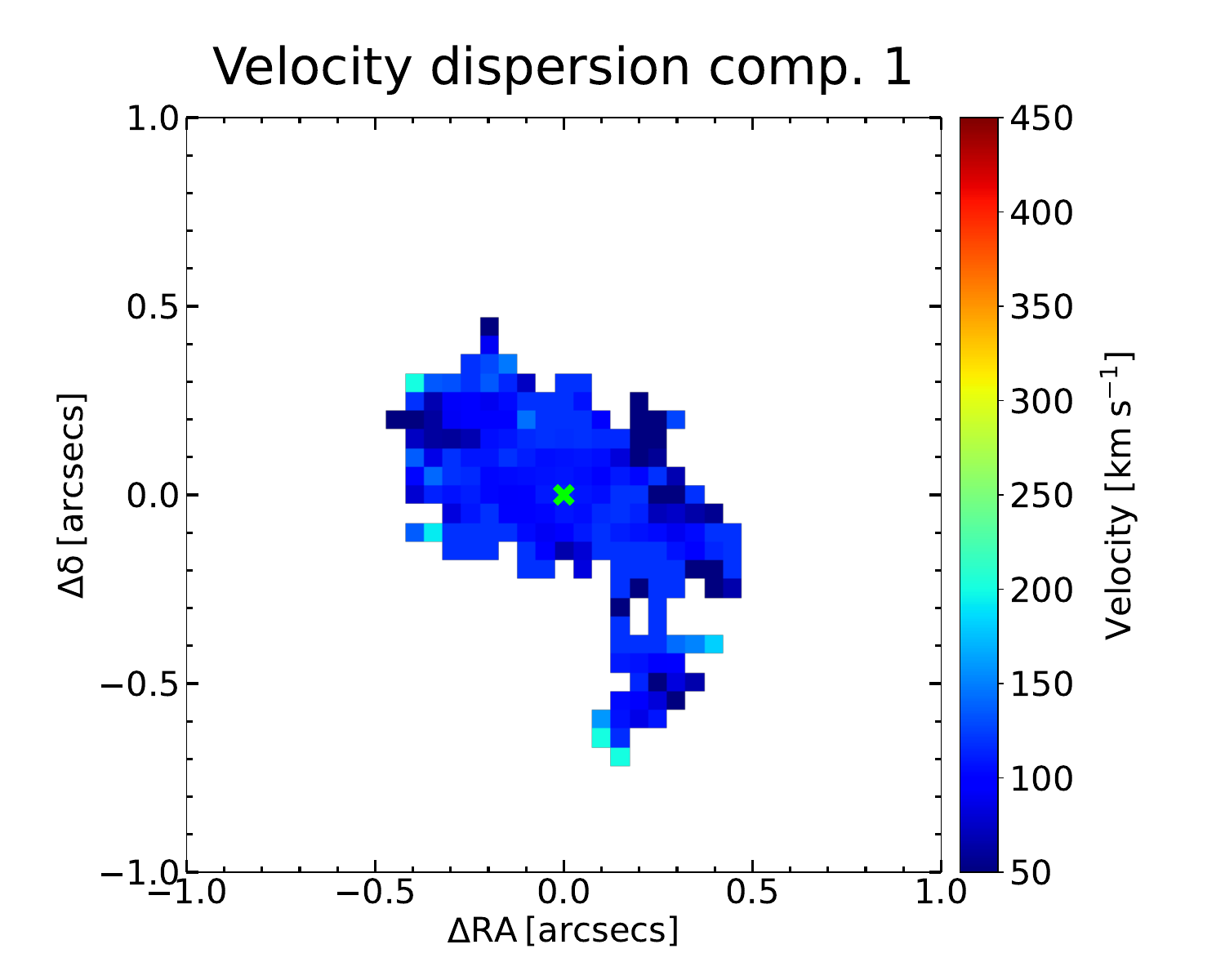}
    \includegraphics[width=0.245\linewidth,trim={1cm 0 1cm 0},clip]{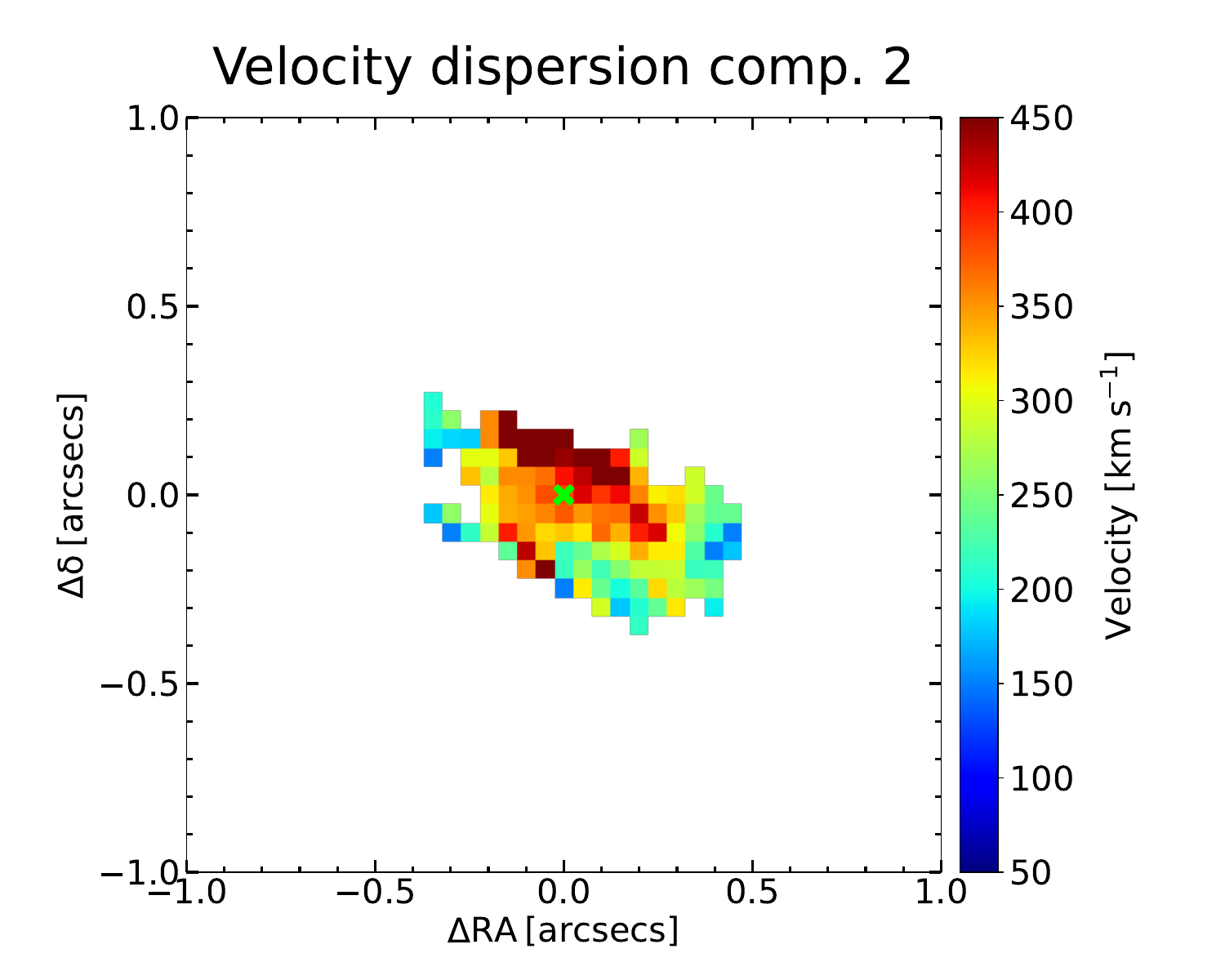}
    \caption{Rest-optical emission maps of GS~133 from JWST NIRSpec $R$2700 observations. 
    \textit{Top row} (from left to right): Flux (moment 0) and velocity (moment 1) of total \oiii$\lambda$5007 line profile and centroid velocity of 1st (narrower) and 2nd (broader) Gaussian components. \textit{Bottom row}: Continuum emission in the rest-frame range 4400--4800~\AA, velocity dispersion (moment 2) of total line profile and of components 1 and 2.
    The black contour on the \oiii velocity dispersion map encompasses the extraction region of the integrated spectrum used for the outflow analysis (see Sect.~\ref{sec:outf_integr_anal} and Figs.~\ref{fig:spectrum_gs133} and \ref{fig:spectra}).
    The white contours on the \oiii and continuum flux maps mark the continuum emission, while the black ones on the \oiii velocity and velocity dispersion maps trace the \oiii emission. In the maps for all the other targets in Figs.~\ref{fig:gs539}-\ref{fig:gs20936}, different colours may be used for the contours to simply enhance the contrast, but they have the same meaning as in this figure (when a different emission line instead of \oiii is reported, the \oiii contours are substituted with those of that line). The lime cross marks the position of the nucleus, as traced by the line emission peak.}
    \label{fig:gs133}
\end{figure*}

We first describe the spaxel-by-spaxel analysis, for which we used custom python scripts. 
The analysis consisted of modelling the emission lines of interest (\hb, \oiii$\lambda\lambda$4950,5008, \nii$\lambda\lambda$6550,85, \ha, \sii$\lambda\lambda$6718,33) with up to two Gaussians, one (hereafter labelled `narrow') aimed at describing the gas at rest in the galaxy, the other (hereafter labelled `outflow') aimed at reproducing the broad emission line wings due to outflowing gas.
This approach is routinely used for the analysis of spectra of galaxies hosting outflows \citep[e.g.][]{Liu2013, Bae2014, Harrison2012, Harrison2016, Brusa2016, Perna2017, Leung2019, Venturi2023, Bessiere2024}.
We added a first-order polynomial (with free amplitude and slope) to reproduce the continuum baseline, including the faint background. 
In case of type-1 AGN (see Table \ref{tab:em_line_fit}), we added an extra Gaussian component in \ha and \hb (plus \hei$\lambda$4923 and \hei$\lambda$6680 in GS~3073) to account for the very broad ($\sigma \gtrsim$ 1000~\kms) emission from the broad line region (BLR), and adopted a power-law to reproduce the continuum emission from the accretion disc. 
We employed the fitting routine \textsc{mpfit} \citep{Markwardt2009} for the spectral modelling described above.
We stress that the spectrally varying spectral resolution of NIRSpec \citep{Jakobsen2022} has been accounted for in the fitting, so that the velocity dispersion of each modelled Gaussian is given by $\sigma = \sqrt{\sigma_\mathrm{intr}^2 + \sigma_\mathrm{LSF}^2}$, where $\sigma_\mathrm{intr}$ and $\sigma_\mathrm{LSF}$ are the intrinsic and the instrumental velocity dispersions (LSF standing for line spread function), respectively. In the rest of the paper, all of the reported velocity dispersions refer to the intrinsic one ($\sigma_\mathrm{intr}$).
For either Gaussian component, we imposed the following constraints on the fit: i) we required all the emission lines to have the same velocity shift and (intrinsic) velocity dispersion, ii) we fixed the flux ratios between the strongest and the faintest lines of the \oiii and \nii doublets to their theoretical value of 3 \citep{Osterbrock2006}, and iii) we allowed the \ha/\hb line flux ratio to vary above the minimum theoretical attenuation-free value of 2.86 (for Case-B recombination and a gas temperature of $10^4$~K; \citealt{Osterbrock2006}) and iv) the \sii$\lambda$6716/$\lambda$6731 ratio to vary between 0.46--1.43, outside which the ratio saturates and stops being sensitive to the electron density \citep{Osterbrock2006}.

We performed two separate fits of the data cube, one with a single Gaussian component and one with two Gaussian components per emission line, as explained above.
To select whether to adopt one or two Gaussians per emission line in each spaxel, we used a combination of a statistically and a physically motivated criterion.
We considered the Bayesian information criterion\footnote{By assuming Gaussian noise, ${\rm BIC} = \chi^2 + k \ln(N)$ where $\chi^2$ is the chi-squared of the fit, $k$ the number of free parameters, and $N$ the number of data points (i.e. the spectral channels in our case).} (BIC; e.g. \citealt{Liddle2007}).
The BIC was calculated in a narrow spectral range encompassing the brightest lines, that is, either \oiii$\lambda$5008 or \nii$\lambda\lambda$6550,85 + \ha, or both, depending on the target.
We adopted the fit with two Gaussian components when $\Delta$BIC = BIC$_1$ -- BIC$_2$  > 4 (BIC$_1$ and BIC$_2$ being the BIC values for the 1- and the 2-Gaussian component fit, respectively), which indicates positive statistical evidence against the 1-component model fit \citep[e.g.][]{Kass1995}.

To this statistical criterion, we added a physically motivated one. The idea behind it is that a second Gaussian component, meant to trace outflowing gas, needs to be added when the line profiles are significantly broader than the galaxy's internal velocity dispersion.
To quantify this, we adopted a second component when the velocity dispersion of the emission lines in the fit with a single component was larger than a given threshold, $\sigma_1 > \sigma_{\rm thr}$. This threshold value was defined to exceed the highest values of velocity dispersion of the narrow component observed throughout the maps obtained from the two-component spectral decomposition of the emission line profiles described above.
This threshold value is therefore different among the objects, reflecting their different host galaxy masses and inclinations relative to the line of sight, resulting in different intrinsic velocity dispersions of the narrow component.
We thus adopted $\sigma_{\rm thr}$ = 150 \kms (GS~133), 250 \kms (GS~539), 150 \kms (GS~511), 250 \kms (GS~774), 300 \kms (GS~811), 150 \kms (GS~3073), 250 \kms (GS~10578), 150 \kms (GS~19293), and 200 \kms (GS~20936).
We also required the signal-to-noise ratio (S/N) on the peak of the second component (obtained from the fit) to be larger than 1.
Summarising, a second component was adopted when (S/N)$_2$ > 1 and either $\Delta$BIC  > 4 or $\sigma > \sigma_{\rm thr}$ for the 1-component fit.
Finally, we calculated the velocity and velocity dispersion of each emission line as the 1st- and 2nd-order moments, respectively, of the total modelled line profile in velocity space (constituted by either the first + second component or just a single Gaussian, depending on whether one or two Gaussians were adopted in the given spaxel).

The maps of GS~133 resulting from the above analysis are shown in Fig.~\ref{fig:gs133} as an example, while those for the other targets are reported in Figs.~\ref{fig:gs539}--\ref{fig:gs20936}, where the emission-line maps of each source are also briefly discussed.
Depending on the source, we only show the maps for the line---either \oiii, \nii, or \ha---for which the outflow signatures are most evident.
For GS~10578, we report the maps of both \nii and \oiii, to highlight different spatial components traced by each of these lines.

\begin{figure*}
    \centering
    \includegraphics[width=0.67\linewidth,trim={0.4cm 0 1cm 0},clip]{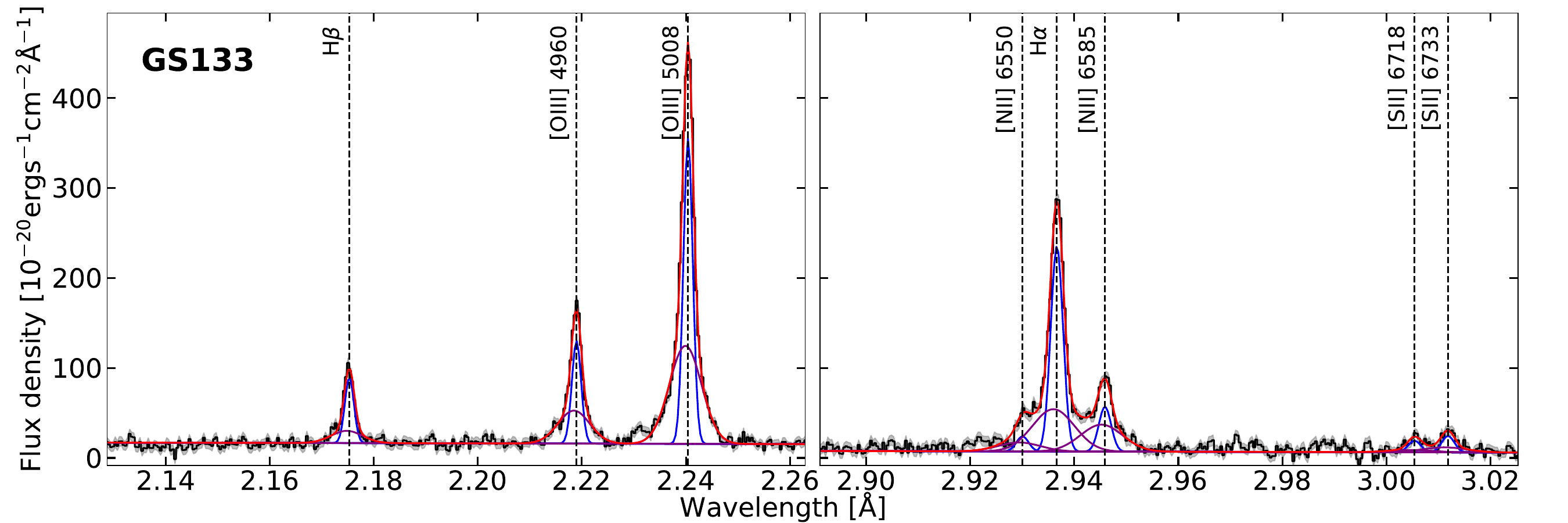}
    \caption{Integrated spectrum for GS~133 (black). The spectrum extraction region is shown by the black contour superimposed on the velocity dispersion map in Fig.~\ref{fig:gs133}. The coloured curves mark the narrow host-galaxy (blue) and broad outflowing (purple) components, while the red curve marks the overall best-fit model. Vertical, dashed labelled lines mark the wavelength of the peak of the fitted narrow component of each fitted emission line, adopted as the systemic velocity of each system. Wavelengths are in the observed frame. Uncertainties on the data are reported as shaded areas. The displayed wavelength range is smaller than the entire spectral range covered by the observations and smaller than the fitted one; the displayed range is [--250, +50] \AA\ around \oiii$\lambda$5008 and [--100, +200] \AA\ around \ha as calculated in rest-frame wavelengths. The modelled integrated spectra for all the other sources are shown in Fig.~\ref{fig:spectra}.}
    \label{fig:spectrum_gs133}
\end{figure*}

\subsection{Integrated outflow spectral analysis}\label{sec:outf_integr_anal}

To derive the outflow properties, we extracted an integrated spectrum for each target from the region encompassing the outflow.
This extraction region was defined by selecting those spaxels whose velocity dispersion of the total spectral line profile exceeds a lower-limit threshold, defined as about two times the typical velocity dispersion of the narrow component in each system, which ensures the selection of only spaxels with a significant outflow component. 
To do this, we use the line which overall has the brightest outflow (i.e. broad) component, the same for which we report the velocity dispersion map in Figs.~\ref{fig:gs133} and \ref{fig:gs539}-\ref{fig:gs20936}, where the spectrum extraction region is marked by a black contour; for GS~10578, for which both \nii and \oiii maps are shown, we adopted the \nii velocity dispersion map to define the outflow extraction region.
The adopted threshold values on the velocity dispersion of the total spectral profile are as follows: GS~133: 200~\kms, GS~511: 250~\kms, GS~10578: 300~\kms, GS~19293: 120~\kms, and GS~20936: 200~\kms.

For some of the targets, namely GS~539, GS~774, GS~811, and GS~3073 (i.e. all the type-1 AGN in the GS sample) the outflow turned out to be spatially unresolved, or at most only marginally resolved, based on the comparison of the spatial profiles of the \oiii emission-line wings with those of the nuclear \ha and \hb BLR components.
In these cases, we extracted the integrated spectrum from a circular aperture, whose radius, $R_\mathrm{ap}$, was defined to encompass the majority of the unresolved nuclear emission, while at the same time minimising the inclusion of noisy background spaxels. We thus adopted $R_\mathrm{ap}$ = 0.23$''$ ($3 \, \sigma_\mathrm{PSF}$) for GS~539, 0.25$''$ ($3.5 \, \sigma_\mathrm{PSF}$) for GS~774, 0.15$''$ ($2.4 \, \sigma_\mathrm{PSF}$) for GS~811, and 0.3$''$ ($3.5 \, \sigma_\mathrm{PSF}$) for GS~3073.

The noise spectrum was first calculated by adding the error spectra of the single spaxels in quadrature, which is equivalent to assuming uncorrelated noise between adjacent spaxels. This was then re-scaled by the ratio between its median and the standard deviation of the signal in the integrated spectrum in regions free of line emission, to account for correlations due to
re-sampling of single dithered exposures (as done e.g. in \citealt{Ubler2023, Jones2024a}). The above re-scaling factor results to be of about 3 to 4, depending on the dataset.

We modelled the continuum and emission lines in the integrated spectra of each target as described in Sect.~\ref{sec:outf_spaxel_anal} for the spaxel-by-spaxel case.
Two Gaussian components (one narrow for the gas at rest and one broad for the outflowing gas) were sufficient to adequately reproduce the line profiles (see Figs.~\ref{fig:spectrum_gs133} and \ref{fig:spectra}).
Only for GS~10578, where two narrow peaks in addition to a broad blueshifted component can be clearly seen in the integrated spectrum (see Fig.~\ref{fig:spectra}), we employed three Gaussians per line to reproduce the profiles.
Moreover, since this target shows the clear presence of the \hb absorption feature due to the stellar populations underlying the \hb emission line, we performed the spectral modelling with the penalized pixel-fitting code (\textsc{ppxf}; \citealt{Cappellari2004,Cappellari2017}) instead of \textsc{mpfit}, to allow for the inclusion of stellar population templates. Specifically, we employed the E-MILES single-stellar population (SSP) model spectra \citep{Koleva2012, Vazdekis2016, LaBarbera2017}, which adopt a unimodal standard Salpeter initial mass function (IMF) with slope 1.3 \citep{Salpeter1955} and scaled-solar `Padova+00' isochrones \citep{Girardi2000}, with M/H metallicities spanning from --1.71 to +0.22 solar (in log) and ages between 0.063 and 2.5 Gyr. Following \cite{DEugenio2024}, the maximum age was set to 0.5 Gyr older than the age of the Universe at the redshift of GS~10578, in order to avoid edge effects.

\section{Incidence of ionised outflows}\label{sec:outf_incid}
Here we discuss the incidence of ionised outflows in our GA-NIFS GS-AGN + COS-AGN sample, and compare it with that estimated in other samples of AGN-host galaxies at different redshifts matched in stellar mass.
Specifically, the stellar masses inferred from SED fitting are in the range $M_* \sim 10^{10-11} M_\odot$ (Table~\ref{tab:outflow_props} for GS-AGN and Table 1 in \citealt{Bertola2025} for COS-AGN).

\begin{table}
\centering
\caption{Outflow incidence from various surveys.}\label{tab_incidence}
\resizebox{\columnwidth}{!}{
\begin{tabular}{ | l | c | c | c | } 
 \hline
  Survey & $z$ & Method & Incidence \\ 
 \hline
 SDSS & <0.4 & FWHM$_\mathrm{Avg} >$ 500~\kms  & 17\% \\ 
 SDSS X-ray & <0.8 & \vmax > 650~\kms & >40\% \\
 KASHz & 0.6--1.1 & $W_\mathrm{80,H\alpha} >$ 600~\kms  & 13\% \\ 
 KASHz & 1.1--2.7 & $W_\mathrm{80,[OIII]} >$ 600~\kms  & 30--70\% \\ 
 KMOS$^\mathrm{3D}$ & 0.6-2.7 & broad comp.  & 60--70\%  \\ 
 MOSDEF & 1.4--3.8 & broad comp.  & 30\%  \\ 
 MOSDEF & 1.4--3.8 & broad comp.; S/N>100  & $>$50\%  \\ 
 SUPER & 2--2.5 & FWHM $>$ 1000~\kms  & 70\%  \\ 
 SUPER & 2--2.5 & $W_\mathrm{80,H\alpha} >$ 600~\kms  & 80--100\% \\
 \hline
 GA-NIFS & 3--6 & broad comp.  & 85\%  \\  
 GA-NIFS & 3--6 & $W_\mathrm{80,[OIII]/H\alpha} >$ 600~\kms & 75\%  \\  
 \hline
\end{tabular}}
\tablefoot{
From left to right: Name of the survey, redshift range, method adopted to identify outflow signature (see Sect.~\ref{sec:outf_incid} for details), and outflow incidence rate.}
\end{table}

Different definitions for identifying outflow signatures in spectra, and consequently various values of outflow incidence rate, are reported in the literature (Table~\ref{tab_incidence}).
At low $z$, \cite{Mullaney2013} report an incidence fraction of $\sim$17\% among all SDSS DR7 optically selected AGN at $z$ < 0.4, defined as the fraction of targets whose \oiii line has FWHM$_\mathrm{Avg}$ > 500~\kms (being the flux-weighted average of the FWHMs of the narrow and the broad modelled Gaussian components), though without any information on the stellar mass.
\cite{Perna2017} instead find an outflow incidence of $\sim$40\% in $z$ < 0.8 X-ray-selected AGN from SDSS, using a threshold of \vmax > 650~\kms (\vmax roughly coincides with $W_\mathrm{80}$; see e.g. \citealt{Fiore2017}), increasing to >50\% for \lbol $\gtrsim$ 10$^{45}$~\ergs, or $L_\mathrm{X} \gtrsim$ 10$^{44}$~\ergs (see also \citealt{Woo2016}).
\citet[][KASHz survey]{Harrison2016} use a threshold of $W_\mathrm{80,\oiii}$ > 600~\kms for the total \oiii or \ha line profile to identify outflows and find an outflow incidence of $\sim$30\% ($L_\mathrm{X} < 6 \times 10^{43}$~\ergs) to 70\% ($L_\mathrm{X} > 6 \times 10^{43}$~\ergs) at $1.1 < z < 1.7$ in a stellar mass range of $M_* \sim 10^{10.3-11.5} M_\odot$\footnote{The stellar masses for a subsample of the KASHz sources are reported in \cite{Bertola2024}.}. When considering a lower-redshift sample targeted with \ha, at $0.6 < z < 1.1$, they find an outflow incidence fraction of only $\sim$13\%.
At a similar redshift range ($0.6 < z < 2.7$), \citet[][KMOS$^\mathrm{3D}$ survey]{ForsterSchreiber2019} report an outflow incidence of approximately 60–70\%, based on the analysis of the \ha-\nii emission lines. The presence of outflows is identified through the requirement of an additional broad Gaussian component (FWHM > 400 km s$^{-1}$) in the line fitting.
\citet[][SUPER survey]{Kakkad2020} analyse a sample of type-1 AGN at Cosmic Noon ($z \sim 2$–2.5) and report an outflow incidence of approximately 70\%, considering only the fastest outflows that require a broad component with FWHM $>$ 1000 km s$^{-1}$. However, we note that if we adopt the threshold $W_\mathrm{80,\oiii}$ $>$ 600~\kms and include all AGN in the SUPER survey (type-1 from \citealt{Kakkad2020};  type-2 from \citealt{Tozzi2024}) the incidence rate is $\sim$80--100\%, where $L_\mathrm{X} \gtrsim 10^{44-45}$~\ergs.
Finally, \citet[][MOSDEF survey]{Leung2019} reports an outflow incidence of $\sim$30\% (17\% after removing possible mergers) in $1.4 < z < 3.8$ AGN, independent on stellar mass, based on the statistical significance of a second Gaussian component in the line fitting.

As can be seen, the different estimates of the outflow incidence reported in the literature rely on different definitions and are not really comparable.
Therefore, we estimate the incidence rate by adopting two methods: detection of a second broad component and $W_\mathrm{80,\oiii}$ $>$ 600~\kms.
The second broad component associated with outflowing gas is detected in all the GS-AGN (Figs.~\ref{fig:spectrum_gs133} and \ref{fig:spectra}), except for the secondary AGN in the two dual systems, namely GS~551-B and GS~10578-B \citep{Perna2025b}. All the COS-AGN, except for one (COS~1656-B), host outflows \citep{Bertola2025}.
With 16 detected outflows out of 19 AGN, the incidence of a broad component in the GA-NIFS GS- and COS-AGN is of $\sim$85\%.
By adopting the threshold of $W_\mathrm{80}$ > 600~\kms (in either \oiii or \ha), we find that all the COS-AGN outflows satisfy it, while two of the GS-AGN outflows do not exceed this threshold.
One is GS~3073, which despite a significantly broad second component (FWHM$_\mathrm{bro} \simeq 2.355\, \sigma_\mathrm{bro} \sim$ 800~\kms; Table~\ref{tab:em_line_fit}), has a very strong narrow line emission core which results in a fairly narrow width of the total line profile. The other is GS~19293, whose asymmetric line wings are just not particularly broad (Fig.~\ref{fig:spectra}).
In this case, the resulting incidence rate is $\sim$75\%.
Interestingly, these two targets, together with the secondary AGN companions above, are the only GS-AGN that are undetected in X-rays.
If we only consider the X-ray selected AGN, excluding the AGN identified from optical diagnostic diagrams---both the two above and the secondary nuclei in the dual systems---the outflow incidence rate becomes 100\%.

We can conclude that the outflow incidence in the GS- and COS-AGN at $z \gtrsim 3$ is similar to that found in the SUPER survey at $z \sim 2-2.5$ \citep{Kakkad2020, Tozzi2024}, while significantly larger than that found in KASHz at at $1.1 < z < 1.7$ \citep{Harrison2016}. 
This might indicate that outflows are more frequent above $z\sim2$.  
Unfortunately, different outflow detection criteria are adopted in other studies, either \cite{Leung2019} in a similar redshift range as those above, or \cite{ForsterSchreiber2019} extending to below cosmic noon, down to \cite{Mullaney2013} in the low-$z$ Universe.
Moreover, the bolometric (or analogously, X-ray) luminosities of the AGN in our GA-NIFS sample and in SUPER \citep{Kakkad2020, Tozzi2024}, typically \lbol~$\gtrsim$~10$^{45-46}$~\ergs ($L_\mathrm{X} \gtrsim 10^{44-45}$~\ergs), lie at the upper end of the luminosity range of the other samples above; this likely contributes to increasing the outflow incidence among the GA-NIFS and SUPER AGN as compared to the other studies, based on the previously mentioned correlation of the outflow incidence rate with \lbol and $L_\mathrm{X}$ found for $z \lesssim 2$ AGN by \cite{Harrison2016} and \cite{Perna2017}.
A study adopting a homogeneous method for the outflow detection, controlling for selection biases (such as in \lbol), would be needed to evaluate the outflow incidence as a function of cosmic time.

We note that our results cannot be extended to the whole population of AGN at $z>3$, in particular those with \lbol < 10$^{45}$~\ergs. Specifically, the new population of X-ray-undetected AGN discovered by JWST at $2 \lesssim z \lesssim 7$ does not show any significant evidence of a broad component in \oiii indicative of outflows, from either NIRSpec MSA \citep{Harikane2023, Maiolino2025} or IFU spectra \citep{Jones2026, Ubler2025, DEugenio2026, Maiolino2026}.
This could be due to the lower \lbol range spanned by these JWST-discovered AGN, generally $\lesssim$10$^{45}$~\ergs \citep{Scholtz2025a, Juodzbalis2026}, compared to that of our GA-NIFS sample and of SUPER.
Also, JWST-detected AGN have different properties compared to X-ray-selected AGN, such as an extreme X-ray weakness \citep{Maiolino2025}, sometimes `V-shaped' rest-frame UV-to-optical SEDs \citep[e.g.][]{Kocevski2025}, and spatially unresolved continuum emission implying very compact physical sizes for these systems ($\lesssim$100~pc; e.g. \citealt{Guia2024}). These properties have been interpreted as due to different black hole accretion properties and/or gas conditions of the AGN-host systems compared to pre-JWST AGN, which may reflect in a reduced capability to launch outflows.

\section{Ionised outflow properties}\label{sec:outf_props}
We calculated the ionised outflow properties by using the parameters for the broad Gaussian component obtained from the fit of the integrated spectrum extracted from the outflow region for each target, as described in Sect.~\ref{sec:outf_integr_anal}. For GS~10578, whose complex line profiles required the use of three Gaussians (see Fig.~\ref{fig:spectra}), we conservatively assumed that only the broadest one of them traces outflowing gas.
We obtained the mass of the outflow from the attenuation-corrected luminosity of the broad component of \ha and \oiii, as follows (as in \citealt{Carniani2015}):
\begin{equation}
    M_{\mathrm{out,H}\alpha}/\mathrm{M}_\odot = 0.6 \times 10^9 
    \left( \dfrac{L_{\mathrm{H}\alpha}}{10^{44}~\mathrm{erg~s}^{-1}} \right) 
    \left( \dfrac{n_\mathrm{e}}{500~\mathrm{cm}^{-3}} \right)^{-1} 
\end{equation}
\begin{equation}
    M_\mathrm{out,\oiii}/\mathrm{M}_\odot =
    0.8 \times 10^8 
    \left( \dfrac{L_\mathrm{\oiii}}{10^{44}~\mathrm{erg~s}^{-1}} \right) 
    \left( \dfrac{n_\mathrm{e}}{500~\mathrm{cm}^{-3}} \right)^{-1} , \label{eq:Mout_oiii}
\end{equation}
under the assumption of fully ionised gas with an electron temperature $T_\mathrm{e}$ = 10$^4$~K, $\langle n_\mathrm{e} \rangle^2 / \langle n_\mathrm{e}^2 \rangle$ = 1 and, for Eq.~\ref{eq:Mout_oiii}, solar oxygen abundance.
We used the \ha/\hb ratio of the broad component to estimate the dust attenuation of the outflow, adopting a \cite{Calzetti2000} starburst attenuation law with $R_V$ = 4.05 and an attenuation-free \ha/\hb ratio of 2.86 (for $T_\mathrm{e}$ = 10$^4$~K and $n_\mathrm{e}$ $\sim$ 10$^{2-3}$~cm$^{-3}$; \citealt{Osterbrock2006}). We note that the dust attenuation correction was possible only for some of the targets, when S/N $>$ 3 for both \ha and \hb broad component; for the other sources, the line fluxes were not corrected for dust attenuation.
The electron density could not be measured for the outflow due to the low S/N of the \sii doublet lines in the broad components.
We attempted performing a stack of the spectra of the targets (the same reported in Figs.~\ref{fig:spectrum_gs133} and \ref{fig:spectra}), in order to constrain the outflow electron density; several approaches were probed, specifically normalising by the emission peak (i.e. either \oiii or \ha depending on the source), by the \ha peak (as e.g. in \citealt{Perna2017a}), or by the \sii peak, as well as weighting (or not) by the uncertainties on the spectrum. None of these approaches could constrain the ratio of the broad \sii components (and thus the density), due to the low S/N of the \sii lines at the broad component level even in the stacked spectrum and/or to degeneracies in the de-blending of the doublet lines. Therefore, we adopt a value of 1000~cm$^{-3}$ for the electron density of the outflowing gas, which is a typical value for AGN outflows, as done in \cite{Bertola2025}.
This value is found by \cite{ForsterSchreiber2019} from the stacked \sii spectrum of 30 AGN-driven outflows at 0.6 < $z$ < 2.6 and by \cite{Perna2017a} at $z$ < 0.8 from both the \sii stacked spectrum of 90 AGN and the median from $\sim$30 individual-object spectra, and is consistent with the \sii-based values found in other single AGN outflows at similar redshifts \citep[e.g.][]{Brusa2016, Husemann2019}.
However, densities of $\sim$100--500~cm$^{-3}$ are also found in other AGN outflows \citep[e.g.][]{Nesvadba2006, Perna2015, Cresci2023, Bessiere2024} and values of 100--200~cm$^{-3}$ are often assumed \citep[e.g.][]{Harrison2014, Brusa2015, Cresci2015, Kakkad2016, Fiore2017}, or even lower (10~cm$^{-3}$; \citealt{Rupke2013, Liu2013}).
Recent theoretical work (\citealt{Huang2026}, QED simulations; \citealt{Howatson2025}, FIRE-2 simulations) indicate that the actual density of the outflowing gas emitting \ha and \oiii is $\sim$1--2~dex smaller than the one estimated from the \sii doublet, the routinely adopted tracer of outflow density (see above).
Therefore, all in all, our adopted value of 1000~cm$^{-3}$ can be considered as a conservative assumption on the inferred outflow mass.
We stress that a different assumption on the electron density just (inversely) changes the outflow mass and related properties by the same factor.

We thus calculate the mass outflow rate as
\begin{equation}
    \dot{M}_\mathrm{out} = C \dfrac{M_\mathrm{out} v_\mathrm{out}}{R_\mathrm{out}}, \label{eq:Moutrate}
\end{equation}
valid for a spherical or (multi-)conical outflow, where \vout is the outflow velocity, $R_\mathrm{out}$ its radius, and $C$ depends on the outflow history. $C$ is equal to 1 in case of a constant mass outflow rate in time and average volume density of the outflowing gas decreasing radially as $R^{-2}$. It is instead equal to 3 in case of constant average volume density and a mass outflow rate that started from a maximum in the past---to which the value thus calculated corresponds---and then decreased in time \citep[e.g.][]{Lutz2020}.
We adopt $C = 3$ for consistency with the COS-AGN study in \cite{Bertola2025} and with the outflow compilation of \cite{Fiore2017}. 

We calculated the outflow velocity as 
$v_\mathrm{max} = |v_\mathrm{bro} - v_\mathrm{nar}| + 2 \sigma_\mathrm{bro}$ 
under the assumption that the extreme wing of the line profile traces the actual outflow velocity, while the rest of the line emission corresponds to lower line-of-sight (l.o.s.) velocities due to projection effects (as in e.g. \citealt{Rupke2013, Fiore2017}). 
This is justified by considering that outflows, as a consequence of having (bi-)conical geometries with wide opening angles (as often observed; e.g. \citealt{Fischer2013, Venturi2018, Lopez-Coba2020, Juneau2022, Marconcini2023}), will have a large fraction of their emitting gas moving along directions different from the l.o.s.
Other definitions of the outflow velocity are also used, based on non-parametric approaches or on the assumption that the line emission at lower l.o.s. velocities arises from intrinsically slower gas rather than from projection effects (see e.g. \citealt{Harrison2018, HervellaSeoane2023, Bessiere2024, Bertola2025}).

\begin{table*}[]
    \centering
    \caption{Results of the Gaussian emission-line fitting of the integrated spectra from Figs.~\ref{fig:spectrum_gs133} and \ref{fig:spectra} extracted from the outflow region (see Sect.~\ref{sec:data_anal}).}
    \resizebox{\textwidth}{!}{
    \setlength{\tabcolsep}{2pt} 
    \renewcommand{\arraystretch}{1.3} 
    \begin{tabular}{l|ccccccccccccc}
    \hline\hline
        Name & Type & $z$ & $F_\mathrm{\oiii,nar}$ & $F_\mathrm{H\beta,nar}$ & $F_\mathrm{\nii,nar}$ & $F_\mathrm{H\alpha,nar}$ & $F_\mathrm{\oiii,bro}$ & $F_\mathrm{H\beta,bro}$ & $F_\mathrm{\nii,bro}$ & $F_\mathrm{H\alpha,bro}$ & $\sigma_\mathrm{nar}$ & $v_\mathrm{bro}$ & $\sigma_\mathrm{bro}$  \\
    \hline
        GS~133   & 2 & 3.47339$_{-0.00002}^{+0.00002}$  & 7.7$_{-0.3}^{+0.3}$  & 1.60$_{-0.10}^{+0.10}$ & 1.48$_{-0.10}^{+0.10}$ & 6.8$_{-0.2}^{+0.2}$    & 8.4$_{-0.2}^{+0.3}$  & 1.01$_{-0.16}^{+0.16}$ & 3.0$_{-0.2}^{+0.2}$ & 4.7$_{-0.3}^{+0.3}$  & 120$_{-3}^{+3}$ &  --67$_{-9}^{+8}$ & 414$_{-16}^{+17}$ \\
        GS~539   & 1 & 4.7556$_{-0.0003}^{+0.0005}$  & --    & --   & 0.31$_{-0.16}^{+0.08}$ & 0.28$_{-0.05}^{+0.04}$     & --    & --   & 0.80 $_{-0.10}^{+0.12}$ & 0.16$_{-0.10}^{+0.15}$   & 160$_{-30}^{+20}$  & --394$_{-120}^{+180}$  & 590$_{-200}^{+90}$  \\
        GS~551   & 2 & 3.703291$_{-0.000015}^{+0.000016}$  & 7.25$_{-0.17}^{+0.19}$  & 0.49$_{-0.04}^{+0.04}$ & 1.23$_{-0.06}^{+0.07}$ & 1.83$_{-0.09}^{+0.08}$    & 9.33$_{-0.17}^{+0.14}$  & 1.00$_{-0.07}^{+0.08}$ & 1.40$_{-0.16}^{+0.15}$ & 3.99$_{-0.15}^{+0.15}$  & 131$_{-2}^{+2}$ &  --26$_{-6}^{+5}$ & 489$_{-11}^{+11}$ \\
        GS~774   & 1 & 3.58442$_{-0.00004}^{+0.00005}$  & 10.6$_{-0.6}^{+0.6}$ & 1.44$_{-0.13}^{+0.13}$ & 0.55$_{-0.16}^{+0.19}$ & 5.17$_{-0.3}^{+0.3}$    & 15.4$_{-0.5}^{+0.5}$ & --   & 10.9$_{-0.4}^{+0.4}$ & 5.2$_{-0.5}^{+0.5}$ & 205$_{-7}^{+7}$ & --137$_{-8}^{+9}$ & 594$_{-15}^{+17}$ \\
        GS~811   & 1 & 3.4678$_{-0.0005}^{+0.0005}$  & 0.27$_{-0.10}^{+0.12}$  & --   & 0.52$_{-0.13}^{+0.10}$ & 0.59$_{-0.14}^{+0.12}$    & 0.63$_{-0.13}^{+0.11}$  & --   & -- & 0.29$_{-0.13}^{+0.15}$  & 240$_{-40}^{+30}$ & --490$_{-100}^{+80}$ & 590$_{-60}^{+80}$ \\
        GS~3073  & 1 & 5.553302$_{-0.000005}^{+0.000005}$  & 16.30$_{-0.07}^{+0.06}$ & 2.02$_{-0.02}^{+0.02}$ & 0.158$_{-0.016}^{+0.016}$ & 5.91$_{-0.05}^{+0.05}$    & 1.92$_{-0.04}^{+0.05}$  & 0.64$_{-0.03}^{+0.03}$ & 0.18$_{-0.04}^{+0.04}$ & 3.09$_{-0.05}^{+0.05}$  & 90.0$_{-0.3}^{+0.3}$  &   +66$_{-4}^{+4}$ & 341$_{-8}^{+8}$ \\
        GS~10578 & 2 & 3.064\tablefootmark{(a)}  & 3.6$_{-0.4}^{+0.4}$  & 0.87$_{-0.16}^{+0.17}$ & 12.9$_{-0.8}^{+0.9}$ & 3.9$_{-0.4}^{+0.4}$  & 11.5$_{-0.5}^{+0.5}$ & -- & 12.0$_{-1.1}^{+1.0}$ & 2.1$_{-0.8}^{+0.7}$  & 314$_{-12}^{+12}$ & --390$_{-30}^{+20}$ & 800$_{-30}^{+30}$ \\
                 &   &        & 1.64$_{-0.14}^{+0.15}$  & 0.39$_{-0.10}^{+0.10}$ & 0.8$_{-0.2}^{+0.3}$ & 1.15$_{-0.13}^{+0.15}$    & & & &                      & 124$_{-8}^{+9}$ & &           \\
        GS~19293 & 2 & 3.11875$_{-0.00005}^{+0.00005}$  & 0.032$_{-0.012}^{+0.011}$  & 0.083$_{-0.012}^{+0.011}$ & 0.22$_{-0.02}^{+0.02}$ & 0.46$_{-0.03}^{+0.02}$    & 0.128$_{-0.015}^{+0.017}$  & --   & 0.26$_{-0.02}^{+0.03}$ & 0.15$_{-0.02}^{+0.02}$  & 94$_{-4}^{+4}$ & --200$_{-30}^{+20}$ & 213$_{-13}^{+13}$ \\
        GS~20936 & 2 & 3.24255$_{-0.00007}^{+0.00007}$  & 0.26$_{-0.03}^{+0.03}$  & 0.050$_{-0.016}^{+0.016}$ & 0.08$_{-0.02}^{+0.03}$ & 0.38$_{-0.04}^{+0.04}$    & 0.78$_{-0.03}^{+0.04}$  & 0.25$_{-0.03}^{+0.03}$ & 0.58$_{-0.03}^{+0.04}$ & 1.71$_{-0.06}^{+0.06}$  & 83$_{-5}^{+6}$  &  +296$_{-11}^{+11}$ & 262$_{-8}^{+8}$ \\
    \hline
    \end{tabular}}
    \label{tab:em_line_fit}
    \tablefoot{
    From left to right: Target name, AGN type (1 and 2 mean presence or absence, respectively, of BLR Balmer lines), redshift (given by the spectral shift of the narrow component centroid), flux, offset velocity, and velocity dispersion of the narrow and broad components of \oiii$\lambda$5008, \hb, \nii$\lambda$6585, and \ha. Fluxes are in units of $10^{-17}$ \ergscm; velocities and velocity dispersions are in units of \kms. For GS~10578, the two values reported for each `nar' quantity belong to the two narrower components (whose velocity shifts relative to the systemic redshift from stellar absorption lines are $v_\mathrm{nar}$ = --71$_{-10}^{+10}$~\kms and --502$_{-11}^{+11}$~\kms).\\
    \tablefoottext{a}{From \cite{DEugenio2024}, based on stellar absorption lines.}
    }
\end{table*}

\begin{table*}[]
    \centering
    \caption{Ionised outflow properties inferred from the fitted emission-line parameters (Table~\ref{tab:em_line_fit}) as explained in Sect.~\ref{sec:outf_props}, and AGN and host-galaxy properties from SED fitting (Circosta et al., in prep.).}
    \resizebox{\textwidth}{!}{
    \setlength{\tabcolsep}{2pt} 
    \renewcommand{\arraystretch}{1.3} 
    \begin{tabular}{l|cccccc|c|ccc}
    \hline\hline
    Name & \vmax & \rmax & \mdot & \edot & $\eta$ = \mdot/SFR & $A_{V,\,\mathrm{out}}$ & \rweig\tablefootmark{(*)} & $L_\mathrm{bol}$ & SFR & $M_*$  \\
    & [\kms] & [kpc] & [\Msunyr] & [$10^{42}$\,\ergs] & & & [kpc] & [$10^{45}$\,\ergs] & [\Msunyr] & [$10^{10}$\,\Msun] \\
    & & & \ha | \oiii & \ha | \oiii & \ha | \oiii & & & & \\
    \hline
    GS~133 & 890$_{-40}^{+40}$ & 3.9$\pm$0.5 & 41$_{-2}^{+2}$ | 45.7$_{-1.4}^{+1.5}$ & 10.5$_{-1.0}^{+1.2}$ | 11.5$_{-1.2}^{+1.4}$ & $>$1.4 | $>$1.5 & 1.7$_{-0.5}^{+0.6}$ & 1.3$\pm$0.5 & 2.0$\pm$1.3 & $<$31 & 3.0$\pm$0.3 \\
    GS~539 & 1600$_{-600}^{+200}$  & $<$1.2 & $>$4.5 | -- & $>$3.6 | -- & $>$0.014 | -- & -- & <0.6 & 1.6$\pm$0.2 & 323$\pm$16 & 9$\pm$3 \\
    GS~551 & 1000$_{-20}^{+30}$ & 2.1$\pm$0.5 & 56$_{-2}^{+3}$ | 70.5$_{-1.3}^{+1.3}$ & 17.8$_{-1.4}^{+1.5}$ | 22.4$_{-1.4}^{+1.5}$ & $>$6.6 | $>$8.3 & 1.2$_{-0.3}^{+0.3}$ & 0.7$\pm$0.5 & 6.6$\pm$0.3 & $<$8 & 13.0$\pm$0.7 \\
    GS~774 & 1320$_{-40}^{+40}$ & $<$1.3 & $>$62 | $>$74 & $>$35 | $>$41 & $>$1.3 | $>$1.5 & -- & <0.7 & 8.2$\pm$1.3 & $<$48 & 2.2$\pm$1.5 \\
    GS~811 & 1670$_{-160}^{+200}$ & $<$1.1 & $>$4.5 | $>$3.9 & $>$3.9 | $>$3.4 & $>$0.14 | $>$0.12 & -- & <0.6 & 1.37$\pm$0.12 & $<$32 & 5.7$\pm$1.9 \\
    GS~3073 & 747$_{-17}^{+18}$ & $<$1.3 & $>$224 | $>$89 & $>$39 | $>$16 & $>$15.5 | $>$6.1 & 1.82$_{-0.16}^{+0.16}$ & <0.7 & <0.8 & $<$14 & 6.1$\pm$0.4 \\
    GS~10578 & 1980$_{-70}^{+80}$ & 2.7$\pm$0.6 & 12$_{-4}^{+4}$ | 26.9$_{-0.9}^{+0.9}$ & 15$_{-5}^{+4}$ | 33$_{-3}^{+3}$ & $>$0.13 | $>$0.28 & -- & 1.1$\pm$0.6 & 6.5$\pm$0.7 & $<$95 & 14.5$\pm$0.7 \\
    GS~19293 & 630$_{-40}^{+40}$ & 2.3$\pm$0.6 & 0.33$_{-0.04}^{+0.05}$ | 0.111$_{-0.014}^{+0.017}$ & 0.041$_{-0.006}^{+0.008}$ | 0.014$_{-0.003}^{+0.003}$ & $>$0.0043 | $>$0.0015 & -- & 0.9$\pm$0.6 & <1.1 & $<$77 & 2.0$\pm$0.6 \\
    GS~20936 & 819$_{-14}^{+14}$ & 2.8$\pm$0.6 & 45.5$_{-1.9}^{+1.9}$ | 18.2$_{-0.9}^{+0.9}$ & 9.6$_{-0.7}^{+0.7}$ | 3.9$_{-0.3}^{+0.3}$ & 1.11$_{-0.04}^{+0.04}$ | 0.44$_{-0.02}^{+0.02}$ & 3.0$_{-0.4}^{+0.4}$ & 1.1$\pm$0.6 & 5.3$\pm$0.5 & 41$\pm$2 & 1.2$\pm$0.3 \\
    \hline
    \end{tabular}}
    \label{tab:outflow_props}
    \tablefoot{
    From left to right: Target name, maximum outflow velocity, maximum outflow radius, mass outflow rate, kinetic rate, and mass loading factor obtained from \ha and \oiii broad component flux (left and right columns, respectively), visual dust attenuation (from \ha/\hb ratio of the broad component), flux-weighted radius (weighted by the \oiii broad component flux), AGN bolometric luminosity, and host-galaxy SFR and stellar mass. Upper limits on \rmax and \rweig correspond to the PSF FWHM and $\simeq$FWHM/2, respectively, for unresolved outflows; $1\,\sigma$ upper limits on \lbol and SFR are reported when these are unconstrained in the SED fitting.\\
    \tablefoottext{*}{The flux-weighted radius (\rweig) was not employed for the calculation of \mdot, \edot, and $\eta$, for which the maximum radius (\rmax) was used instead.}
    }
\end{table*}

For the outflow radius, we considered the maximum projected distance reached by the outflow as traced by the broad component, \rmax. We stress that \rmax depends on S/N and spatial resolution of the observations. Using another definition of the outflow radius, such as the flux-weighted radius (the sum of the distances of each spaxel from the nucleus weighed by the flux of the broad component in each spaxel; see e.g. \citealt{Bertola2025}) or a 2D-Gaussian fitting to the broad component, would give a more meaningful description of the outflow mass spatial distribution. 
We thus also report the flux-weighted outflow radius, \rweig, in Table~\ref{tab:outflow_props}.
Nevertheless, we adopted \rmax to estimate the outflow properties, to allow them to be compatible with those in the literature, because \rmax is the only definition of outflow radius which can be retrieved for all of the literature sources. 
We note that \rmax is a conservative assumption on the outflow radius for what concerns the estimate of the mass outflow rate (and related quantities), since the bulk of the outflowing material is generally closer to the nucleus, while this definition is driven by the faint outskirts of the outflowing material. The resulting mass outflow rates when using \rweig would be a factor of $\sim$ 2 to 3 larger than those estimated through \rmax.
We stress, though, that both \rmax and \rweig are projected radii, since the outflow inclination in the line of sight is unknown, and the intrinsic outflow radius could be larger.

For GS~539, GS~774, GS~811, and GS~3073, whose outflows are spatially unresolved, we defined an upper limit on \rmax based on the PSF, which implies a lower limit on \mdot. For consistency with the definition of \rmax in the spatially resolved case as the maximum observed radius, here we considered the radius encompassing 98\% of the PSF flux, corresponding to $2.33 \, \sigma_\mathrm{PSF}$ $\sim$ FWHM$_\mathrm{PSF}$.
Specifically, for GS~539 we adopted the value estimated from the BLR PSF in \cite{Parlanti2024} for this source from the same NIRSpec IFU data used in the present work, of FWHM$_\mathrm{PSF}$ $\sim$ 1.2~kpc.
For GS~811, we inferred the PSF from a 2D Gaussian fitting to the unresolved spatial profile of the modelled \ha BLR ($\sim$2.9~$\mu$m), which gives FWHM$_\mathrm{PSF}$ $\sim$ 0.15$''$ (1.13~kpc), consistent with the estimates of \cite{DEugenio2024} and \cite{Marshall2025a} for other targets.
For GS~774, we found FWHM$_\mathrm{PSF}$ $\sim$ 0.17$''$ $\times$ 0.15$''$ from a 2D fitting to the \hb BLR spatial profile, and similarly for the \oiii outflowing wings; we thus conservatively adopted 0.17$''$ (1.26~kpc) as an upper limit on $R_\mathrm{max}$.
Finally, for GS~3073, we obtained a FWHM $\sim$ 0.21$''$ $\times$ 0.18$''$ from a 2D fitting of the outflowing broad wing of \oiii, comparable with the PSF FWHM$_\mathrm{PSF}$ of $\sim$ 0.20$''$ $\times$ 0.17$''$ from the \ha BLR spatial profile. Therefore, the outflow is at most very marginally resolved, if not unresolved. We thus conservatively adopted 0.21$''$ (1.28~kpc) as an upper limit on $R_\mathrm{max}$ for GS~3073.
Since the flux-weighted radius (\rweig) of a 2D Gaussian function roughly corresponds to its FWHM/2, we adopted this as an upper limit on \rweig for the case of the unresolved outflows above.

Summarising, we adopted $v_\mathrm{out} \equiv v_\mathrm{max}$, $R_\mathrm{out} \equiv R_\mathrm{max}$, and $n_\mathrm{e} \sim 1000$~cm$^{-3}$ (as done in \citealt{Bertola2025} for their comparison with outflows from the literature).
The kinetic energy and momentum rates are then obtained as
    $\dot{E}_\mathrm{out} = \dot{M}_\mathrm{out}  v_\mathrm{out}^2 / 2$
and
    $\dot{p}_\mathrm{out} = \dot{M}_\mathrm{out}  v_\mathrm{out}$,
respectively. The outflow properties thus obtained for our sample are reported in Table~\ref{tab:outflow_props}.
Here, the outflow masses (and resulting rates) derived from \oiii were multiplied by a factor of 3, to account for the known discrepancy between \ha- (or \hb-) versus \oiii-derived AGN outflow masses (see Appendix~\ref{sec:app_outf_props_lit} for more details).
In any case, in the rest of this work, we adopt the mass (and derived rates) obtained from \ha rather than \oiii emission, because for a given gas mass, the former directly relates to the number of ionising photons, while the latter depends on gas metallicity and on the hardness of the radiation field \citep[e.g.][]{Osterbrock2006, Venturi2023}.

We estimated the uncertainties on the outflow properties through Monte Carlo simulations, by perturbing the integrated spectra of each source (reported in Figs.~\ref{fig:spectrum_gs133} and \ref{fig:spectra}) by their errors. We ran 10,000 iterations for each target, by repeating the fitting procedure described in Sect.~\ref{sec:outf_integr_anal} and calculating the outflow properties as described above, for each iteration. The uncertainties were obtained as the 16th and the 84th percentile of the distribution of each quantity, and are reported in Tables~\ref{tab:em_line_fit} and \ref{tab:outflow_props}. 
Since the dust attenuation estimates for the outflowing gas are available for only a few AGN, we fixed $A_V$ to the best-fit value derived for their integrated spectra (see above and Table~\ref{tab:outflow_props}) and did not propagate its uncertainties into the derived outflow properties in order to homogenise the error estimates across the entire sample.
The outflow radius, \rmax, was also fixed to that adopted for the best-fit calculations, as reported in Table~\ref{tab:outflow_props}; for its uncertainty, we considered a typical spatial resolution of $\sigma_\mathrm{PSF} \simeq \mathrm{FWHM}_\mathrm{PSF}/2.355 = 0.17''/2.355 \simeq 0.07''$ (see above and e.g. \citealt{DEugenio2024}).

As for the COS-AGN in \cite{Bertola2025}, the AGN bolometric luminosity (\lbol) and the star formation rate (SFR) of the host galaxies were derived through SED fitting with CIGALE \citep{Boquien2019, Yang2020_cigale, Yang2022_cigale}, employing rest-frame UV-optical to FIR photometry, which will be presented in Circosta et al., in prep. 
We thus also calculate the mass loading factor, $\eta$ = \mdot/SFR, which describes how quickly the outflow is removing gas relative to its consumption by star formation. These three quantities are also reported in Table~\ref{tab:outflow_props}. 

We scale the outflow properties for the COS-AGN from \cite{Bertola2025} to our framework (\rmax, \vmax, $n_\mathrm{e}$ = 1000~cm$^{-3}$), using the values reported in their Table~2. The only difference is that here we consider the non-PSF-corrected \rmax for consistency with the rest of the literature sample. We note that the difference is negligible for COS-AGN (3\% at most).

We collected a compilation of AGN ionised outflows from the literature to compare them with those from our GS-AGN + COS-AGN GA-NIFS sample. We only considered outflow studies from rest-optical lines, for a consistent comparison. Also, we specifically only considered spatially resolved observations in order to have an estimate of the outflow radius, which is a key quantity for the mass outflow rate and related quantities (see Eq.~\ref{eq:Moutrate}). We re-calculated all of the outflow properties from the literature consistently with those obtained for our sample. Specifically, for all of them we re-calculated the outflow velocity as \vmax (see above), estimated \rmax from the outflow (i.e. broad component) flux maps, when not already provided, and adopted an outflow electron density of 1000~cm$^{-3}$ and $C$ = 3. 
All of the details on these samples and on how the outflow properties for each of them were re-calculated are given in Appendix \ref{sec:app_outf_props_lit}.
Here we just stress that for the QSO from \cite{Marshall2025a}, since the authors did not provide the outflow properties, we calculated them for the first time in our work, by using the results of the fiducial emission line fitting (which employed a double Gaussian for the BLR Balmer lines) to the NIRSpec IFU data from their study. 

The ionised outflow properties of some of the GS-AGN were already analysed in previous single-target studies as part of the GA-NIFS collaboration, as mentioned in Appendix~\ref{sec:app_emline_maps}. 
The estimates of the properties presented in these studies do not necessarily match those reported in the present work, which used uniform, consistent assumptions to calculate the outflow quantities, as explained above.

\section{Outflow properties versus AGN and host galaxy properties}\label{sec:outf_vs_agn+gal_props}

\begin{figure*}
    \centering
    \includegraphics[width=0.65\textwidth]{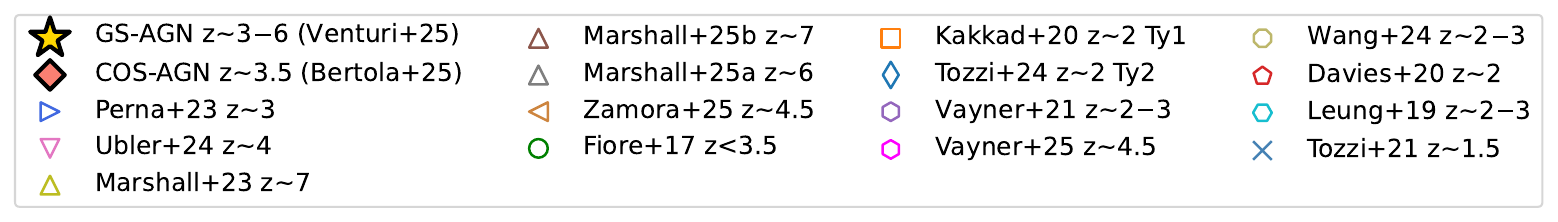}\\
    \includegraphics[width=0.35\linewidth]{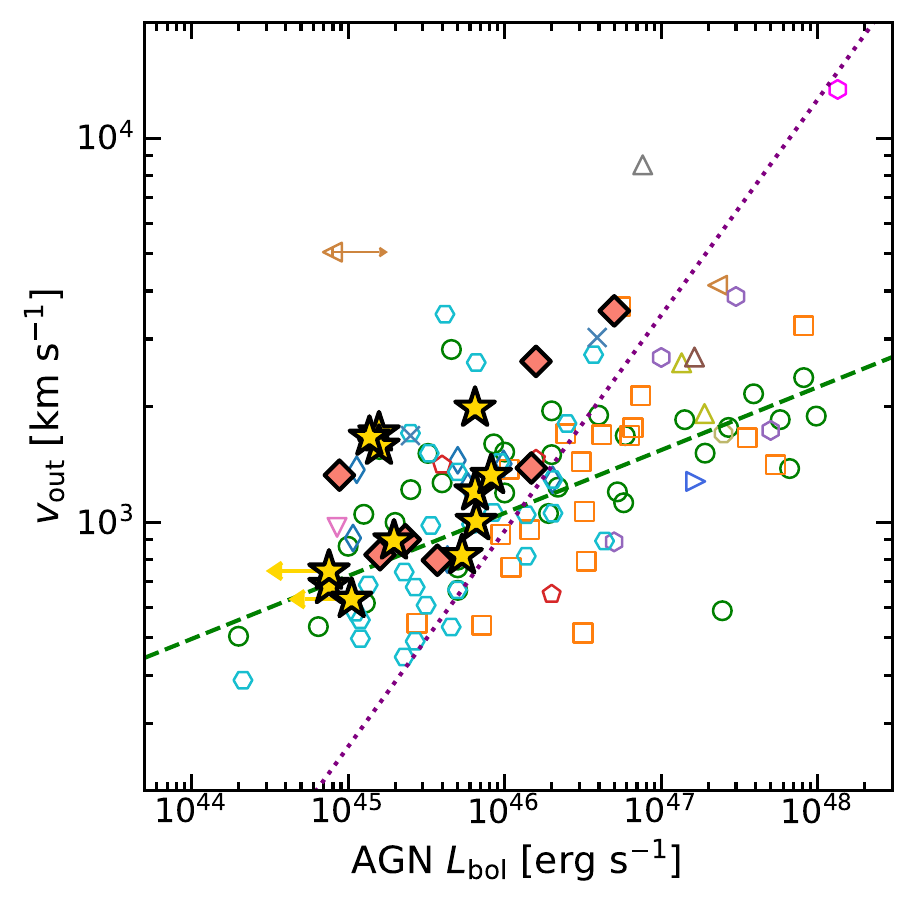}
    \includegraphics[width=0.35\linewidth]{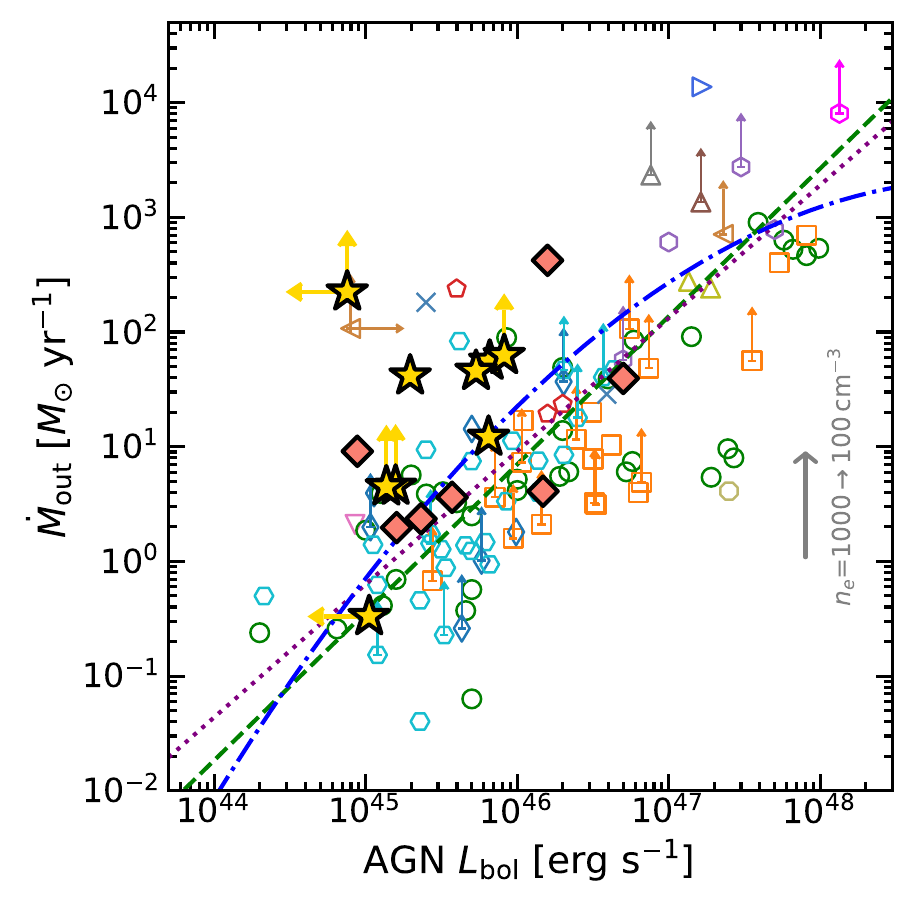}\\
    \includegraphics[width=0.35\linewidth]{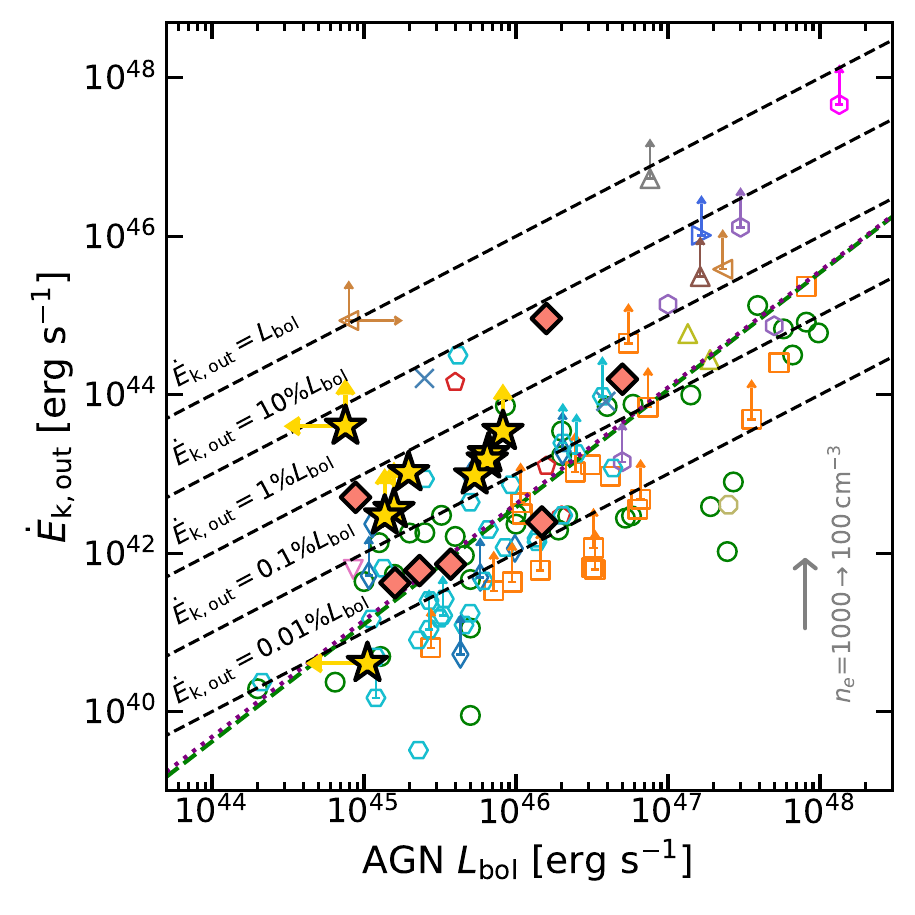}
    \includegraphics[width=0.35\linewidth]{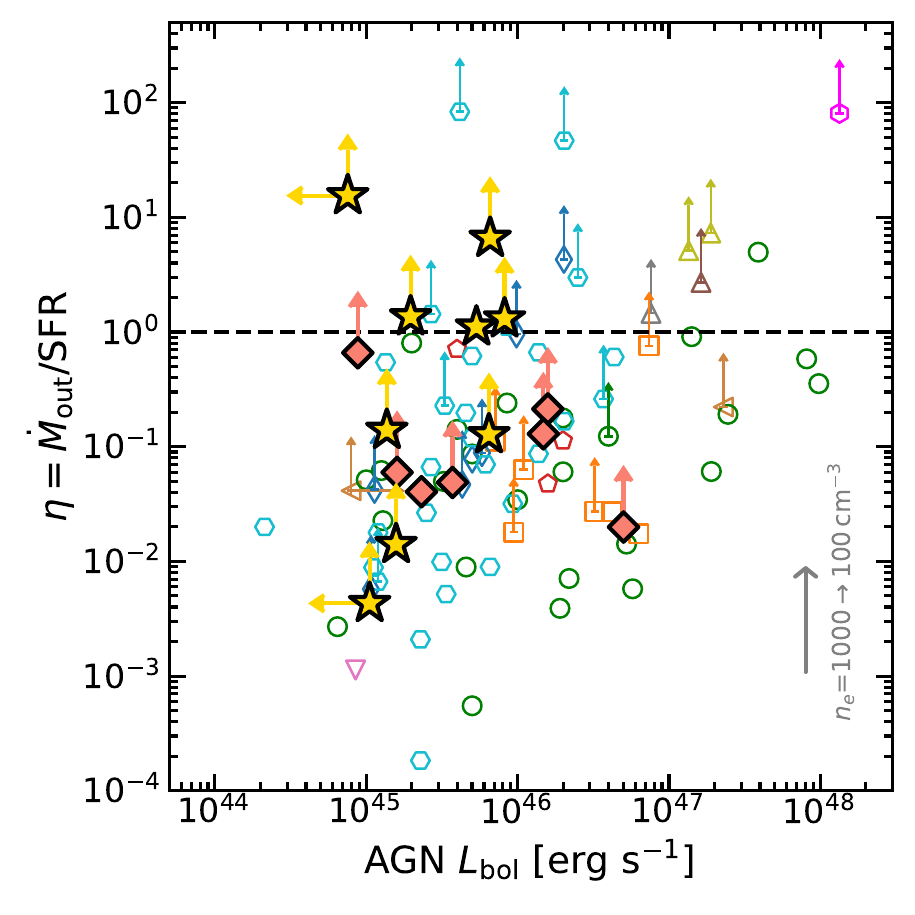}
    \caption{
    Bolometric luminosity (\lbol) versus outflow properties for AGN from GA-NIFS and from the literature. \textit{From left to right}: AGN \lbol versus outflow velocity (\vout), mass outflow rate (\mdot), kinetic energy rate (\edot), and mass loading factor ($\eta$ = \mdot/SFR). Symbols and colours are the same as in Fig.~\ref{fig:Lbolvz}.
    The dashed green, dotted purple, and dash-dot blue lines represent the best-fit relations from \cite{Fiore2017}, \cite{Musiimenta2023}, and \cite{Bischetti2019}, respectively, re-scaled to $n_\mathrm{e}$ = 1000~cm$^{-3}$.
    The diagonal dashed black lines in the lower-left plot mark \edot = 100\%, 10\%, 1\%, 0.1\%, and 0.01\% \lbol. The horizontal dashed black line in the lower-right plot indicates $\eta$ = 1. The grey arrow quantifies the linear scaling factor that would be applied to all of the points when assuming a different value of the electron density, specifically when using 100~cm$^{-3}$ instead of the value of 1000~cm$^{-3}$ assumed.
    }
    \label{fig:Lbolvoutprops}
\end{figure*}

We can now compare the ionised outflow properties obtained for the GS-AGN and COS-AGN with those from existing spatially resolved (mostly IFU) studies from the literature, re-calculated and homogenised according to the definitions and assumptions adopted in our work, as explained in Sect. \ref{sec:outf_props} and detailed in Appendix \ref{sec:app_outf_props_lit}.
This comparison is shown in Fig. \ref{fig:Lbolvoutprops}, which reports the AGN bolometric luminosity versus various outflow properties, namely the velocity, mass outflow rate, kinetic rate, and mass loading factor.

We see that the outflows from the GS- and COS-AGN from GA-NIFS are generally consistent with those from the literature, and with the empirical relations fitted in previous studies (dashed coloured lines), as found in \cite{Bertola2025}. 
That said, it appears that the bulk of our targets (with AGN \lbol between 10$^{45-46}$~\ergs) have larger \mdot than the sources in the literature at the same \lbol, which are also generally at lower redshifts; we investigate this aspect and any possible evolution of the outflow properties with redshift in Sect.~\ref{sec:zevol}.

The mass loading factors of a significant fraction of the GS- and COS-AGN from GA-NIFS are close to the threshold of 1, above which outflows evacuate gas at a faster pace than it is used to form new stars, though different assumptions on the outflow electron density would move the values up and down in the plot.
Specifically, assuming a density of 100~cm$^{-3}$ (close to that of 200~cm$^{-3}$ assumed in various AGN outflow studies; e.g. \citealt{Fiore2017}) instead of 1000~cm$^{-3}$ would move all of the mass outflow rates, and thus mass loading factors, up by one order of magnitude.
Moreover, we stress that, since we only have upper limits on the SFRs of all but two of the GS- and COS-AGN from SED fitting (Table~\ref{tab:outflow_props}), their mass loading factors are lower limits, and therefore the actual values could be higher.

\begin{figure*}
    \centering
    \includegraphics[width=0.33\linewidth]{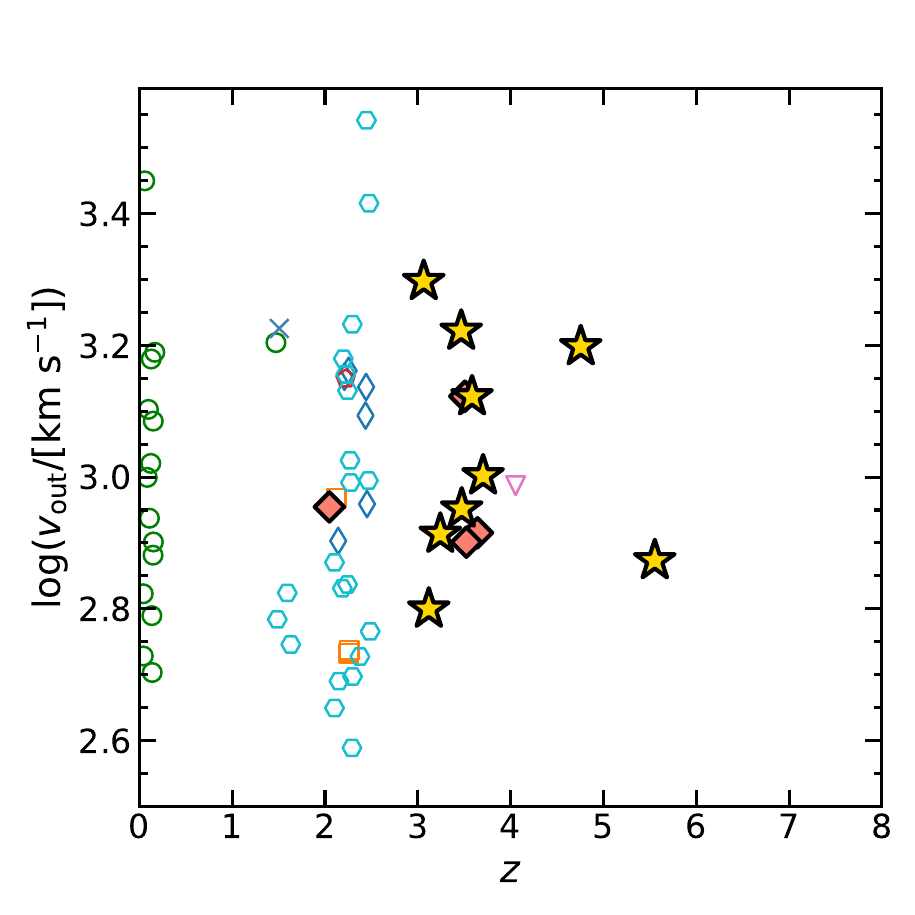}
    \includegraphics[width=0.33\linewidth]{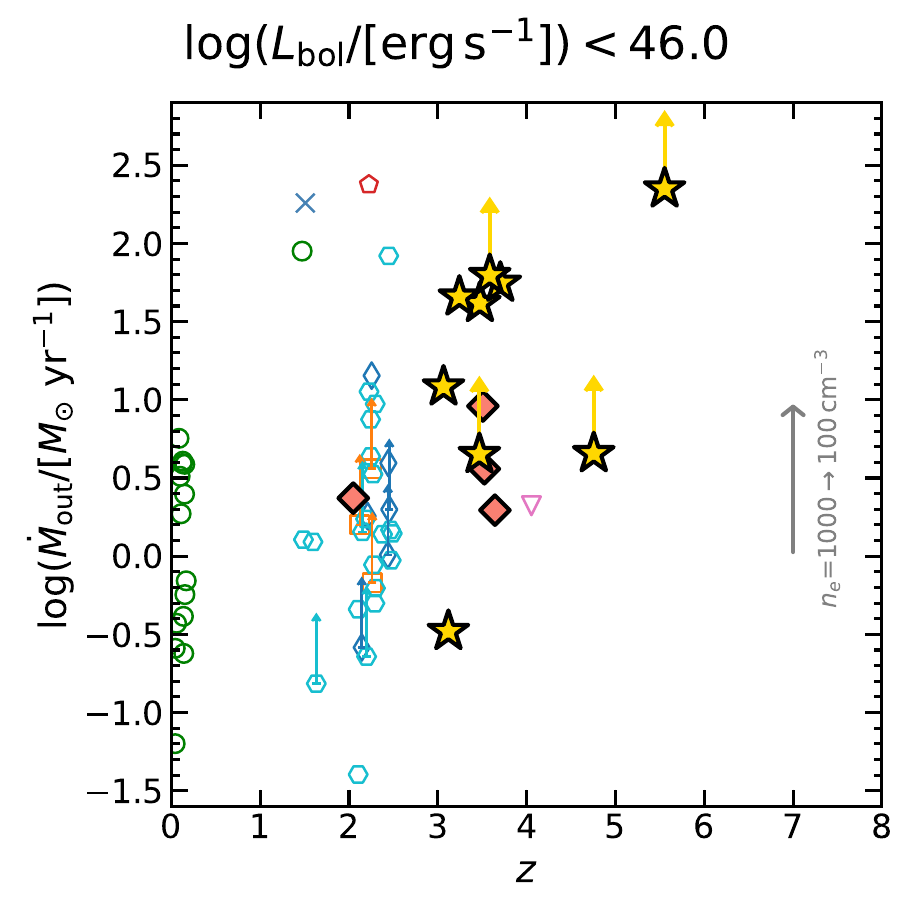}
    \includegraphics[width=0.33\linewidth]{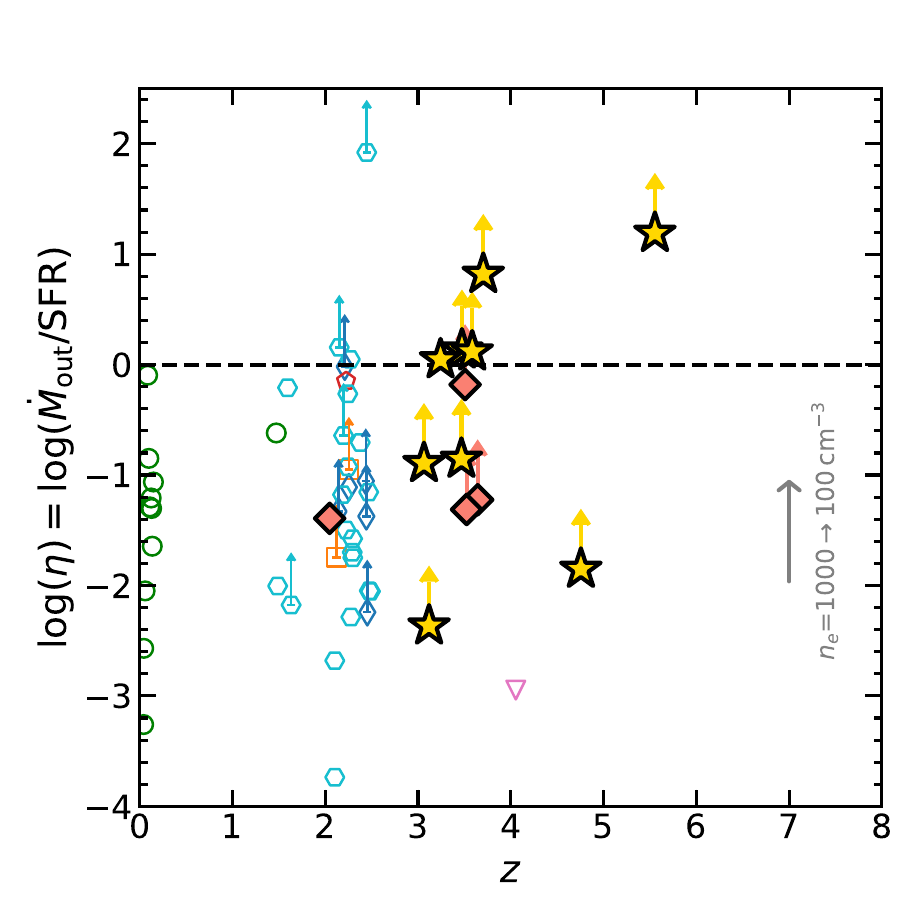}\\
    \includegraphics[width=0.33\linewidth]{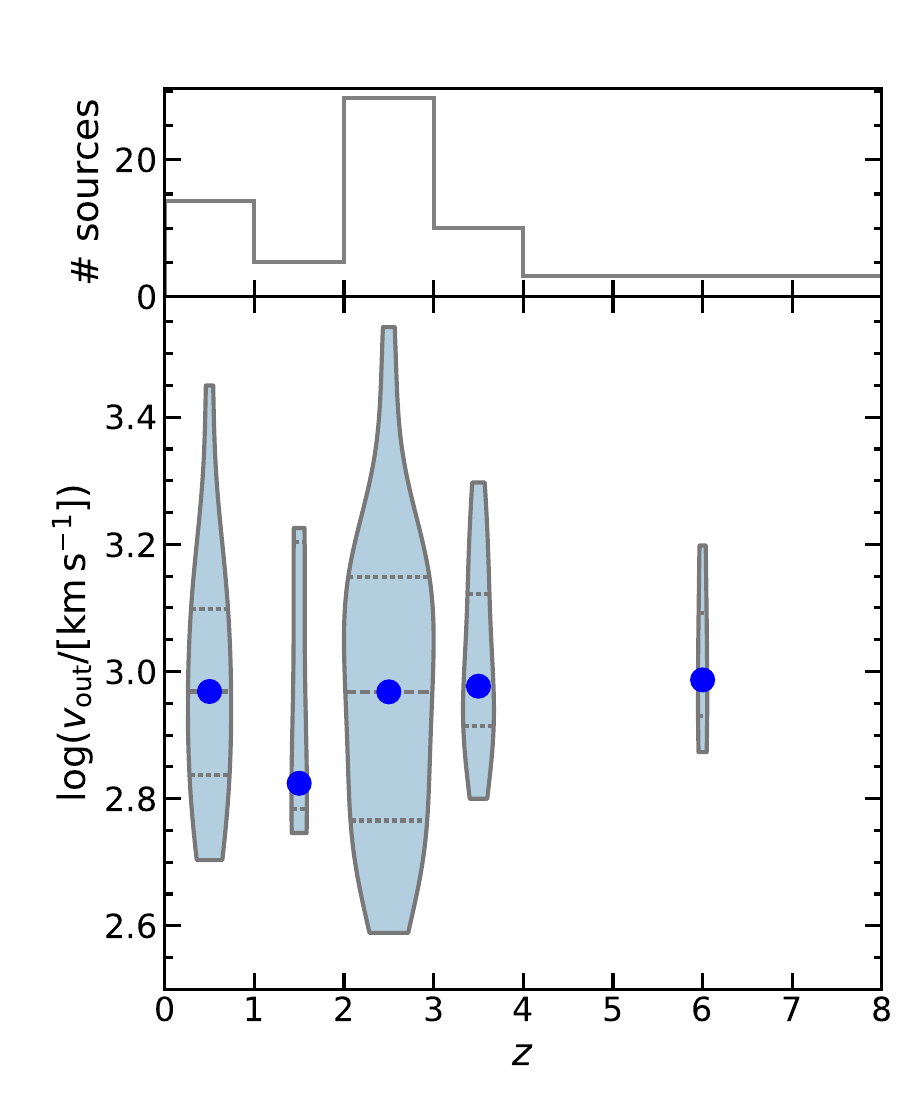}
    \includegraphics[width=0.33\linewidth]{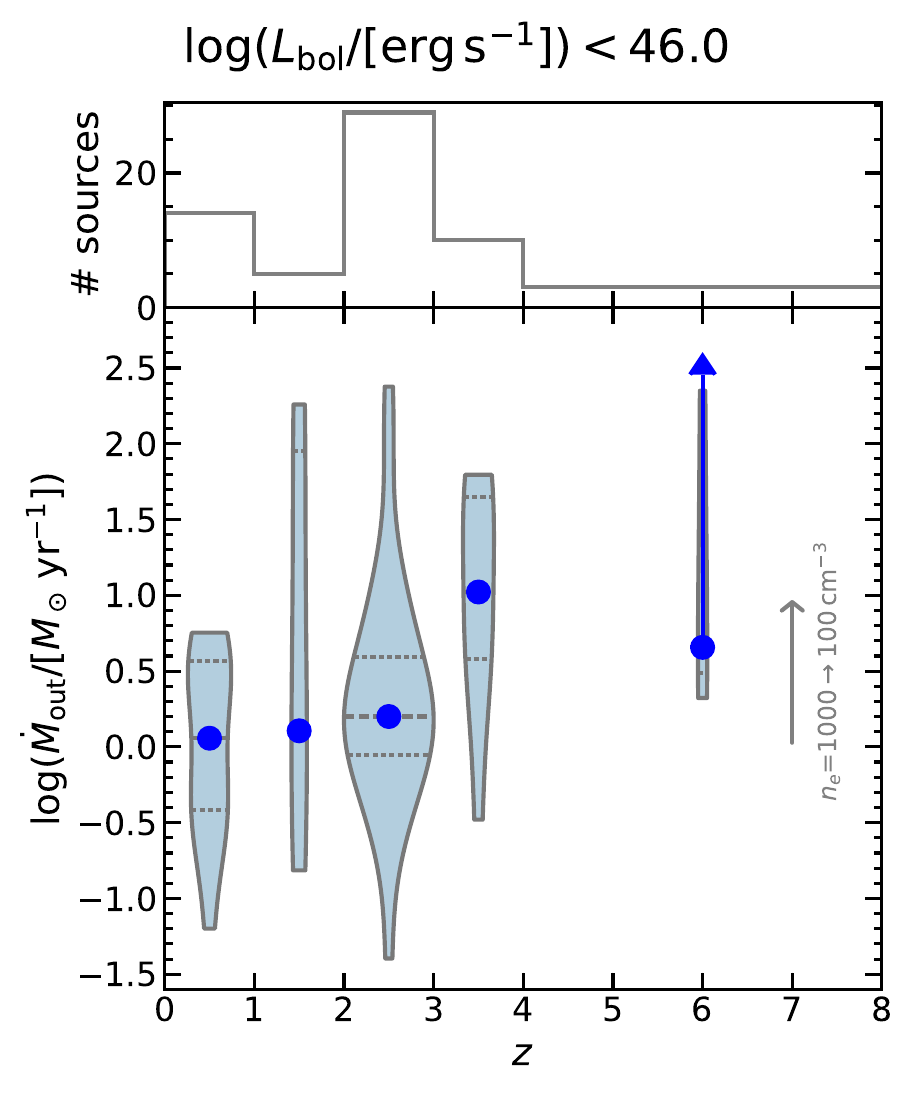}
    \includegraphics[width=0.33\linewidth]{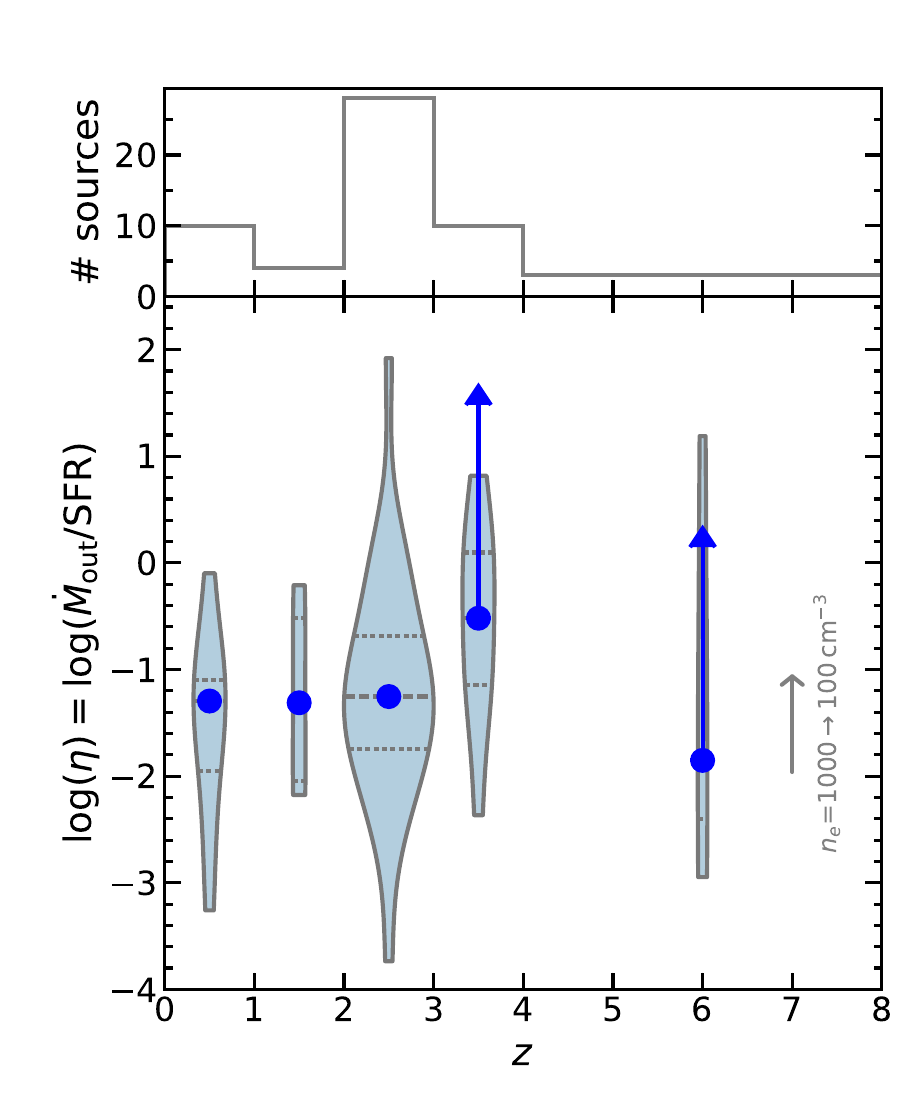}\\
    \caption{Redshift evolution of the outflow properties for sources with AGN \lbol < 10$^{46}$~\ergs, matching the bolometric luminosities of our GA-NIFS GS- and COS-AGN targets. \textit{From left to right}: Outflow velocity (\vout), mass outflow rate (\mdot), and mass loading factor ($\eta$ = \mdot/SFR) versus redshift. Measurements are reported for both single sources (top row) and in redshift bins (bottom row). For the latter, each point is the median (50th percentile) and the first and third quartiles (25th and 75th percentiles, respectively) of the distribution of values at each $z$ bin are marked by the dotted horizontal segments within each violin.
    The bins are $z$ = 0--1, 1--2, 2--3, 3--4, and 4--8. The area of each violin is proportional to the fraction of sources in the given redshift bin out of the total. The upper panels report the number of sources in each redshift bin.}
    \label{fig:zevol_lbolcut}
\end{figure*}

\section{Redshift evolution of outflow properties}\label{sec:zevol}
Given that the AGN targeted by GA-NIFS allow us to probe the ionised outflow properties in a redshift range ($z \gtrsim 3$) which has been poorly probed so far with spatially resolved observations, with the exception of a few luminous QSOs (see Fig.~\ref{fig:Lbolvz}), we can investigate whether there is any evolution of the outflow strength with redshift.
To do so, in Fig.~\ref{fig:zevol_lbolcut} we report \vout, \mdot, and $\eta$ versus $z$ for the GA-NIFS targets and those from literature. 
Given the existing relations between outflow properties and AGN bolometric luminosity (see Fig.~\ref{fig:Lbolvoutprops}), we compare our GS- and COS-AGN to literature targets matched in AGN luminosity (\lbol $<$ $10^{46}$~\ergs), in order to remove any luminosity bias (see the distribution of \lbol versus $z$ in Fig.~\ref{fig:Lbolvz}).
The measurements for the single targets are reported in the top panels, and are binned by redshift in the bottom panels, to better investigate the presence of any trends with redshift.
We used a binning of $\Delta z$ = 1 up to $z$ = 4, above which we adopted a single bin encompassing all of the measurements between $4 < z < 8$, due to the low number statistics at these high redshifts. The violin plots illustrate the spread of the observed values in each redshift bin, each point is the median (50th percentile) and the first and third quartiles (25th and 75th percentiles, respectively) of the distribution are indicated by the dotted segments; we stress that these do not have to be interpreted as statistical errors, but just illustrate the distribution of values in each bin. The upward arrows in the last and second-to-last redshift bins of $\eta$, and last of \mdot, indicate that more than 50\% of the measurements in that bin are lower limits.

On the one hand, the plots indicate an increasing trend with redshift for \mdot, particularly for outflows above cosmic noon ($z \gtrsim 2-3$).
On the other hand, \vout does not seem to show any redshift evolution.
$\eta$ also shows a rise from $z \sim 3$ to 4, though the many lower limits on $\eta$ (for more than 50\% of the sources at $3 < z < 4$) and the limited statistics at the highest redshifts ($>$4) hinder assessing the presence of a clear redshift evolution of the mass loading factor beyond cosmic noon.
In order to quantify this trend, we evaluated the Spearman rank correlation coefficient ($\rho$) and the Pearson correlation coefficient ($r$) between the above quantities and redshift for the single targets reported in the upper panels of Fig.~\ref{fig:zevol_lbolcut}, by also including lower limits as if they were constrained values. We obtained $\rho$ = 0.37 ($p$ = 0.0036) and $r$ = 0.40 ($p$ = 0.0016) for the mass outflow rate, $\rho$ = 0.21 ($p$ = 0.12) and $r$ = 0.27 ($p$ = 0.046) for the mass loading factor, and $\rho$ = 0.11 ($p$ = 0.38) and $r$ = 0.05 ($p$ = 0.68) for the outflow velocity, where $p$-values $<$ 0.05 indicate a significant correlation. This supports the presence of a correlation between the mass outflow rate and redshift, a weaker one for the mass loading factor, and the absence of a correlation for the outflow velocity. We stress that since lower limits on \mdot and $\eta$ were also included and treated as the other measured values for this correlation analysis, the actual correlation coefficients are expected to be higher.
Furthermore, we checked whether excluding GS~20936, whose broad line attribution to an outflow is ambiguous due to the presence of a merger, changes the observed trends. The median value in the $3<z<4$ bin is unaffected for the mass outflow rate and mildly lowered for the mass loading factor, though still higher than at $z<3$; the correlation coefficients are only lowered by 0.02. Therefore, we conclude that the presence or absence of GS~20936 does not significantly change our conclusions.

The same plots done for all sources (without any cut in \lbol), reported in Fig.~\ref{fig:zevol_all} (top and middle rows), show instead a clear increasing trend with redshift for all of the three outflow quantities. This is driven by a luminosity bias, with increasingly higher-\lbol AGN --- driving faster and more powerful outflows --- being observed at higher $z$, which is accounted for in Fig.~\ref{fig:zevol_lbolcut} by only considering sources at similar AGN luminosities.
To account for the luminosity bias, while keeping all of the sources in order to maximise the statistics (which is instead lowered when cutting by \lbol as in Fig.~\ref{fig:zevol_lbolcut}), we produced plots where the outflow quantities are normalised by \lbol (Fig.~\ref{fig:zevol_all}, bottom row), as done in \cite{Bertola2025}.
While these plots also show no clear redshift trend for \vout, they indicate a smoother rise of \mdot and $\eta$ at $z \gtrsim 2-3$ than that observed for the \lbol-matched subsample in Fig.~\ref{fig:zevol_lbolcut}.
However, as discussed in Appendix~\ref{sec:zevol_appendix}, the normalisation of the quantities by \lbol does not necessarily remove their dependence on \lbol; therefore, we consider the \lbol-matched plots in Fig.~\ref{fig:zevol_lbolcut} more robust against a luminosity bias, though they have more limited statistics.

In \cite{Bertola2025} we found a flattening of the mean mass outflow rate normalised by \lbol above $z\sim2$, when considering only COS-AGN plus a literature sample of IFU outflow studies substantially overlapping with the one considered here and including also other works using spatially integrated data (see Appendix~\ref{sec:app_outf_props_lit}). 
While in this work we build on and further develop the analysis shown in \citet{Bertola2025}, we note that these results are not in contradiction with each other. 
As can be seen in Fig.~\ref{fig:zevol_all}, bottom panels, we only observe a very mild enhancement of \mdot/\lbol between $z \sim 2$ and 3, likely driven by the fact that the GS-AGN included in this work generally have higher \mdot than the COS-AGN (see Figs.~\ref{fig:zevol_lbolcut} and \ref{fig:zevol_all}, top panels). Moreover, the evolution at low-$z$ observed in \citet{Bertola2025} disappears when considering a single low-redshift bin for $z<1$, as done here.  All in all, \mdot/\lbol versus $z$ between this work and \cite{Bertola2025} is overall consistent.

To summarise, the increasing trend with redshift of \mdot found above cosmic noon may indicate that high-$z$ outflows are more powerful than those in the local Universe and around cosmic noon.
However, more statistics of spatially resolved outflows would be needed to confirm and further investigate this trend, especially at the highest redshifts ($z > 4$). Moreover, mostly only lower limits on the mass loading factors are available at $3 < z < 4$ at the moment, which prevents us from robustly assessing whether there is an increase in the impact of AGN outflows on the host galaxy SFR above cosmic noon.
We also stress that a constant electron density for the outflowing gas has been assumed in our analysis. If this increases with redshift, the observed trend could be significantly reduced or even disappear. For instance, an increase in $n_\mathrm{e}$ of about a factor of 3 from $z\sim2$ to 6 is found for the host galaxy ISM in \citet[see also \citealt{Isobe2023}]{Marconcini2024a}, which if also affecting the outflowing gas, would flatten the \mdot versus $z$ trend by the same amount.
A thorough study of the cosmic evolution of the outflowing gas electron density would be needed to address this aspect.

\section{Conclusions}\label{sec:concl}
In this work, we have presented a spatially resolved study of ionised outflows in nine AGN (two of which are dual) at $z$ $>$ 3--6 from the GOODS-S field exploiting JWST NIRSpec IFU R2700 spectroscopy, as part of the GA-NIFS GTO survey.
Together with its twin subsample of six AGN (two of which are dual) from COSMOS \citep{Bertola2025}, also part of GA-NIFS, this constitutes the largest spatially resolved sample of AGN outflows at these redshifts to date, numbering a total of 19 sources, 16 of which with detected outflows. 
This allowed us to probe the so far virtually unexplored regime of AGN outflows at $z$ $\sim$ 3--6 and AGN bolometric luminosities \lbol $\sim$ 10$^{45-46}$~\ergs. 

We modelled the rest-optical emission lines (\oiii, \hb, \nii, \ha, \sii) with multiple Gaussian profiles (maximum of two except for one case), including an additional Gaussian component to model the \ha and \hb (plus \hei$\lambda$4923 and \hei$\lambda$6680 in GS~3073) emission from the BLR for the case of type-1 AGN, and spectrally isolated the outflows traced by broad wings from the gas at rest in the galaxy. We produced maps of the flux, velocity, and velocity dispersion of the ionised gas in the outflow and at rest, which revealed different outflow morphologies, from mostly blueshifted and spatially symmetric (the majority of cases) to bipolar in one case (see also \citealt{Perna2025a}); all the outflows in type-1 AGN are spatially unresolved.

By exploiting the spatially resolved information, we obtained the integrated outflow velocity, mass outflow rate, and kinetic rate, as well as the outflow mass loading factor, that is, the ratio between mass outflow rate and SFR, obtained from SED fitting (Circosta et al., in prep.).
We find outflow velocities of $\sim$600-2000~\kms, maximum outflow radii of $\lesssim$1~kpc in unresolved sources up to $\sim$4~kpc in spatially resolved ones, and mass outflow rates ranging from $\sim$0.1~\Msunyr up to >100~\Msunyr, which in some cases may exceed the SFR and thus significantly affect the galaxy star formation (though the uncertainty on the gas electron density, assumed to be 1000~cm$^{-3}$, prevents us from drawing firm conclusions).

The incidence of ionised outflows in the GOODS-S and COSMOS GA-NIFS sample is high (>75\%), comparable with what is found in some $z \sim 2$ AGN samples with similar luminosities (\lbol $\gtrsim$ 10$^{45-46}$~\ergs) and significantly higher than at lower $z$. However, a homogeneous outflow detection method and sample selection would be required to draw firm conclusions on the redshift evolution of the outflow incidence.
We note that our results cannot be extended to the whole population of AGN at $z>3$, in particular to those with \lbol~$\lesssim$~10$^{45}$~\ergs undetected in X-rays discovered by JWST.

In order to compare the properties of the outflows in GOODS-S and COSMOS AGN from GA-NIFS with those from the literature, we re-calculated the outflow properties of the latter in a homogeneous way by using the same assumptions and method adopted in our work and, to this end, we only considered spatially resolved studies.
We find that the outflow properties from the GOODS-S and COSMOS AGN from GA-NIFS are generally consistent with those from the literature, though some of the sources in our sample tend to be at the high end of mass outflow rate, kinetic rate, and mass loading factor, given their (moderate) AGN \lbol.
The mass loading factors of many of them are close to or above 1, and could even be higher since they are all lower limits, or if the actual outflow density is smaller than the conservative value of 1000~cm$^{-3}$ assumed; this implies that these outflows could significantly affect the star formation processes in the host galaxy.

To further investigate this aspect, we checked for the presence of any evolution with redshift of the outflow properties. 
We removed the luminosity bias by only considering sources at \lbol $<$ 10$^{46}$~\ergs, without which all of the outflow properties show a steep increase with redshift, due to the fact that more luminous AGN, which drive more powerful outflows, are preferentially observed at higher redshifts.
Nevertheless, even after removing the luminosity bias, our analysis suggests an increase in the mass outflow rate with redshift, especially above cosmic noon ($z\gtrsim3$). There is also evidence of an increase in mass loading factor at $z\gtrsim3$; the large number of upper limits on SFR (lower limits on mass loading factor) makes this less clear, but it also implies that the actual mass loading factors at $z\gtrsim3$ could be significantly larger. At $z>4$, the limited statistics on outflows at these moderate AGN luminosities makes the inferred evolution of the outflow properties at this upper redshift range very uncertain.
The outflow velocity shows no evolution with redshift.

All in all, the results of our work indicate that AGN outflows were stronger in the early Universe than at later times, and potentially more capable of affecting their host galaxy.
More statistics at $z > 3-4$ would be beneficial to confirm these results, while robust and homogeneous measurements of SFR, most of which are upper limits at $z$ $>$ 3 (especially for \lbol $<$ 10$^{46}$~\ergs), would help constrain the mass loading factor and its cosmic evolution, which are critical to quantify the role of AGN feedback in the early Universe.

\begin{acknowledgements}
We acknowledge support from European Union's HE ERC Starting Grant No. 101040227 - WINGS (G.V., S.C., and S.Z.); the Italian National Institute for Astrophysics (INAF) under the IAF - Astrophysics Fellowships in Italy grant CUP C59J21034720001 - ``AD MAJORA'' (G.V.); European Union's ERC APEX No. 101164796 and the Max Planck Society through the Lise Meitner Excellence Program (H.\"U.); INAF Large Grant 2022 ``The metal circle: a new sharp view of the baryon cycle up to Cosmic Dawn with the latest generation IFU facilities'' and INAF GO grant ``A JWST/MIRI MIRACLE: Mid-IR Activity of Circumnuclear Line Emission'' (E.B. and G.C.); INAF ``Ricerca Fondamentale 2024'' program: mini-grant 1.05.24.07.01 (E.B.); grants PID2021-127718NB-I00 (S.A., M.P., and B.R.P.), PID2024-159902NA-I00 (M.P.), PID2024-158856NA-I00 (B.R.P.), and RYC2023-044853-I (M.P.), funded by the Spain Ministry of Science and Innovation/State Agency of Research MCIN/AEI/10.13039/501100011033; El Fondo Social Europeo Plus FSE+ (M.P.); ``ERDF A way of making Europe'' (B.R.P.); ``FirstGalaxies'' Advanced Grant from the European Research Council (ERC) under the European Union’s Horizon 2020 research and innovation program (Grant agreement No. 789056; A.J.B.); the Science and Technology Facilities Council (STFC), the ERC through Advanced Grant 695671 ``QUENCH'', and the UKRI Frontier Research grant RISEandFALL (R.M., F.D.E., G.C.J., and J.S.); a research professorship from the Royal Society (R.M.); PRIN-MUR project ``PROMETEUS''  financed by the European Union -  Next Generation EU, Mission 4 Component 1 CUP B53D23004750006 (I.L.).

Views and opinions expressed are however those of the authors only and do not necessarily reflect those of the European Union or the European Research Council Executive Agency. Neither the European Union nor the granting authority can be held responsible for them.
This work is based on observations made with the NASA/ESA/CSA James Webb Space Telescope. The data were obtained from the Mikulski Archive for Space Telescopes at the Space Telescope Science Institute, which is operated by the Association of Universities for Research in Astronomy, Inc., under NASA contract NAS 5-03127 for JWST. These observations are associated with program \#1216. The specific observations analysed can be accessed via \url{https://doi.org/10.17909/vn0z-4j48}.
This research has made use of NASA’s Astrophysics Data System Bibliographic Services, of the FITS files visualisation tool QFitsView (\url{https://www.mpe.mpg.de/~ott/QFitsView/}), of the cosmology calculator by \cite{Wright2006}, and of the Python packages NumPy \citep{numpy2020}, SciPy \citep{scipy2020},  IPython \citep{ipython2007}, Matplotlib \citep{matplotlib2007}, seaborn \citep{Waskom2021}, and Astropy (\url{http://www.astropy.org}), a community-developed core Python package for Astronomy \citep{astropy:2013, astropy:2018}.
\end{acknowledgements}

\bibliographystyle{aa} 
\bibliography{bibliography_ganifs_out}

\begin{appendix}

\section{Emission line maps}\label{sec:app_emline_maps}

\begin{figure*}
    \centering
    \includegraphics[width=0.245\linewidth,trim={1cm 0 1cm 0},clip]{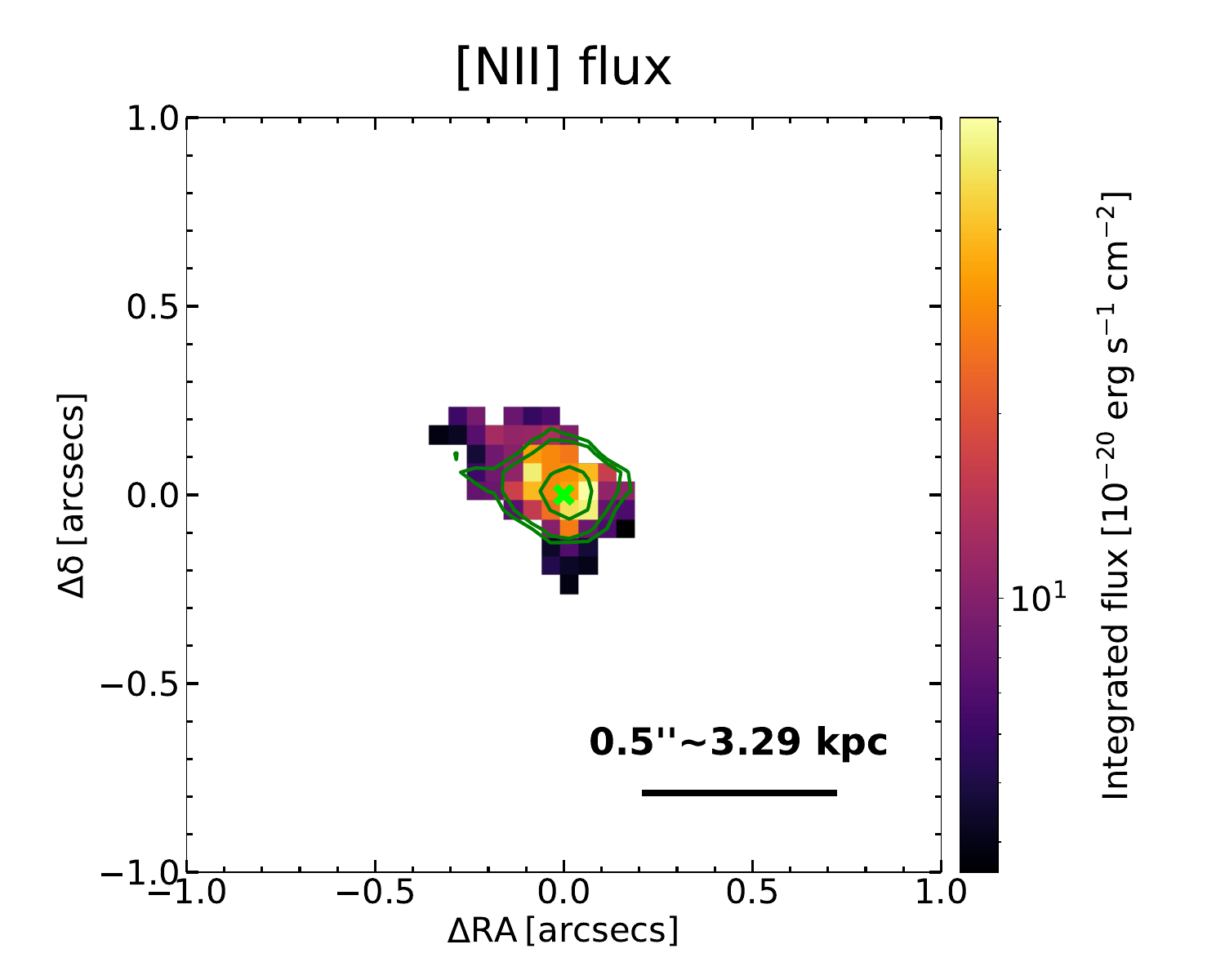}
    \includegraphics[width=0.245\linewidth,trim={1cm 0 1cm 0},clip]{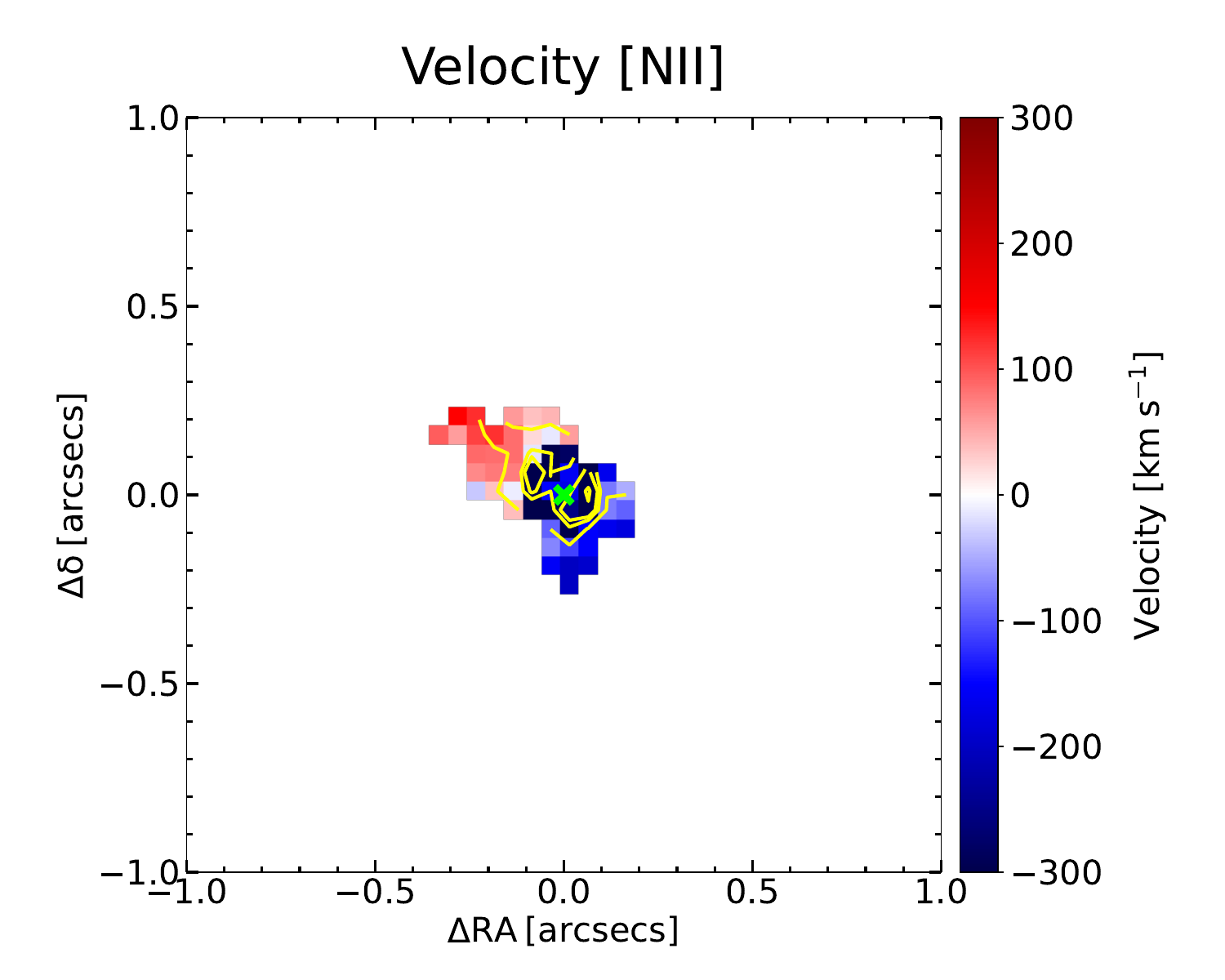}
    \includegraphics[width=0.245\linewidth,trim={1cm 0 1cm 0},clip]{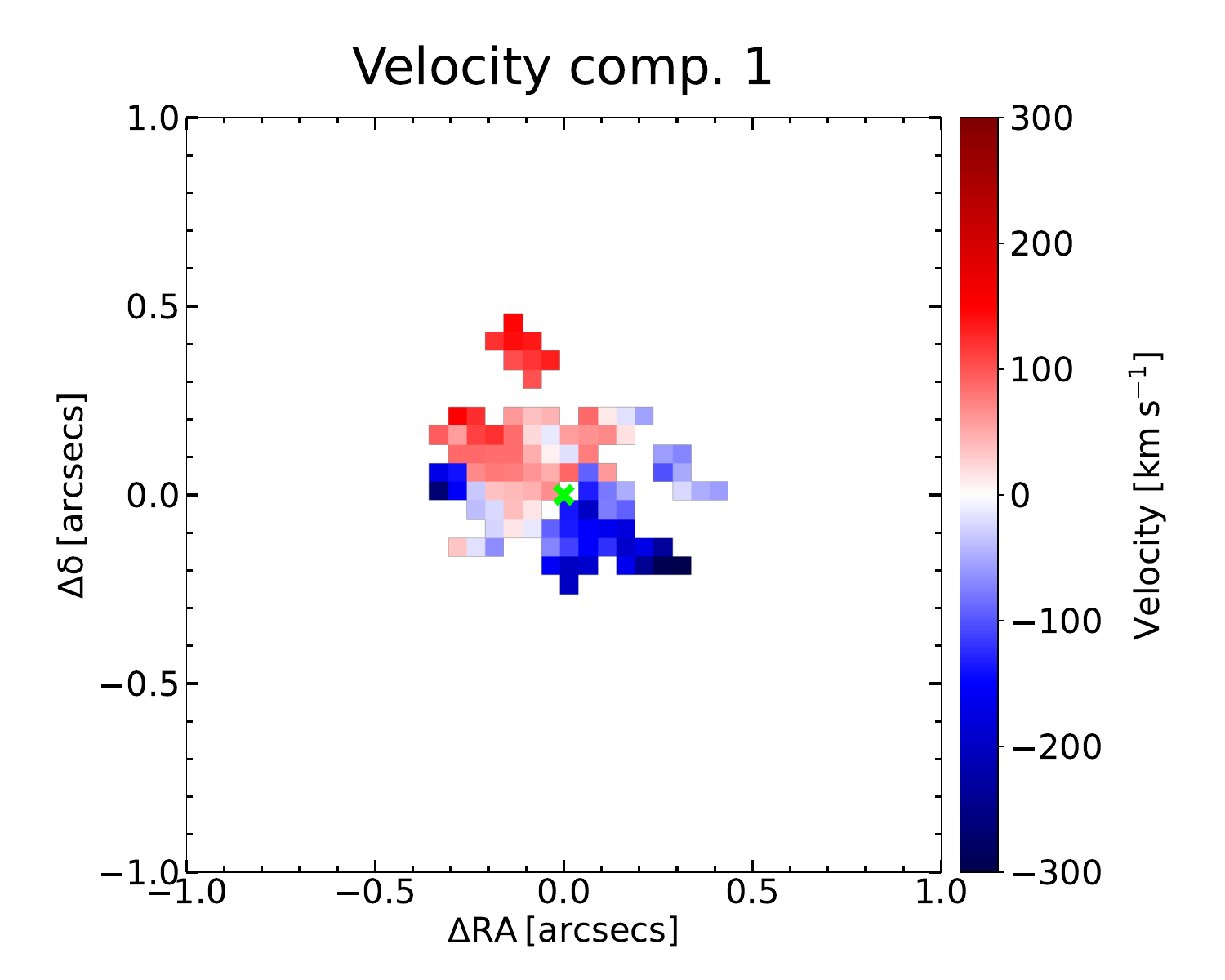}
    \includegraphics[width=0.245\linewidth,trim={1cm 0 1cm 0},clip]{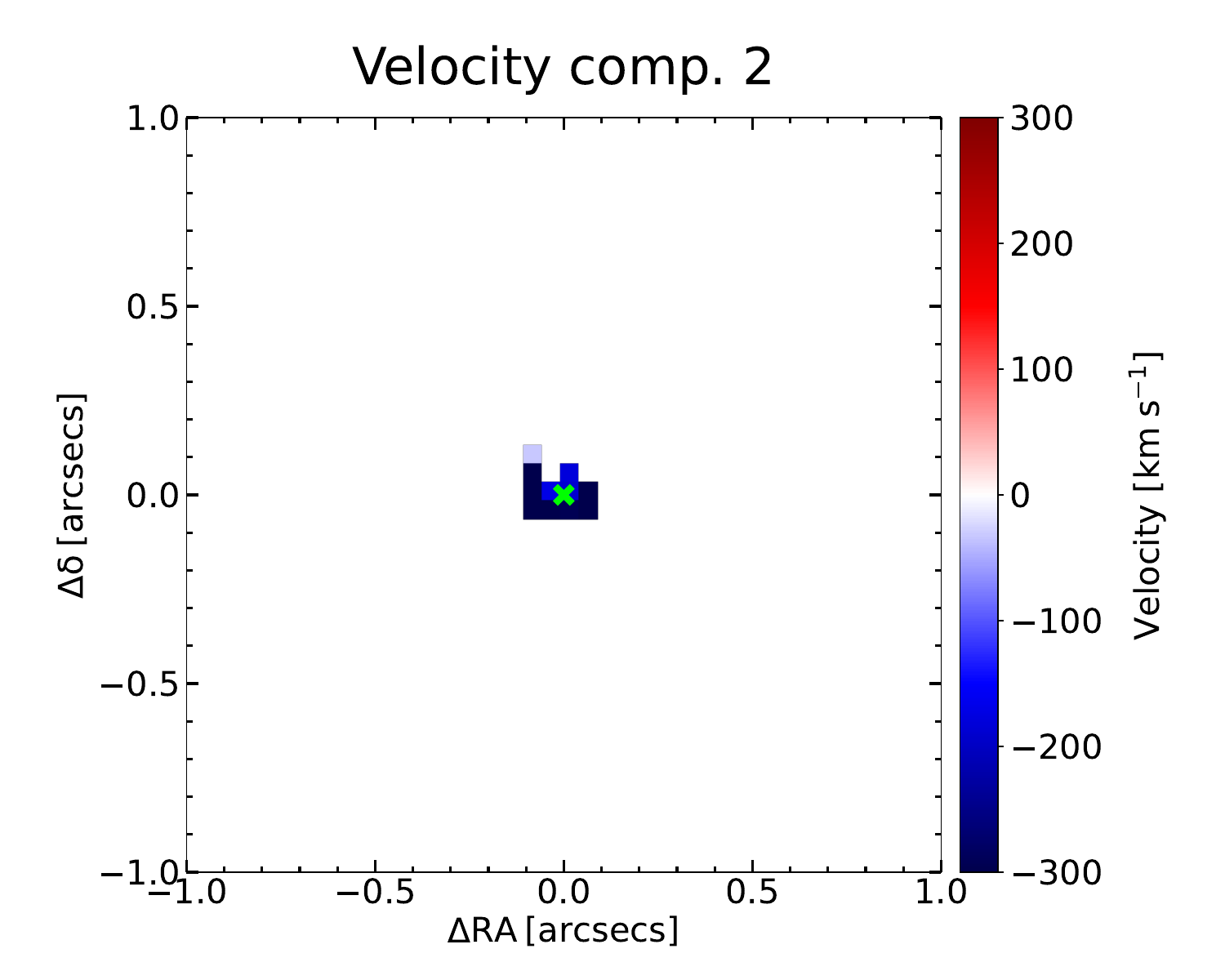}\\
    \includegraphics[width=0.245\linewidth,trim={1cm 0 1cm 0},clip]{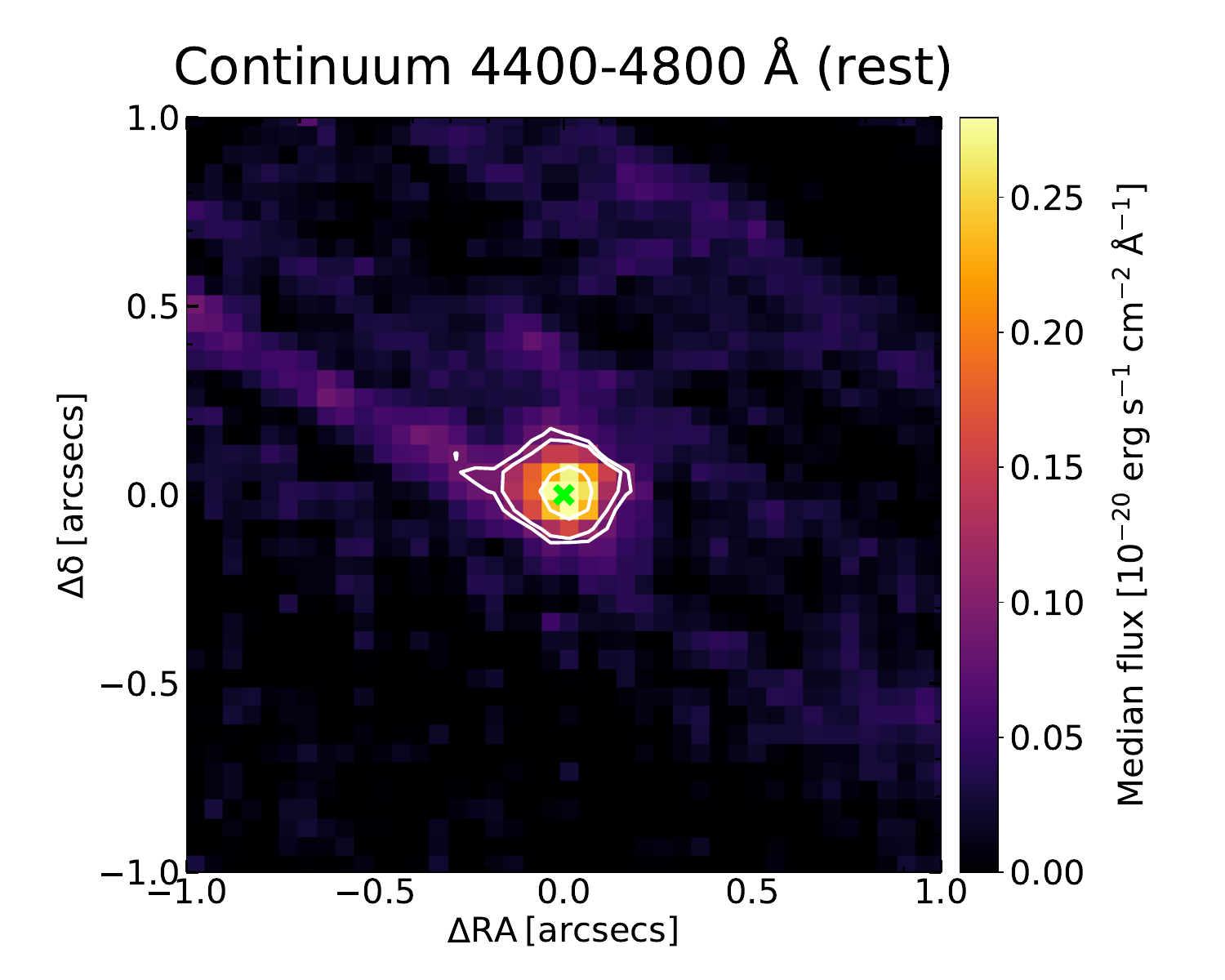}
    \includegraphics[width=0.245\linewidth,trim={1cm 0 1cm 0},clip]{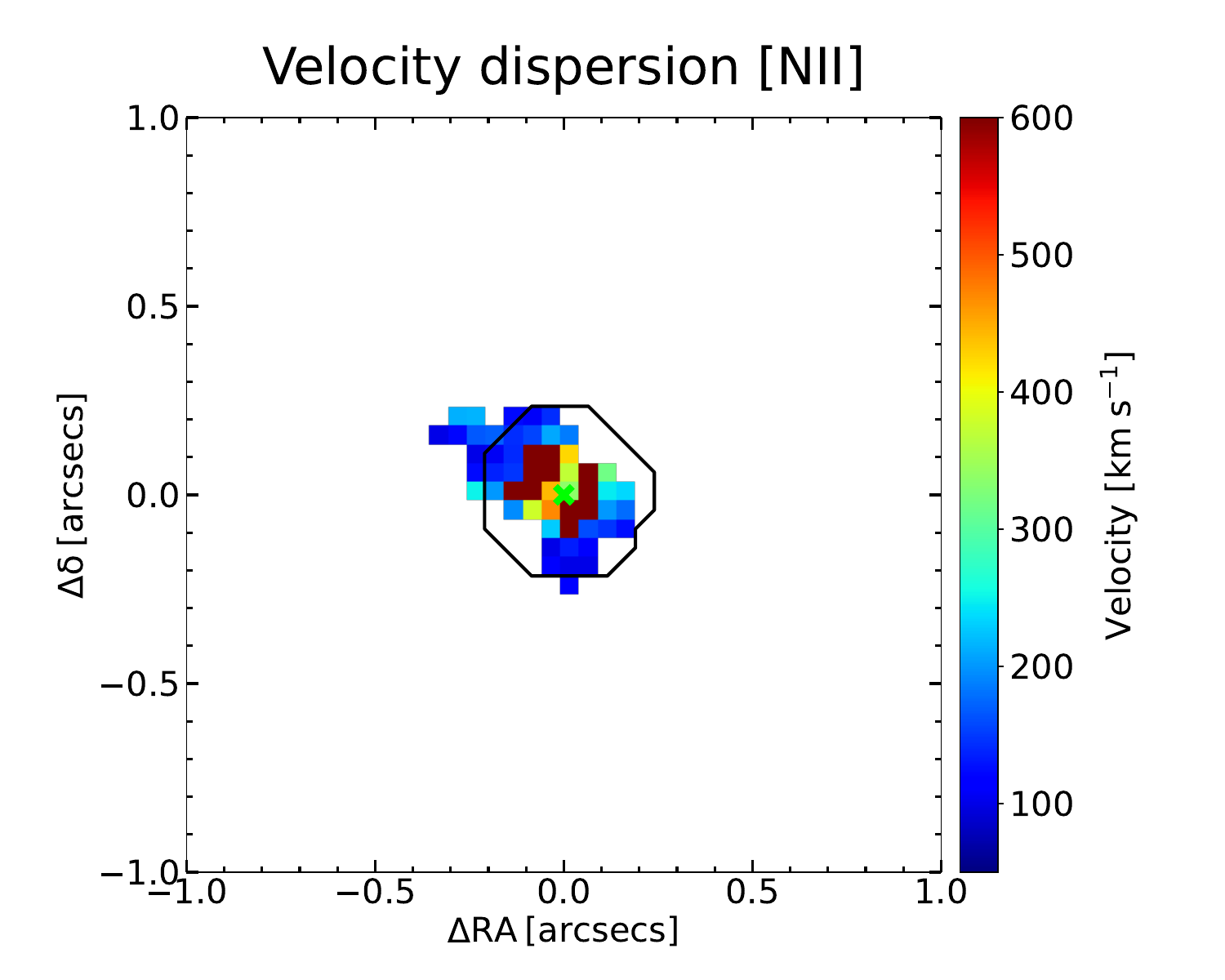}
    \includegraphics[width=0.245\linewidth,trim={1cm 0 1cm 0},clip]{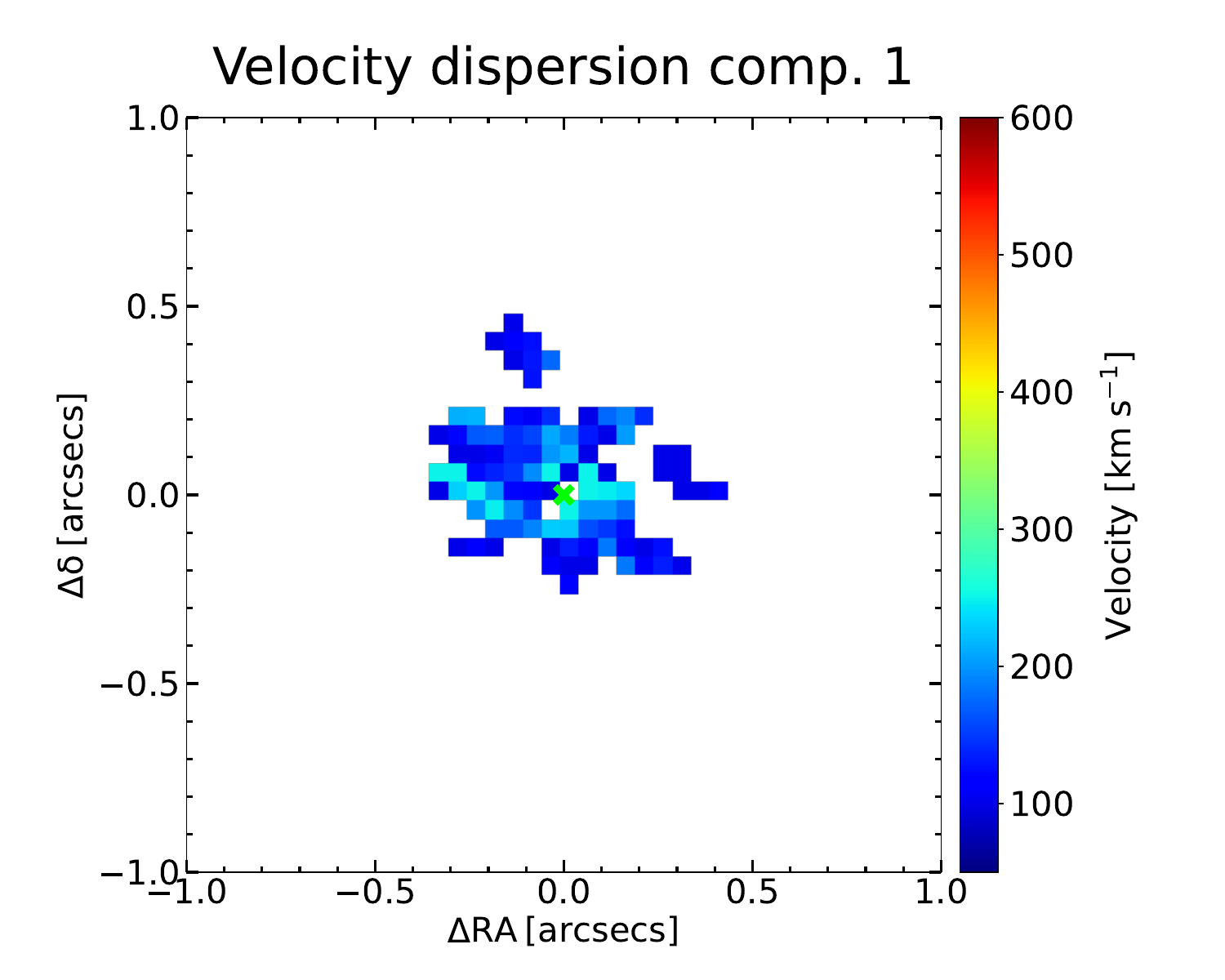}
    \includegraphics[width=0.245\linewidth,trim={1cm 0 1cm 0},clip]{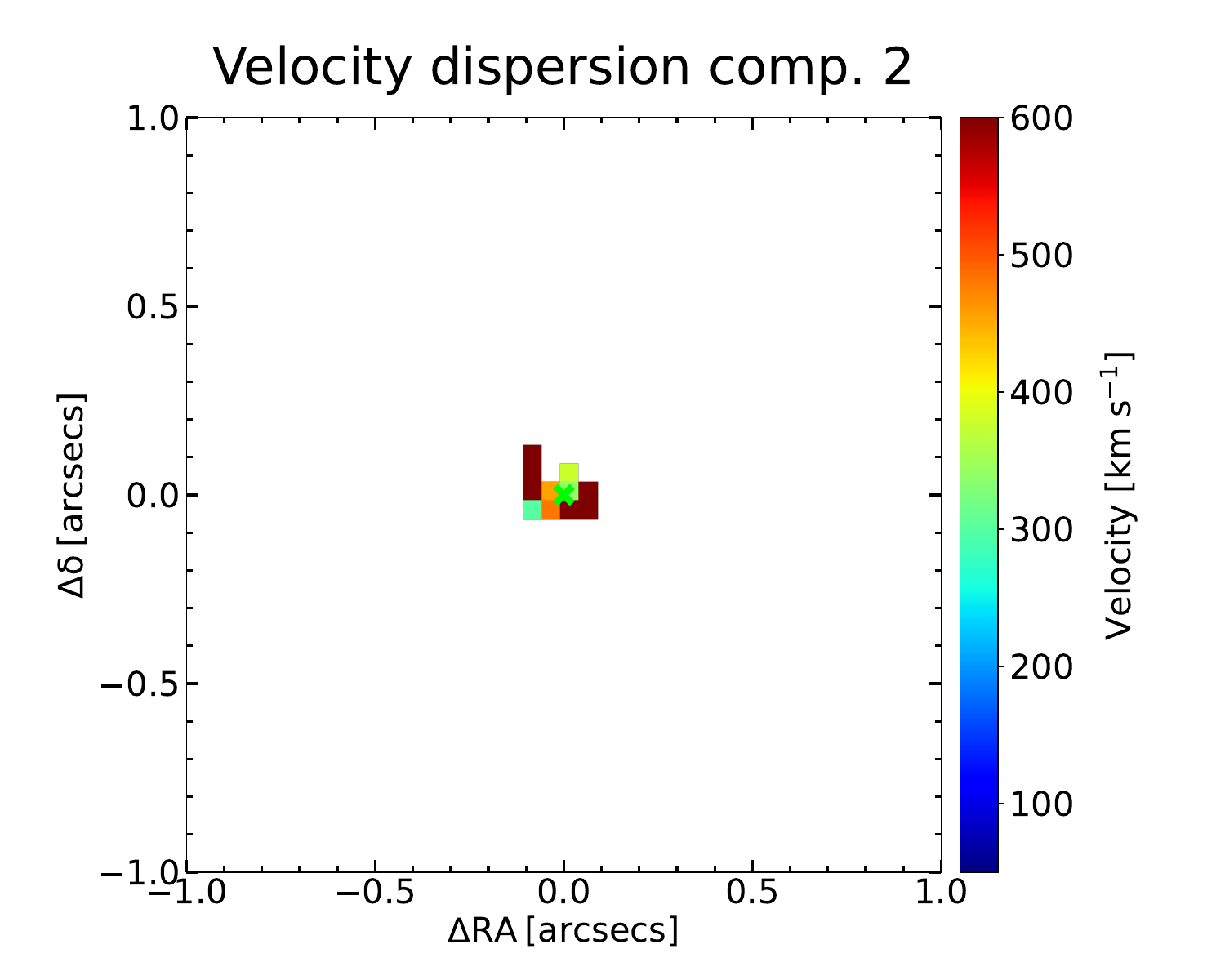}
    \caption{Same as Fig.~\ref{fig:gs133} but for GS~539.}
    \label{fig:gs539}
\end{figure*}

\begin{figure*}
    \centering
    \includegraphics[width=0.245\linewidth,trim={1cm 0 1cm 0},clip]{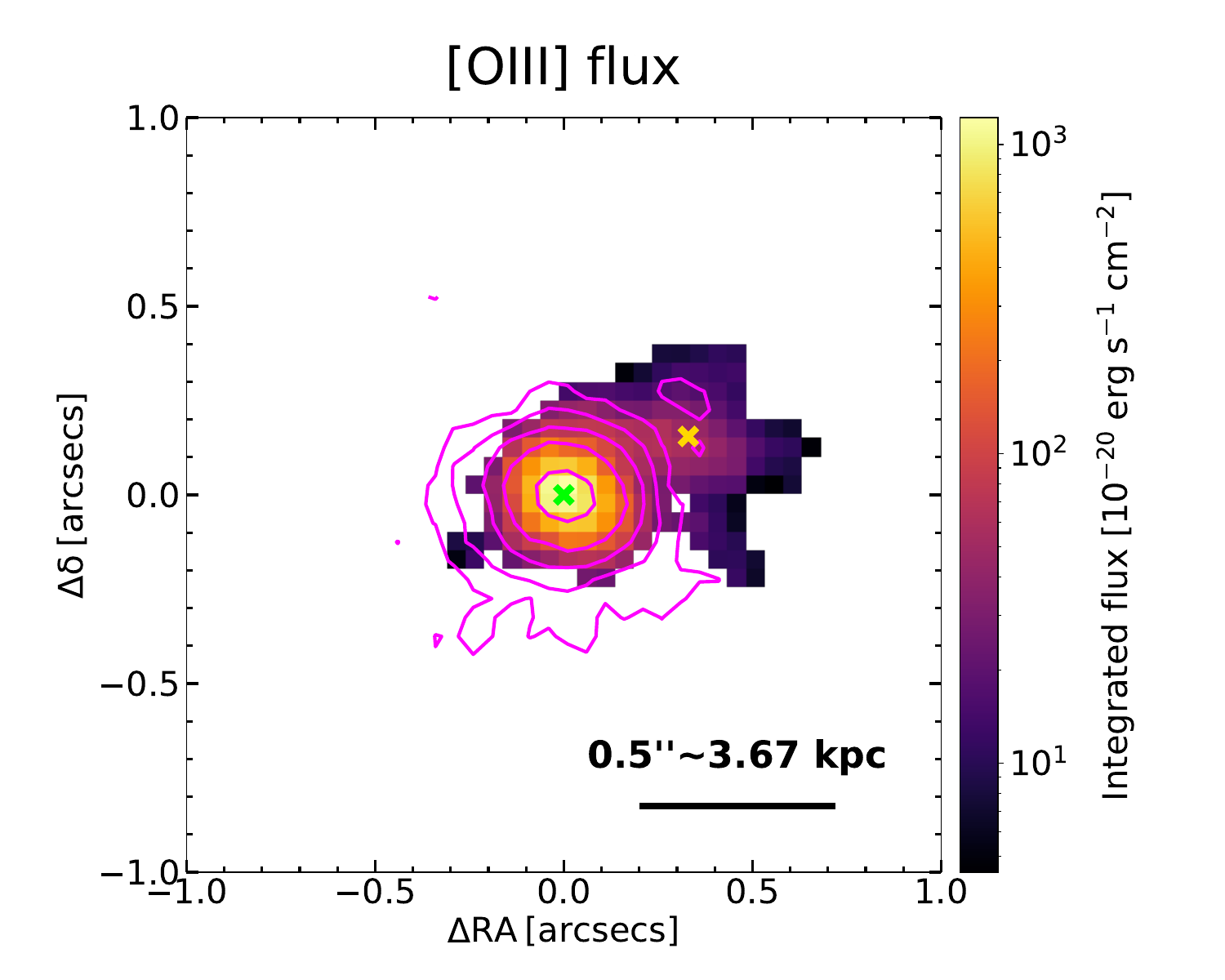}
    \includegraphics[width=0.245\linewidth,trim={1cm 0 1cm 0},clip]{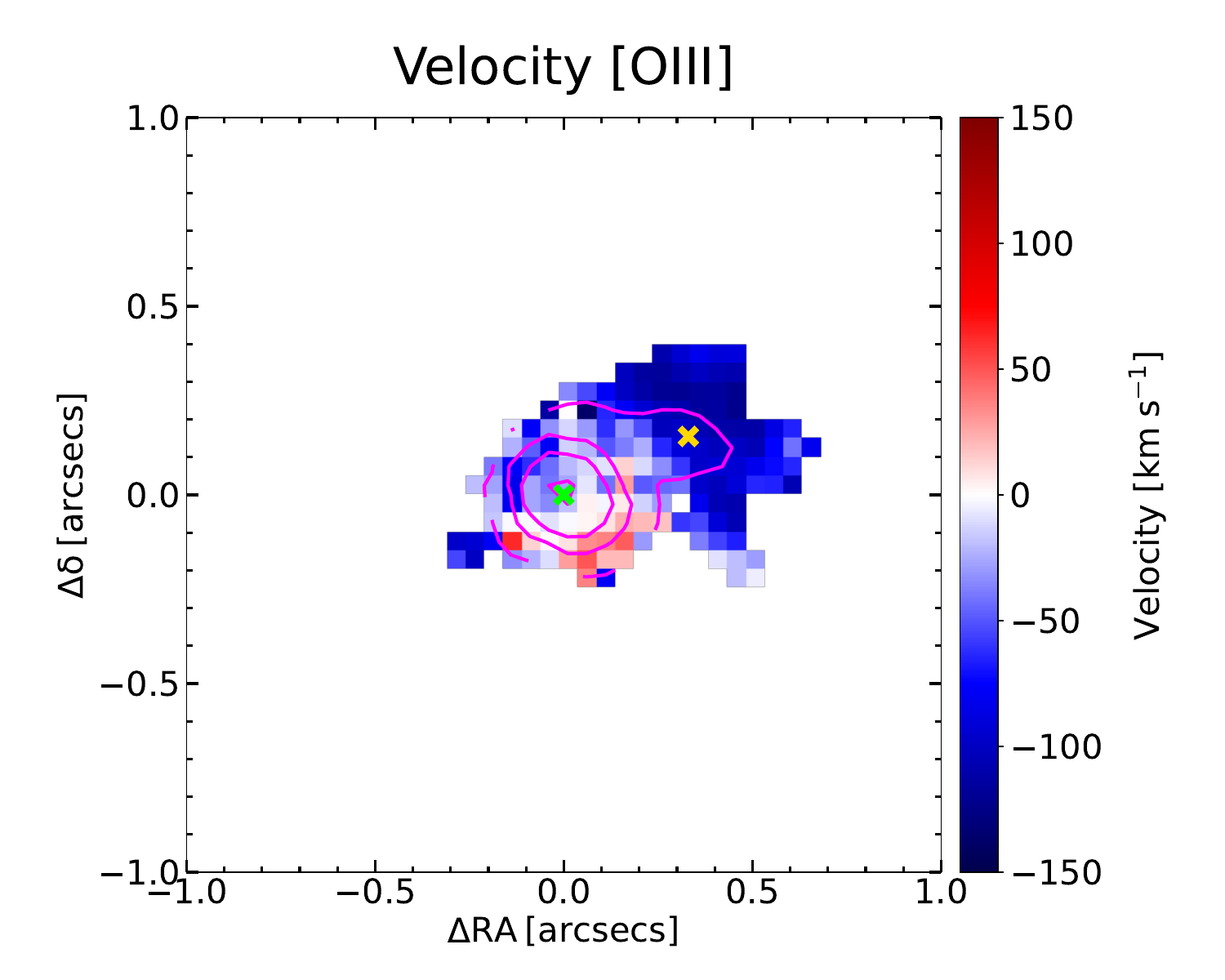}
    \includegraphics[width=0.245\linewidth,trim={1cm 0 1cm 0},clip]{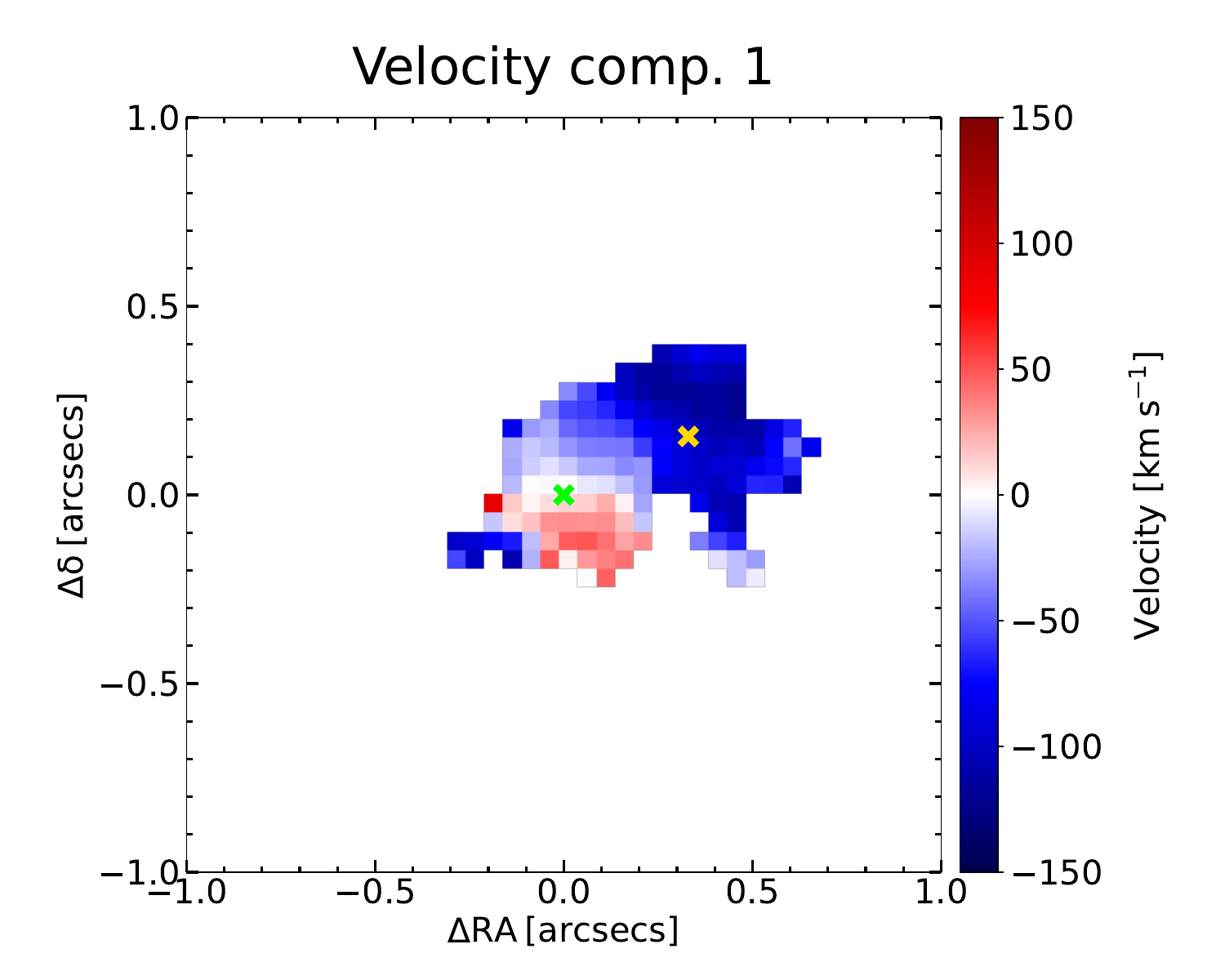}
    \includegraphics[width=0.245\linewidth,trim={1cm 0 1cm 0},clip]{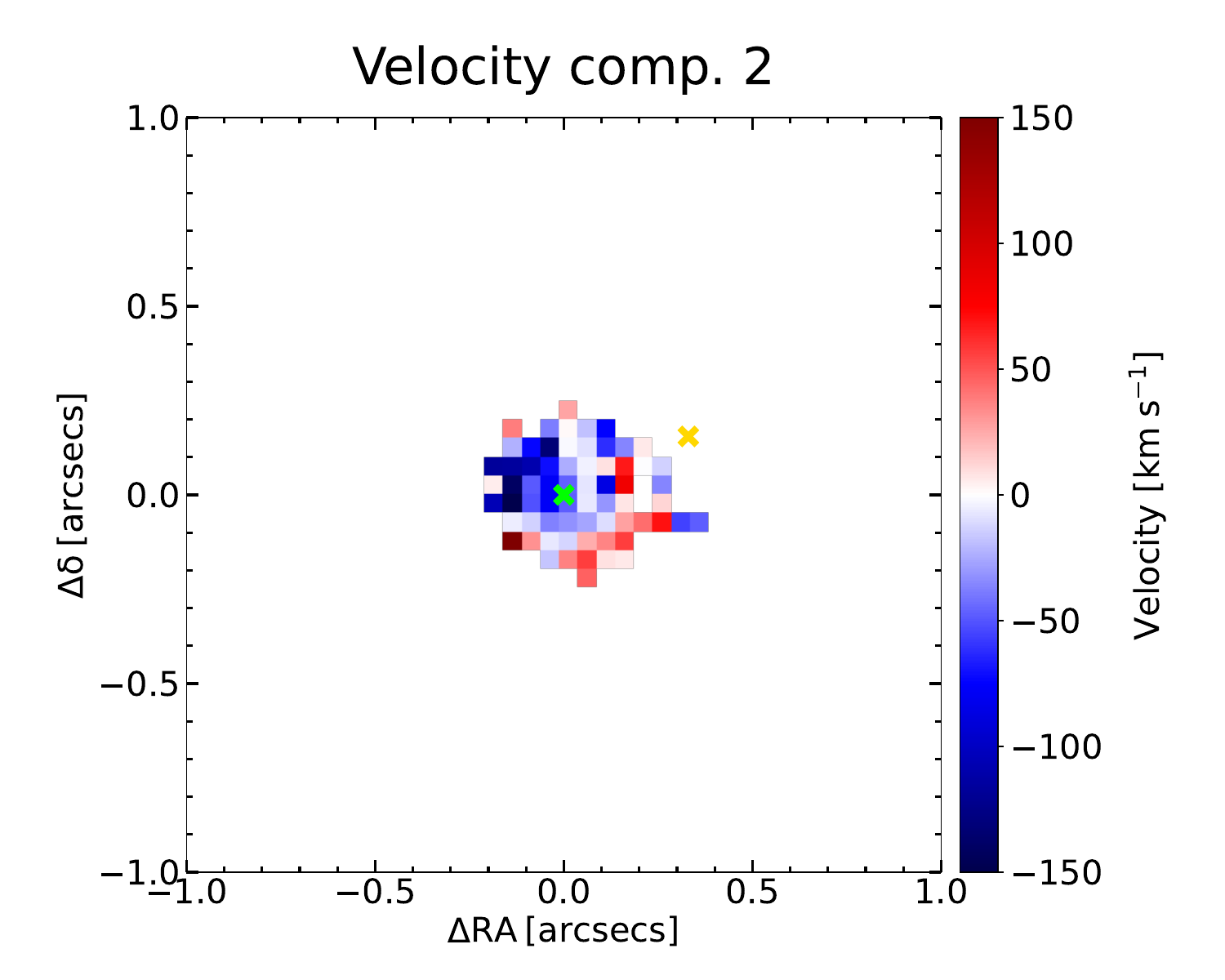}\\
    \includegraphics[width=0.245\linewidth,trim={1cm 0 1cm 0},clip]{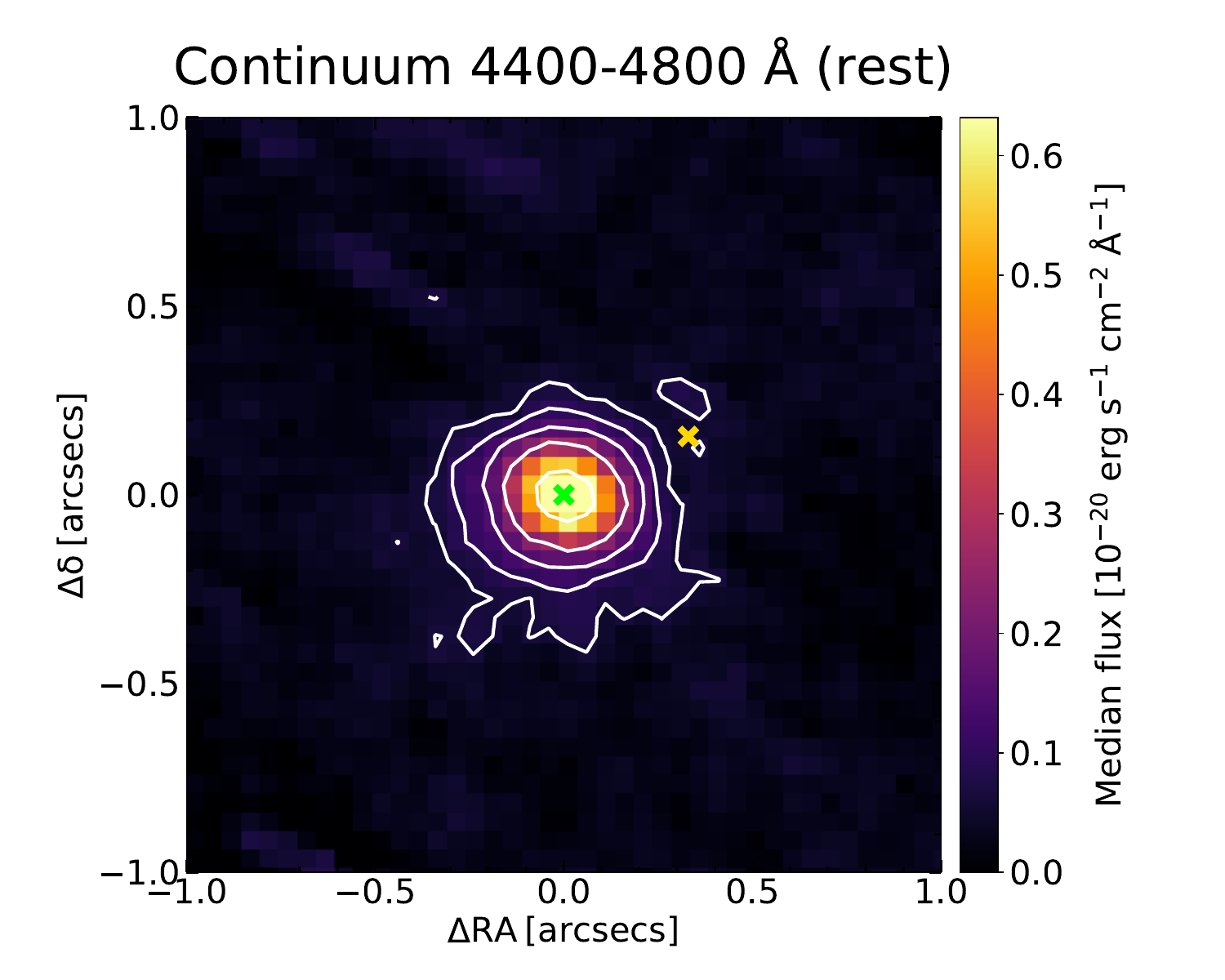}
    \includegraphics[width=0.245\linewidth,trim={1cm 0 1cm 0},clip]{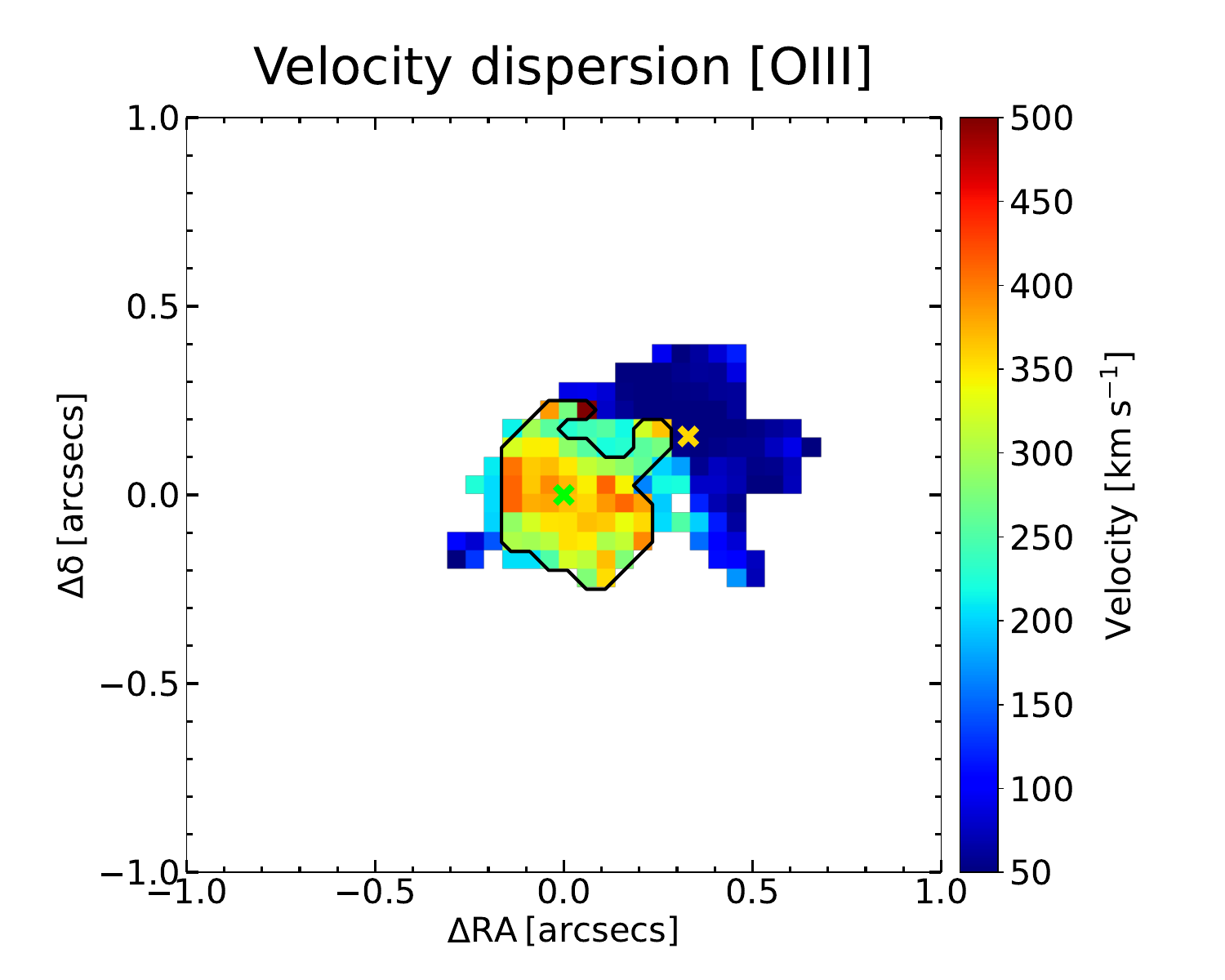}
    \includegraphics[width=0.245\linewidth,trim={1cm 0 1cm 0},clip]{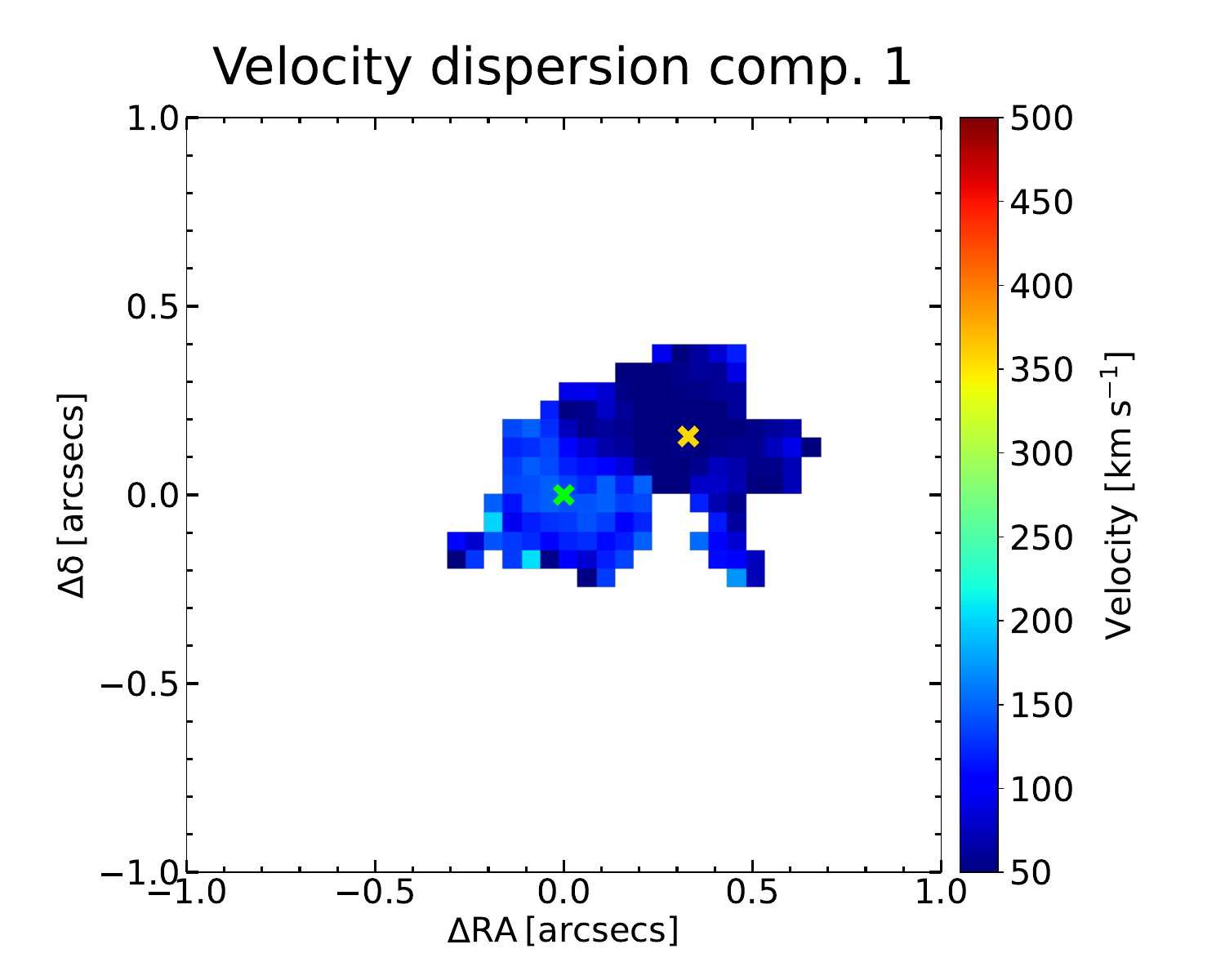}
    \includegraphics[width=0.245\linewidth,trim={1cm 0 1cm 0},clip]{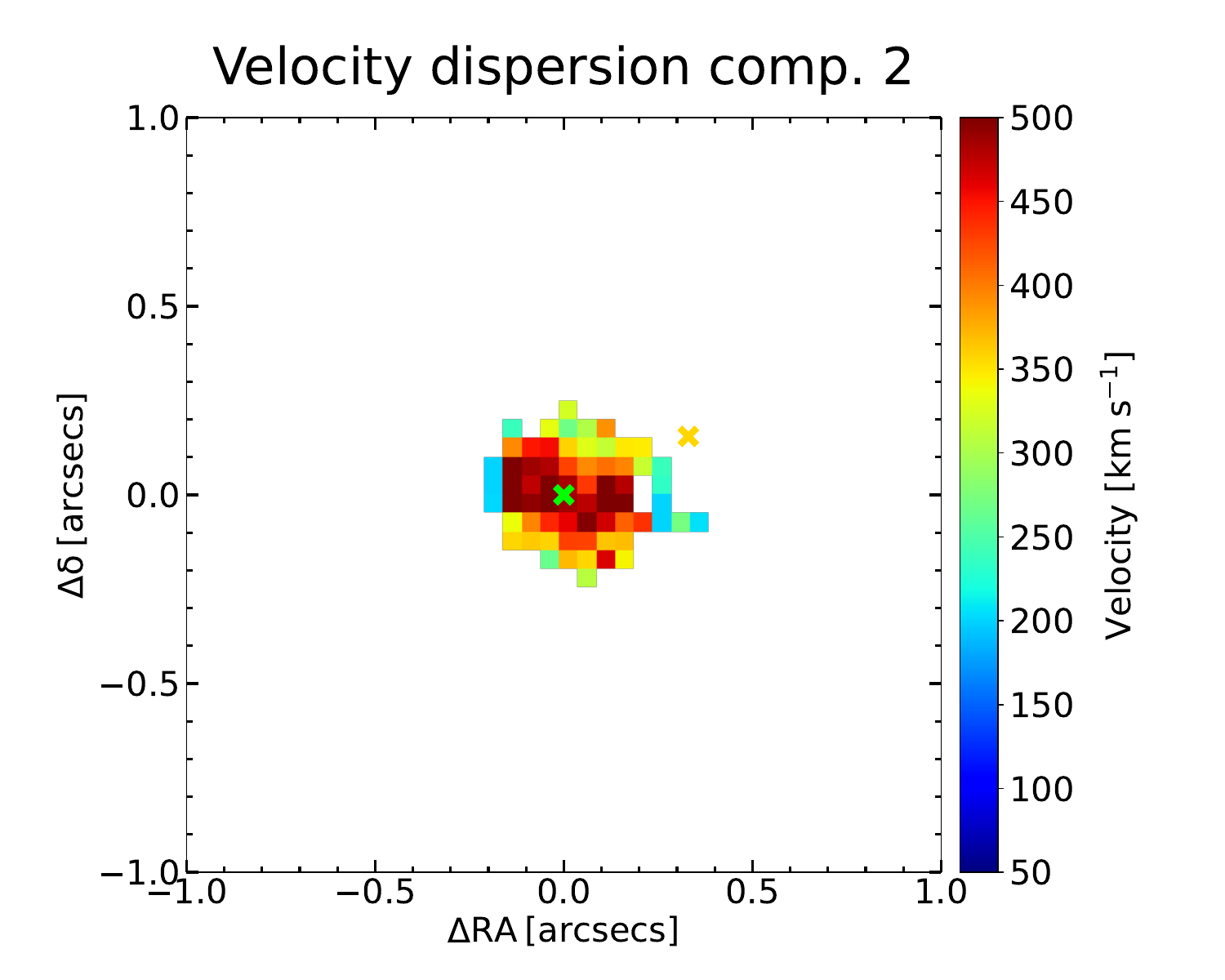}
    \caption{Same as Fig.~\ref{fig:gs133} for but GS~551. The golden cross marks the position of the secondary AGN.}
    \label{fig:gs551}
\end{figure*}

\begin{figure*}
    \centering
    \includegraphics[width=0.245\linewidth,trim={1cm 0 1cm 0},clip]{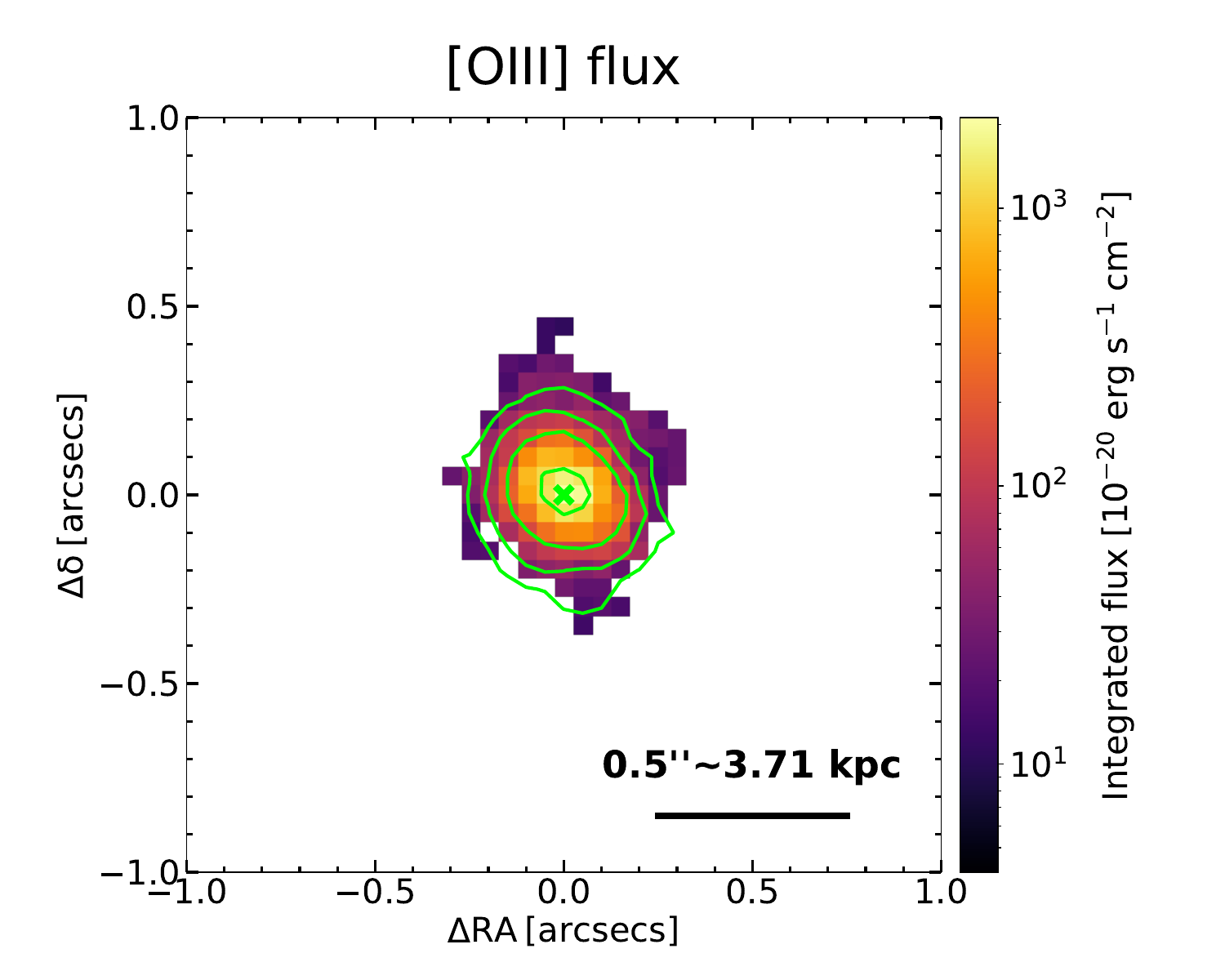}
    \includegraphics[width=0.245\linewidth,trim={1cm 0 1cm 0},clip]{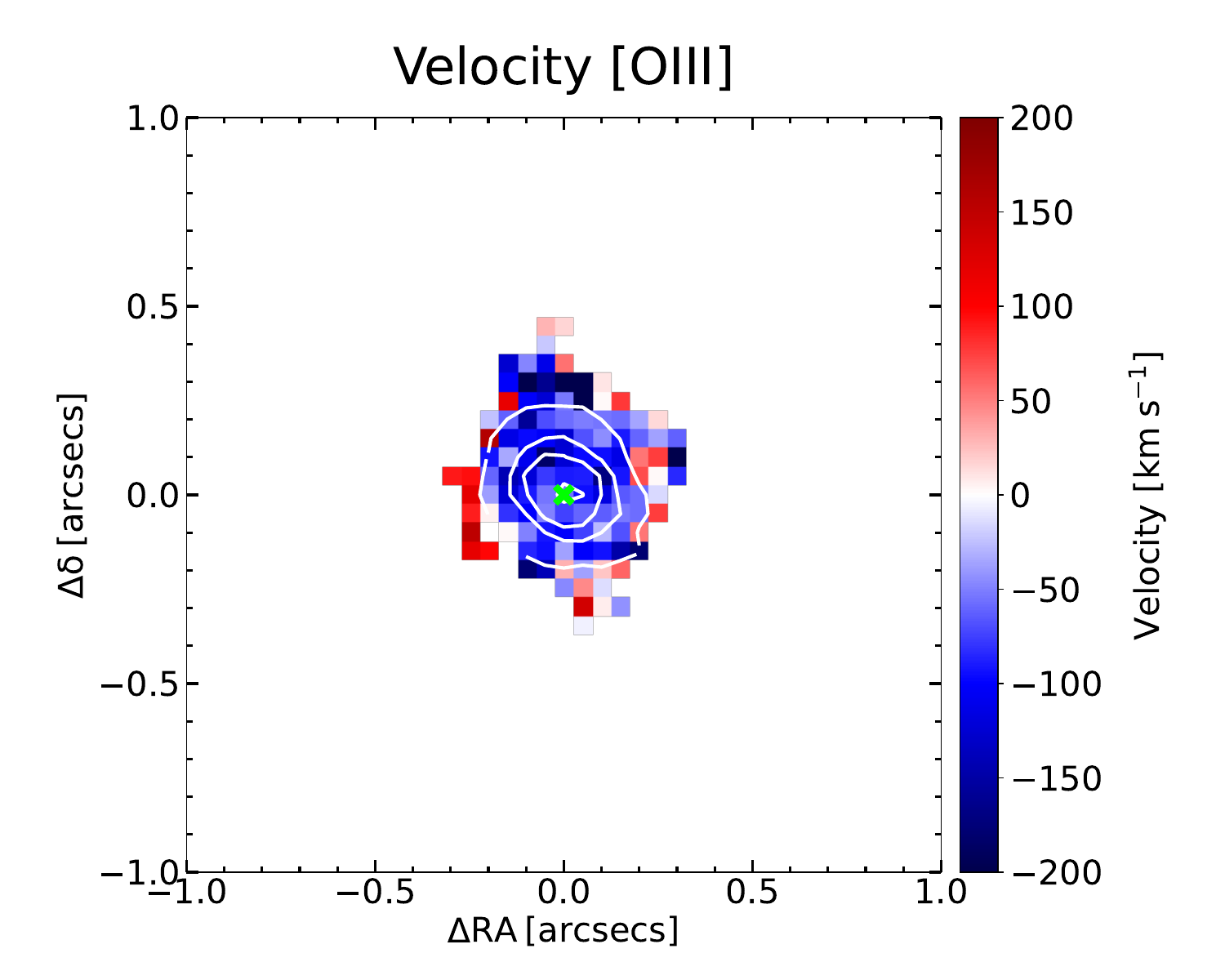}
    \includegraphics[width=0.245\linewidth,trim={1cm 0 1cm 0},clip]{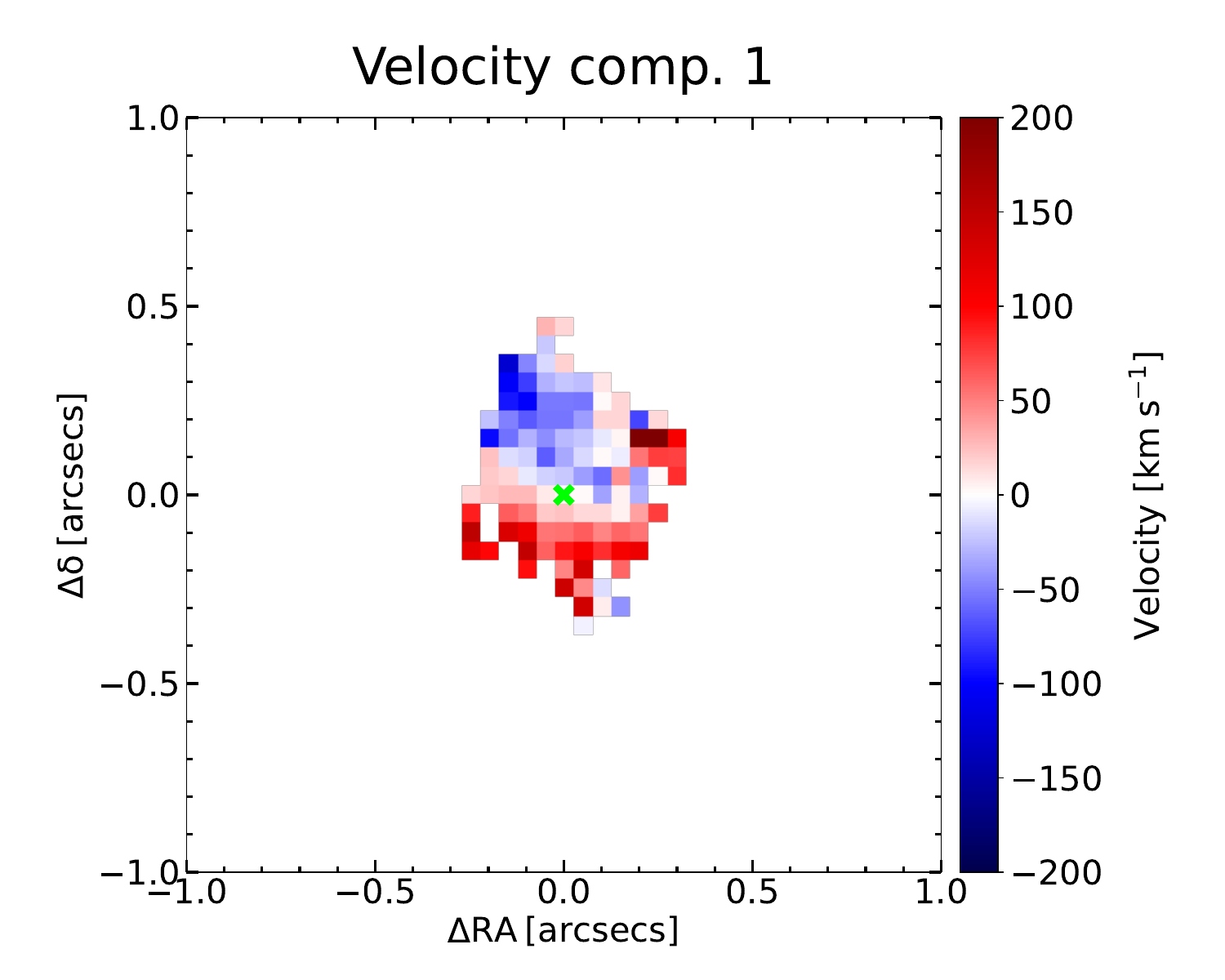}
    \includegraphics[width=0.245\linewidth,trim={1cm 0 1cm 0},clip]{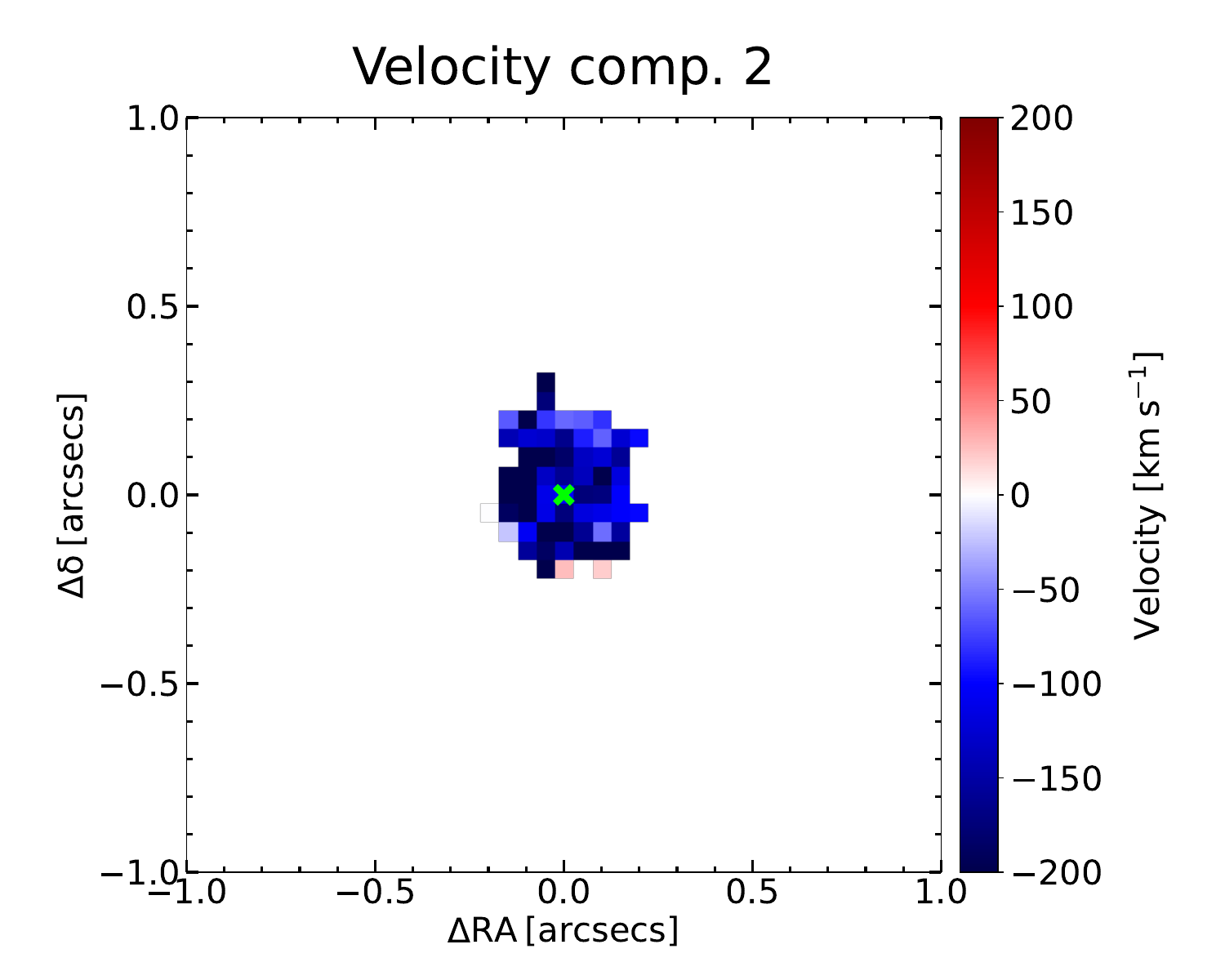}\\
    \includegraphics[width=0.245\linewidth,trim={1cm 0 1cm 0},clip]{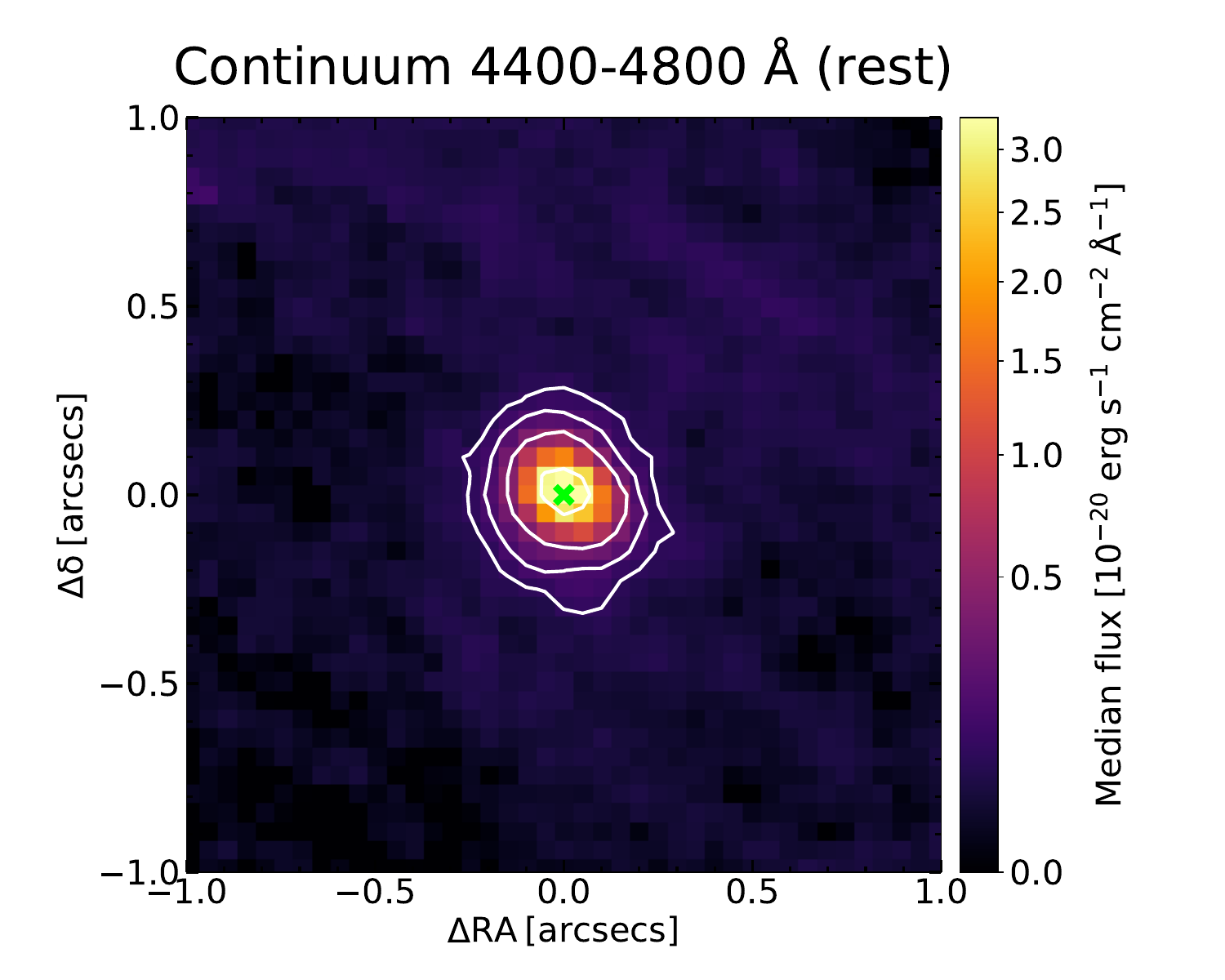}
    \includegraphics[width=0.245\linewidth,trim={1cm 0 1cm 0},clip]{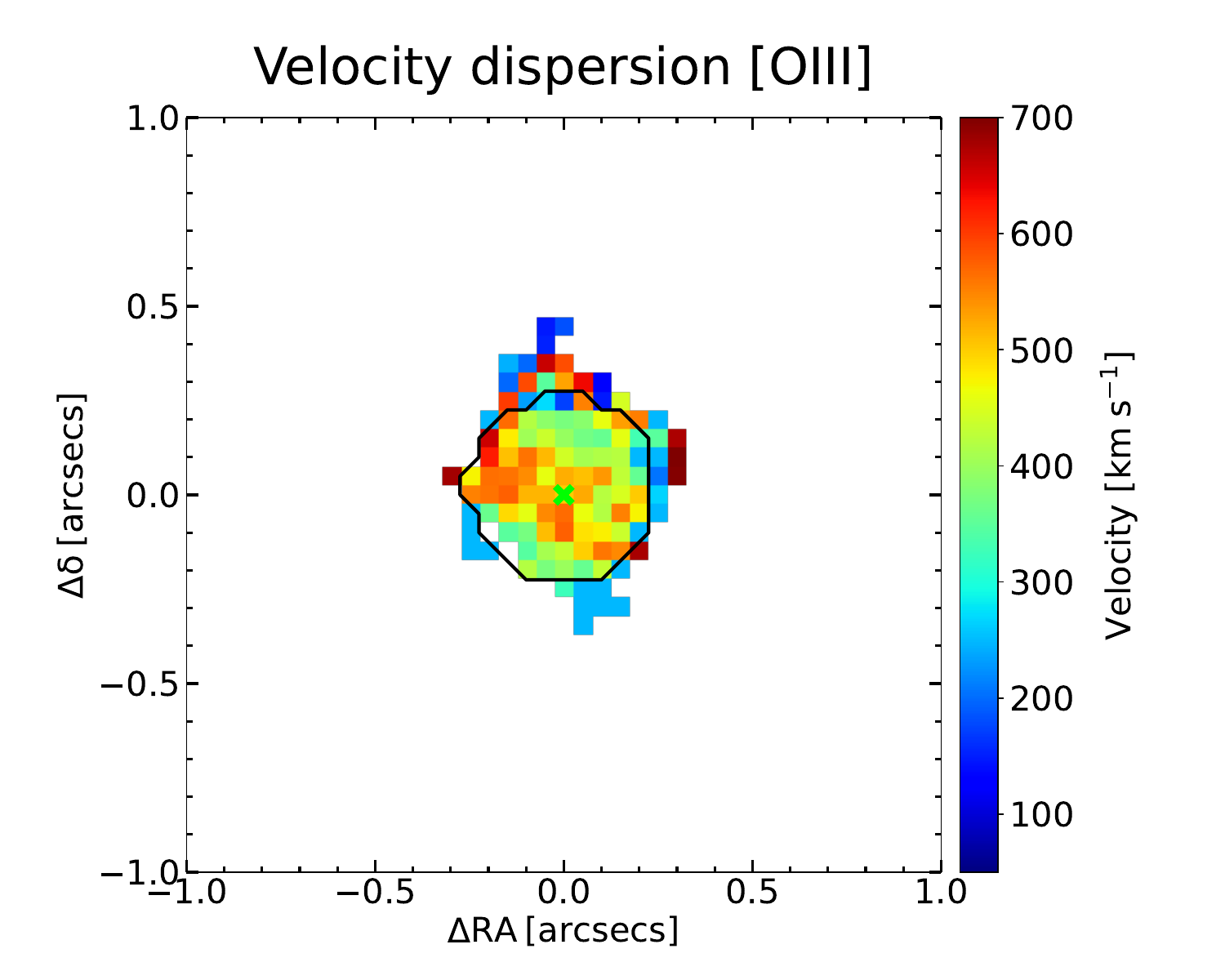}
    \includegraphics[width=0.245\linewidth,trim={1cm 0 1cm 0},clip]{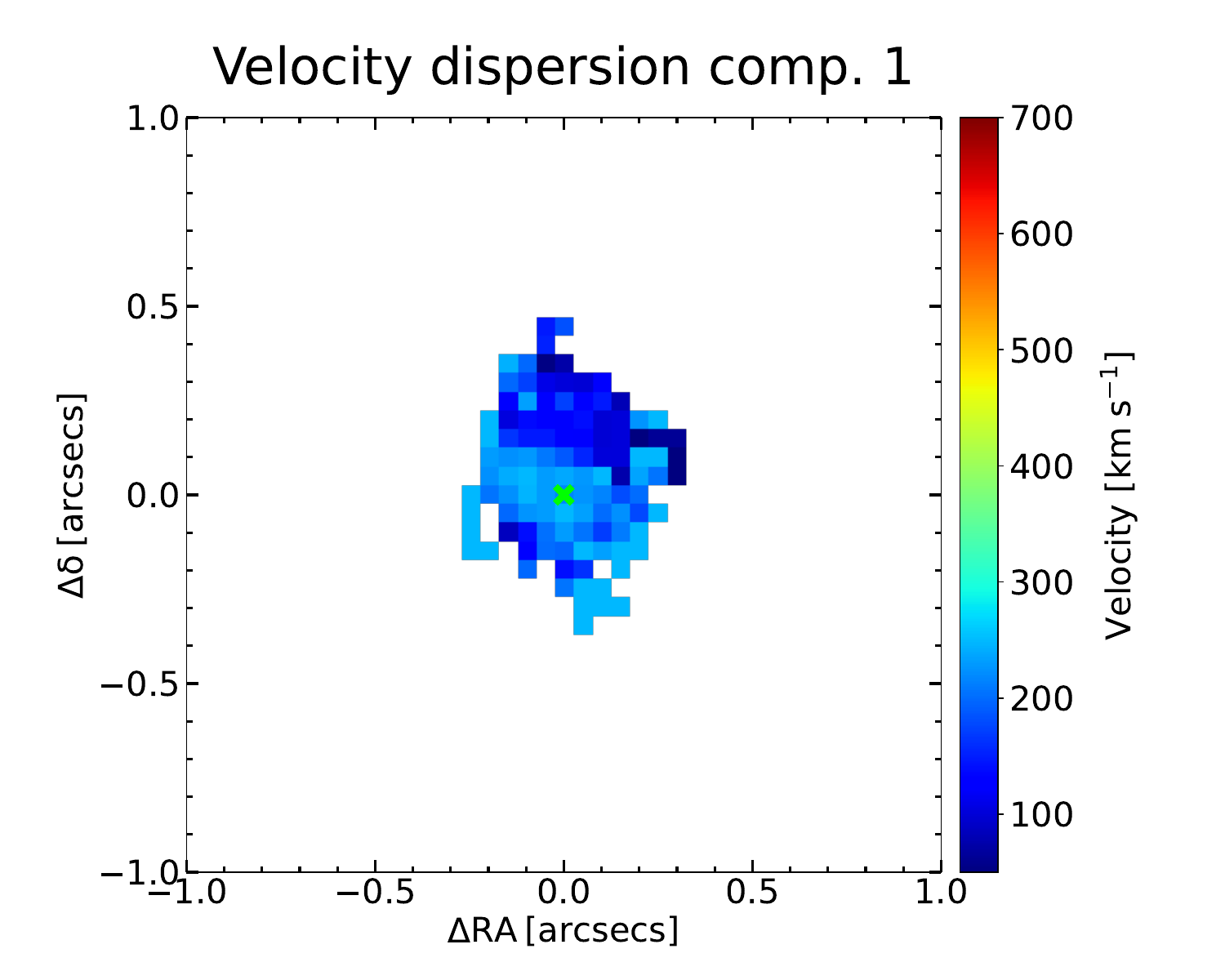}
    \includegraphics[width=0.245\linewidth,trim={1cm 0 1cm 0},clip]{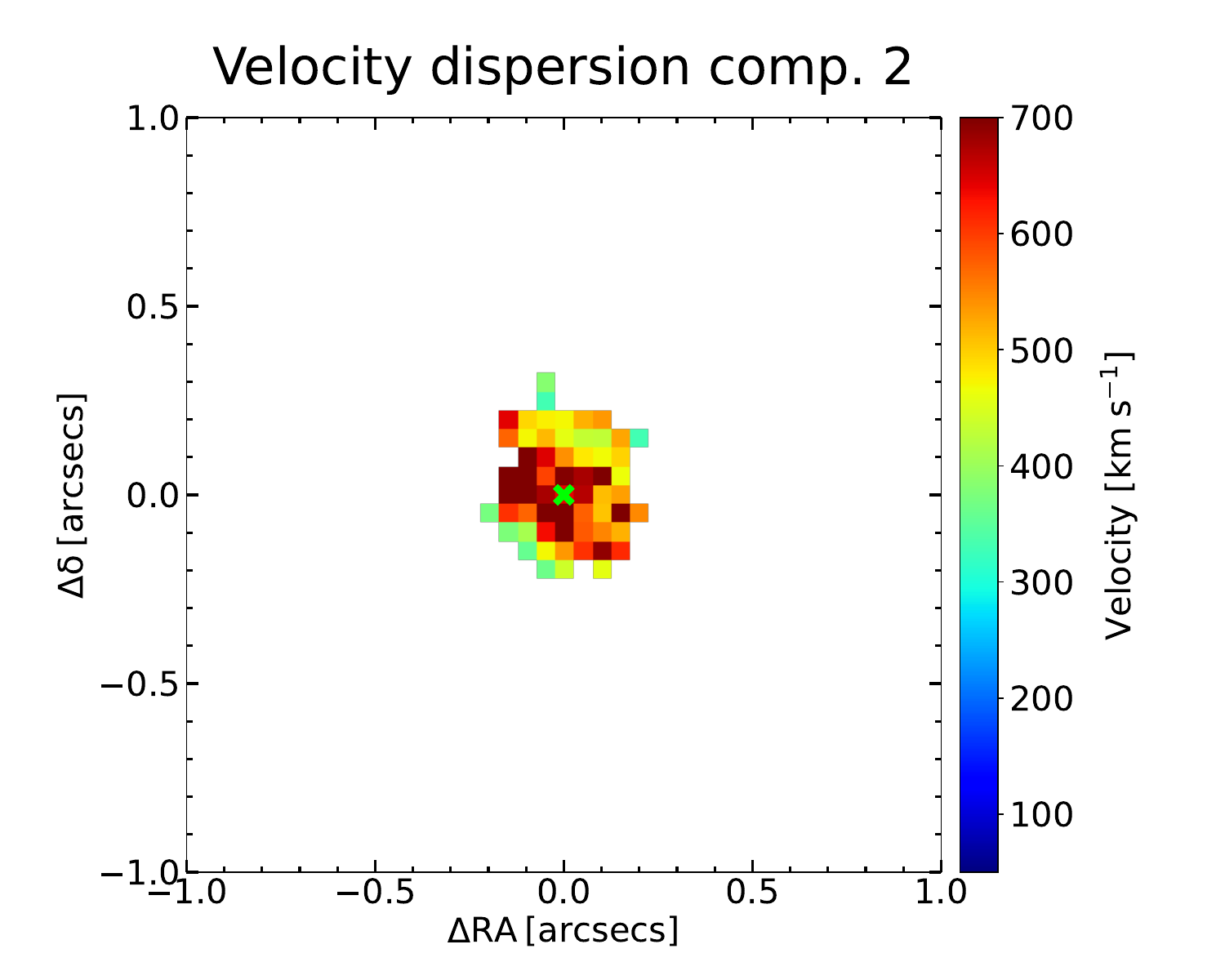}
    \caption{Same as Fig.~\ref{fig:gs133} for but GS~774.}
    \label{fig:gs774}
\end{figure*}

\begin{figure*}
    \centering
    \includegraphics[width=0.245\linewidth,trim={1cm 0 1cm 0},clip]{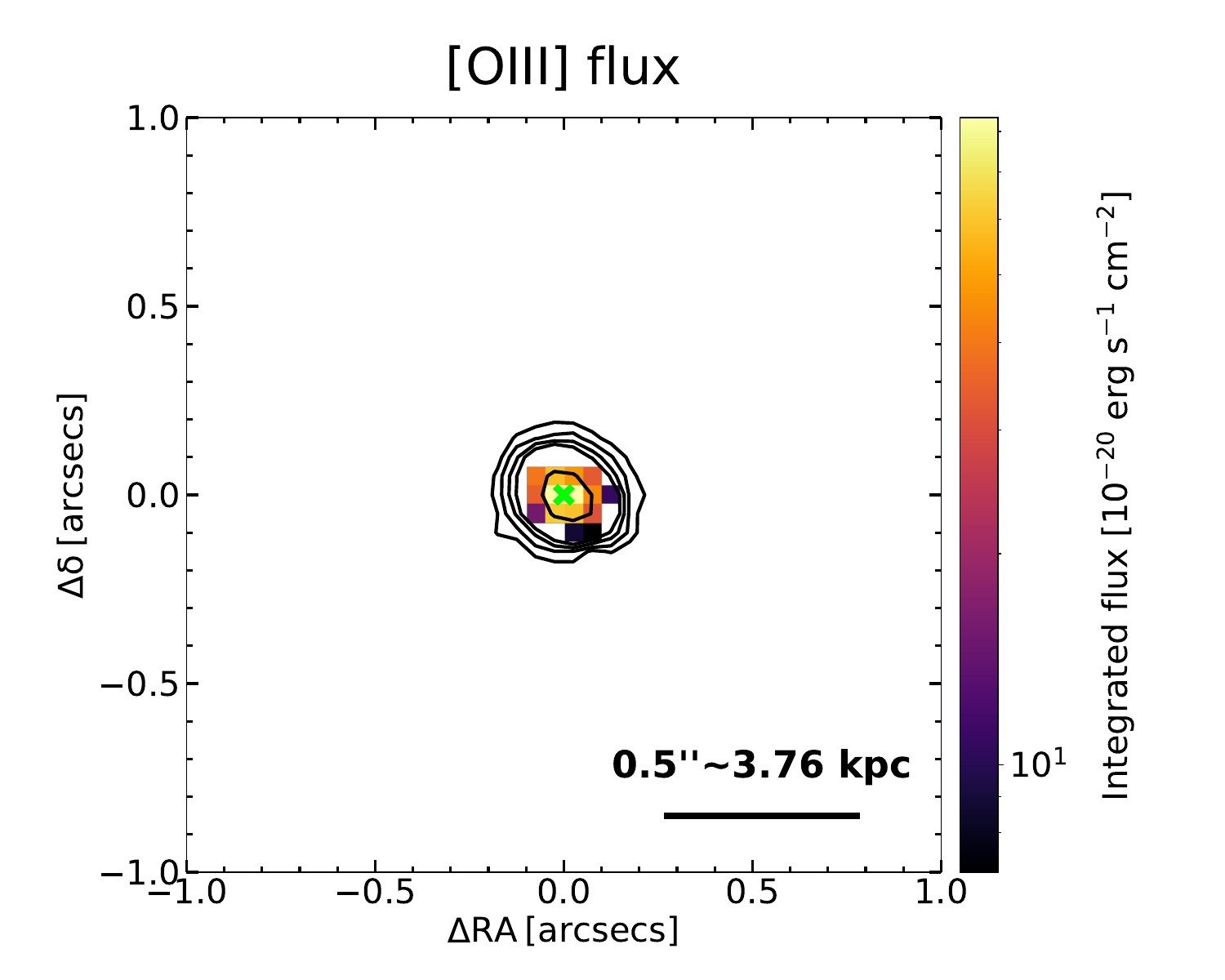}
    \includegraphics[width=0.245\linewidth,trim={1cm 0 1cm 0},clip]{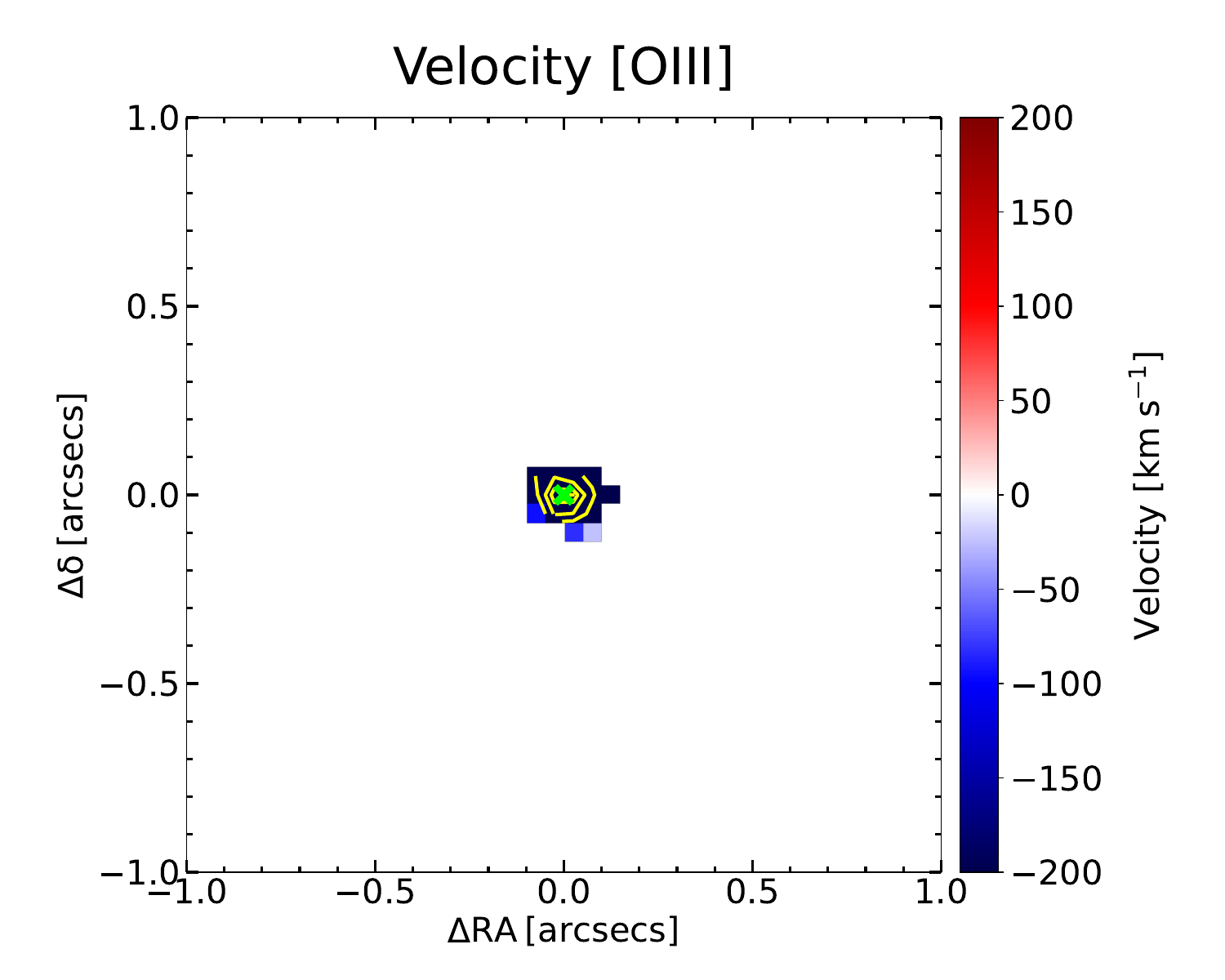}
    \includegraphics[width=0.245\linewidth,trim={1cm 0 1cm 0},clip]{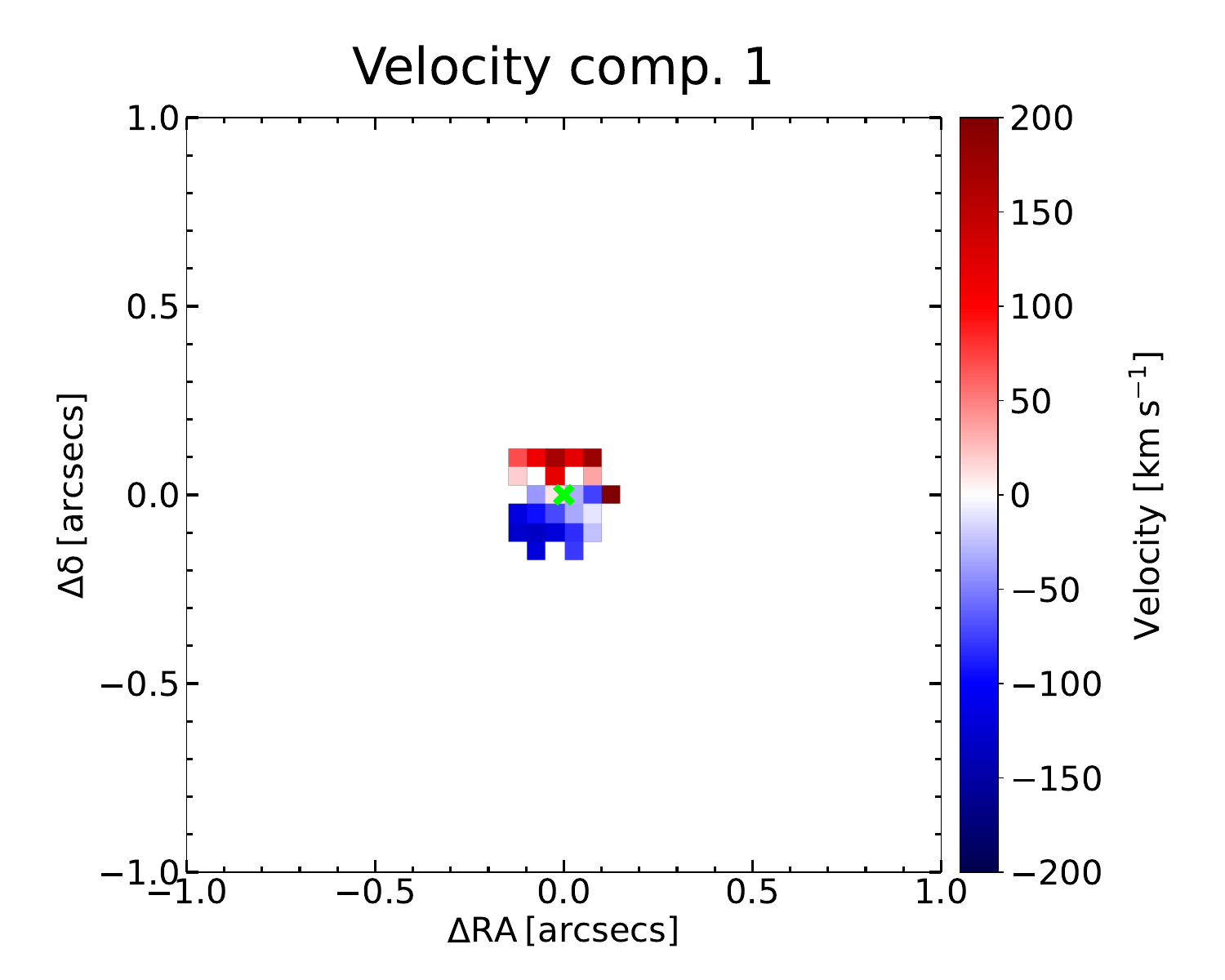}
    \includegraphics[width=0.245\linewidth,trim={1cm 0 1cm 0},clip]{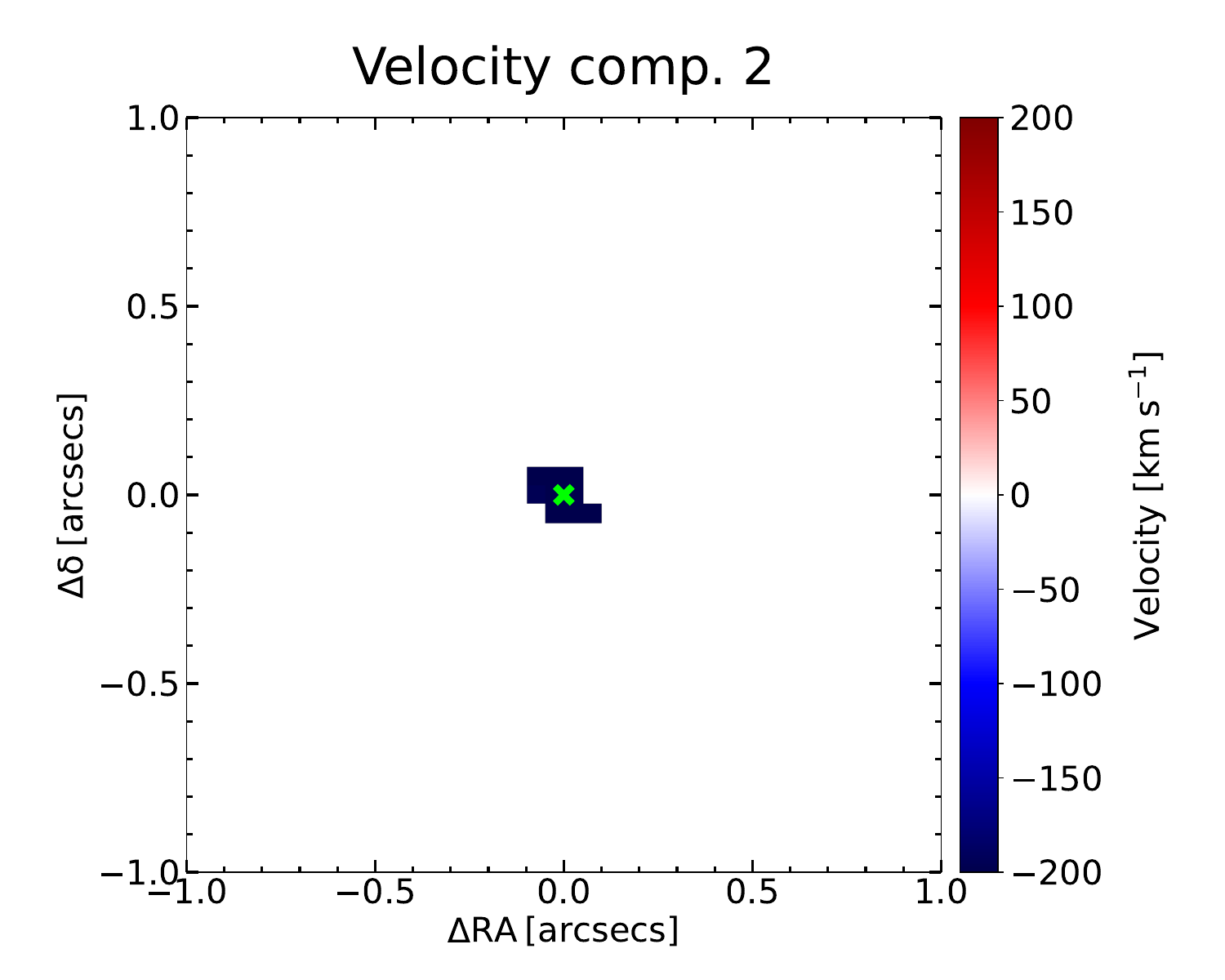}\\
    \includegraphics[width=0.245\linewidth,trim={1cm 0 1cm 0},clip]{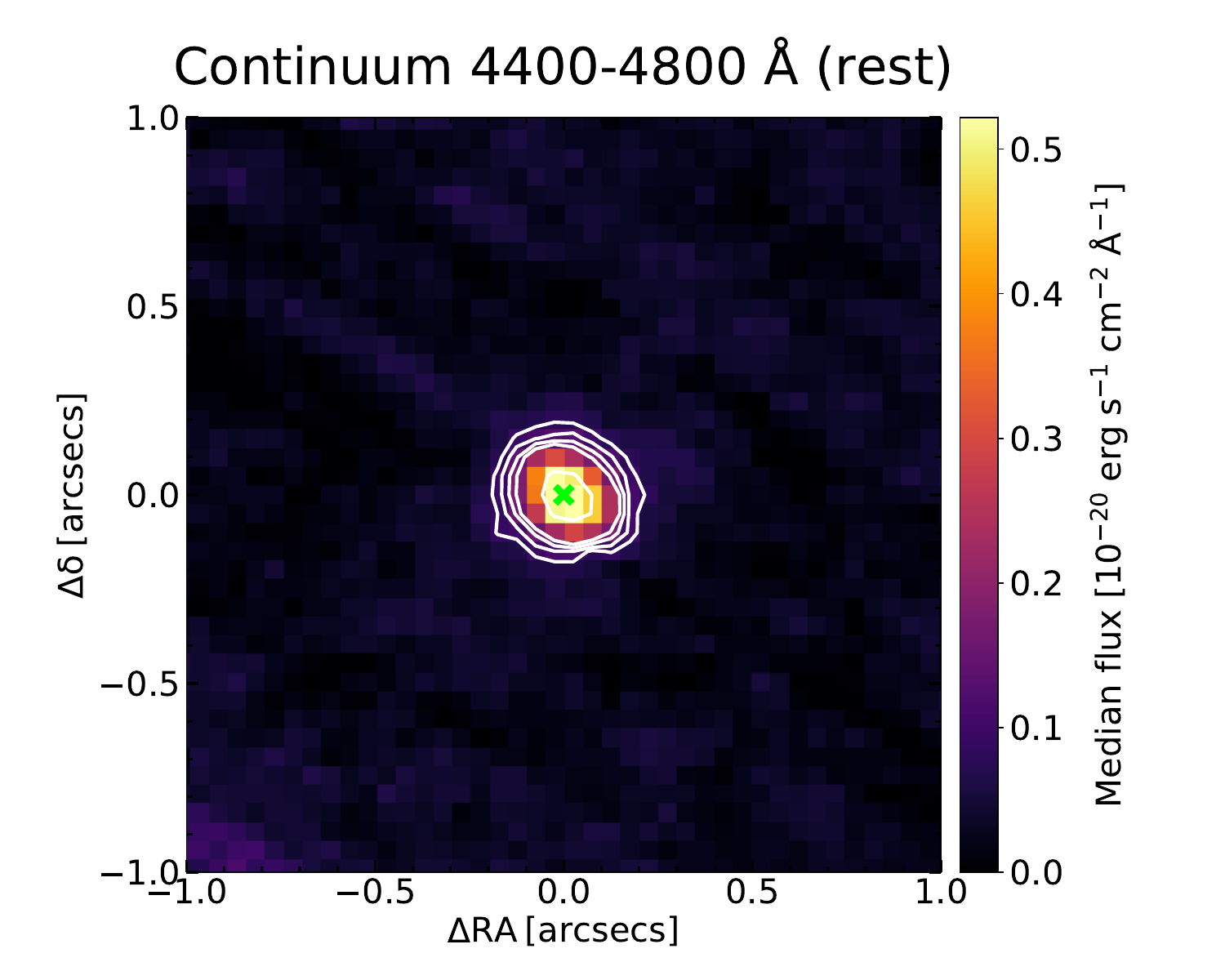}
    \includegraphics[width=0.245\linewidth,trim={1cm 0 1cm 0},clip]{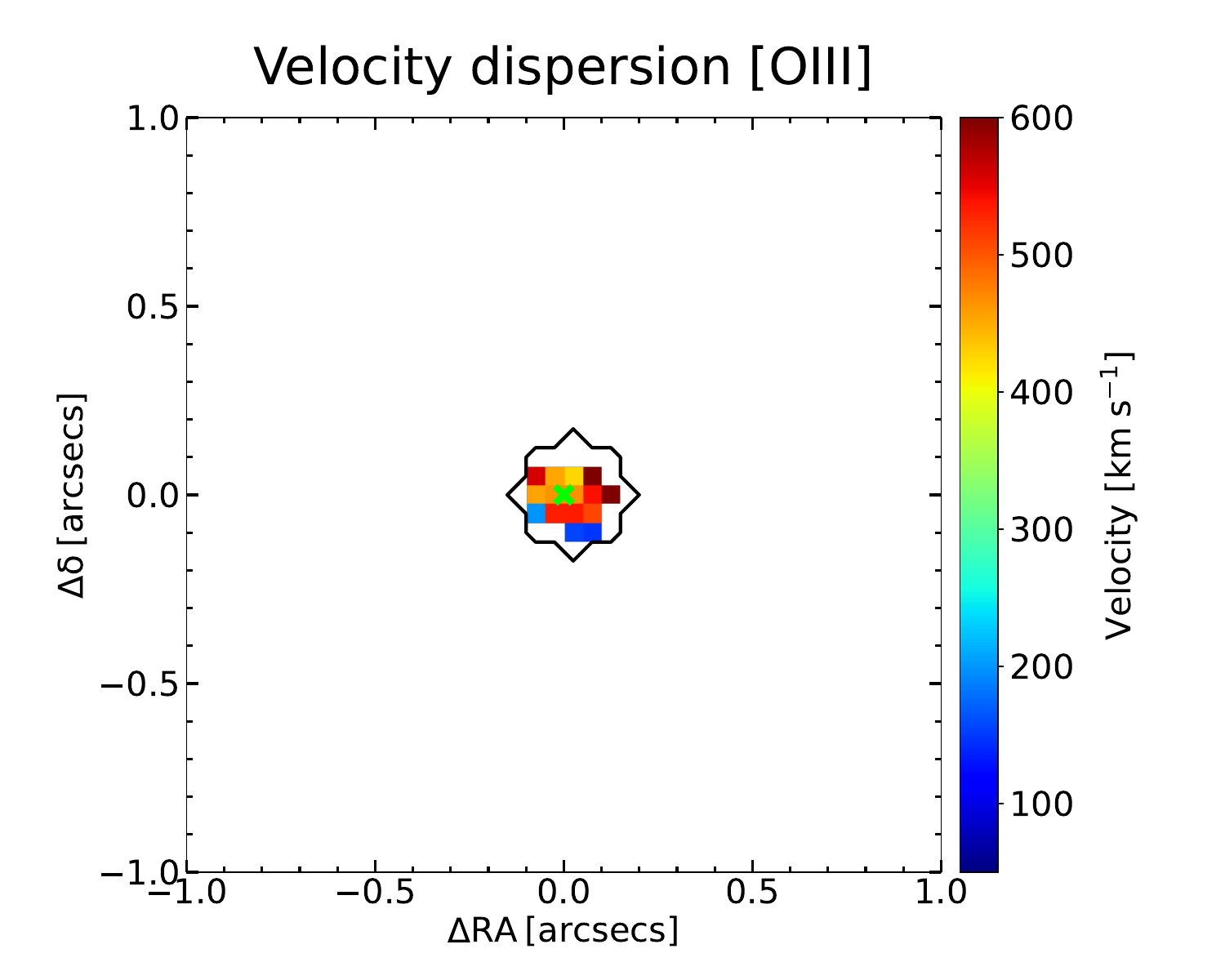}
    \includegraphics[width=0.245\linewidth,trim={1cm 0 1cm 0},clip]{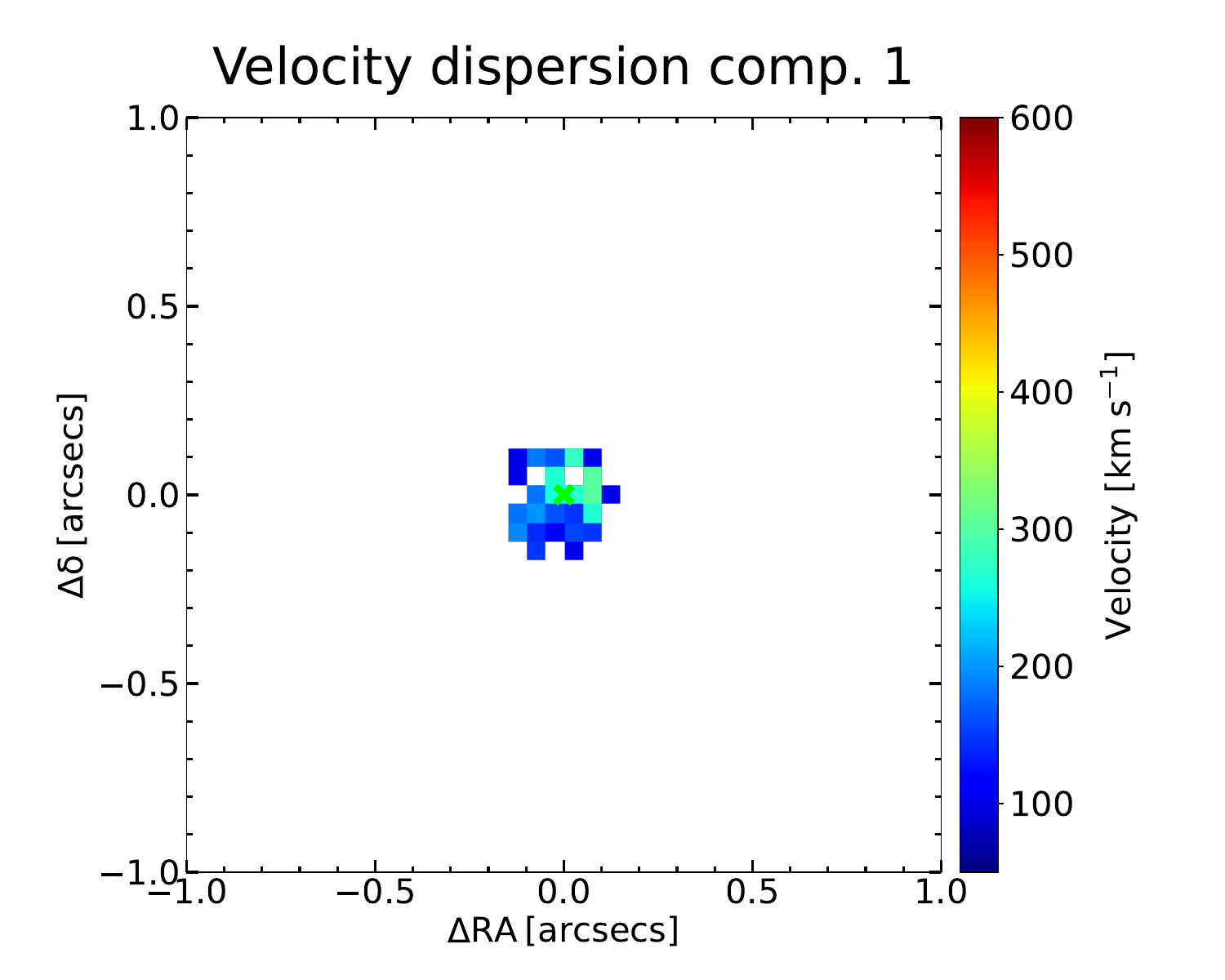}
    \includegraphics[width=0.245\linewidth,trim={1cm 0 1cm 0},clip]{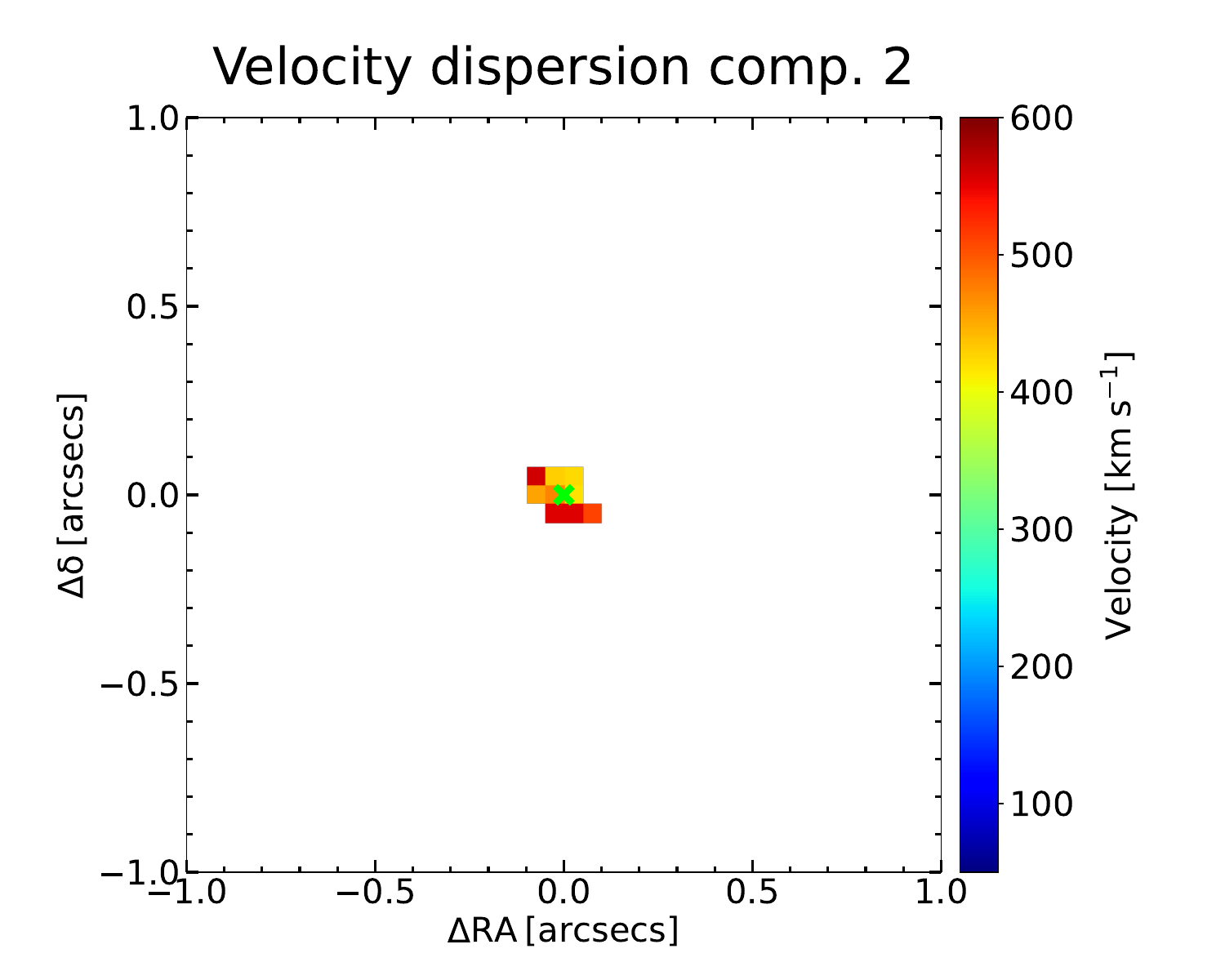}
    \caption{Same as Fig.~\ref{fig:gs133} for but GS~811.}
    \label{fig:gs811}
\end{figure*}

\begin{figure*}
    \centering
    \includegraphics[width=0.245\linewidth,trim={1cm 0 1cm 0},clip]{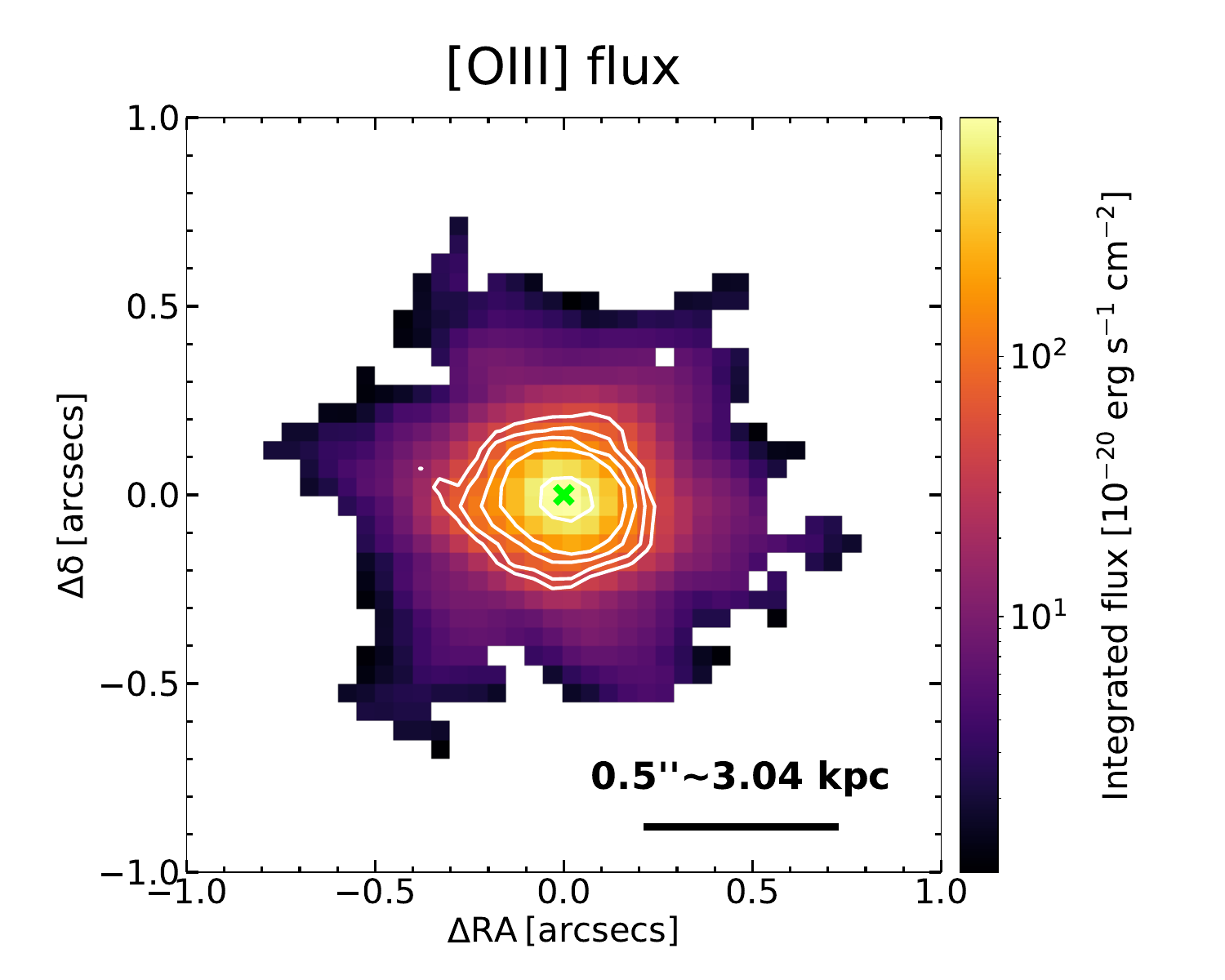}
    \includegraphics[width=0.245\linewidth,trim={1cm 0 1cm 0},clip]{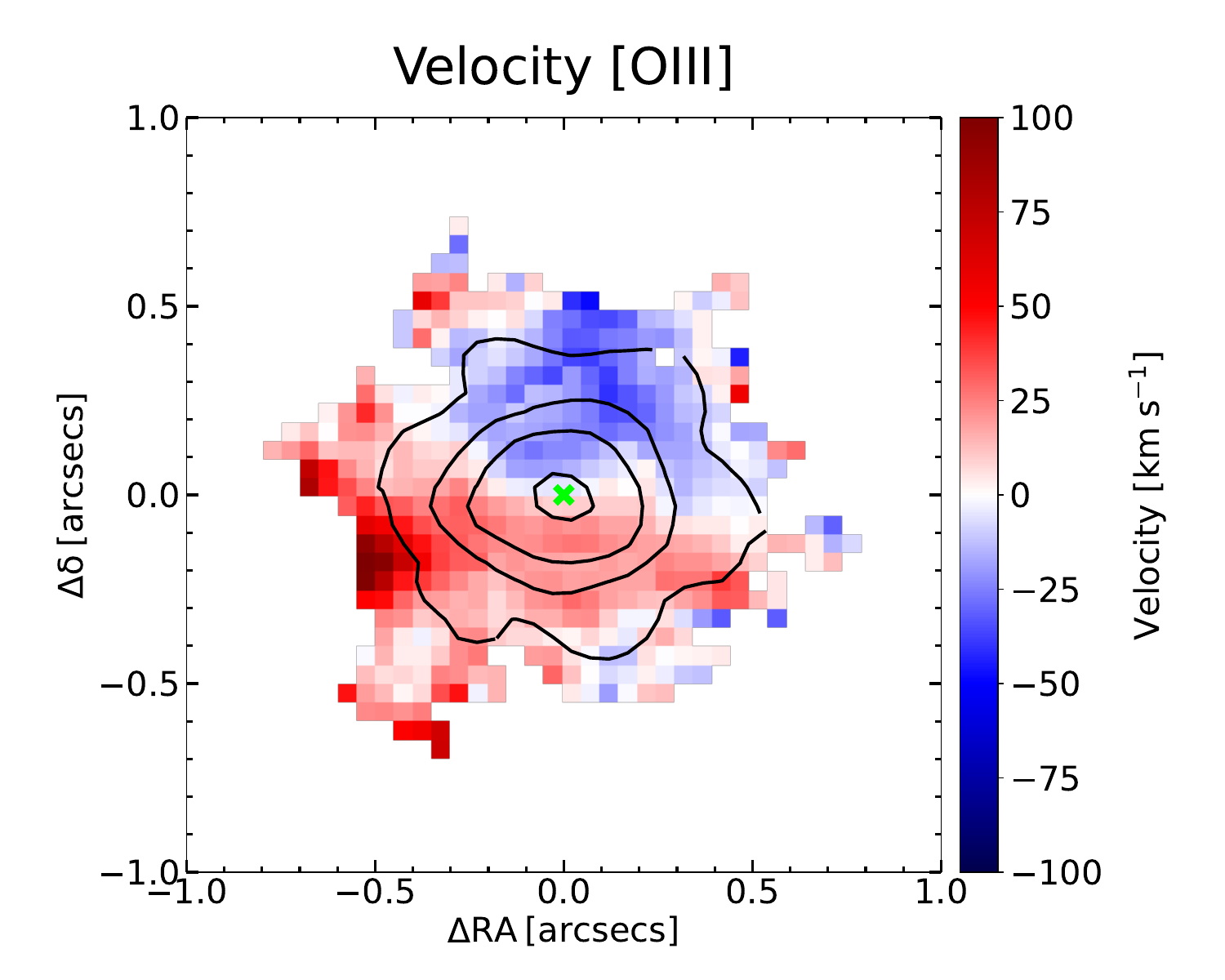}
    \includegraphics[width=0.245\linewidth,trim={1cm 0 1cm 0},clip]{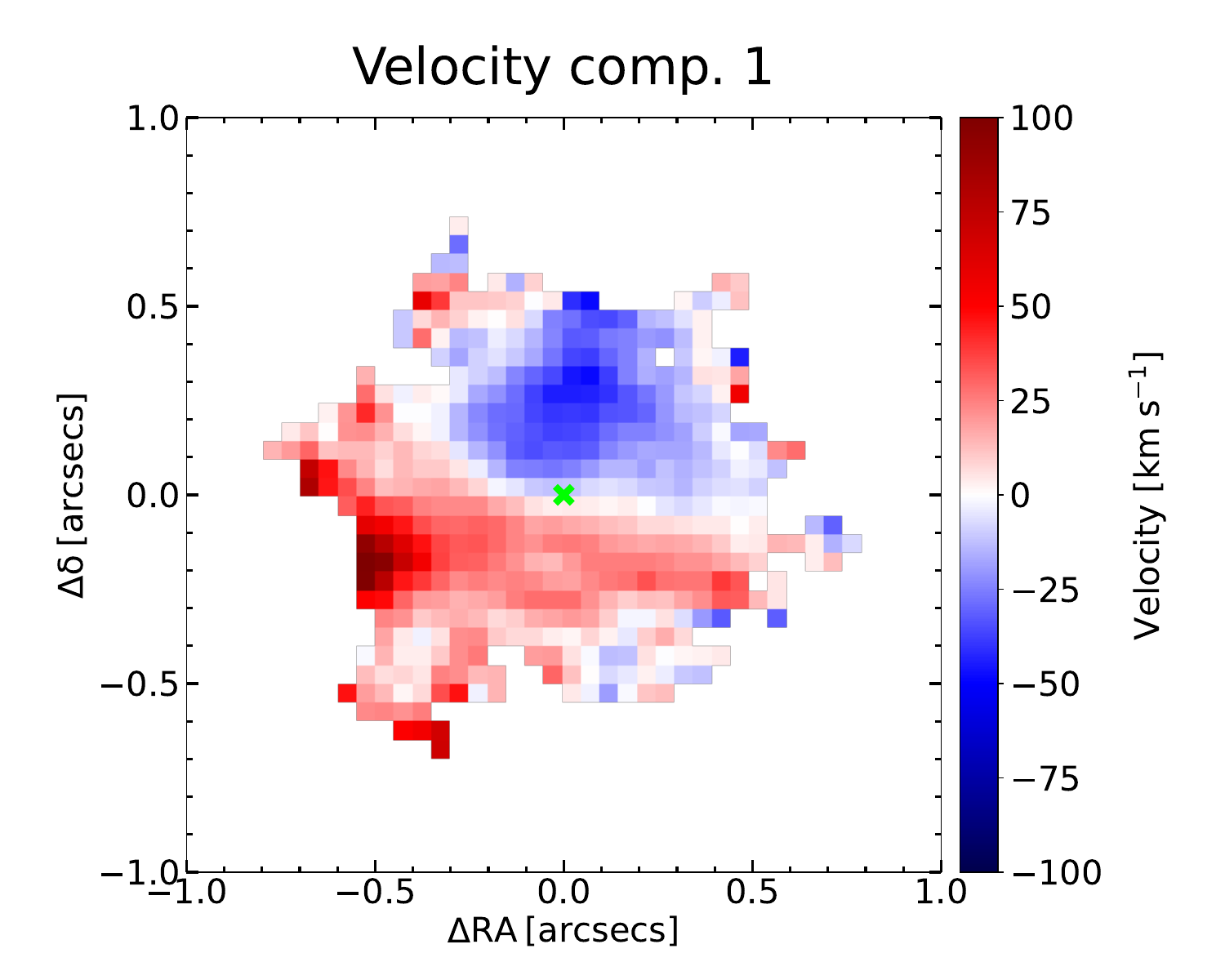}
    \includegraphics[width=0.245\linewidth,trim={1cm 0 1cm 0},clip]{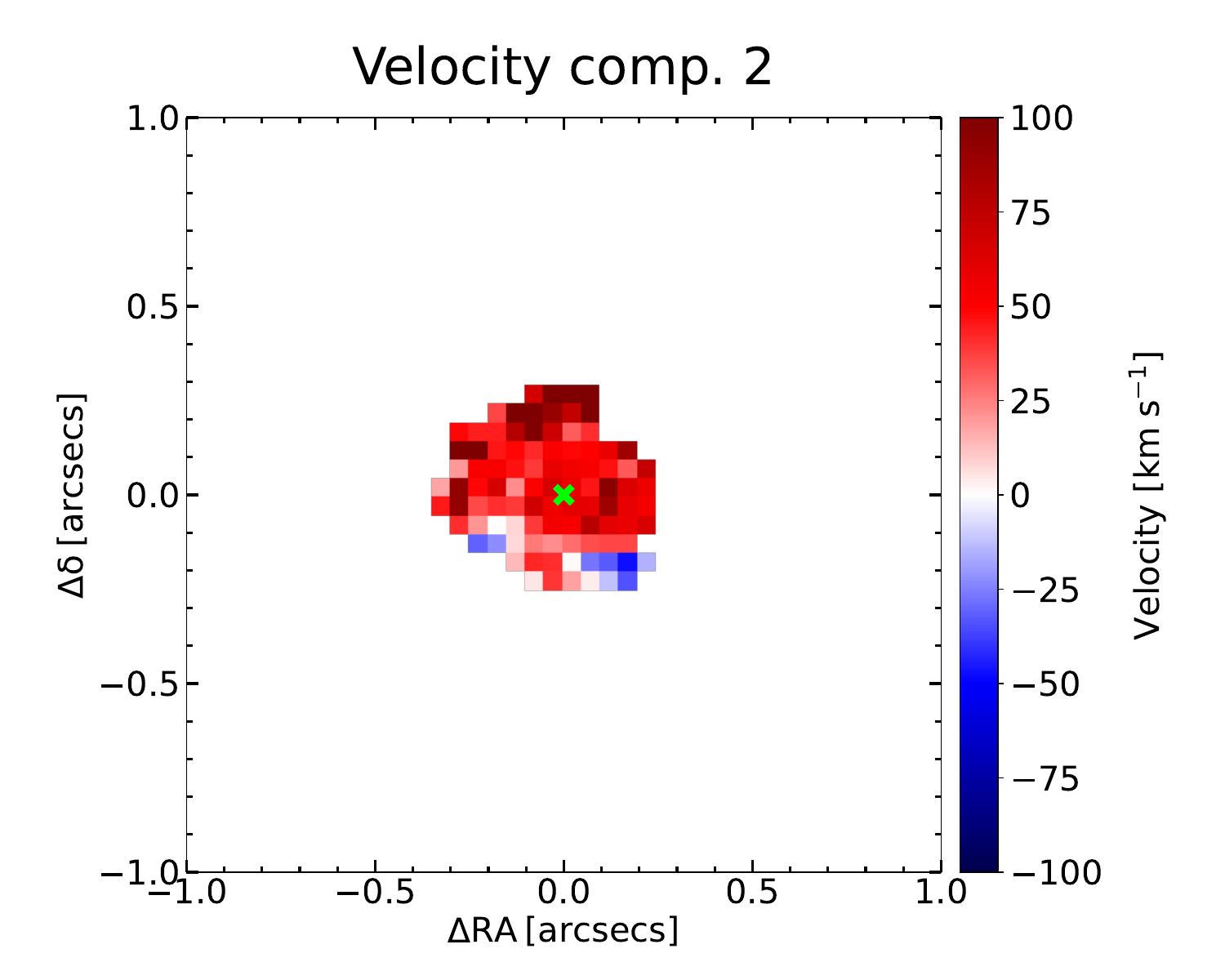}\\
    \includegraphics[width=0.245\linewidth,trim={1cm 0 1cm 0},clip]{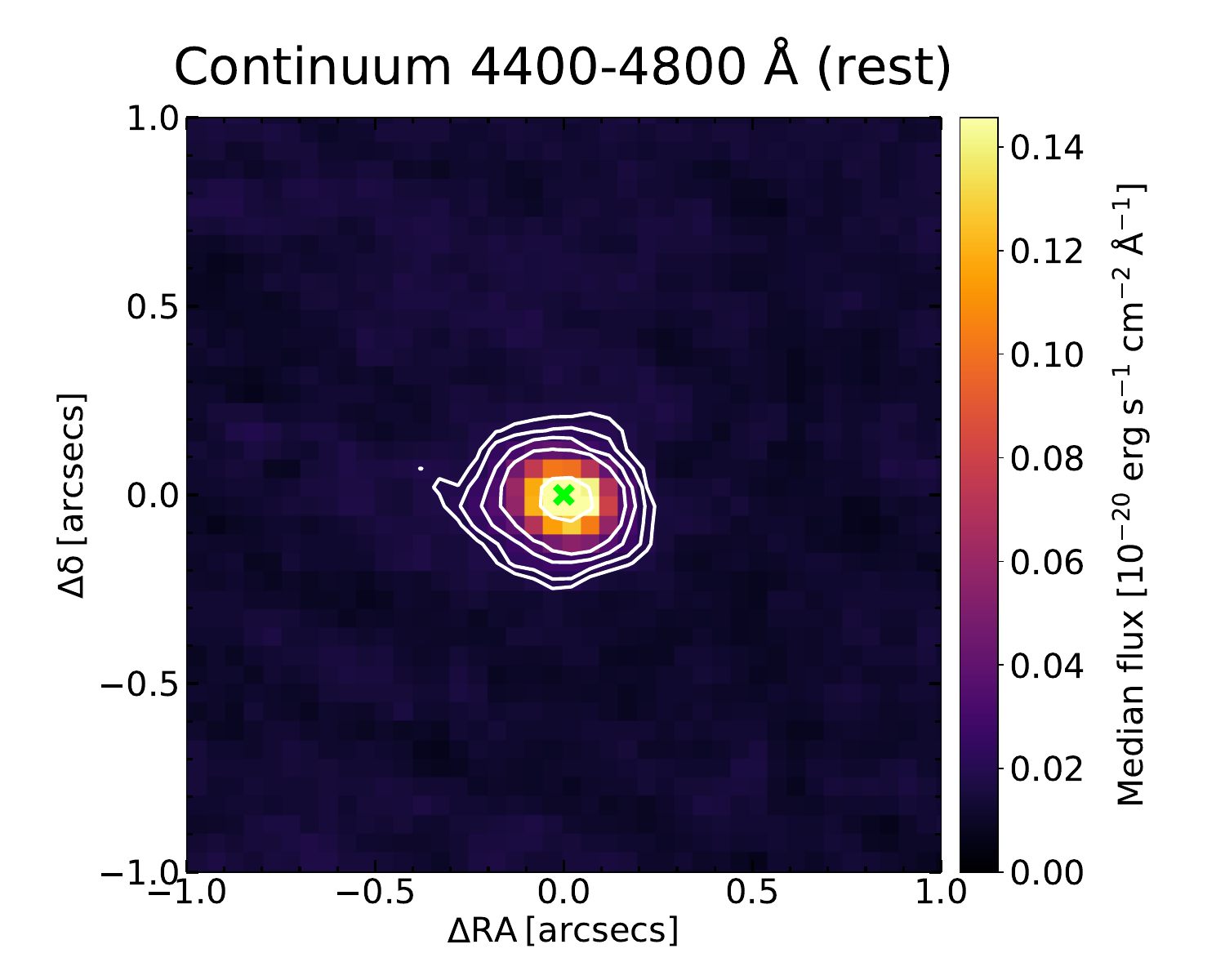}
    \includegraphics[width=0.245\linewidth,trim={1cm 0 1cm 0},clip]{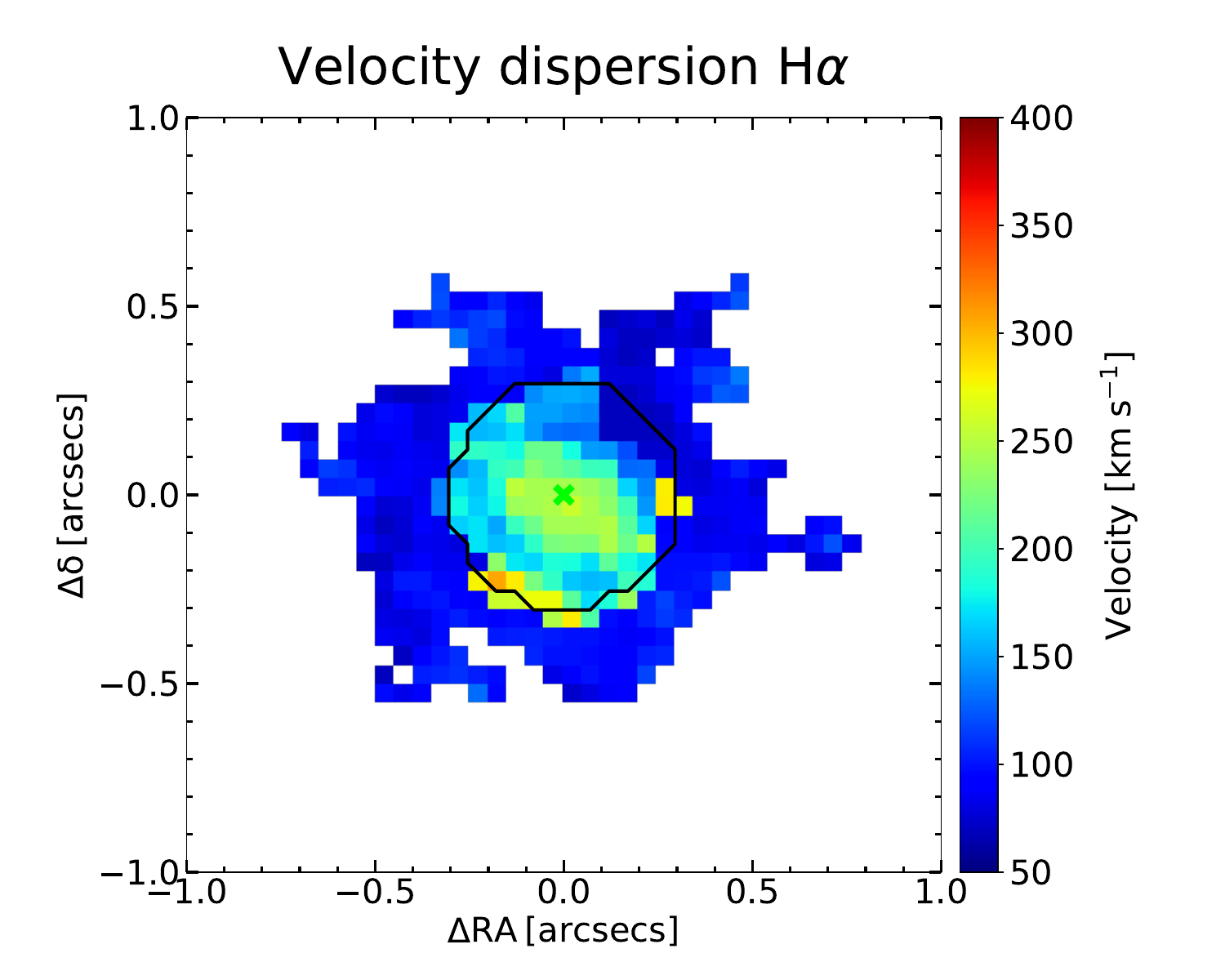}
    \includegraphics[width=0.245\linewidth,trim={1cm 0 1cm 0},clip]{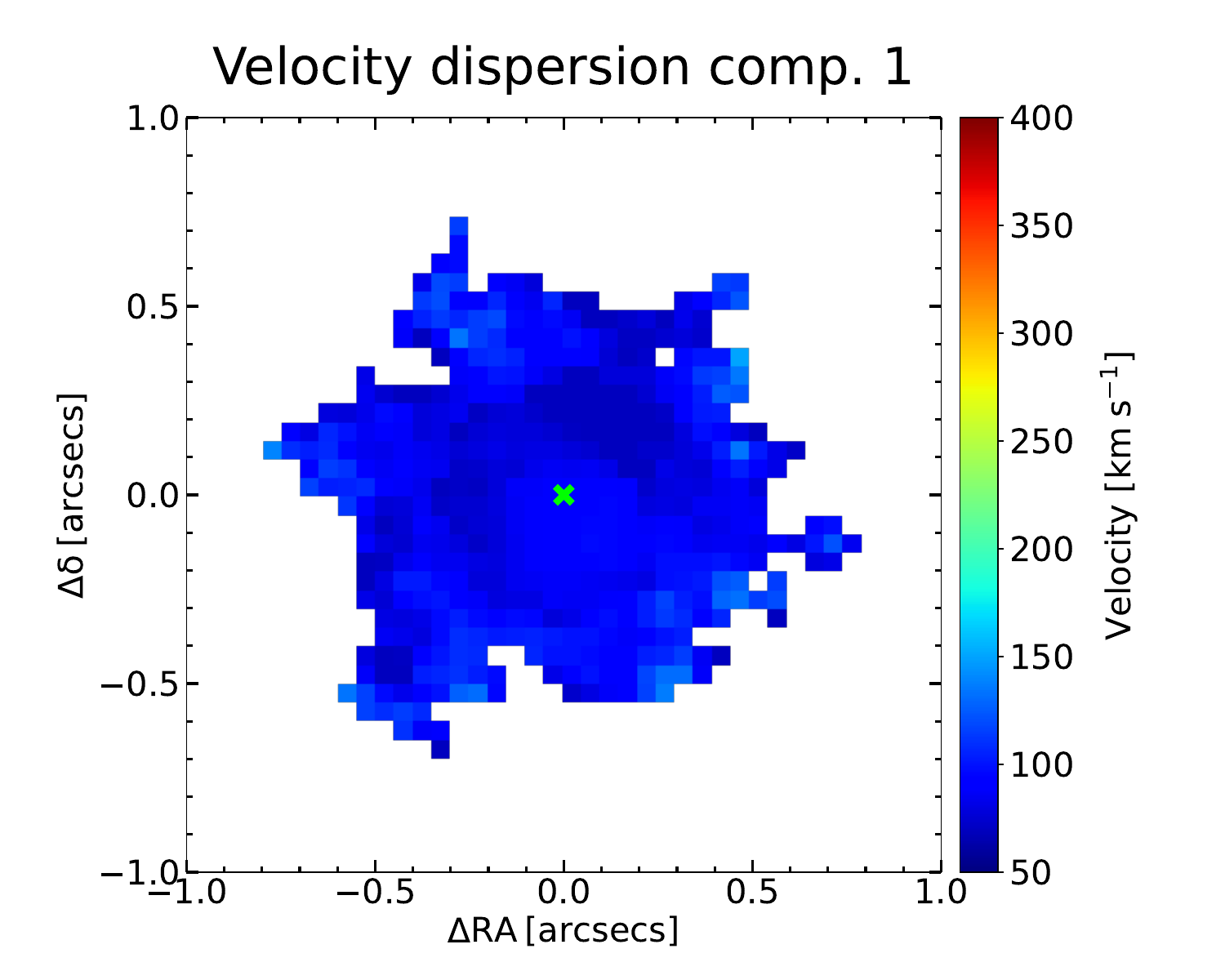}
    \includegraphics[width=0.245\linewidth,trim={1cm 0 1cm 0},clip]{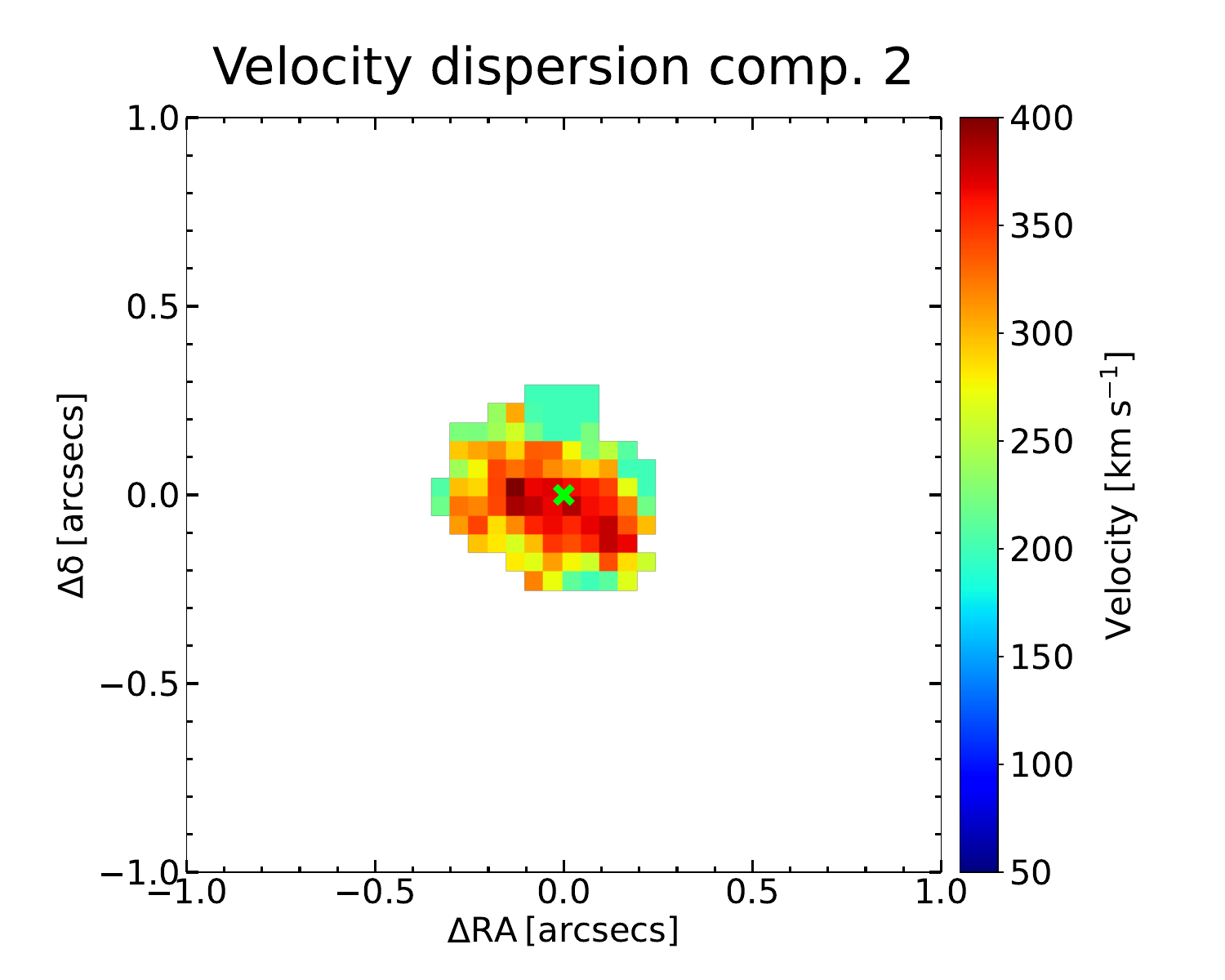}
    \caption{Same as Fig.~\ref{fig:gs133} but for GS~3073. We show the flux and velocity maps of \oiii, being the brightest line, but we report the velocity dispersion of \ha instead, because the broad emission tracing the outflow is more evident in the profile of \ha than of \oiii in this target.}
    \label{fig:gs3073}
\end{figure*}

\begin{figure*}
    \centering
    \includegraphics[width=0.245\linewidth,trim={1cm 0 1cm 0},clip]{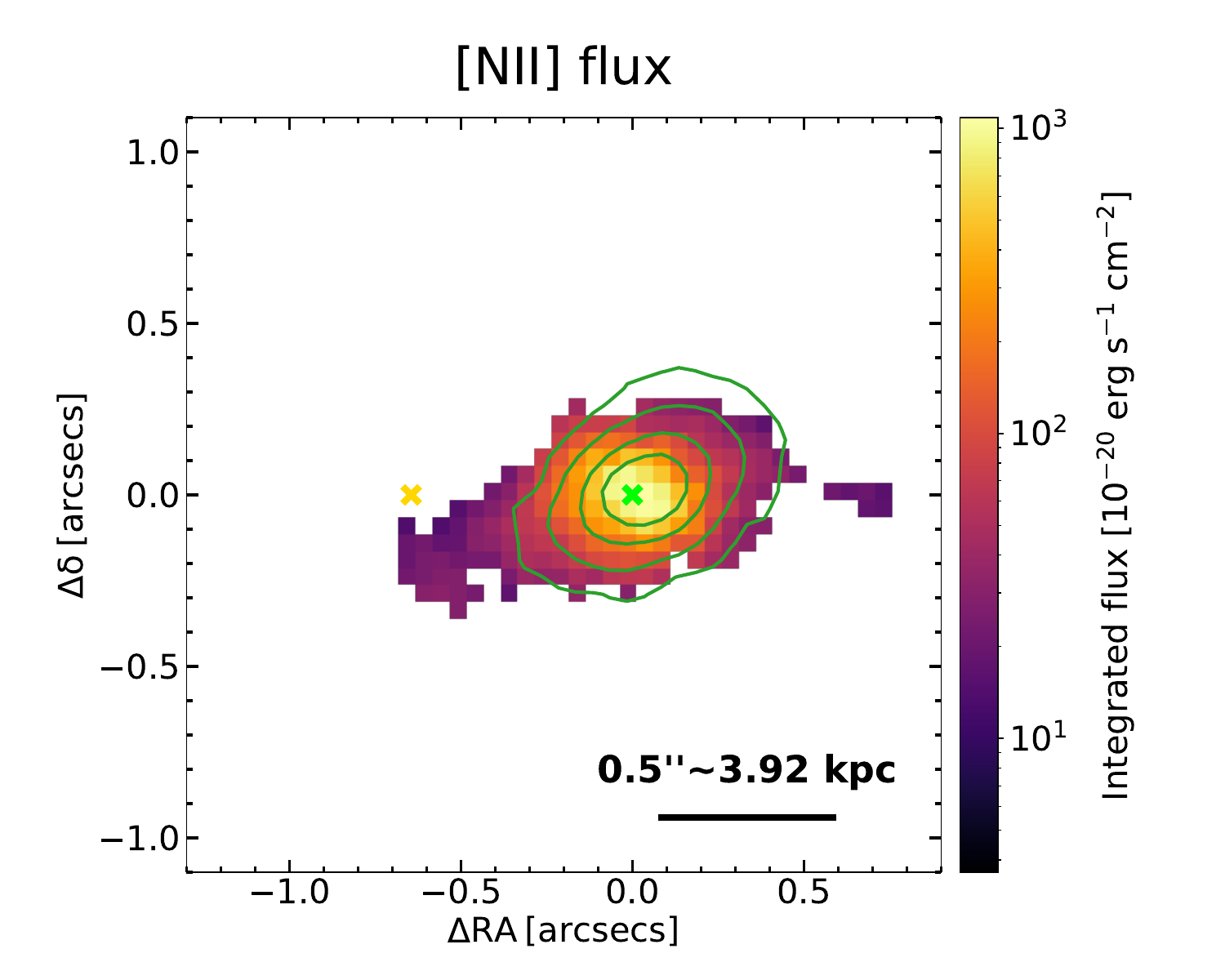}
    \includegraphics[width=0.245\linewidth,trim={1cm 0 1cm 0},clip]{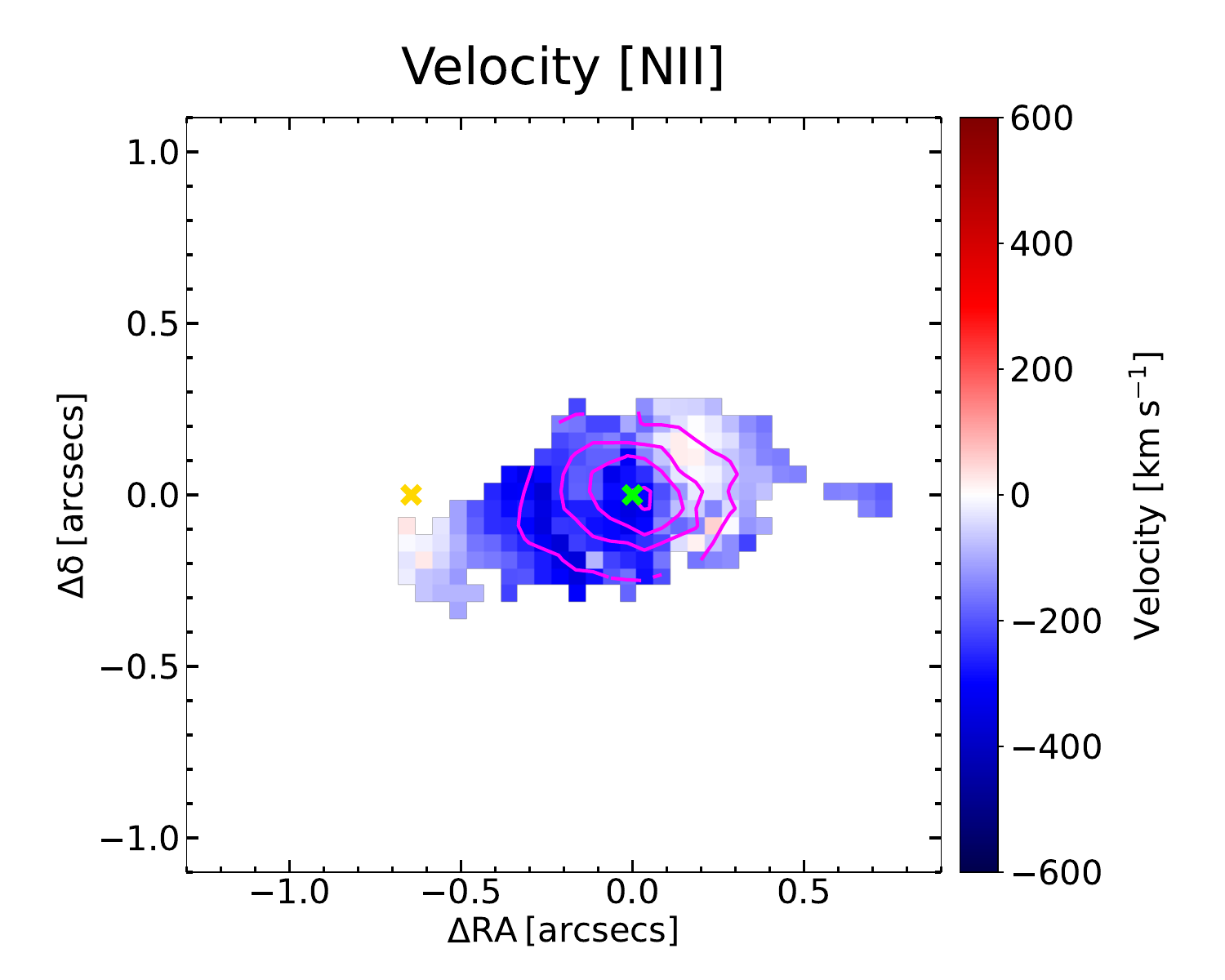}
    \includegraphics[width=0.245\linewidth,trim={1cm 0 1cm 0},clip]{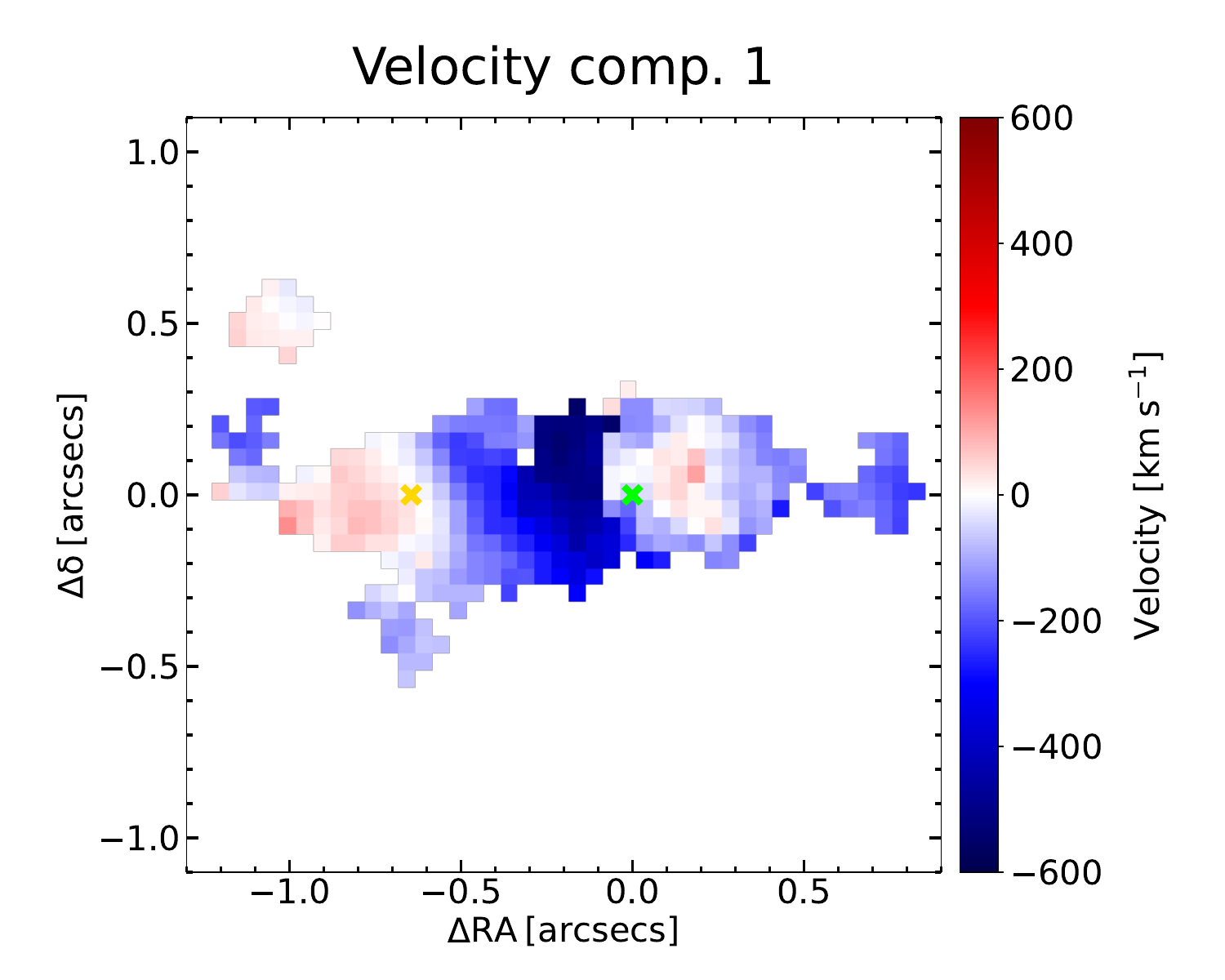}
    \includegraphics[width=0.245\linewidth,trim={1cm 0 1cm 0},clip]{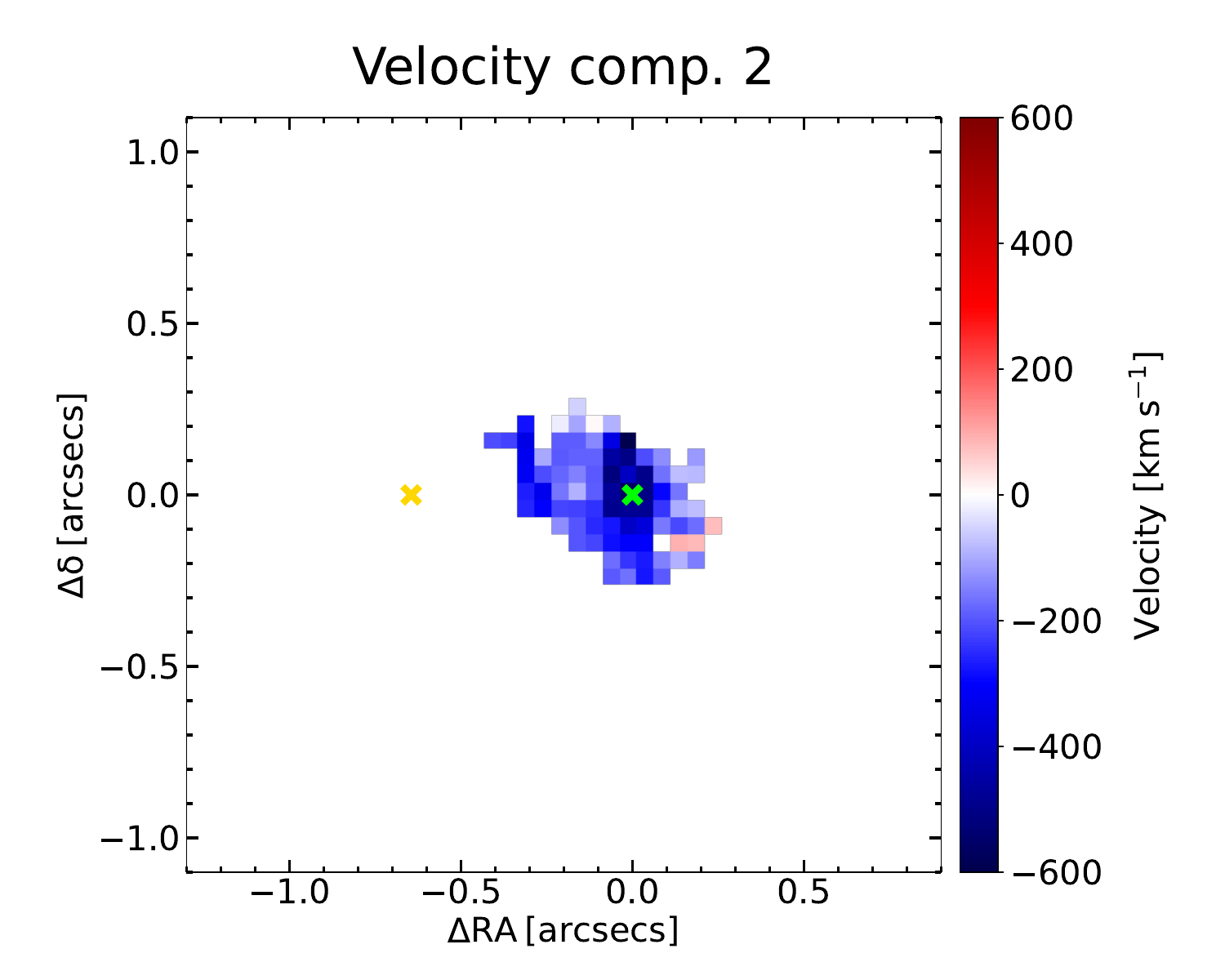}\\
    \includegraphics[width=0.245\linewidth,trim={1cm 0 1cm 0},clip]{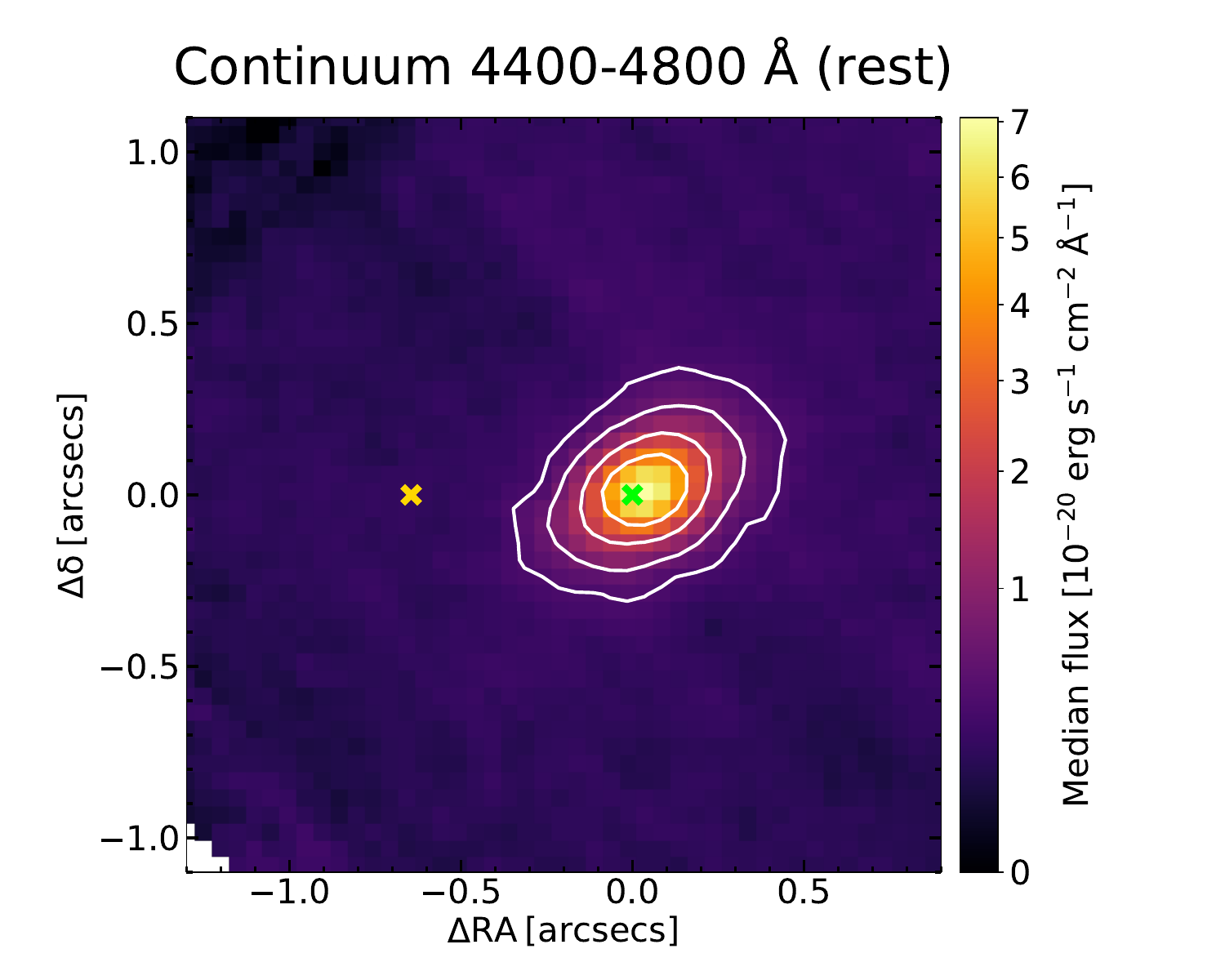}
    \includegraphics[width=0.245\linewidth,trim={1cm 0 1cm 0},clip]{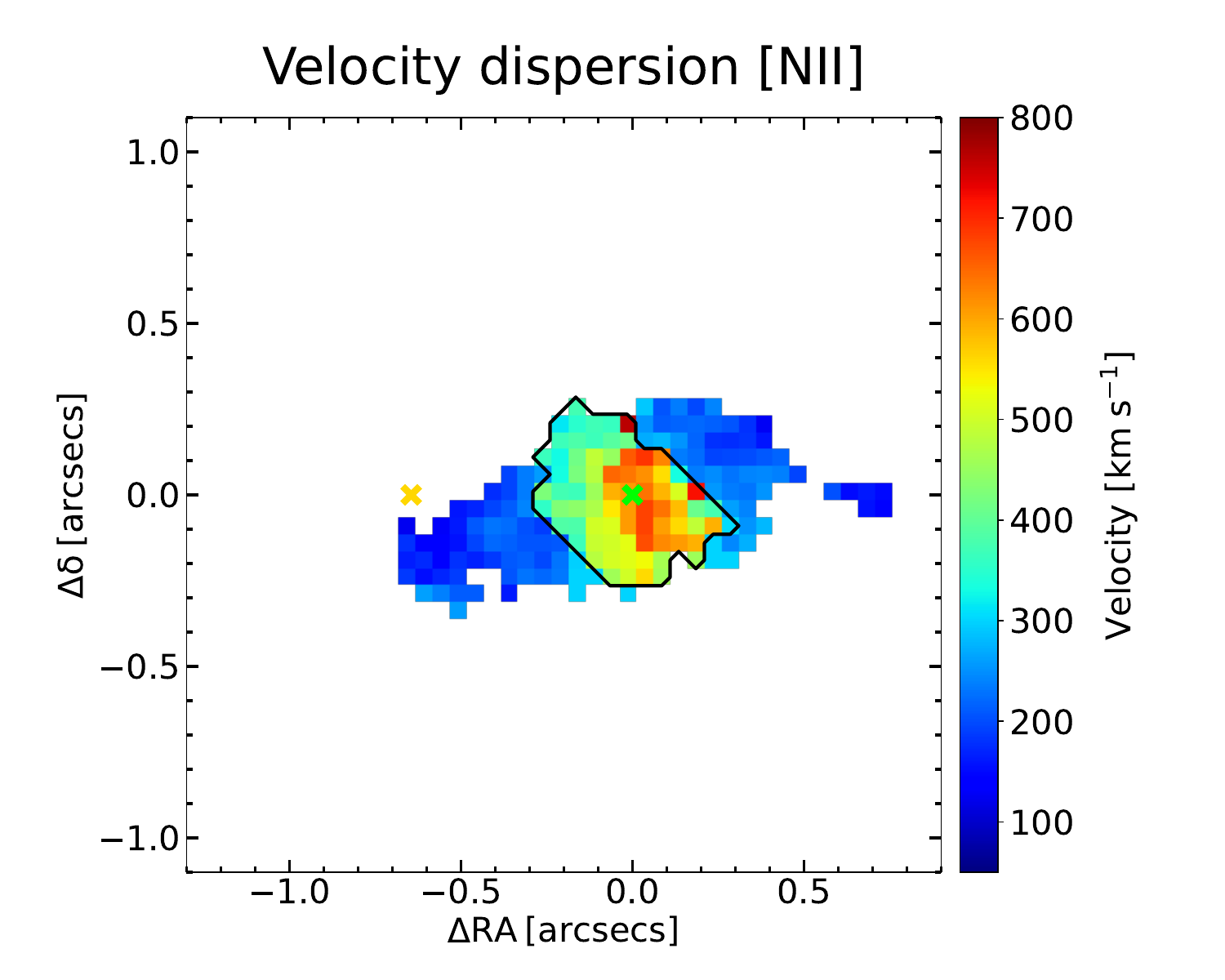}
    \includegraphics[width=0.245\linewidth,trim={1cm 0 1cm 0},clip]{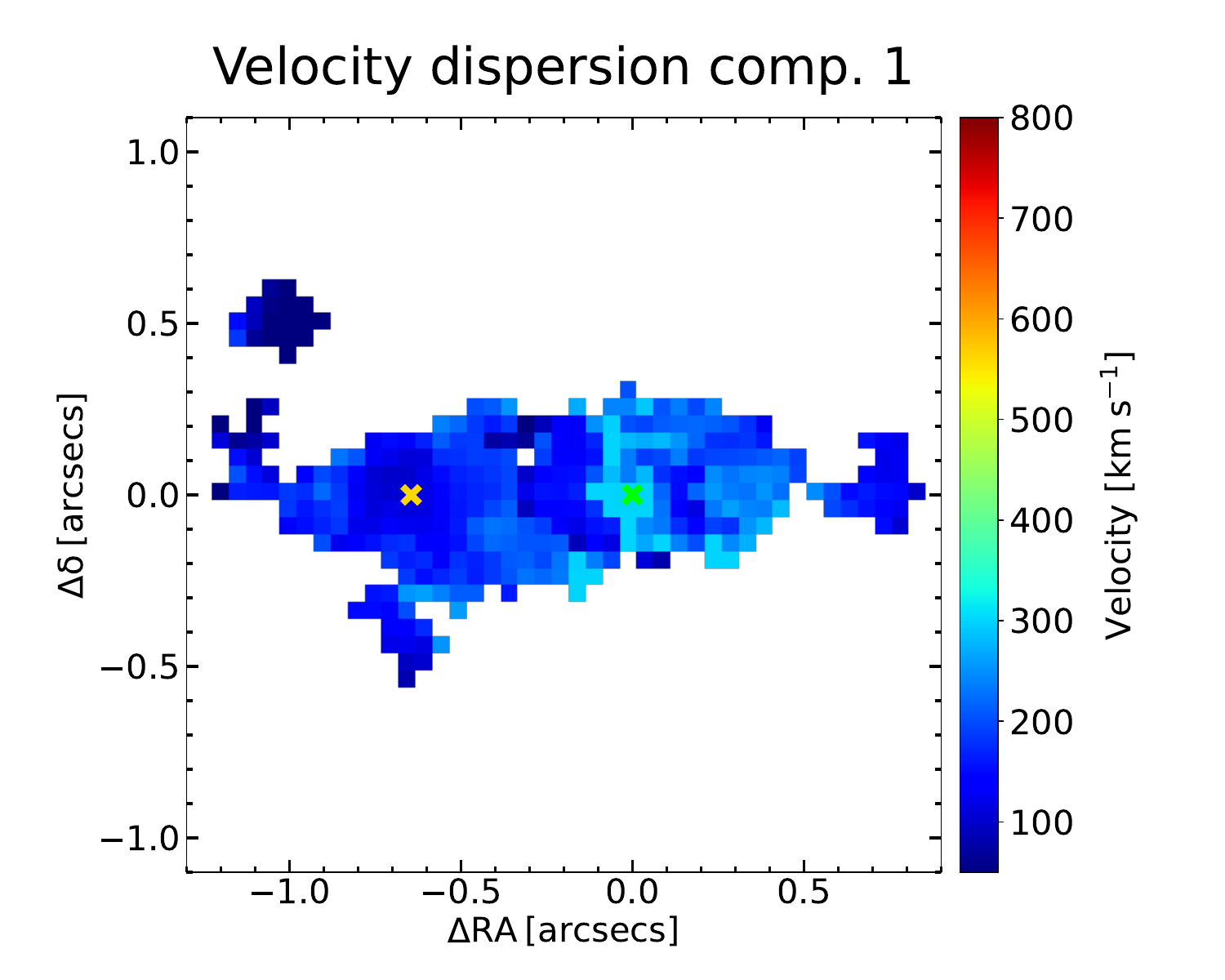}
    \includegraphics[width=0.245\linewidth,trim={1cm 0 1cm 0},clip]{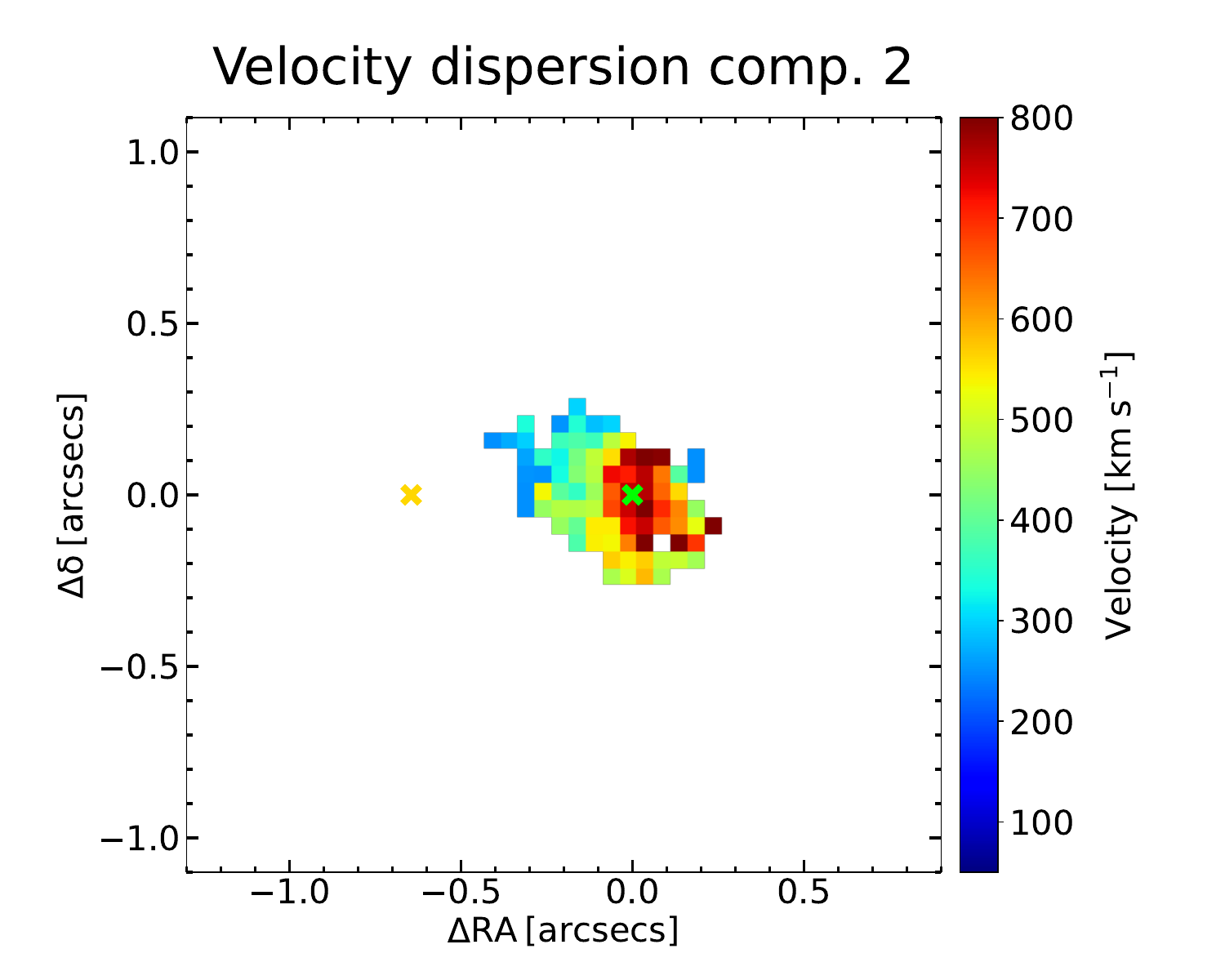}\\
    \includegraphics[width=0.245\linewidth,trim={1cm 0 1cm 0},clip]{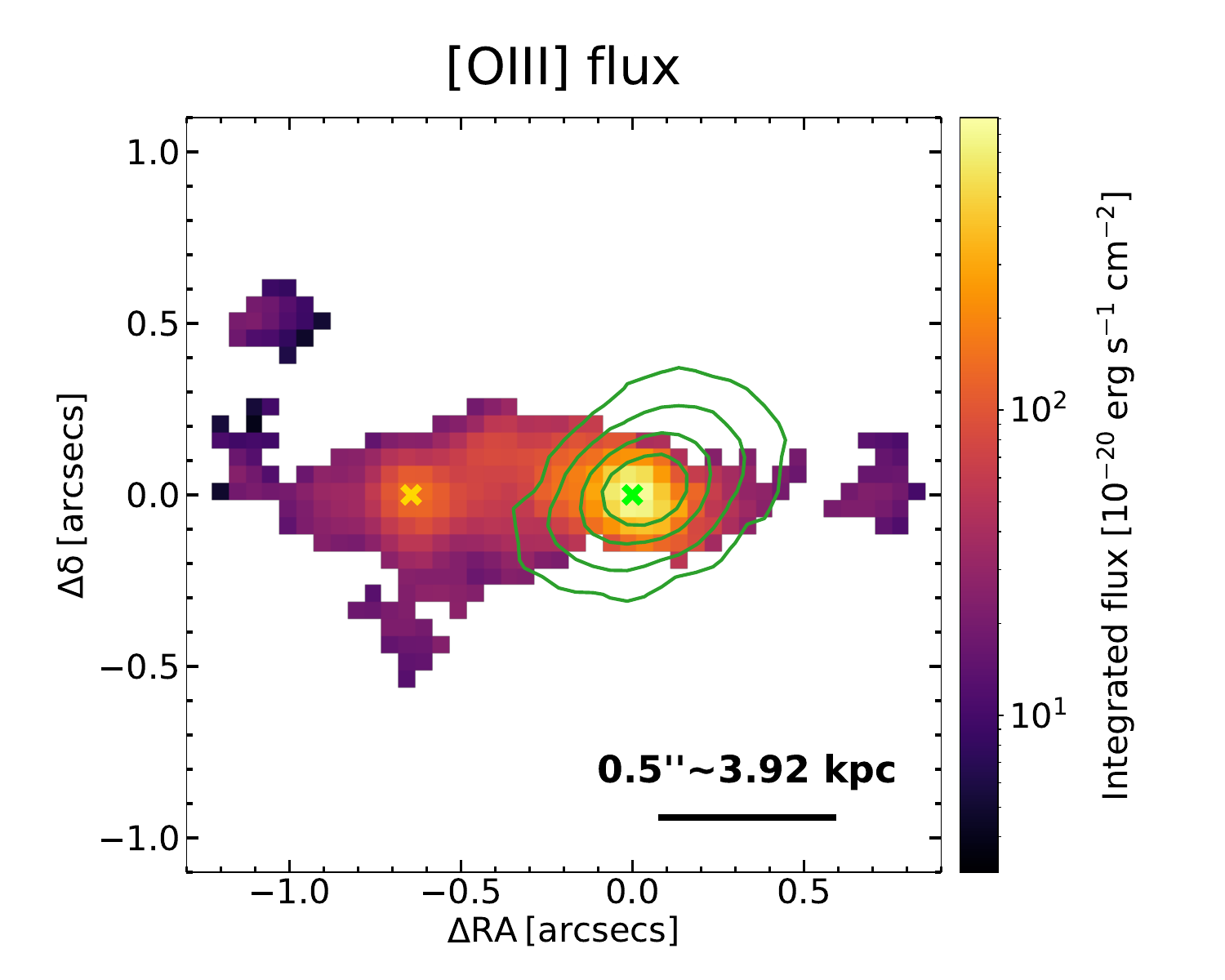}
    \includegraphics[width=0.245\linewidth,trim={1cm 0 1cm 0},clip]{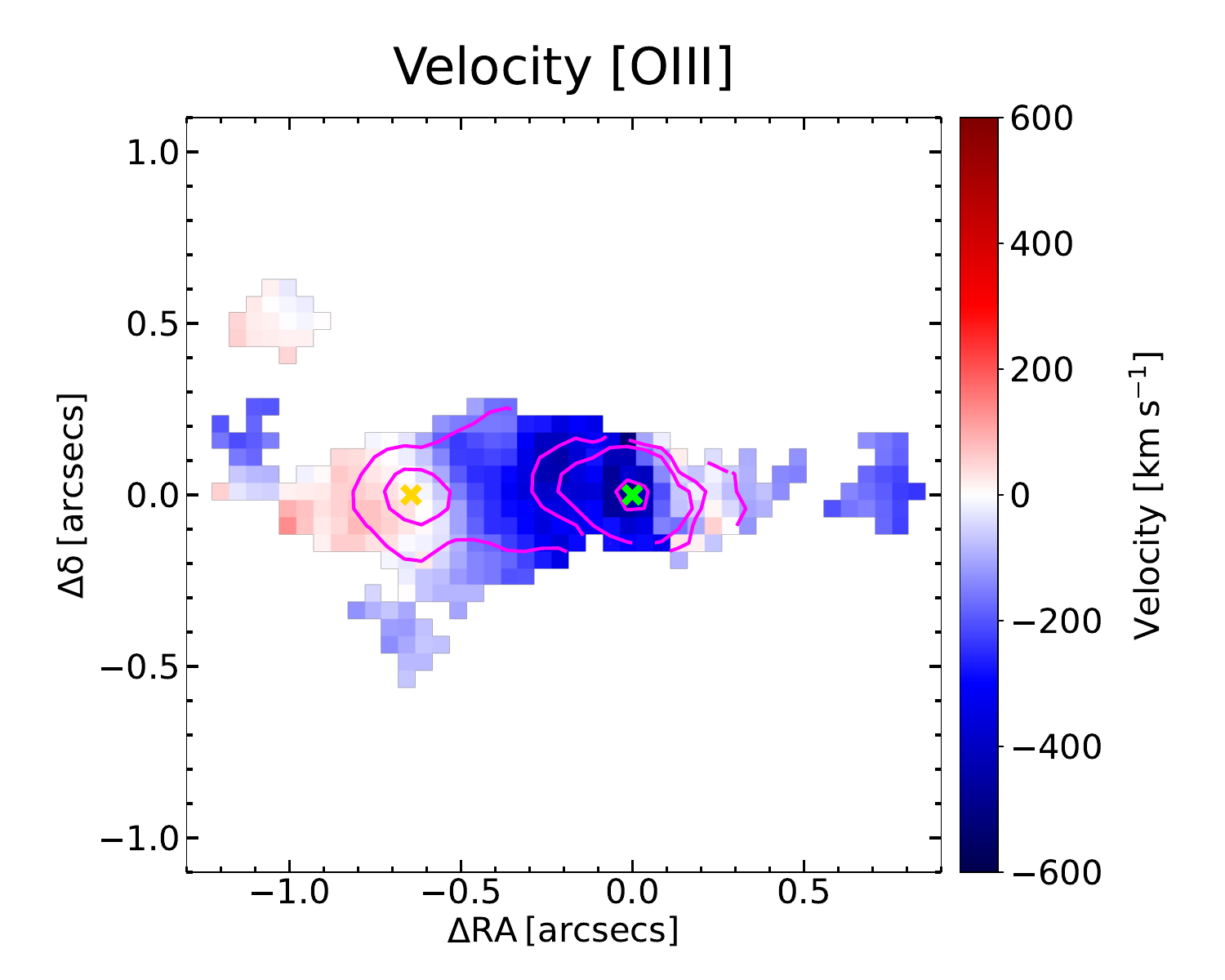}
    \includegraphics[width=0.245\linewidth,trim={1cm 0 1cm 0},clip]{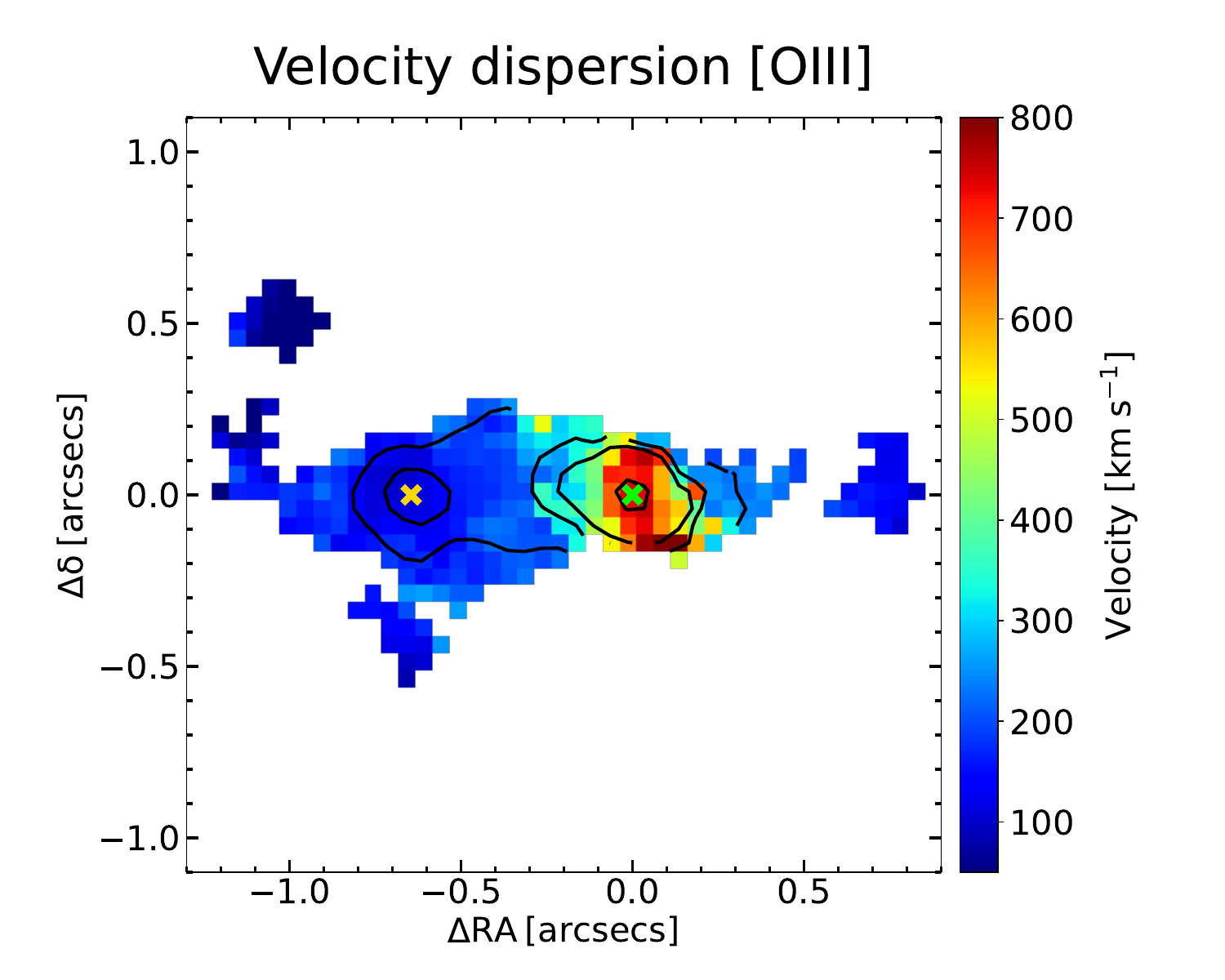}
    \caption{Same as Fig.~\ref{fig:gs133} but for GS~10578. Here we also report the maps of the flux, velocity, and velocity dispersion of the total fitted profile of \oiii, due to their different appearance as compared to those of \nii. The golden cross marks the position of the secondary AGN.}
    \label{fig:gs10578}
\end{figure*}

\begin{figure*}
    \centering
    \includegraphics[width=0.245\linewidth,trim={1cm 0 1cm 0},clip]{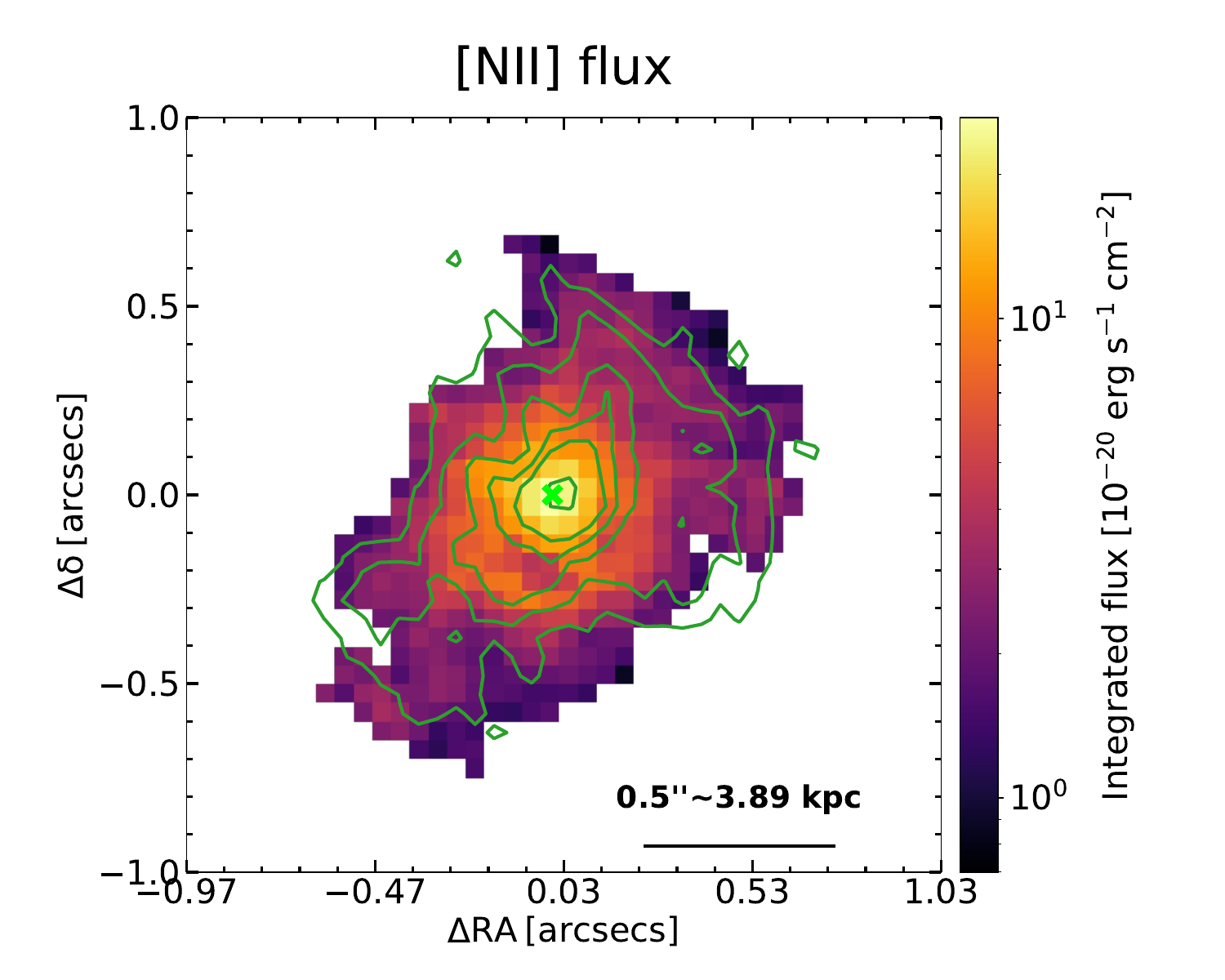}
    \includegraphics[width=0.245\linewidth,trim={1cm 0 1cm 0},clip]{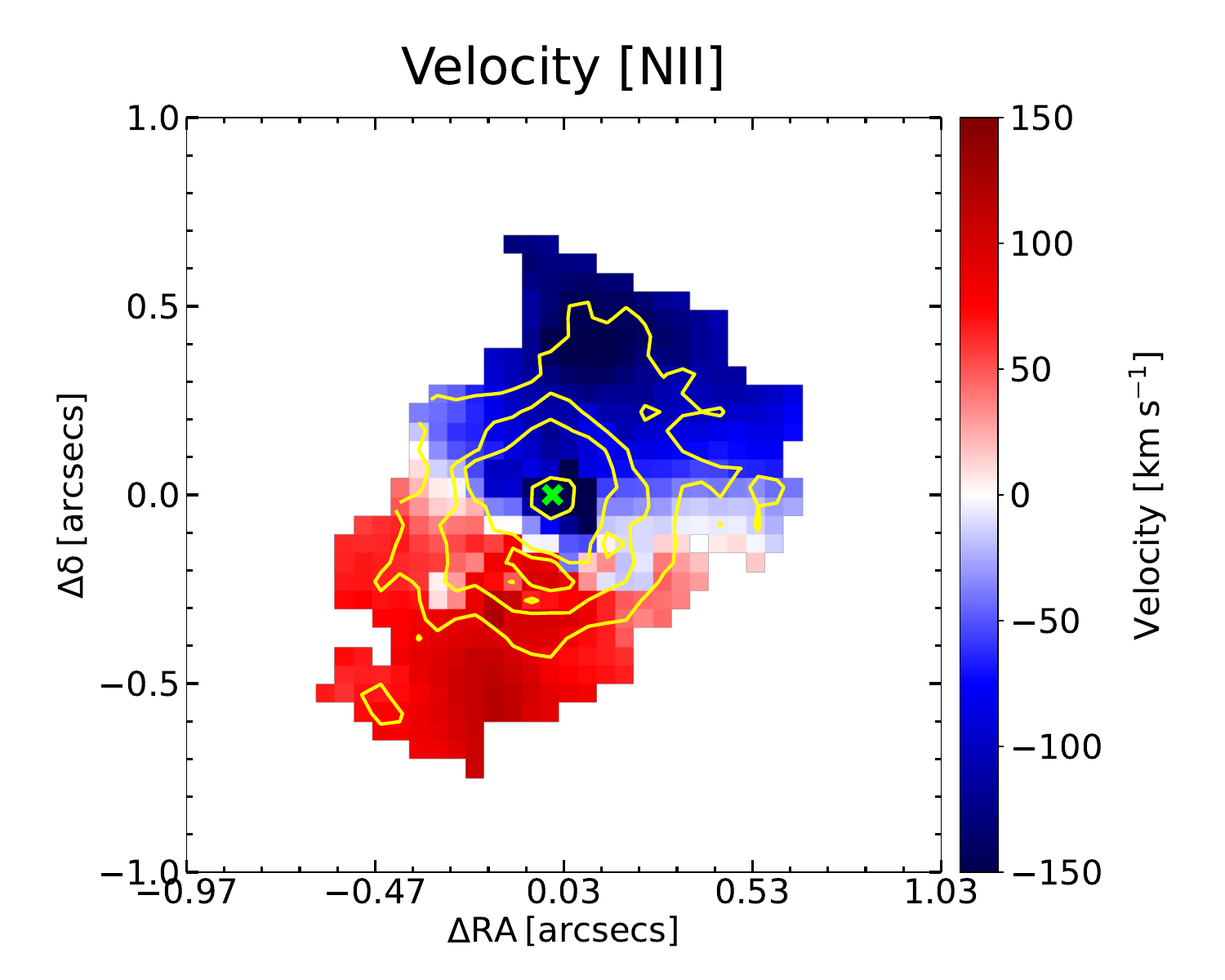}
    \includegraphics[width=0.245\linewidth,trim={1cm 0 1cm 0},clip]{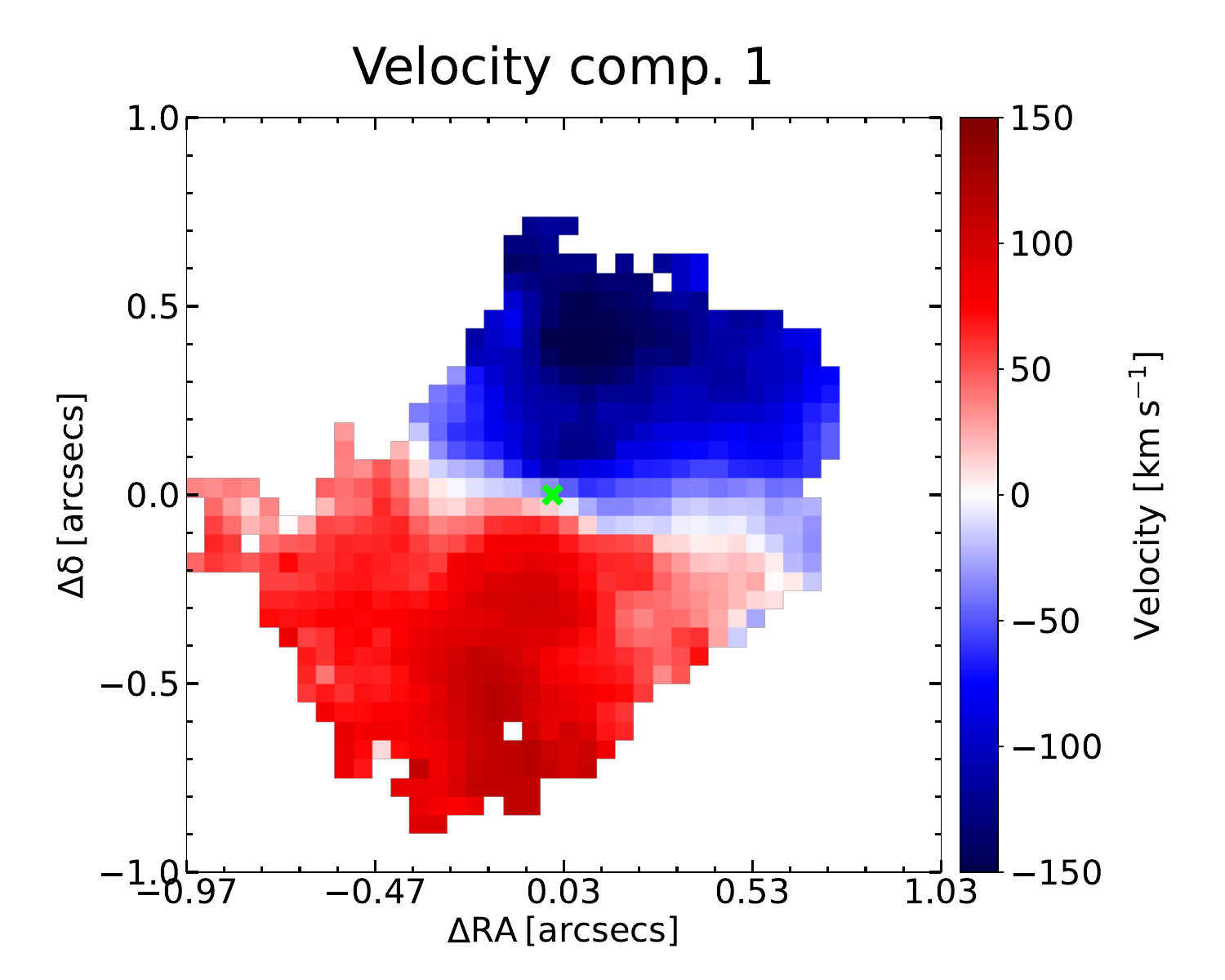}
    \includegraphics[width=0.245\linewidth,trim={1cm 0 1cm 0},clip]{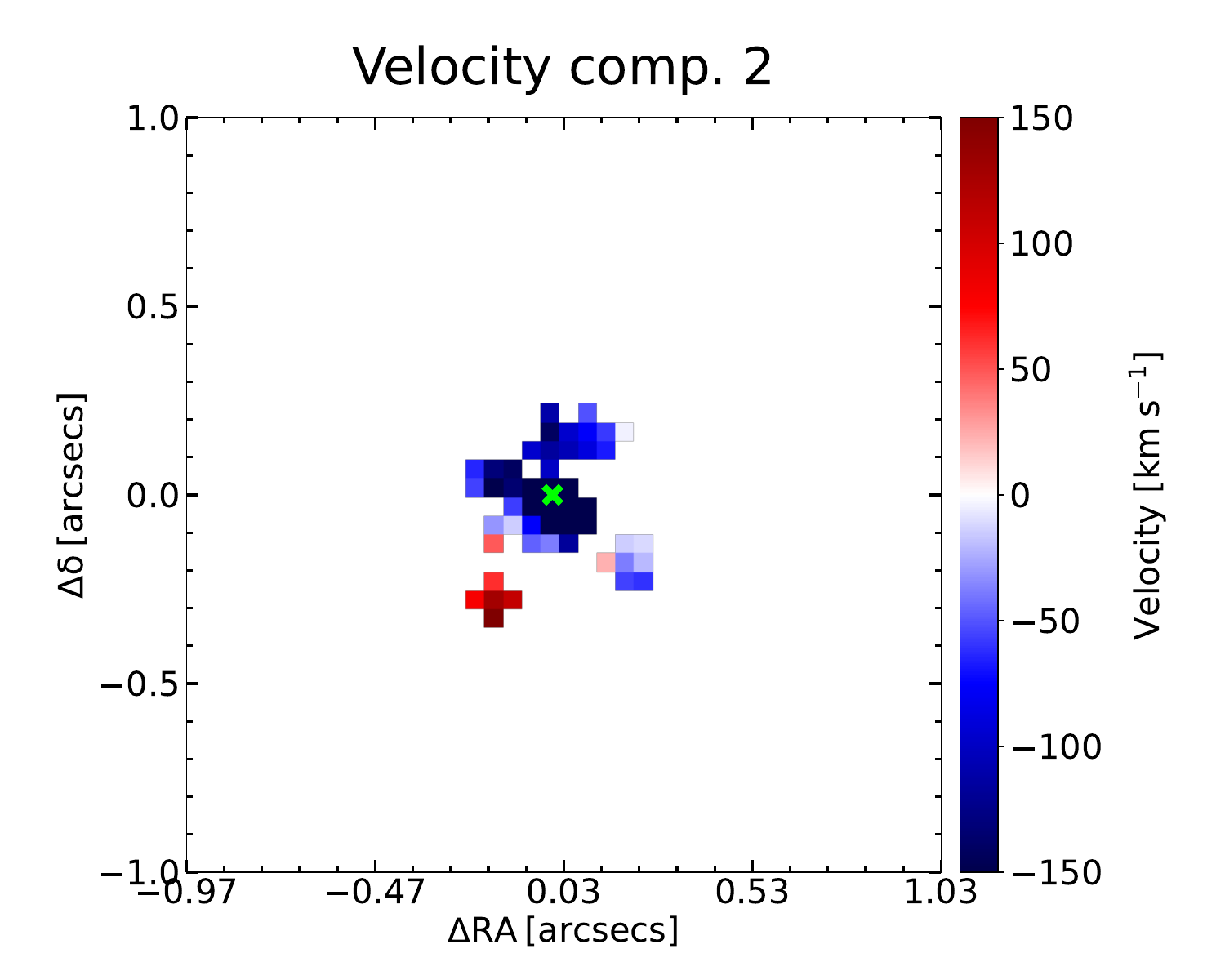}\\
    \includegraphics[width=0.245\linewidth,trim={1cm 0 1cm 0},clip]{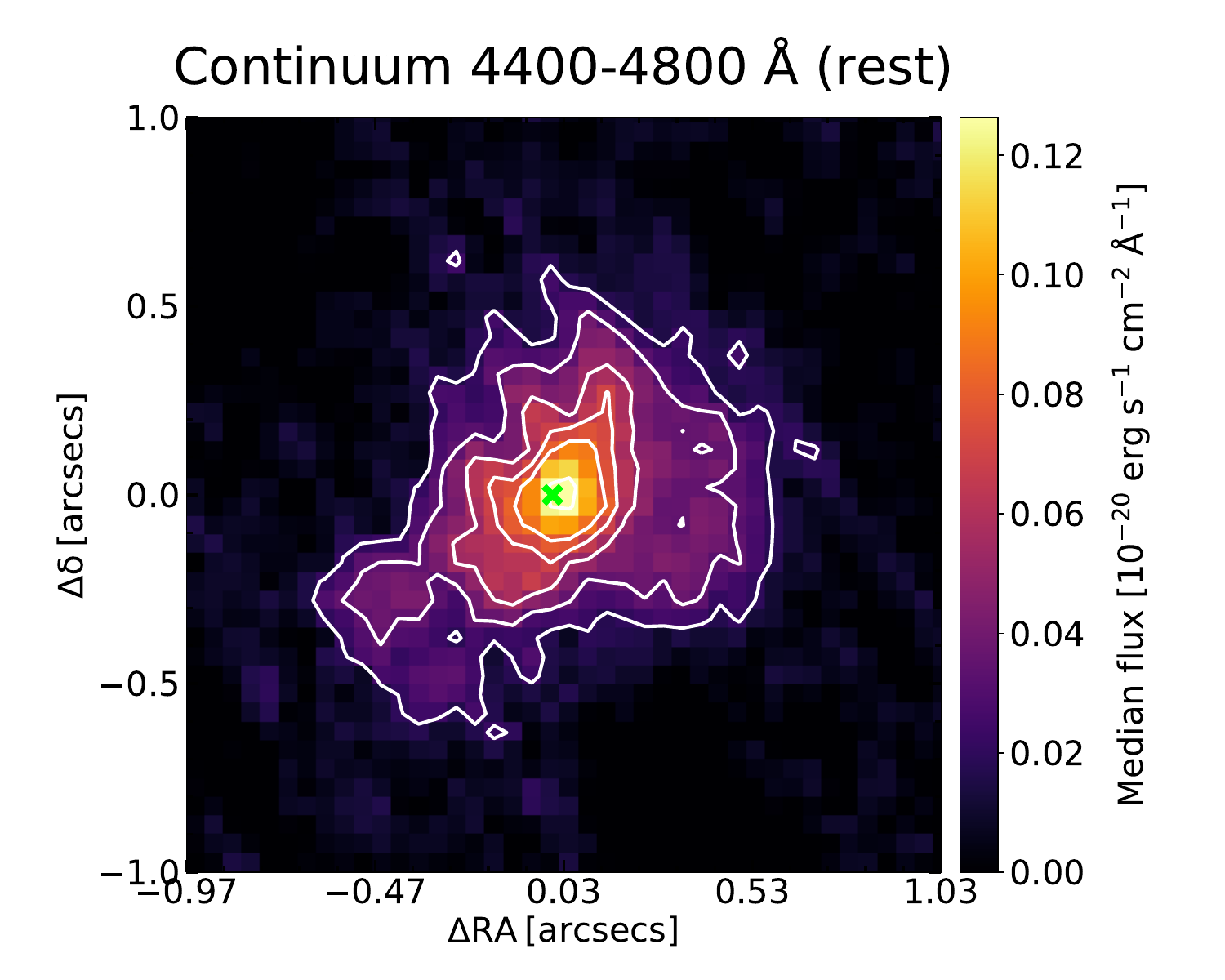}
    \includegraphics[width=0.245\linewidth,trim={1cm 0 1cm 0},clip]{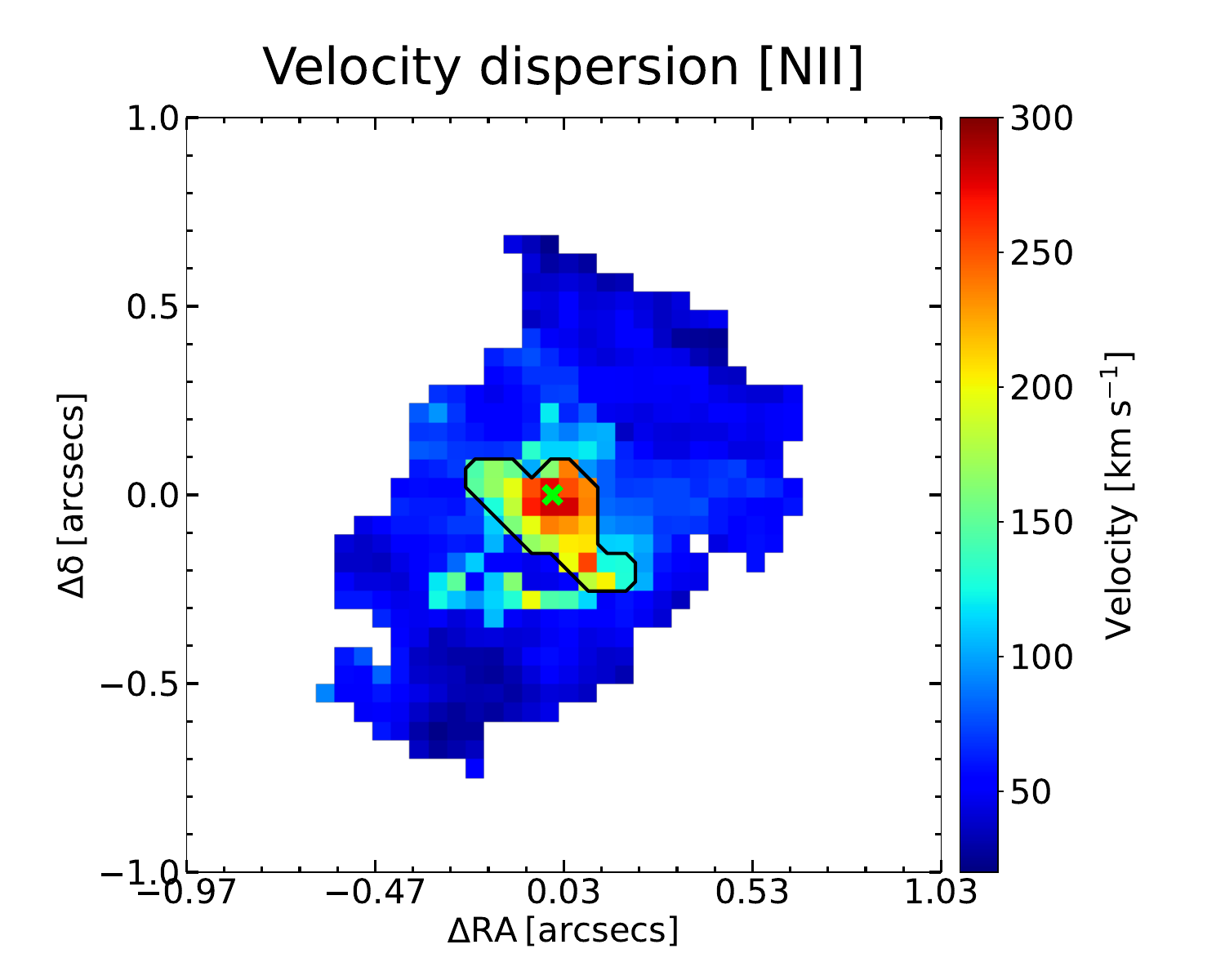}
    \includegraphics[width=0.245\linewidth,trim={1cm 0 1cm 0},clip]{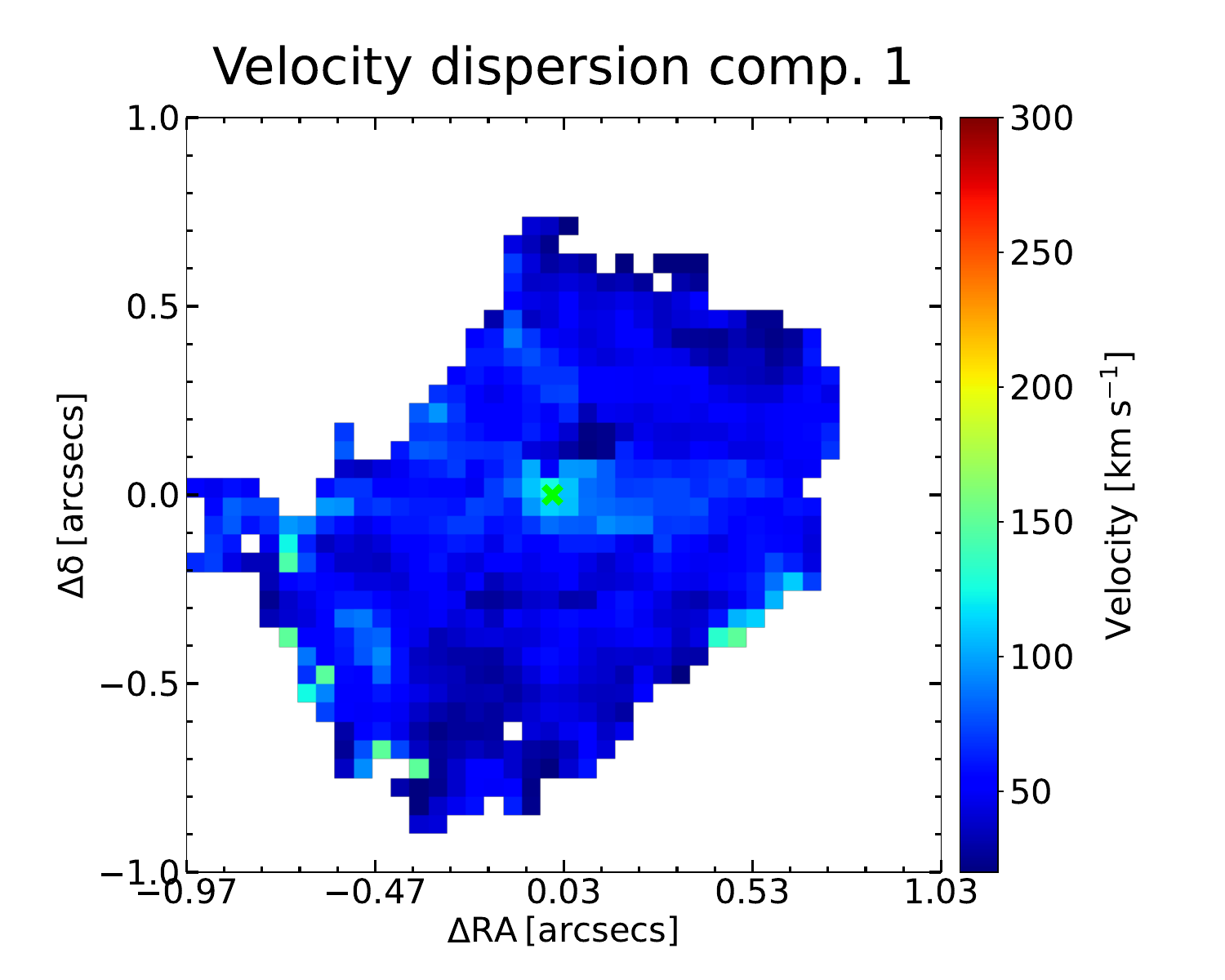}
    \includegraphics[width=0.245\linewidth,trim={1cm 0 1cm 0},clip]{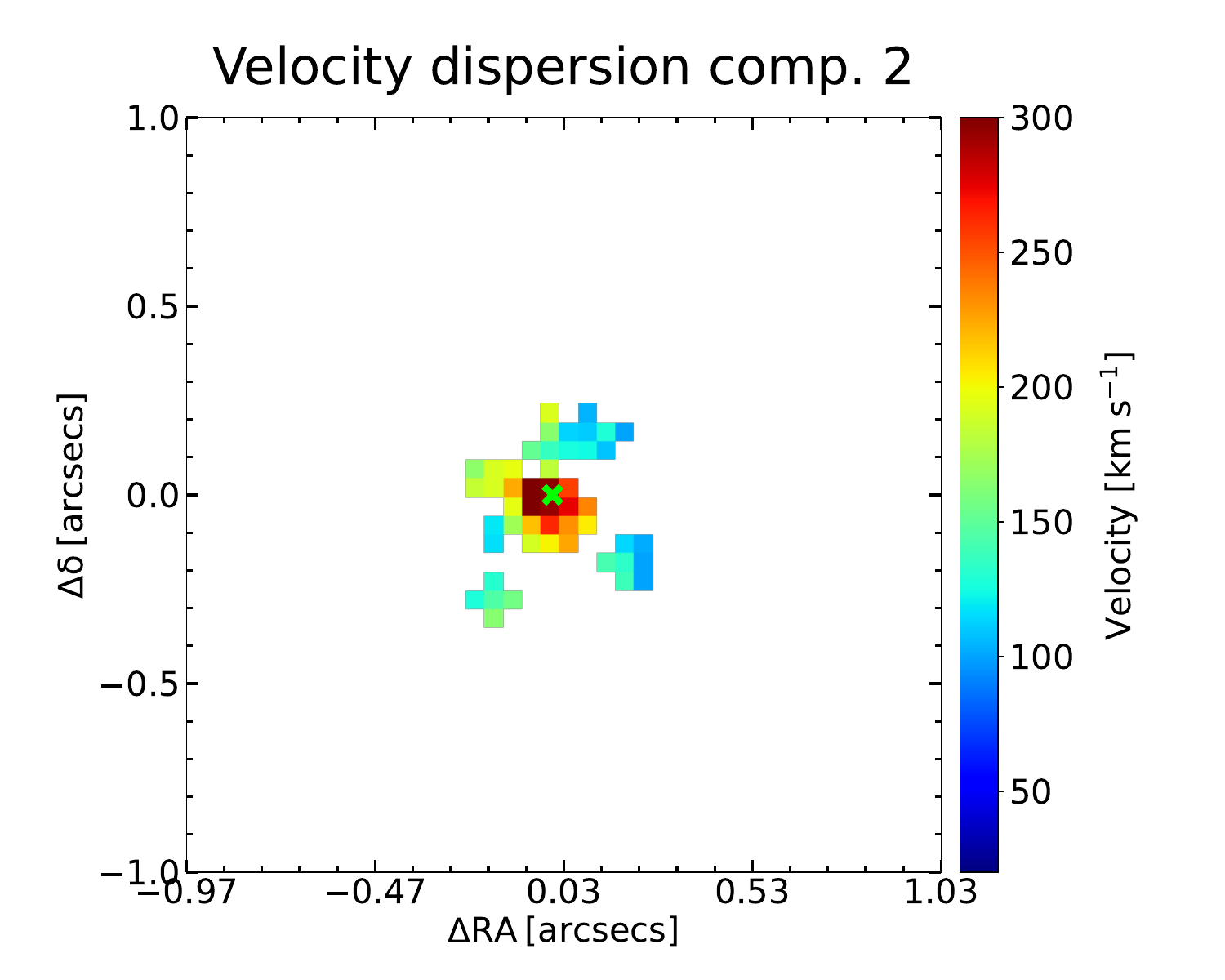}
    \caption{Same as Fig.~\ref{fig:gs133} but for GS~19293.}
    \label{fig:gs19293}
\end{figure*}

\begin{figure*}
    \centering
    \includegraphics[width=0.245\linewidth,trim={1cm 0 1cm 0},clip]{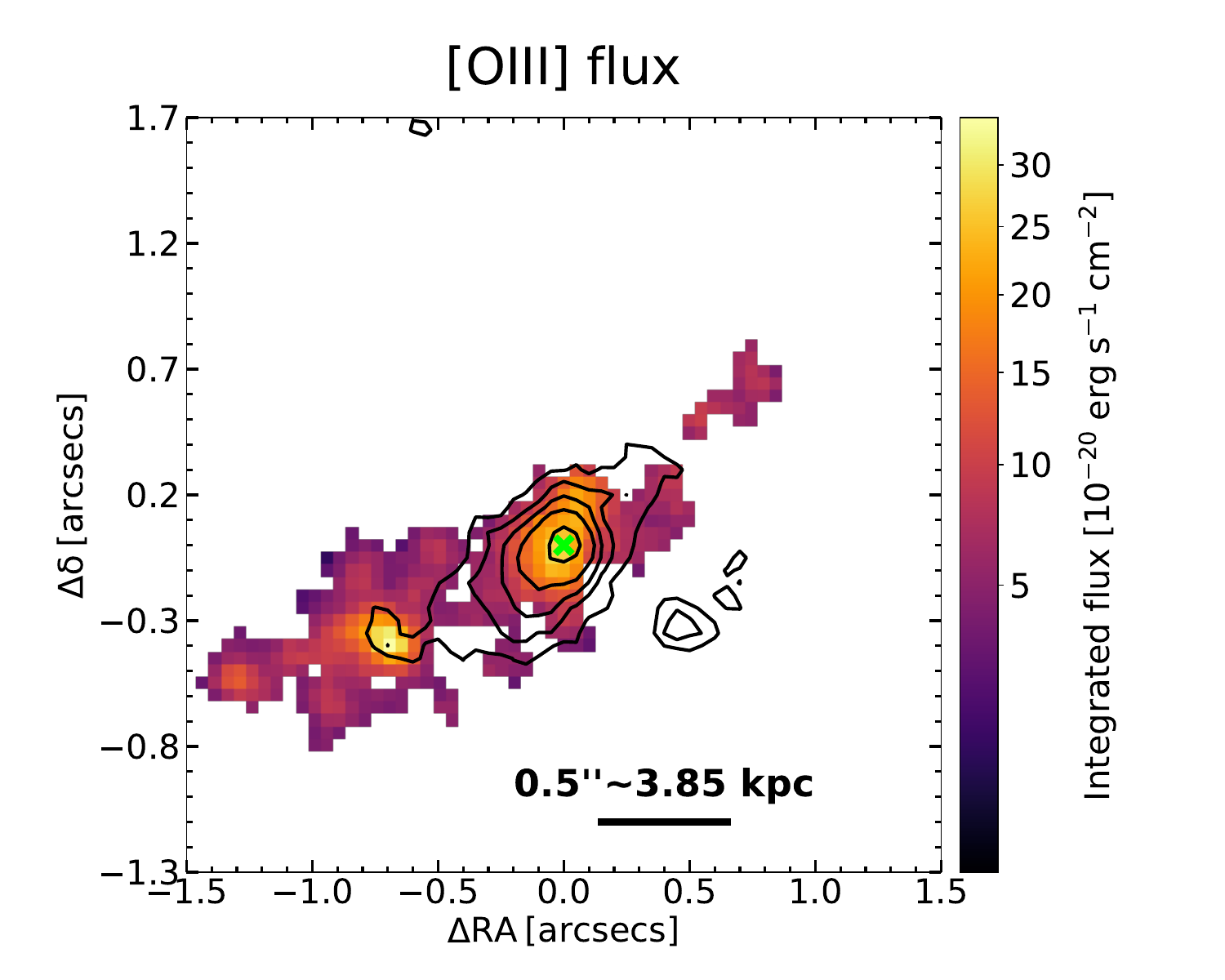}
    \includegraphics[width=0.245\linewidth,trim={1cm 0 1cm 0},clip]{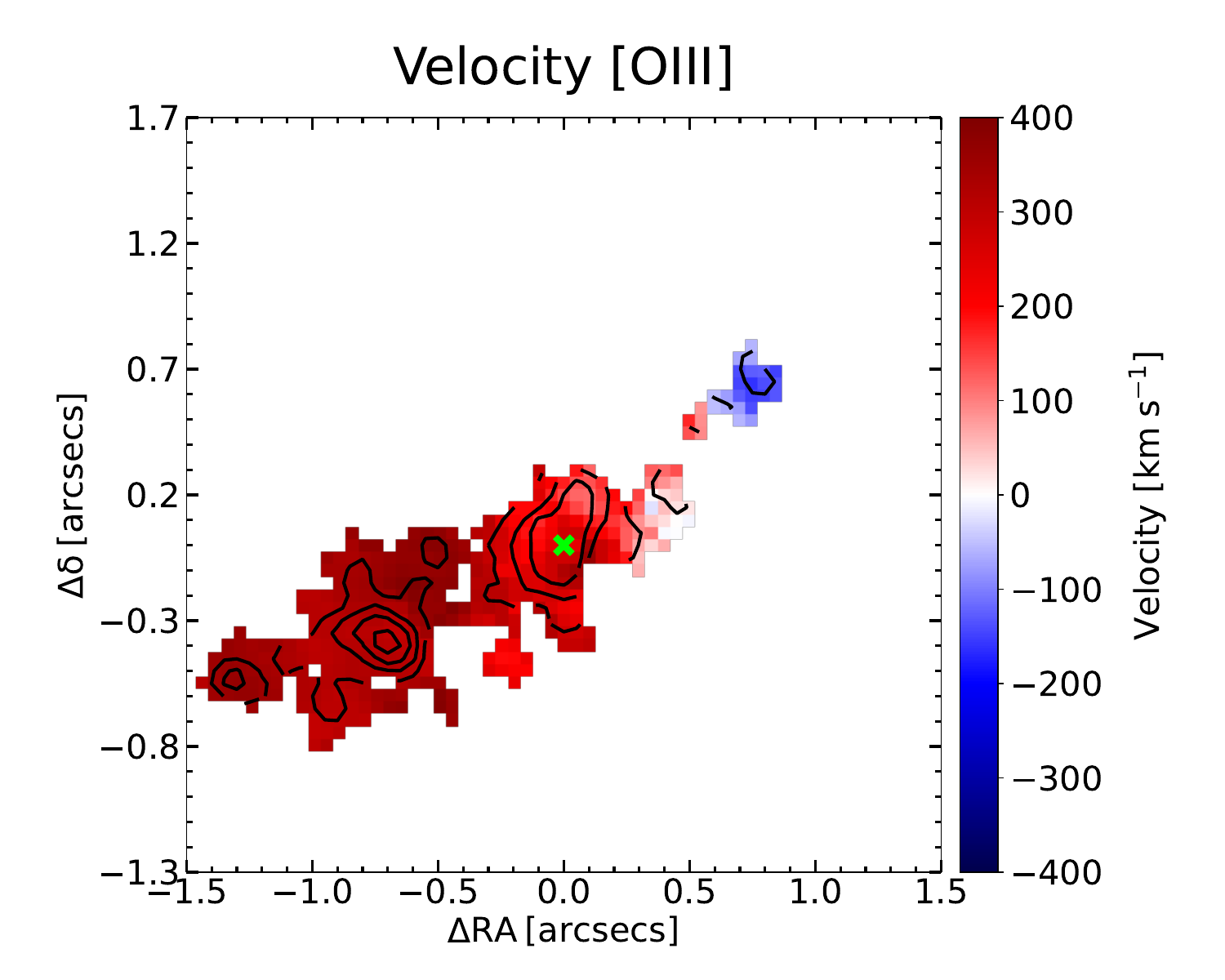}
    \includegraphics[width=0.245\linewidth,trim={1cm 0 1cm 0},clip]{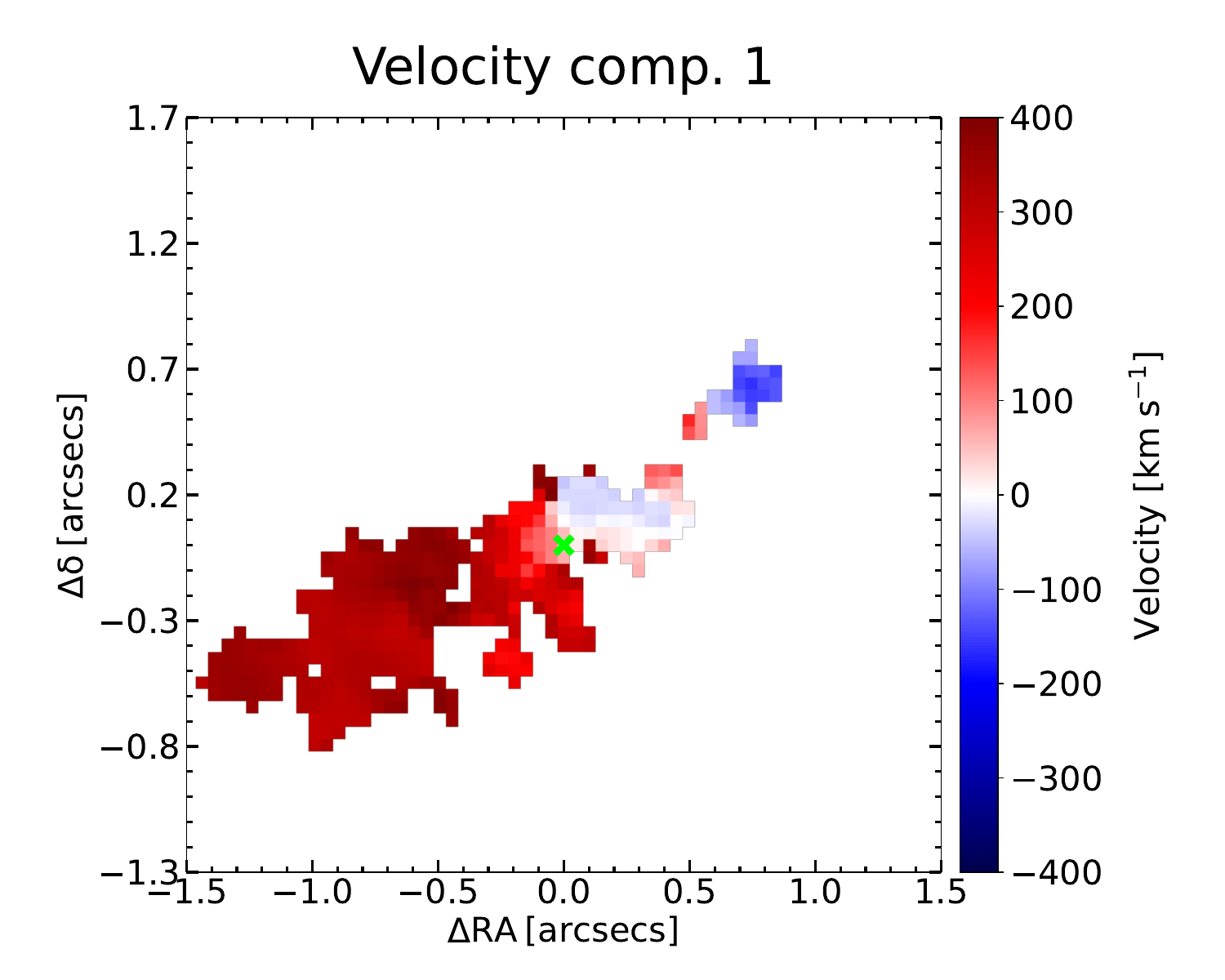}
    \includegraphics[width=0.245\linewidth,trim={1cm 0 1cm 0},clip]{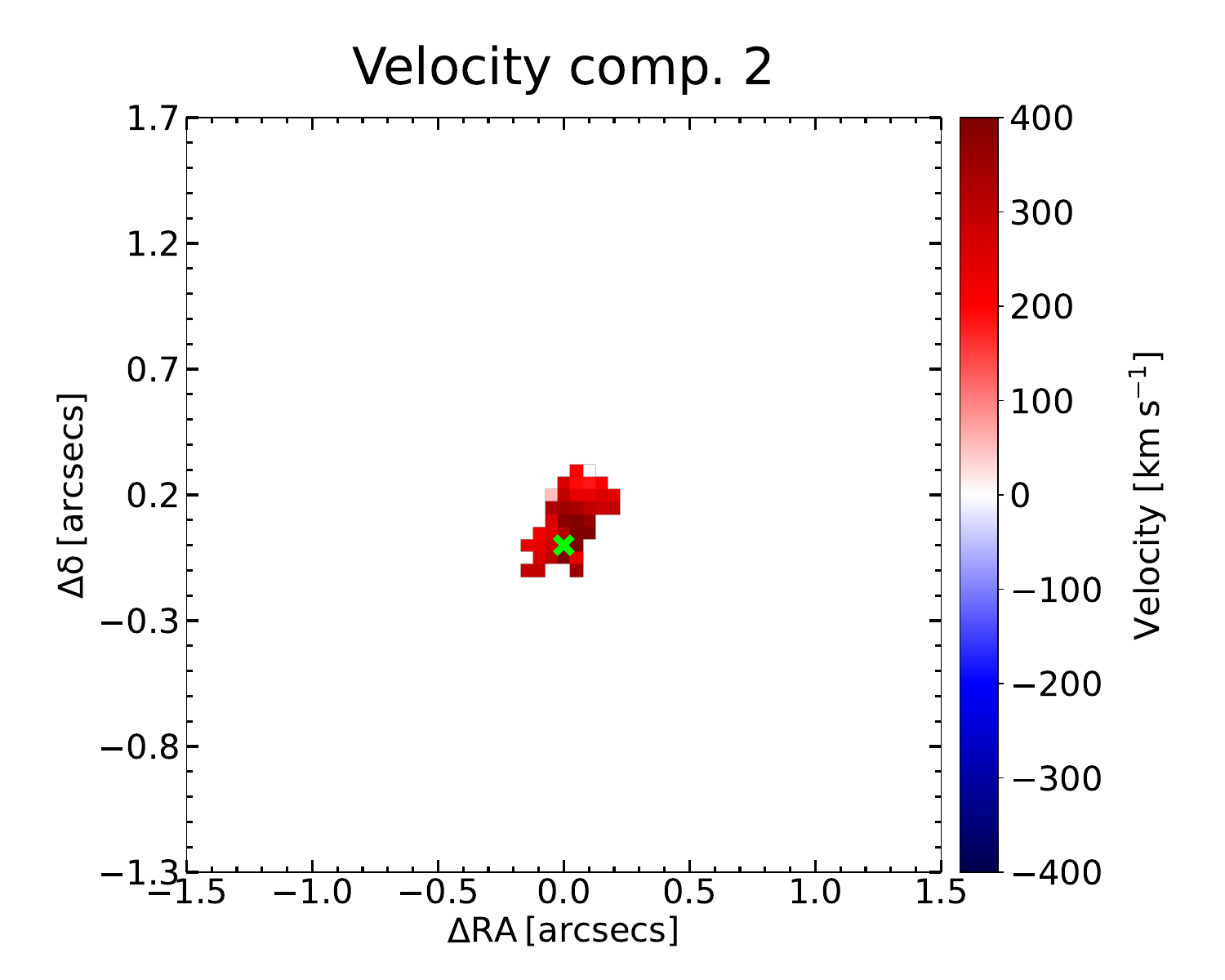}\\
    \includegraphics[width=0.245\linewidth,trim={1cm 0 1cm 0},clip]{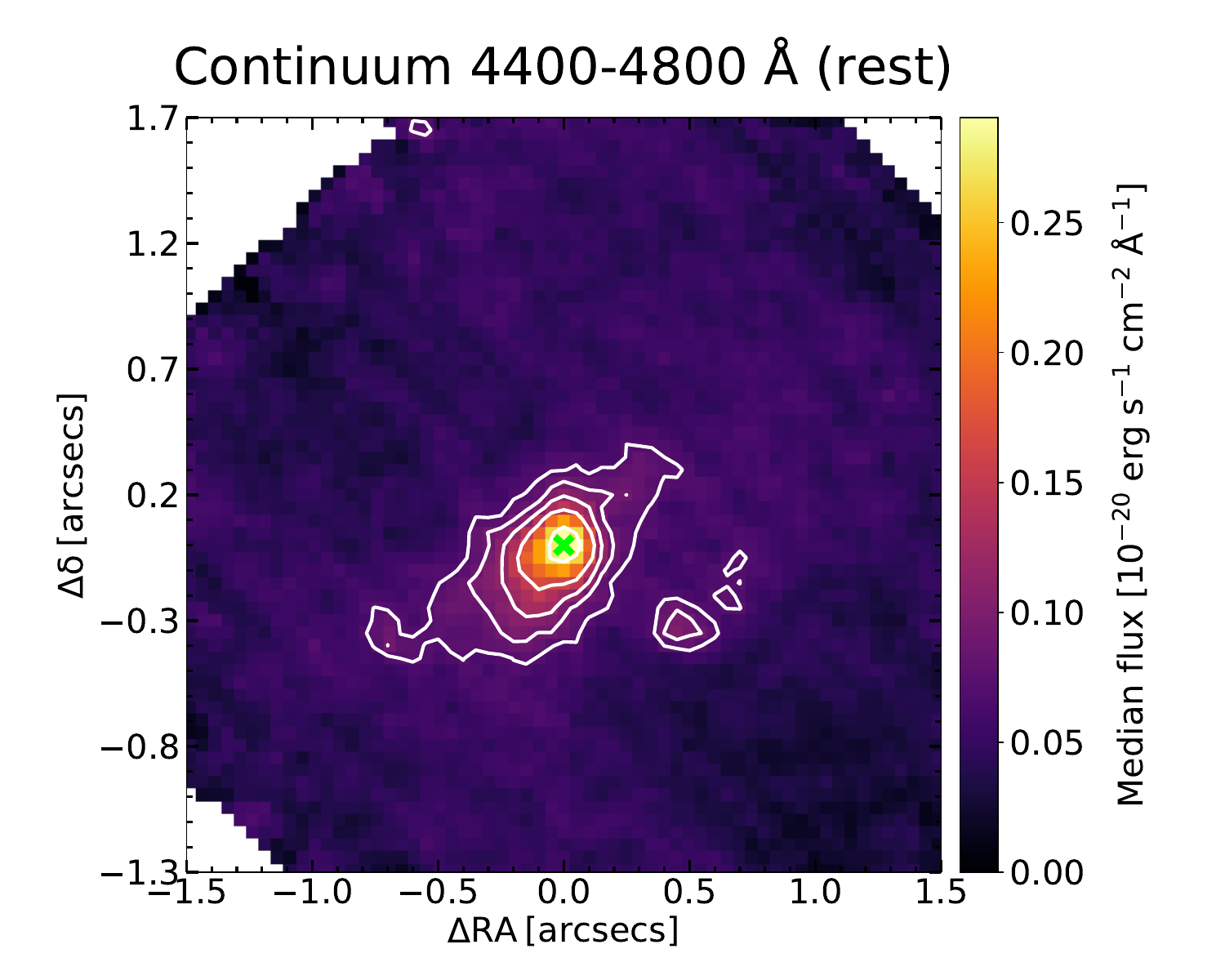}
    \includegraphics[width=0.245\linewidth,trim={1cm 0 1cm 0},clip]{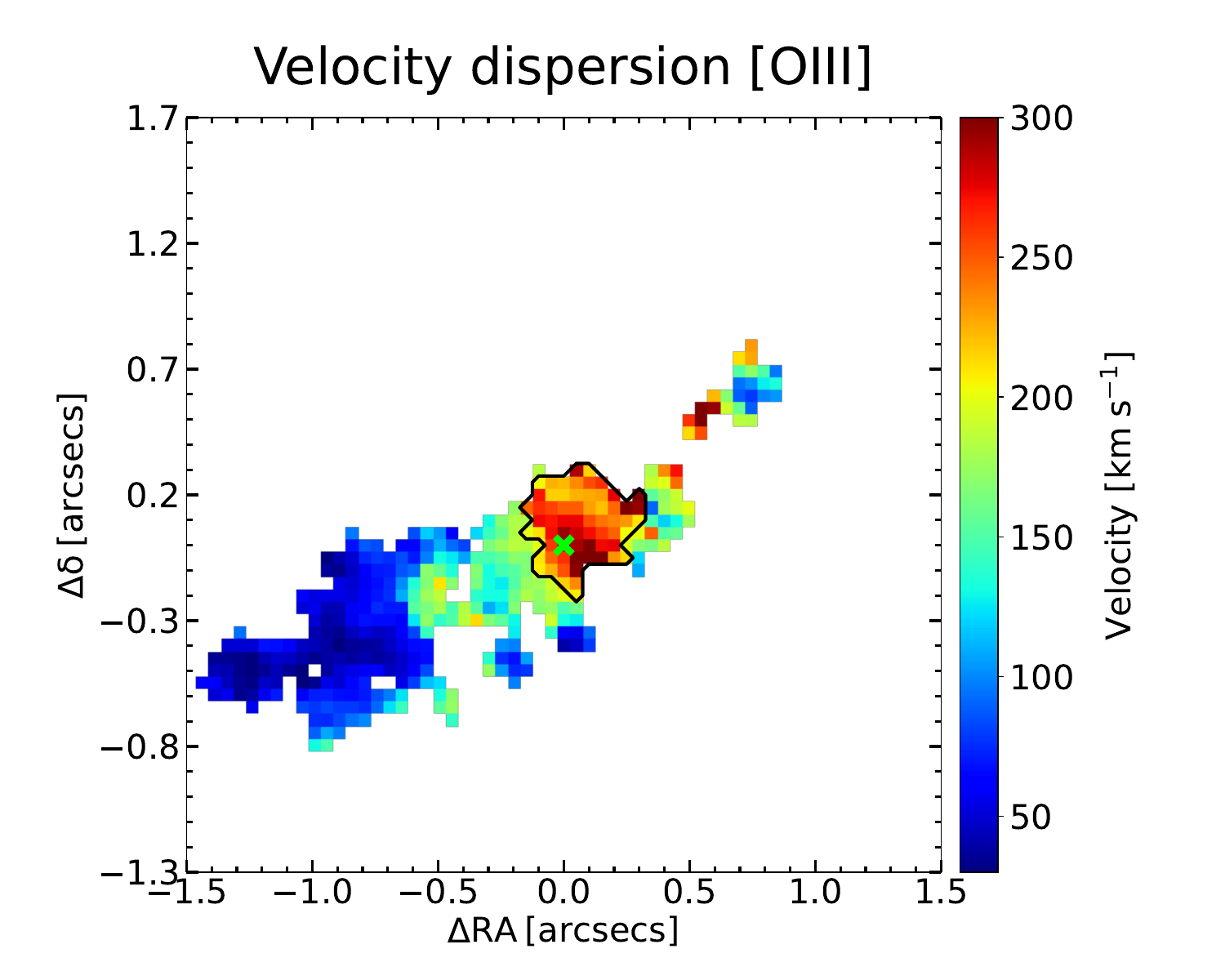}
    \includegraphics[width=0.245\linewidth,trim={1cm 0 1cm 0},clip]{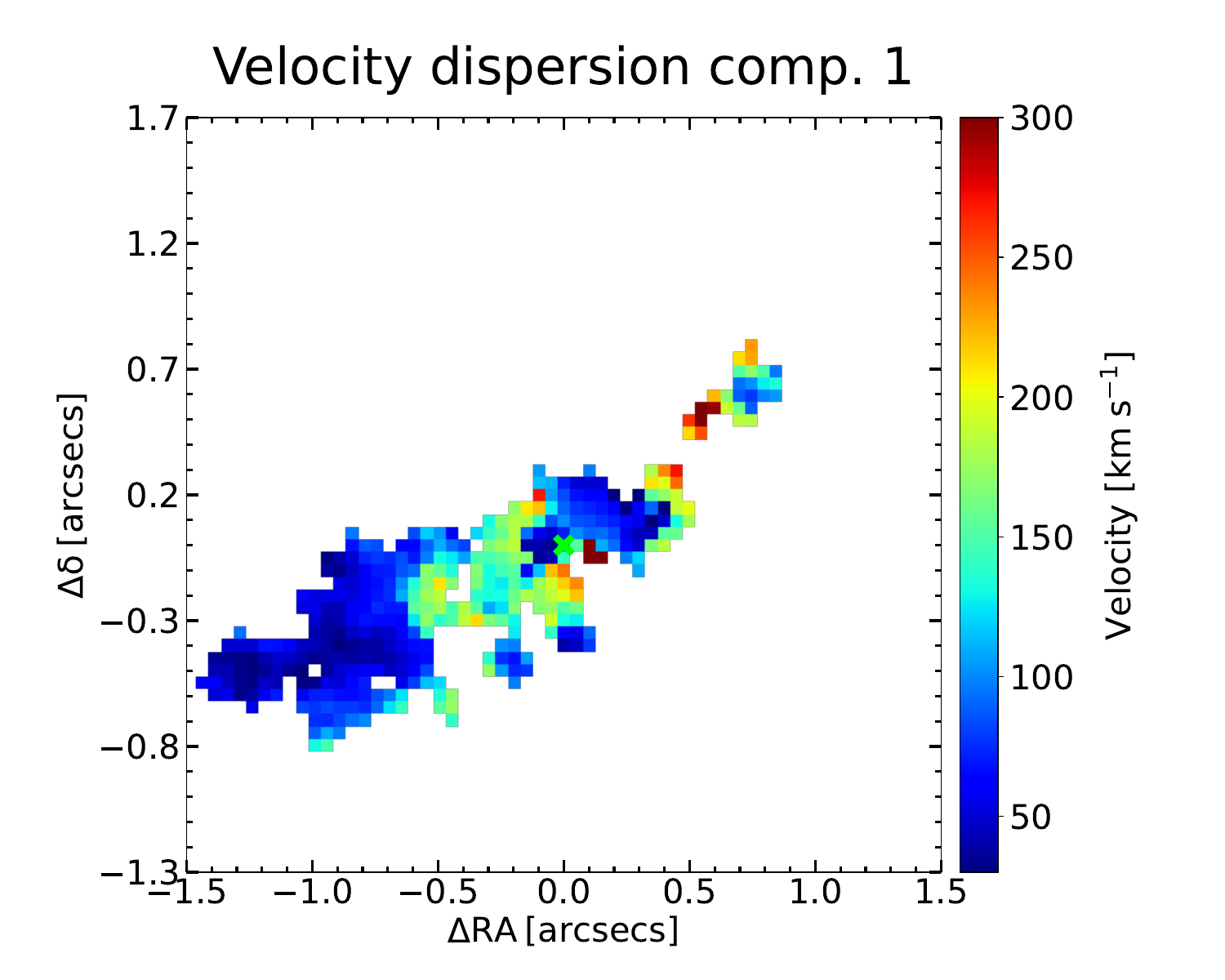}
    \includegraphics[width=0.245\linewidth,trim={1cm 0 1cm 0},clip]{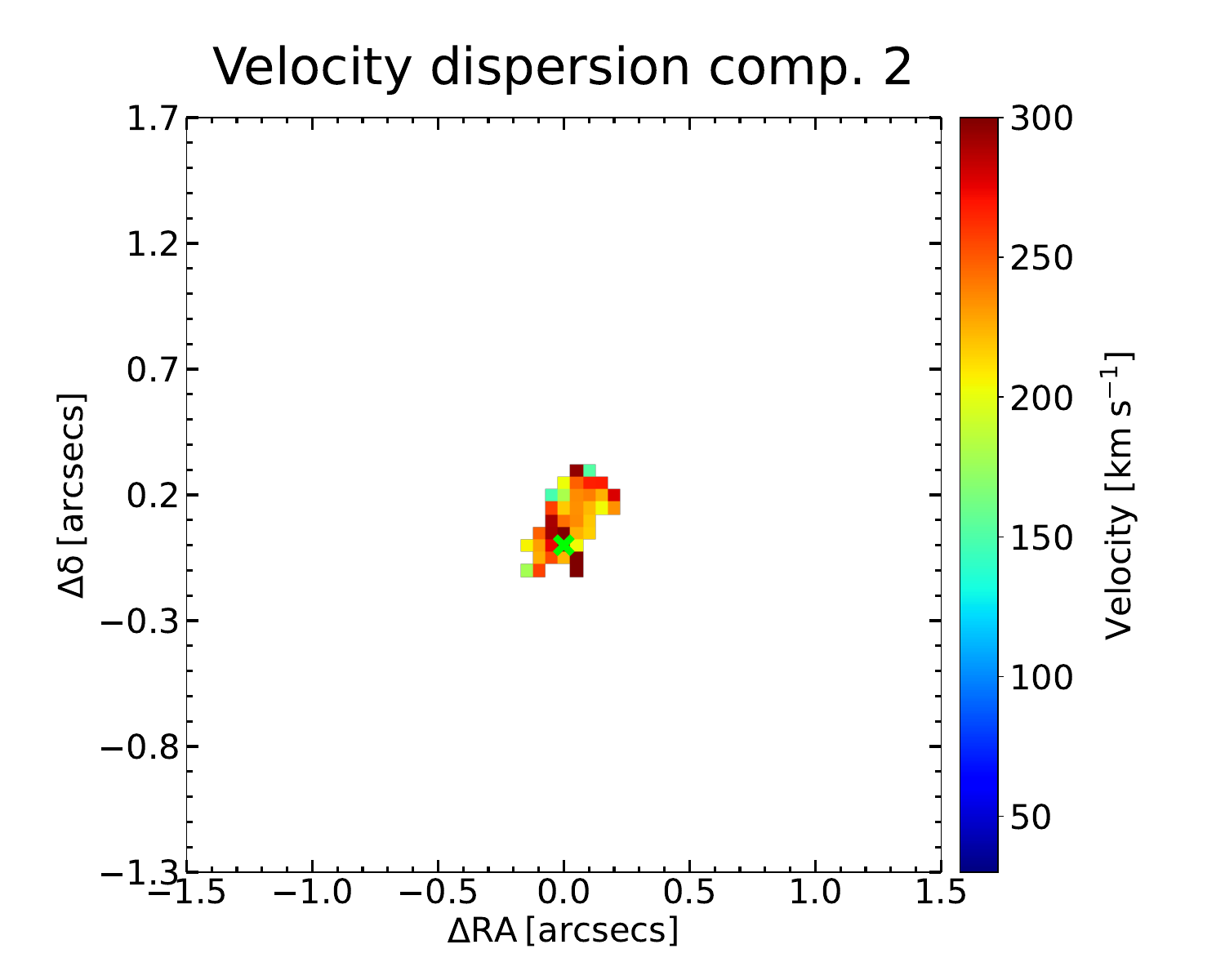}
    \caption{Same as Fig.~\ref{fig:gs133} but for GS~20936.}
    \label{fig:gs20936}
\end{figure*}

In this section we briefly discuss the emission line maps obtained for each source from the spaxel-by-spaxel analysis described in Sect. \ref{sec:data_anal} and how they relate to previous work.
The maps of GS~539, GS~551, GS~774, GS~811, GS~3073, GS~10578, GS~19293, and GS~20293 are reported in Figs.~\ref{fig:gs539}--\ref{fig:gs20936}, respectively, while the maps of GS~133 are shown in Fig.~\ref{fig:gs133}.

\subsection{GS~133}
GS~133 ($z \simeq 3.473$; Fig.~\ref{fig:gs133}) is a Compton-thick ($N_\mathrm{H} > 1.5 \times 10^{24}$~cm$^{-2}$; \citealt{Guo2021}) type-2 AGN, which has been the focus of another study from the GA-NIFS collaboration \citep{Perna2025a} employing the same NIRSpec IFU data used in this work. Its maps (Fig.~\ref{fig:gs133}) show extended \oiii emission in the NE-SW direction (spanning $\sim$7.5~kpc), while the continuum is radially symmetric around the galaxy centre and concentrated on smaller scales. The velocity field is complex, but the single narrow and broad component maps reveal that the former traces a gradient along the NW-SE direction with low velocity dispersions ($\lesssim$100~\kms), likely associated with rotation, while the latter shows a clear gradient in the NE-SW direction and large velocity dispersions ($\gtrsim$300-400~\kms), consistent with a bipolar outflow (see also \citealt{Perna2025a}).

\subsection{GS~539}
GS~539 ($z \simeq 4.756$; Fig.~\ref{fig:gs539}), also known as ALESS073.1, is a sub-millimetre galaxy (SMG) hosting a Compton-thick ($N_\mathrm{H} \sim 1.7 \times 10^{24}$~cm$^{-2}$; \citealt{Circosta2019}), type-1 AGN and has been the subject of another study from the GA-NIFS collaboration \citep{Parlanti2024}. The ionised gas traced by \nii (\oiii is too weak at the spaxel level due to the high obscuration) is elongated and the narrow component traces a symmetric gradient consistent with rotation, while the broad component indicates an unresolved, blueshifted (i.e. approaching) outflow (see also \citealt{Parlanti2024}).

\subsection{GS~551}
GS~551 ($z \simeq 3.703$; Fig.~\ref{fig:gs551}) is a candidate dual AGN, in which both nuclei are type-2. The most luminous of the two was selected from X-rays (Compton-thick, $N_\mathrm{H} \sim 1.2 \times 10^{24}$~cm$^{-2}$; \citealt{Circosta2019}) while the secondary nucleus, at a separation of 2.9~kpc to the NW, was discovered on the basis of rest-optical emission line diagnostics in \cite{Perna2025b} as part of GA-NIFS through the same NIRSpec IFU data used in this work.
As stressed in \cite{Perna2025b}, the secondary core is a candidate AGN, since ionising radiation from the primary AGN could in principle be responsible for its gas excitation and line emission.
The secondary nucleus shows a different, blueshifted velocity offset and much lower velocity dispersions ($\sim$50 \kms) compared to the primary AGN. The narrow component of the primary nucleus has velocity dispersions of $\sim$100--150 \kms and shows a mild gradient in the NS direction possibly due to rotation. The broad component has velocity dispersions of $\sim$300--500 \kms and mostly blueshifted velocity offsets. In the rest of this work, when referring to GS~551, we intend the primary nucleus, since the second, fainter one does not show any evidence of outflows.

\subsection{GS~774}
GS~774 ($z \simeq 3.584$; Fig.~\ref{fig:gs774}) is an unobscured ($N_\mathrm{H} \sim 7.4 \times 10^{22}$~cm$^{-2}$; \citealt{Luo2017}), type-1 AGN. It exhibits a fully blueshifted, unresolved outflow with velocity dispersions in the range $\sim$450--700~\kms as traced by the broad component, while the narrow component shows a velocity gradient in the NS direction, possibly due to rotation. 
The approaching outflow velocities, being spatially unresolved, are consistent with the fact that this source is a type 1 AGN, so we might be observing the galaxy almost face on and the outflow being launched mostly in our line of sight.

\subsection{GS~811}
GS~811 ($z \simeq 3.468$; Fig.~\ref{fig:gs811}) is an obscured ($N_\mathrm{H} \sim 2.3 \times 10^{23}$~cm$^{-2}$; \citealt{Luo2017}), type-1 AGN. The low signal-to-noise due to the target faintness allows the ionised gas emission to be traced only in a handful of spaxels, especially in the broad component, which shows blueshifted velocities only, consistent with the target being a type-1 AGN. The narrow component velocity map instead shows a tentative resolved gradient in the NS direction.

\subsection{GS~3073}
GS~3073 ($z \simeq 5.553$; Fig.~\ref{fig:gs3073}), originally selected for GA-NIFS observations as a star-forming galaxy due to its lack of X-ray emission in deep \textit{Chandra} observations \citep[7 Ms;][]{Grazian2020}, turned out to be an AGN based on the detection of prominent BLR components in permitted (hydrogen and helium) lines and a large equivalent width of \heii$\lambda$4686 \citep{Ubler2023}. The maps (Fig.~\ref{fig:gs3073}) show the characteristic star-shaped PSF of JWST, due to the strong unresolved line emission of this source. There is also spatially extended emission though, with the narrow component (having $\sigma \lesssim 100$~\kms) showing a gradient suggestive of rotation in the NS direction. The broad, spatially unresolved (or at most only marginally resolved) component shows a redshifted velocity offset (despite the source being a type-1 AGN) and velocity dispersions up to $\sim$350~\kms.
We note that the stellar mass from the UV-optical to FIR photometric SED fitting (Circosta et al., in prep.) reported in Table~\ref{tab:outflow_props}, of log($M_*/M_\odot$) $\sim$ 10.8 (consistent with previous photometric SED fitting estimates; e.g. \citealt{Barchiesi2023}), is significantly larger than that reported by \cite{Ubler2023}, of $\sim$ 9.5, obtained from the SED fitting of the NIRSpec prism spectrum with \textsc{beagle} masking the emission lines, and even larger than the dynamical mass, between 9.0--10.3, also inferred by \cite{Ubler2023} from the same NIRSpec IFU data used in this work.
The reported stellar mass for this galaxy should thus be taken with caution.

\subsection{GS~10578}
GS~10578 ($z \simeq 3.064$; Fig.~\ref{fig:gs10578}) is an obscured ($N_\mathrm{H} \sim 5.5 \times 10^{23}$~cm$^{-2}$; \citealt{Circosta2019}), type-2 AGN. It resides in a massive ($M_* \sim 1.6 \times 10^{11}$~\Msun), post-starburst galaxy hosting a rotating stellar disc as well as powerful atomic ionised and neutral outflows as found from NIRSpec IFU data in GA-NIFS \citep{DEugenio2024}. It has low cold molecular gas content ($<$0.8\% of its stellar mass) and evidence of the galaxy having evolved with gas inflows being balanced by AGN feedback, either ejecting material out of the galaxy or preventing its accretion onto it \citep{Scholtz2026}.

\oiii and \nii emission show quite different morphologies (Fig.~\ref{fig:gs10578}). \oiii is elongated in the E-W direction, especially eastward of the nucleus, and only partially follows the continuum major axis, which is mostly directed NW-SE. 
It also peaks at the location of a secondary clump to the E of the primary system, not traced in continuum, which is a type-2 second AGN, as revealed by the analysis in \cite{Perna2025b}. 
\nii shows instead no emission at the location of the second AGN (at least at the spaxel level; see \citealt{Perna2025b} for the \nii detection in its integrated spectrum), and its overall morphology is more oriented in the NW-SE direction and has an elongation to the SE, which terminates southward of the second AGN, possibly tracing an additional gas clump or weak companion. \nii is also more puffed up around the primary nucleus and has stronger emission to the NW of it, unlike \oiii.

The velocity field in the primary system is complex, as can be seen in the narrow component map, possibly due to multiple overlapping kinematic and spatial components in the ionised gas (see also \citealt{DEugenio2024}).
The second AGN has a separate velocity offset from the rest of the system.
The velocity dispersion is enhanced in an extended region around the primary nucleus and we are able to isolate this broad component, which reaches centroid velocities faster than --600~\kms and dispersions $>$700~\kms. 
The secondary AGN does not show evidence of broad wings, therefore it is not considered in our outflow analysis. In the rest of this work, when mentioning GS~10578, we refer to the primary nucleus.

\subsection{GS~19293}
GS~19293 ($z \simeq 3.064$; Fig.~\ref{fig:gs19293}), undetected in deep \textit{Chandra} X-ray observations \citep[7 Ms;][]{Luo2017}, was originally selected as a star-forming galaxy for GA-NIFS observations, but turned out to be an AGN based on emission-line diagnostic ratios \citep{Perna2025b}, for which we find \oiii/\hb $\sim$4.6 and \nii/\ha $\sim$ 1.0 in a 1-spaxel radius aperture centred on the line peak. 
As Fig.~\ref{fig:gs19293} shows, the galaxy has extended emission in both line emission and stellar continuum and shows a large-scale velocity gradient in the moment 1 map, though with a strongly blueshifted excess on the nucleus, associated with larger velocity dispersions. Disentangling between narrow and broad components isolates the outflow, mostly blueshifted and with dispersions of $\sim$200--300~\kms, leaving a smooth rotating disc with dispersions $\lesssim$50~\kms ($\sim$100~\kms in the core) in the narrow component.

\subsection{GS~20936}
GS~20936 ($z \simeq 3.243$; Fig.~\ref{fig:gs20936}), selected due to its nature of distant red galaxy, was not initially included in the AGN sample of GA-NIFS, since its absorption-corrected X-ray luminosity ($1.3 \times 10^{43}$~\ergs) did not satisfy the adopted threshold of $10^{44}$~\ergs. However, its emission line ratios (\oiii/\hb $\sim$ 3.3, \nii/\ha $\sim$ 0.46, and \sii/\ha $\sim$ 0.40; from an aperture radius of 2 spaxels centred on the line peak), above the star formation versus AGN demarcation curve from \cite{Kewley2001} in the \oiii/\hb versus \nii/\ha and \sii/\ha diagnostic diagrams \citep{Baldwin1981, Veilleux1987}, indicate the presence of an AGN.
The morphology of both line emission and continuum is very complex, with clumpy, asymmetric and disturbed emission, indicating ongoing interactions and merging. The kinematic maps generally trace a gradient in the NW-SE direction, with enhanced velocity dispersions ($\gtrsim$250 up to 300~\kms) in a large part of the system. Our spectral disentangling isolates a broad ($\sigma \gtrsim 220$~\kms) component with redshifted centroid velocity offsets ($\sim$150--300~\kms), while leaving areas with fairly high velocity dispersion in the narrow component ($\sim$150--200~\kms).
It is thus not obvious whether the observed broadening, even the highest-velocity dispersion regions around the nucleus in the broad component, are actually due to an outflow or to disturbances and tidal interactions due to the merging. We therefore include this source in the outflow sample, but caution about this aspect.

\section{Integrated spectra}
In Fig.~\ref{fig:spectra}, we report the integrated spectra for all the targets analysed in this work (for GS~133, see Fig.~\ref{fig:spectrum_gs133}), extracted from the black contoured region in the velocity dispersion map in Figs.~\ref{fig:gs539}-\ref{fig:gs20936}, as described in Sect.~\ref{sec:outf_integr_anal}. These were used to derive the ionised outflow properties, as detailed in Sect.~\ref{sec:outf_props}. 

\begin{figure*}
    \centering
    \includegraphics[width=0.325\linewidth,trim={1cm 0 0.3cm 0},clip]{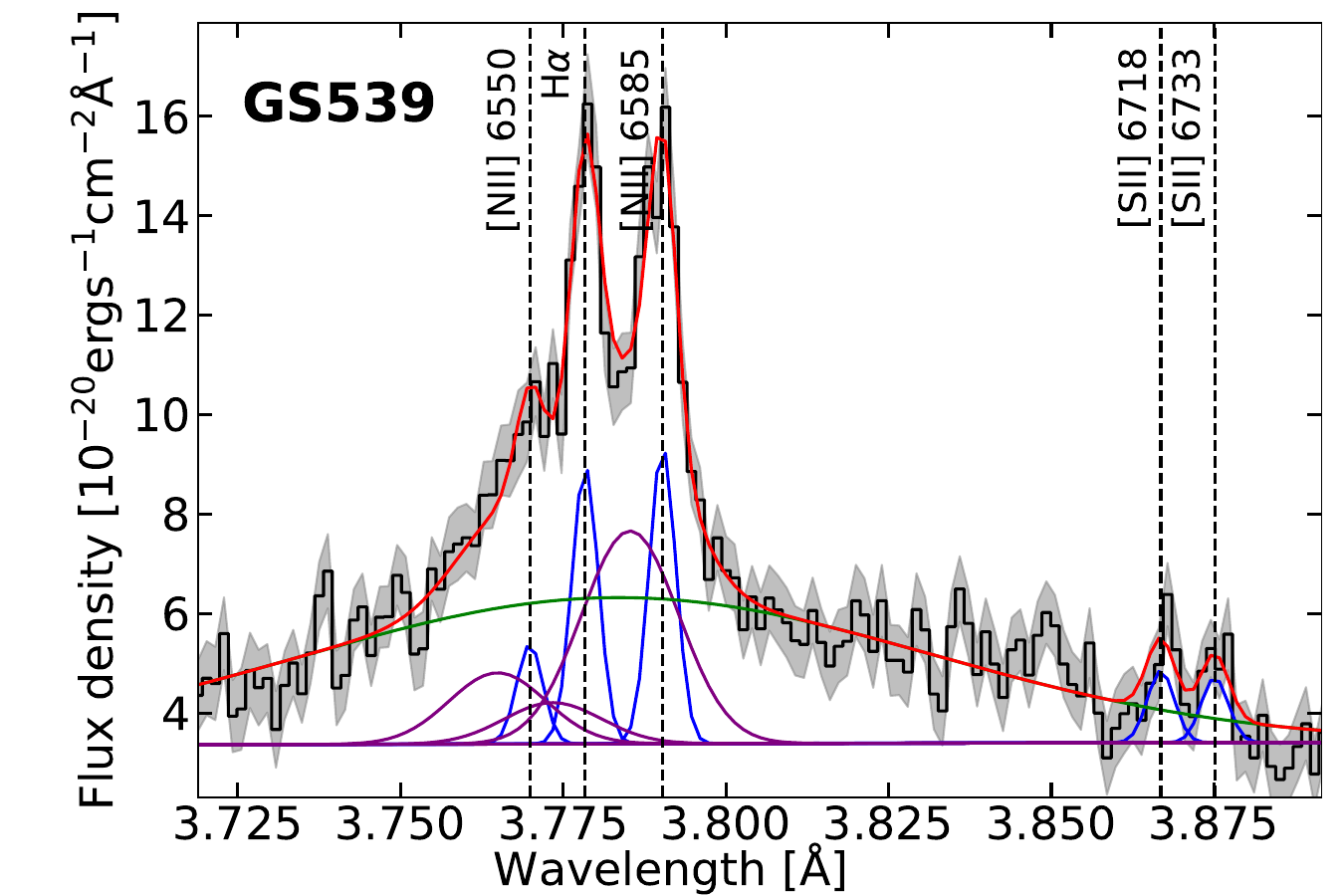}\\
    \includegraphics[width=0.7\linewidth,trim={0.4cm 0 0cm 0},clip]{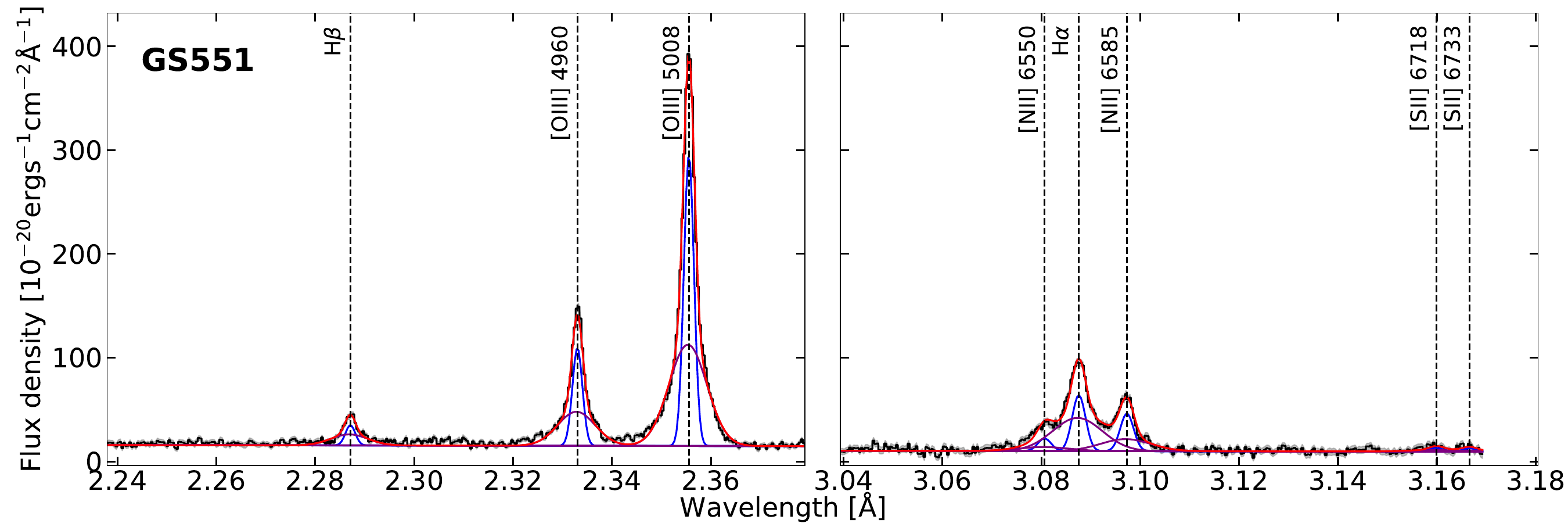}\\
    \includegraphics[width=0.7\linewidth,trim={0.4cm 0 0cm 0},clip]{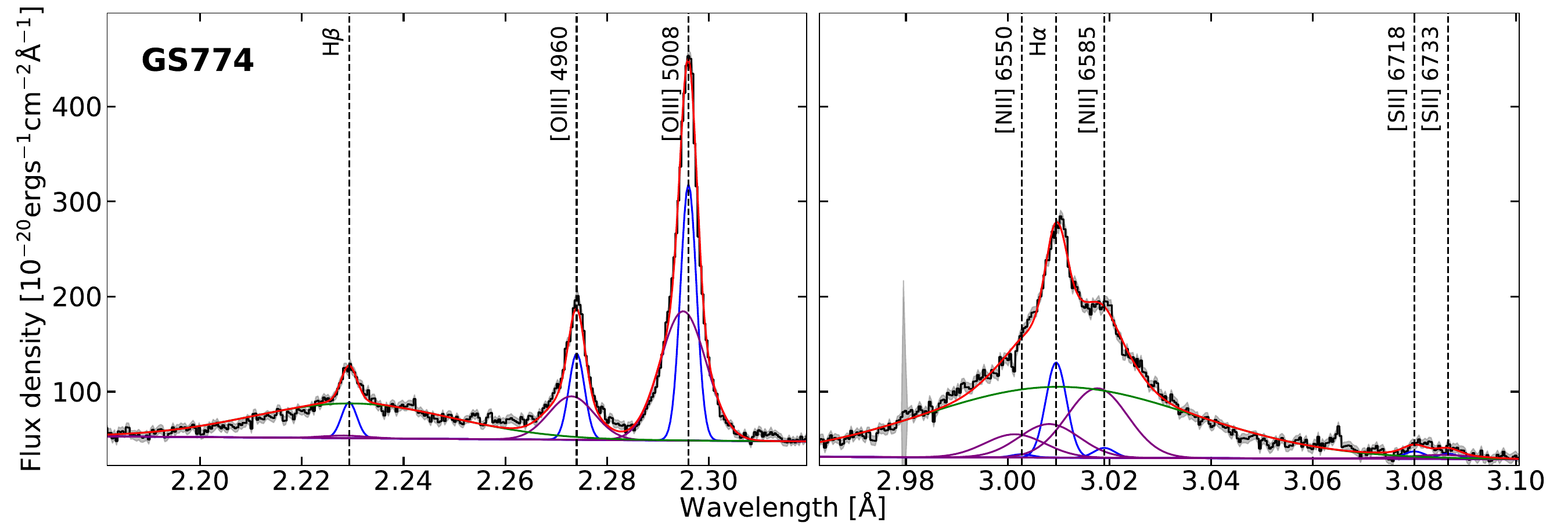}\\
    \includegraphics[width=0.7\linewidth,trim={0.4cm 0 0cm 0},clip]{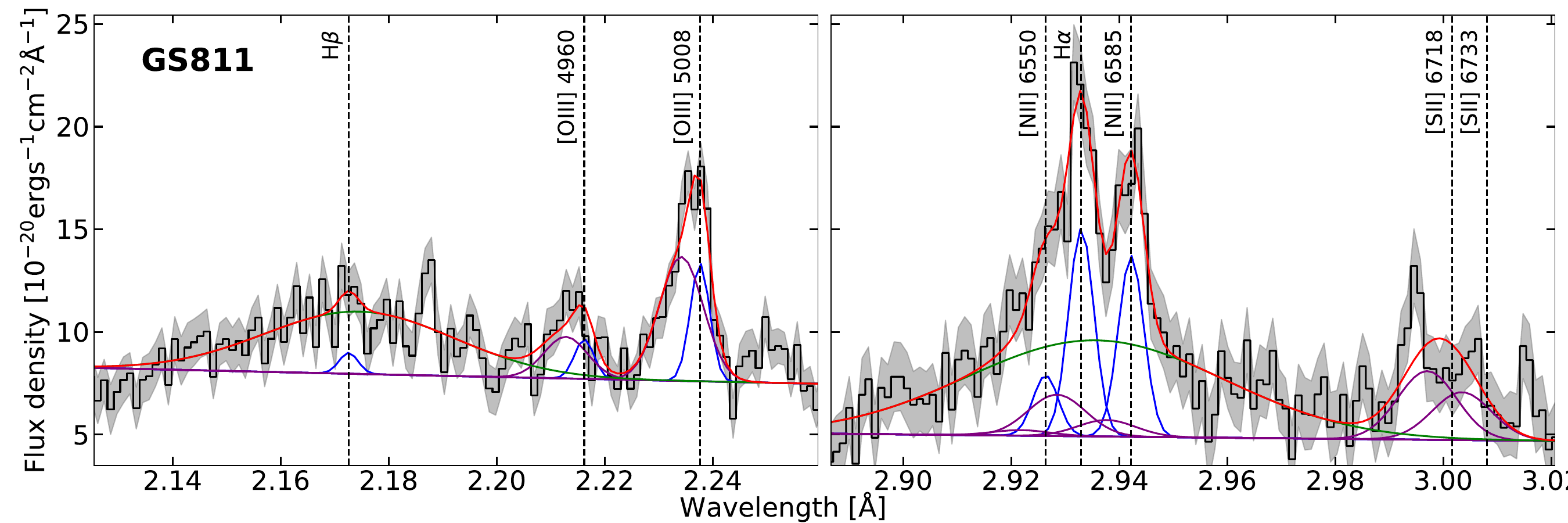}\\
    \caption{Integrated spectra for all of the sources (black), except GS~133 whose spectrum is shown in Fig.~\ref{fig:spectrum_gs133}. The spectrum extraction region is shown by the black contour superimposed on the velocity dispersion map of each source in Figs.~\ref{fig:gs539}-\ref{fig:gs20936}. The coloured curves mark the narrow host-galaxy (blue; blue and orange for GS~10578), broad outflowing (purple), and BLR (green; for type-1 AGN) components, while the red curve marks the overall best-fit model. Vertical, dashed labelled lines mark the wavelength of the peak of the fitted narrow component of each fitted emission line, adopted as the systemic velocity of each system (except for GS~10578, for which we adopted the systemic redshift based on stellar absorption lines from \citealt{DEugenio2024}). Wavelengths are in the observed frame. Uncertainties on the data are reported as shaded areas. The displayed wavelength range is smaller than the entire spectral range covered by the observations and than the fitted one; the displayed range is the same for all the objects, being [--250, +50] \AA\ around \oiii$\lambda$5008 and [--100, +200] \AA\ around \ha as calculated in rest-frame wavelengths. The figure continues on the next page.}
    \label{fig:spectra}
\end{figure*}

\begin{figure*}
    \ContinuedFloat
    \captionsetup{list=off,format=cont}
    \centering
    \includegraphics[width=0.7\linewidth,trim={0.4cm 0 0cm 0},clip]{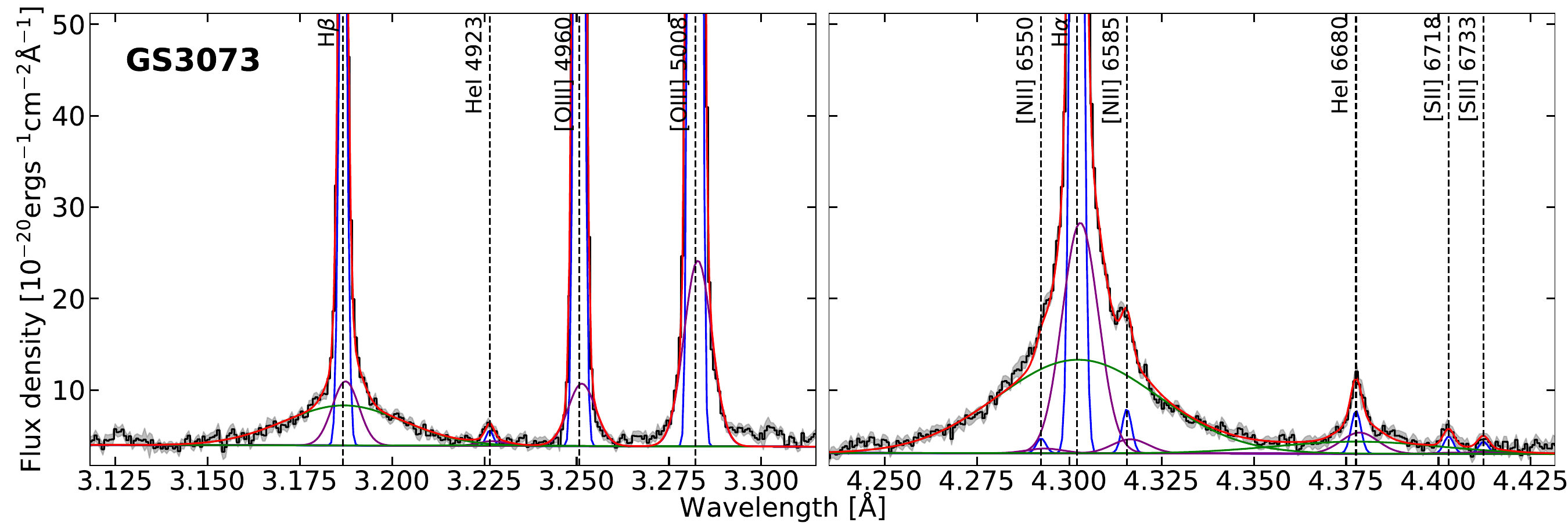}\\
    \includegraphics[width=0.7\linewidth,trim={0.4cm 0 0cm 0},clip]{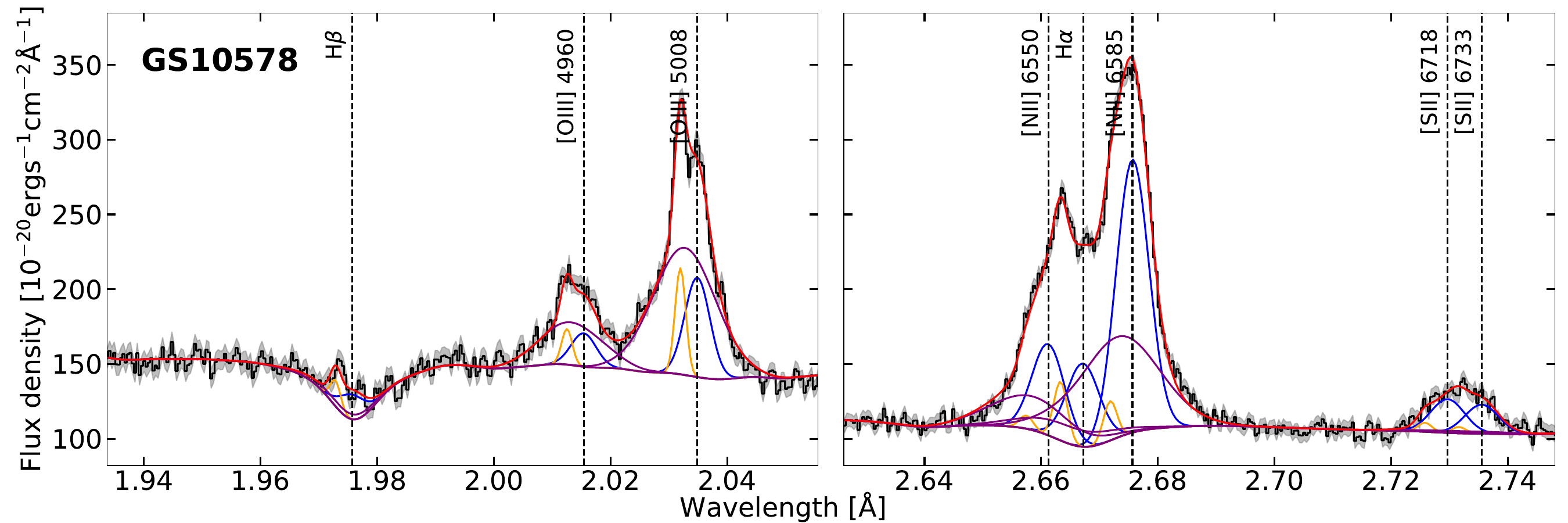}\\
    \includegraphics[width=0.7\linewidth,trim={0.4cm 0 0cm 0},clip]{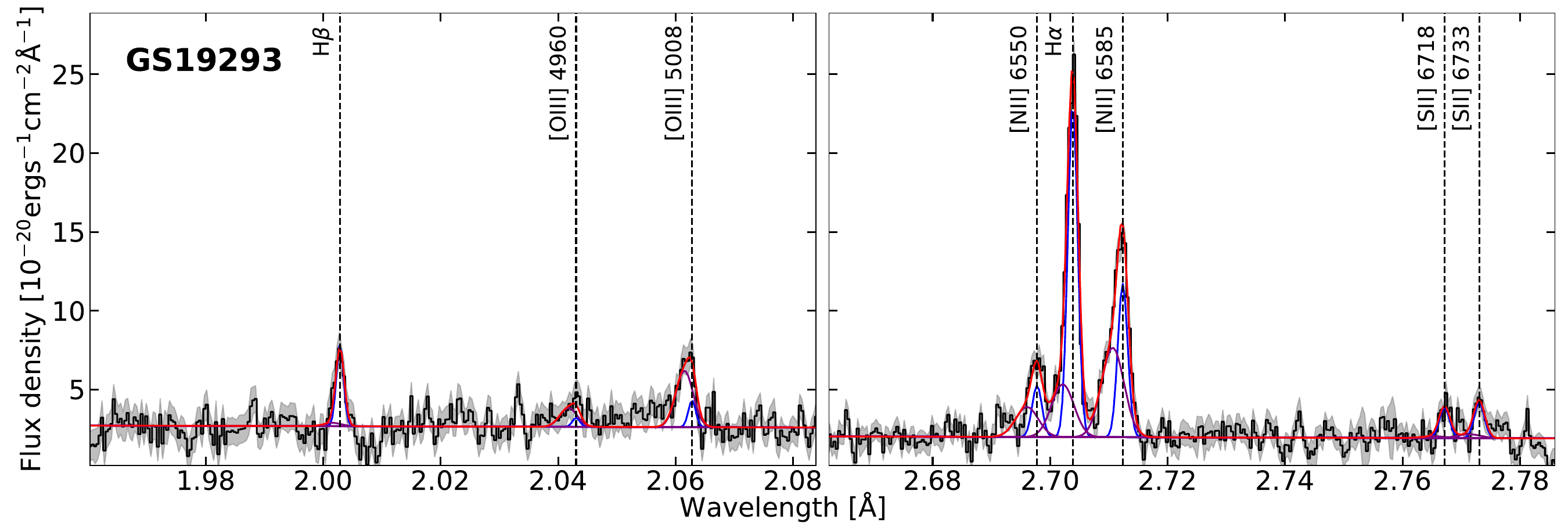}\\
    \includegraphics[width=0.7\linewidth,trim={0.4cm 0 0cm 0},clip]{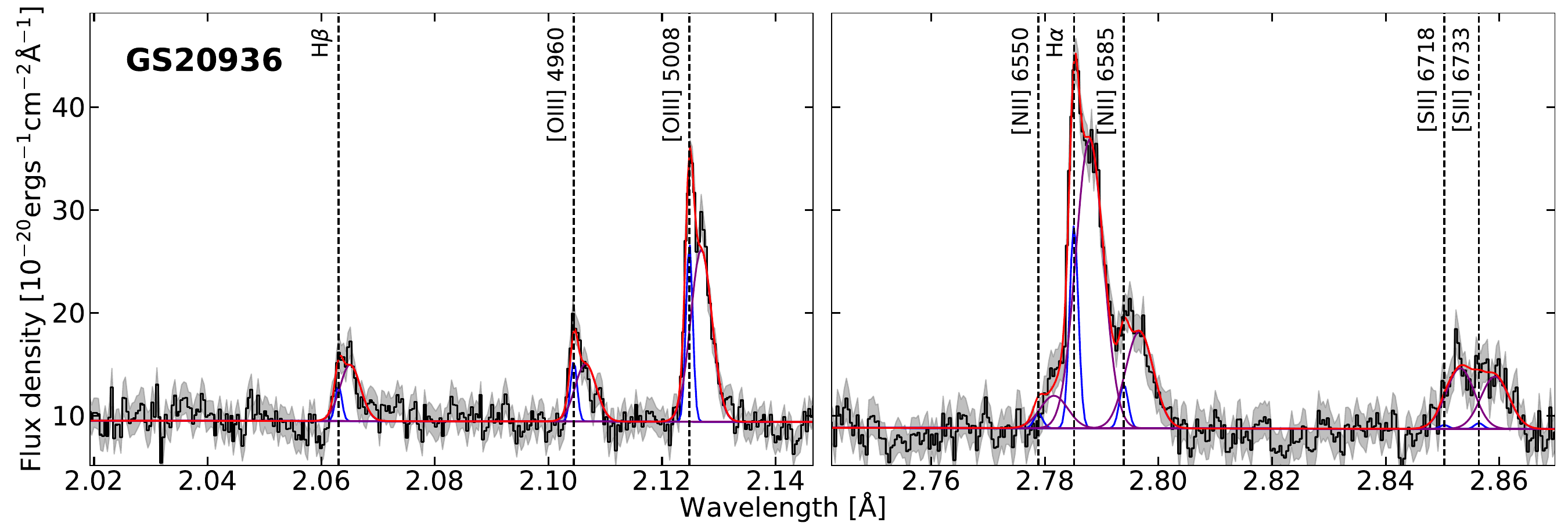}
    \caption{continued.}
\end{figure*}

\section{Re-calculation of outflow properties from the literature}\label{sec:app_outf_props_lit}
Here we describe how the properties of the AGN ionised outflows from the literature were re-calculated to be directly comparable with the analysis presented in this work (Sect.~\ref{sec:outf_props}). 
As a general rule, we only considered outflow measurements for which the flux of a broad, modelled Gaussian component was employed for the outflow mass and \rmax was provided or could be inferred from the reported broad component emission maps. 
The outflow mass and related quantities have in all cases been scaled to the assumed density of 1000~cm$^{-3}$ and a value of $C$ = 3 was adopted for the mass outflow rate; 
also, when \oiii instead of \ha (or \hb) was used, we multiplied the derived outflow masses (and resulting rates) by a factor of 3, which is the \ha- (or \hb-) versus \oiii-derived mass ratio commonly found on average for AGN outflows \citep[e.g.][]{Carniani2015, Fiore2017, Venturi2023}, due to the different volumes from which these lines are emitted.

When a different definition of the outflow velocity was used, the centroid offset velocity and the velocity dispersion of the broad component, when available, were used to calculate $v_\mathrm{max} = |v_\mathrm{bro} - v_\mathrm{nar}| + 2 \sigma_\mathrm{bro}$ (as in Sect.~\ref{sec:outf_props}). When instead only non-parametric velocities (e.g. $v_{02}$, $v_{05}$, and $v_{10}$, being the 2nd-, 5th-, and 10th-percentile velocities of the observed line profile, respectively) were given, we adopted $|v_{02}|$ (to be more precise, the maximum between it and $|v_{98}|$) as tracer of \vmax \citep[see e.g.][]{Rupke2013}. 
This is the case for \cite{Rupke2013},  \cite{Harrison2014}, \cite{Perna2019}, and \cite{Vayner2025}; the latter two include in the compilation by \cite{Fiore2017}.
In the cases in which $v_{10}$ was adopted \citep{Vayner2021a, Tozzi2021}, we re-scaled this to \vmax by a multiplicative factor of $\sim$1.7 or 1.4 when obtained from the total \oiii line profile or the broad component only, respectively, 
which are the median ratios between \vmax and $v_{10}$ that we find for our GS-AGN sample (consistent with those found by \citealt{Bertola2025} for the GA-NIFS COS-AGN sample and by \citealt{Tozzi2024}). For \cite{Tozzi2024}, who adopted $v_{10}$ to infer the outflow quantities, these were re-scaled by using the values of \vmax provided by the authors.
The outflow velocity of the targets from \cite{Carniani2015} reported in the \cite{Fiore2017} compilation was obtained from spectro-astrometry, not compatible with the method used in our work; therefore, for these targets, we considered the \oiii $v_{10}$ in the outflow region from the maps in \cite{Carniani2015} as a proxy of the outflow velocity and re-scaled it to \vmax as above.
\cite{Tozzi2021} instead adopted the maximum \oiii $v_{10}$ observed in the maps as the outflow velocity, which is an extreme definition and also incompatible with our approach; therefore also in this case we adopted the typical $v_{10}$ from the outflow region observed in the maps, and re-scaled it to \vmax.
Finally, \cite{Davies2020} used the full width at zero power (FWZP) of the entire \nii+\ha complex minus the velocity separation of the \nii doublet lines, divided by two; this gives the total width of the line where its intensity is non-zero, which is roughly compatible with \vmax.

For the outflow radius, when a different estimate other than the maximum radius \rmax was provided, we re-estimated \rmax from the broad outflowing component maps (or 2D spectra, for the case of the long-slit observations in \citealt{Bischetti2017}) reported in each study. This is the case of \citet[][who adopted the semi-major axis of an ellipse fitted to the \oiii emission line image]{Harrison2014}, \citet[][whose radius was obtained through spectro-astrometry, giving much smaller values than \rmax]{Carniani2015}, and \citet[][who adopted a single value for all the targets, as obtained from the most robust measurement among their 2D spectra]{Bischetti2017}, among the compilation in \cite{Fiore2017}.
It also applies to \citet[][who adopted the PSF-de-convolved HWHM of a 2D-Gaussian fit to the broad component emission]{Davies2020} and \citet[][who adopted the 90\% of the broad component flux radius]{Vayner2021a}.
We stress that in all cases the reported \rmax are not de-convolved by the spatial resolution of the observations, since this was not provided in the studies from the literature. In \cite{Leung2019}, where the PSF-de-convolved outflow radii were given, we re-convolved them by the PSF in order for them to be comparable with the other measurements.
In all the cases in which the outflow (broad component) was spatially unresolved, we adopted the FWHM$_\mathrm{PSF}$ as an upper limit on \rmax (see Sect.~\ref{sec:outf_props} for the choice of FWHM$_\mathrm{PSF}$ over HWHM$_\mathrm{PSF}$).
This is also the case of the spatially unresolved outflow in \cite{Marshall2025a}, whose properties were calculated for the first time in this work, as mentioned in Sect.~\ref{sec:outf_props}.

To calculate the mass outflow rate and related quantities, \cite{Kakkad2020} used a fixed radius of 2~kpc for all targets, the $W_{80}$ at 2~kpc for resolved targets (or the integrated $W_{80}$ value for unresolved targets) as the outflow velocity, and the emission line flux at $|v| > 300$~\kms.
Since these are incompatible with our approach, we re-calculated all of the outflow properties by using the flux of the \oiii broad component (multiplied by 3, as explained before) and \vmax provided by \cite{Kakkad2020}. For the outflow radius, we used the maximum projected spatial extent where $W_{80}$ $>$ 600~\kms, $D_{600}$; when not available, we used the PSF FWHM as an upper limit on \rmax.
For the AGN outflows from the SUPER survey \citep{Kakkad2020, Tozzi2024}, for the values of \lbol and SFR we adopted the SED fitting results as updated in \citet[which for \citealt{Kakkad2020} only applies to the targets in the COSMOS field]{Bertola2024}. 

We also re-calculated the outflow properties of \cite{Vayner2021a}, who separated nuclear and extended outflow components. In order to get outflow quantities comparable with the approach we used for the other targets, we  summed the masses of these nuclear and extended outflow components, obtained a single mass-weighted outflow velocity from the outflow velocities of the different components, and adopted the maximum radius reached by the outflow in each source.

\cite{Marshall2025b} reported two values of AGN bolometric luminosity for their target, one obtained by the authors from the 5100~\AA\ luminosity, and the other by \cite{Shen2019} from the 3000~\AA\ luminosity. We adopt the mean between these two values, giving \lbol $\sim$ 7.6 $\times$ 10$^{46}$~\ergs.

From the \cite{Fiore2017} compilation, we excluded the targets from \cite{Liu2013} since they considered the surface brightness of (the whole) \oiii line at the break radius between radiation-bounded and matter-bounded clouds (where the \oiii/\hb ratio drops), which is smaller than the \rmax equivalents they report, and cannot be re-scaled to be compatible to the method used in our work due to their use of surface brightness.
Out of the targets from \cite{Fiore2017}, we also removed the $z \sim 6$ QSO SDSS~J1148 given the disputed detection of an ionised outflow from \cii \citep{Maiolino2012, Cicone2015, Meyer2022}.
We also removed the three targets of \cite{Nesvadba2006, Nesvadba2008} from the \cite{Fiore2017} compilation, because the outflow flux is obtained for the whole emission line profile instead of the broad components only and the parameters needed to infer \vmax are not available; as a consequence, we could not re-scale their properties in accordance with our approach.
For \cite{Genzel2014}, who used the average HWHM of the broad component as outflow radius for all of the targets, it was not possible to estimate \rmax, due to the absence of maps of the broad component, except for one target for which such a map was reported in \cite{ForsterSchreiber2014}; moreover, the outflow velocities in \cite{Genzel2014} are defined as $|v_{\mathrm{bro}} - \rm{FWHM}_{\mathrm{bro}}/2|$ (where the subscripts identifies the broad component), which is incompatible with our adopted \vmax definition and cannot be re-scaled to it since the authors did not provide the values of $v_{\mathrm{bro}}$ and $\rm{FWHM}_{\mathrm{bro}}$ that would have allowed for its calculation. Therefore, we excluded all of the \cite{Genzel2014} targets present in \cite{Fiore2017}. 
We excluded the outflows in the radio-loud QSOs from \cite{Vayner2021b}, since they are obtained with a method which cannot be re-scaled to the one used in this work.
For the same reason, we also did not employ the outflow measurements for the target in \cite{Vayner2024a} obtained with NIRSpec data, but we instead kept the outflow properties for the same target presented in the original \cite{Vayner2021a} work (re-calculated as explained before).
We did not include the $z\sim4$ AGN outflow from \cite{Suh2025} since the assumed electron density, needed to re-scale the outflow mass by the electron density value assumed in our work, is not provided.

Many studies in the literature assumed an outflow radius to calculate the mass outflow rate instead of estimating it from the data, due to the lack of spatially resolved observations (e.g. assumed to be $R_\mathrm{out}$ = 1~kpc in \citealt{Perrotta2019}, equal to the host galaxy half-light radius in \citealt{ForsterSchreiber2019}, \citealt{Brusa2022}, and \citealt{Musiimenta2023}, or equal to the median outflow radius found from another AGN sample at similar \lbol and $z$ in \citealt{Bessiere2024}); other studies only reported the outflow velocities \citep{Yang2023, Loiacono2024, Bugiani2025} or missed some of the quantities needed to calculate the mass outflow rate as in Eq.~\ref{eq:Moutrate} \citep{Nesvadba2017, Lau2024}.
Therefore, we excluded these targets from our literature sample as their outflow properties are not compatible with those obtained from our adopted method, and information on the quantities (e.g. \rmax or line fluxes) that would allow us to re-scale them is not available.

\cite{Speranza2024} and \cite{Musiimenta2024} obtained the outflowing gas flux with a method incompatible with ours, by considering only the fraction of the total line profile at a velocity higher than a certain value, instead of the total flux of a broad modelled Gaussian; therefore, we did not include their values in our work. 
Similarly, we did not include the outflow estimates for the $z$ < 0.3 radio galaxies from \cite{Speranza2021}, who considered as the outflowing flux the asymmetric blueshifted excess of the \oiii after mirroring the line around the peak and masking the line core.
\cite{Kakkad2022} also adopted different methods to infer the outflow velocity and the mass outflow rate in $z<0.1$ AGN from IFU data, by adopting small apertures encompassing only the nuclear emission or pseudo-slits covering only part of the outflows, or a spatially resolved approach which used the size of each spaxel as the outflow radius instead of the whole outflow size, as well as outflow electron densities estimated at the spaxel level instead of for the entire outflow; these approaches cannot be re-scaled to comply with our method, therefore we did not consider the targets from \cite{Kakkad2022} in this work.
Similarly, the estimates of the outflow properties for radio galaxies at $z \sim 3.5-4$ from \cite{Roy2026} were not included in this work, because they adopted recipes which cannot be re-scaled to those adopted in our work, specifically \vout = $[ (v_\mathrm{bro} - v_\mathrm{nar})^2 + W_{50}^2 ]^{0.5}$, a spatially resolved outflow density by fitting a single Gaussian to each \sii doublet line (instead of from broad components), and the total mass outflow rate as the average of resolved values from radial shells; in any case, their targets host powerful AGN, with \lbol $\gtrsim 10^{46.3-48.3}$~\ergs, which would not change our results since they are obtained for AGN with \lbol $\lesssim 10^{46}$~\ergs (Sect.~\ref{sec:zevol} and Fig.~\ref{fig:zevol_lbolcut}).
Our final spatially resolved outflow sample from the literature, excluding the GS-AGN from this work and the COS-AGN from \cite{Bertola2025}, comprises 107 targets.
Their properties---together with those of the COS-AGN---re-calculated as described above are reported in Table~\ref{tab:outflow_props_lit}.

\begin{table*}
\centering
\caption{Properties of the ionised outflows from the GA-NIFS COS-AGN sample \citep{Bertola2025} and from the spatially resolved literature studies considered in this work, re-calculated in a homogeneous way as detailed in Sect.~\ref{sec:outf_props} and Appendix \ref{sec:app_outf_props_lit}.}
\begin{adjustbox}{max totalheight=0.86\textheight}
\begin{tabular}{p{0.2\linewidth}ccccccccc}
\hline\hline
Name & $z$ & log(\mdot) & log(\edot) & log($\eta$) & \vmax & \rmax & log(\lbol) & log(SFR) & Ref.\\
 &  & [\Msunyr] & [\ergs] &  & [\kms] & [kpc] & [\ergs] & [\Msunyr] & \\
\hline
COS1118 & 3.643 & $0.29$ & 41.62 & $>$$-1.2$ & 823 & 2.6 & 45.20 & $<$$1.52$ & 1 \\
COS1638-A & 3.506 & $1.60$ & 44.20 & $>$$-1.7$ & 3551 & 3 & 46.70 & $<$$3.30$ & 1 \\
COS1638-B & 3.511 & $2.63$ & 44.96 & $>$$-0.67$ & 2623 & 1.9 & 46.20 & $<$$3.30$ & 1 \\
COS1656-A & 3.507 & $0.96$ & 42.70 & $>$$-0.18$ & 1326 & 2.6 & 44.94 & $<$$1.14$ & 1 \\
COS349 & 3.509 & $0.61$ & 42.39 & $>$$-0.89$ & 1384 & 2.6 & 46.17 & $<$$1.50$ & 1 \\
COS590 & 3.524 & $0.56$ & 41.86 & $>$$-1.3$ & 796 & 2.6 & 45.57 & $<$$1.87$ & 1 \\
COS2949 & 2.048 & $0.37$ & 41.78 & $-1.4$ & 901 & 2.1 & 45.36 & $1.76$ & 1 \\
\hline
SDSSJ0945 & 0.128 & $0.61$ & 42.48 & $-1.3$ & 1511 & 5.6 & 45.51 & $1.91$ & 2 \\
SDSSJ0958 & 0.109 & $0.27$ & 41.64 & $-1.3$ & 866 & 3.5 & 45.00 & $1.56$ & 2 \\
SDSSJ1000 & 0.148 & $0.40$ & 41.67 & $-1.1$ & 761 & 5 & 45.70 & $1.46$ & 2 \\
SDSSJ10101 & 0.199 & $0.62$ & 42.49 & $-1.5$ & 1523 & 12 & 46.00 & $2.08$ & 2 \\
SDSSJ10100 & 0.098 & $0.51$ & 42.21 & $-0.85$ & 1267 & 2.8 & 45.60 & $1.36$ & 2 \\
SDSSJ1100 & 0.101 & $0.72$ & 42.37 &  & 1192 & 3.3 & 46.00 &  & 2 \\
SDSSJ1125 & 0.167 & $-0.16$ & 41.73 &  & 1547 & 4.6 & 45.20 &  & 2 \\
SDSSJ1130 & 0.135 & $-0.38$ & 40.70 & $-1.6$ & 616 & 2.7 & 45.11 & $1.26$ & 2 \\
SDSSJ1316 & 0.150 & $0.59$ & 42.26 &  & 1216 & 4.8 & 45.40 &  & 2 \\
SDSSJ1339 & 0.139 & $-0.62$ & 40.29 &  & 505 & 3.5 & 44.30 &  & 2 \\
SDSSJ1355 & 0.152 & $-0.25$ & 41.05 &  & 797 & 4.6 & 45.70 &  & 2 \\
SDSSJ1356 & 0.124 & $0.59$ & 42.13 & $-1.2$ & 1049 & 6.3 & 45.10 & $1.80$ & 2 \\
SDSSJ1430 & 0.086 & $0.75$ & 42.25 & $-0.096$ & 999 & 3.2 & 45.30 & $0.85$ & 2 \\
SDSSJ1326 & 3.304 & $2.96$ & 45.13 & $0.7$ & 2160 & 10 & 47.59 & $2.26$ & 2 \\
SDSSJ1549 & 2.367 & $2.72$ & 44.50 &  & 1380 & 7 & 47.82 &  & 2 \\
SDSSJ1201 & 3.512 & $2.80$ & 44.83 &  & 1850 & 7 & 47.76 &  & 2 \\
SDSSJ0745 & 3.220 & $2.73$ & 44.78 & $-0.45$ & 1890 & 15 & 47.99 & $3.18$ & 2 \\
SDSSJ0900 & 3.297 & $2.67$ & 44.92 & $-0.23$ & 2380 & 10 & 47.91 & $2.90$ & 2 \\
LBQS0109 & 2.350 & $0.90$ & 42.90 &  & 1761 & 6.6 & 47.43 &  & 2 \\
2QZJ0028 & 2.401 & $1.96$ & 44.00 & $-0.042$ & 1845 & 5.7 & 47.15 & $2.00$ & 2 \\
HB8905 & 2.480 & $1.93$ & 43.88 &  & 1677 & 4.6 & 46.77 &  & 2 \\
HE0109 & 2.407 & $0.98$ & 42.02 & $-0.72$ & 587 & 7.5 & 47.39 & $1.70$ & 2 \\
HB8903 & 2.440 & $0.73$ & 42.59 & $-1.2$ & 1509 & 4.2 & 47.28 & $1.95$ & 2 \\
RGJ0302 & 2.239 & $0.78$ & 42.47 & $-2.1$ & 1234 & 8 & 46.34 & $2.93$ & 2 \\
SMMJ0943 & 3.351 & $0.87$ & 42.47 & $-2.2$ & 1124 & 15 & 46.76 & $3.11$ & 2 \\
SMMJ1237 & 2.060 & $0.78$ & 42.44 & $-1.8$ & 1200 & 7 & 46.72 & $2.63$ & 2 \\
SMMJ1636 & 2.385 & $0.74$ & 42.29 & $-2.4$ & 1054 & 7 & 46.28 & $3.15$ & 2 \\
XID2028 & 1.593 & $1.69$ & 43.54 & $-0.75$ & 1500 & 13 & 46.30 & $2.44$ & 2 \\
XID5321 & 1.470 & $1.14$ & 43.23 & $-1.2$ & 1950 & 11 & 46.30 & $2.36$ & 2 \\
XID5395 & 1.472 & $1.95$ & 43.86 & $-0.62$ & 1600 & 4.3 & 45.93 & $2.57$ & 2 \\
MIRO20581 & 2.450 & $1.59$ & 43.85 & $>$$-0.91$ & 1900 & 4.8 & 46.60 & $<$$2.50$ & 2 \\
I08572 & 0.058 & $-0.43$ & 41.97 & $-2$ & 2817 & 2 & 45.66 & $1.62$ & 2 \\
I10565 & 0.043 & $-0.59$ & 40.37 & $-2.6$ & 535 & 5 & 44.81 & $1.98$ & 2 \\
Mrk 231 & 0.042 & $-1.20$ & 39.95 & $-3.3$ & 665 & 3 & 45.70 & $2.06$ & 2 \\
X\_N\_160\_22 & 2.445 & $>$$2.03$ & $>$44.65 &  & 3637 & $<$0.4 & 46.74 &  & 3 \\
X\_N\_81\_44 & 2.311 & $0.60$ & 42.55 & $-1.8$ & 1682 & 0.9 & 46.80 & $2.36$ & 3 \\
X\_N\_66\_23 & 2.386 & $1.23$ & 42.49 & $>$$-1.2$ & 764 & 0.4 & 46.04 & $<$$2.43$ & 3 \\
X\_N\_35\_20 & 2.261 & $>$$-0.17$ & $>$40.80 &  & 546 & $<$2.2 & 45.44 &  & 3 \\
X\_N\_12\_26 & 2.471 & $>$$0.49$ & $>$41.79 &  & 792 & $<$0.3 & 46.52 &  & 3 \\
X\_N\_4\_48 & 2.317 & $>$$0.32$ & $>$41.78 &  & 956 & $<$2.9 & 46.16 &  & 3 \\
X\_N\_102\_35 & 2.190 & $>$$0.69$ & $>$42.68 &  & 1761 & $<$7.6 & 46.82 &  & 3 \\
X\_N\_115\_23 & 2.342 & $1.30$ & 43.11 &  & 1437 & 0.9 & 46.49 &  & 3 \\
cid\_166 & 2.448 & $>$$1.69$ & $>$43.84 & $>$$-0.12$ & 2136 & $<$0.7 & 46.87 & $<$$1.81$ & 3 \\
cid\_1605 & 2.121 & $>$$0.20$ & $>$41.64 & $>$$-1.7$ & 929 & $<$5.8 & 45.97 & $<$$1.95$ & 3 \\
cid\_346 & 2.219 & $1.02$ & 42.97 & $-1.6$ & 1689 & 2.8 & 46.62 & $2.58$ & 3 \\
cid\_1205 & 2.255 & $>$$0.56$ & $>$41.52 & $>$$-0.95$ & 540 & $<$2.5 & 45.85 & $1.51$ & 3 \\
cid\_467 & 2.288 & $>$$0.51$ & $>$42.06 & $>$$-1.6$ & 1067 & $<$9 & 46.51 & $<$$2.07$ & 3 \\
\hline
\end{tabular}
\end{adjustbox}
    \label{tab:outflow_props_lit}
    \tablefoot{
    From left to right: target name, redshift ($z$), mass outflow rate (\mdot), kinetic energy rate (\edot), mass loading factor ($\eta$ = \mdot/SFR), maximum outflow velocity (\vmax), maximum outflow radius (\rmax), AGN bolometric luminosity (\lbol), host galaxy star formation rate (SFR), and reference for the measurements.}
    \tablebib{
    1: \cite{Bertola2025}; 2: \cite{Fiore2017}; 3: \cite{Kakkad2020}; 4: \cite{Tozzi2024}; 5: \cite{Vayner2021a}; 6: \cite{Vayner2025}; 7: \cite{Wang2024}; 8: \cite{Davies2020}; 9: \cite{Leung2019}; 10: \cite{Perna2023b}; 11: \cite{Ubler2024b}; 12: \cite{Marshall2023}; 13: \cite{Marshall2025a}; 14: \cite{Marshall2025b}; 15: \cite{Zamora2025}; 16: \cite{Tozzi2021}.
    }
\end{table*}

\begin{table*}
\ContinuedFloat
\captionsetup{list=off,format=cont}
\centering
\caption{}
\begin{tabular}{lccccccccc}
\hline\hline
Name & $z$ & log(\mdot) & log(\edot) & log($\eta$) & \vmax & \rmax & log(\lbol) & log(SFR) & Ref.\\
 &  & [\Msunyr] & [\ergs] & & [\kms] & [kpc] & [\ergs] & [\Msunyr] & \\
\hline
J1333+1649 & 2.089 & $2.84$ & 45.37 &  & 3248 & 0.3 & 47.91 &  & 3 \\
J1441+0454 & 2.059 & $>$$1.75$ & $>$43.69 &  & 1661 & $<$2.8 & 47.55 &  & 3 \\
J1549+1245 & 2.365 & $2.60$ & 44.40 &  & 1413 & 0.2 & 47.73 &  & 3 \\
S82X1905 & 2.263 & $0.89$ & 41.82 &  & 515 & 2.7 & 46.50 &  & 3 \\
S82X1940 & 2.351 & $>$$0.86$ & $>$42.64 &  & 1373 & $<$0.6 & 46.03 &  & 3 \\
S82X2058 & 2.308 & $>$$1.06$ & $>$43.02 &  & 1701 & $<$2.2 & 46.39 &  & 3 \\
cdfs\_36 & 2.255 & $1.15$ & 42.98 & $-1.1$ & 1450 & 3.7 & 45.70 & $2.26$ & 4 \\
cdfs\_419 & 2.143 & $>$$-0.58$ & $>$40.72 & $>$$-1.3$ & 800 & $<$4.1 & 45.63 & $<$$0.75$ & 4 \\
cdfs\_614 & 2.453 & $>$$0.30$ & $>$41.72 & $>$$-2.2$ & 910 & $<$2.4 & 45.03 & $2.54$ & 4 \\
cid\_1057 & 2.210 & $0.26$ & 42.06 & $>$$-0.019$ & 1420 & 3.4 & 45.99 & $<$$0.27$ & 4 \\
cid\_1143 & 2.443 & $0.60$ & 42.37 & $>$$-1.4$ & 1370 & 3.3 & 45.05 & $<$$1.97$ & 4 \\
cid\_2682 & 2.437 & $>$$0.008$ & $>$41.69 & $>$$-1.1$ & 1240 & $<$2.4 & 45.76 & $<$$1.06$ & 4 \\
cid\_451 & 2.444 & $>$$1.57$ & $>$43.29 & $>$$0.63$ & 1290 & $<$2.9 & 46.31 & $<$$0.93$ & 4 \\
0812N & 2.379 & $>$$1.76$ & $>$43.15 &  & 889 & $<$1.9 & 46.70 &  & 5 \\
0826N & 2.577 & $>$$3.44$ & $>$46.11 &  & 3870 & $<$2.5 & 47.48 &  & 5 \\
1652E & 2.948 & $2.89$ & 44.87 &  & 1734 & 4.6 & 47.70 &  & 5 \\
2323E & 2.369 & $2.78$ & 45.14 &  & 2685 & 2.4 & 47.00 &  & 5 \\
J224607.57-052635.0 & 4.601 & $>$$3.91$ & $>$47.66 & $>$$1.9$ & 13392 & $<$1.0 & 48.13 & $<$$2.00$ & 6 \\
4C+19.71 & 3.589 & $0.61$ & 42.61 & $-1.3$ & 1702 & 0.7 & 47.40 & $1.92$ & 7 \\
K20-ID5 & 2.224 & $2.38$ & 44.16 & $-0.15$ & 1410 & 3.3 & 45.60 & $2.53$ & 8 \\
COS4-11337 & 2.096 & $1.29$ & 43.10 & $-1.3$ & 1459 & 8.5 & 46.20 & $2.60$ & 8 \\
J0901 & 2.259 & $1.37$ & 42.47 & $-0.95$ & 650 & 1.5 & 46.30 & $2.30$ & 8 \\
3D-HST\_1606 & 2.475 & $-0.03$ & 42.30 & $-2$ & 2605 & 5.6 & 45.82 & $2.02$ & 9 \\
3D-HST\_5095 & 2.295 & $0.97$ & 42.94 & $-1.6$ & 1706 & 5.8 & 45.40 & $2.55$ & 9 \\
3D-HST\_5130 & 2.377 & $0.14$ & 41.09 & $-0.7$ & 534 & 5.8 & 45.66 & $0.85$ & 9 \\
3D-HST\_5224 & 2.151 & $>$$0.16$ & $>$41.04 & $>$$0.16$ & 490 & $<$5.5 & 45.43 & $<$$0.00$ & 9 \\
3D-HST\_6743 & 2.487 & $>$$1.61$ & $>$43.98 & $>$$-0.58$ & 2731 & $<$5.4 & 46.57 & $2.19$ & 9 \\
3D-HST\_8388 & 2.198 & $>$$-0.64$ & $>$41.21 & $>$$-0.64$ & 1513 & $<$5.5 & 45.52 & $<$$0.00$ & 9 \\
3D-HST\_10769-1 & 2.103 & $-0.34$ & 40.90 & $-2.7$ & 742 & 10 & 45.36 & $2.34$ & 9 \\
3D-HST\_10769-2 & 2.103 & $-1.40$ & 39.51 & $-3.7$ & 446 & 12 & 45.36 & $2.34$ & 9 \\
3D-HST\_11487 & 3.408 & $>$$1.25$ & $>$43.26 & $>$$0.47$ & 1807 & $<$4.5 & 46.40 & $0.78$ & 9 \\
3D-HST\_12302 & 2.276 & $-0.05$ & 41.42 & $-2.3$ & 981 & 5.6 & 45.53 & $2.23$ & 9 \\
3D-HST\_12615 & 3.179 & $0.93$ & 42.48 & $-0.78$ & 1056 & 6.4 & 46.31 & $1.71$ & 9 \\
3D-HST\_13429 & 3.475 & $0.88$ & 42.19 & $-1.1$ & 816 & 4.5 & 46.14 & $1.94$ & 9 \\
3D-HST\_14596 & 2.447 & $1.92$ & 44.50 & $>$$1.9$ & 3481 & 5.6 & 45.62 & $<$$0.00$ & 9 \\
3D-HST\_15359 & 1.594 & $0.09$ & 41.24 & $-0.21$ & 667 & 7.3 & 45.70 & $0.30$ & 9 \\
3D-HST\_17664 & 2.187 & $0.24$ & 41.40 & $-1.2$ & 678 & 7.8 & 45.43 & $1.41$ & 9 \\
3D-HST\_17754 & 2.297 & $-0.20$ & 40.69 & $-1.7$ & 498 & 6.1 & 45.08 & $1.54$ & 9 \\
3D-HST\_19082 & 2.487 & $0.15$ & 41.17 & $-2.1$ & 583 & 11 & 45.05 & $2.20$ & 9 \\
3D-HST\_21290 & 2.215 & $0.60$ & 42.14 & $-0.18$ & 1046 & 6.3 & 46.14 & $0.78$ & 9 \\
3D-HST\_21492 & 2.472 & $>$$1.67$ & $>$43.39 & $>$$1.7$ & 1292 & $<$5.4 & 46.31 & $0.00$ & 9 \\
3D-HST\_22995 & 2.468 & $0.17$ & 41.65 & $-1.2$ & 988 & 5.9 & 45.79 & $1.32$ & 9 \\
3D-HST\_24192 & 2.243 & $0.87$ & 42.64 & $-0.92$ & 1353 & 5.9 & 45.70 & $1.80$ & 9 \\
3D-HST\_26009 & 3.434 & $1.67$ & 43.08 & $-0.22$ & 894 & 4.6 & 46.64 & $1.89$ & 9 \\
3D-HST\_26304 & 1.632 & $>$$-0.81$ & $>$40.17 & $>$$-2.2$ & 557 & $<$5.8 & 45.08 & $1.36$ & 9 \\
3D-HST\_30014 & 2.291 & $-0.30$ & 40.37 & $-1.7$ & 388 & 5.8 & 44.33 & $1.40$ & 9 \\
3D-HST\_30274 & 2.225 & $1.05$ & 42.86 & $-1.5$ & 1432 & 5.8 & 45.97 & $2.55$ & 9 \\
3D-HST\_32856 & 2.270 & $0.52$ & 42.07 & $0.048$ & 1060 & 6.2 & 45.93 & $0.48$ & 9 \\
3D-HST\_33691 & 2.243 & $0.64$ & 41.81 & $-0.26$ & 687 & 7 & 45.13 & $0.90$ & 9 \\
3D-HST\_38195 & 1.487 & $0.11$ & 41.18 & $-2$ & 608 & 10 & 45.50 & $2.11$ & 9 \\
LBQS0302-0019 & 3.287 & $>$$4.14$ & $>$46.01 &  & 1274 & $<$1.2 & 47.22 &  & 10 \\
GN20 & 4.054 & $0.32$ & 41.79 & $-2.9$ & 970 & 2 & 44.93 & $3.27$ & 11 \\
DELS\_J0411-0907 & 6.818 & $2.38$ & 44.44 & $>$$0.86$ & 1919 & 4 & 47.28 & $<$$1.52$ & 12 \\
VDES\_J0020-3653 & 6.855 & $2.44$ & 44.77 & $>$$0.71$ & 2601 & 3.3 & 47.13 & $<$$1.73$ & 12 \\
J1120+0641 & 7.080 & $>$$3.13$ & $>$45.49 & $>$$0.43$ & 2694 & $<$0.81 & 47.21 & $<$$2.70$ & 13 \\
NDWFS\_J1425+3254 & 5.890 & $>$$3.37$ & $>$46.73 & $>$$0.16$ & 8523 & $<$1.1 & 46.88 & $<$$3.20$ & 14 \\
BR1202-0725\_QSO & 4.694 & $>$$2.85$ & $>$45.58 & $>$$-0.65$ & 4140 & $<$1.0 & 47.36 & $3.51$ & 15 \\
BR1202-0725\_SMG & 4.694 & $>$$2.03$ & $>$44.94 & $>$$-1.4$ & 5050 & $<$1.0 & 44.90 & $3.41$ & 15 \\
HS0810+2554 & 1.508 & $2.26$ & 44.21 &  & 1677 & 8.7 & 45.40 &  & 16 \\
SDSSJ1353+1138 & 1.632 & $1.46$ & 43.92 &  & 3018 & 9.2 & 46.59 &  & 16 \\
\hline
\end{tabular}
\end{table*}

\section{Redshift evolution of outflow properties for all sources}\label{sec:zevol_appendix}

\begin{figure*}
    \centering
    \includegraphics[width=0.33\linewidth]{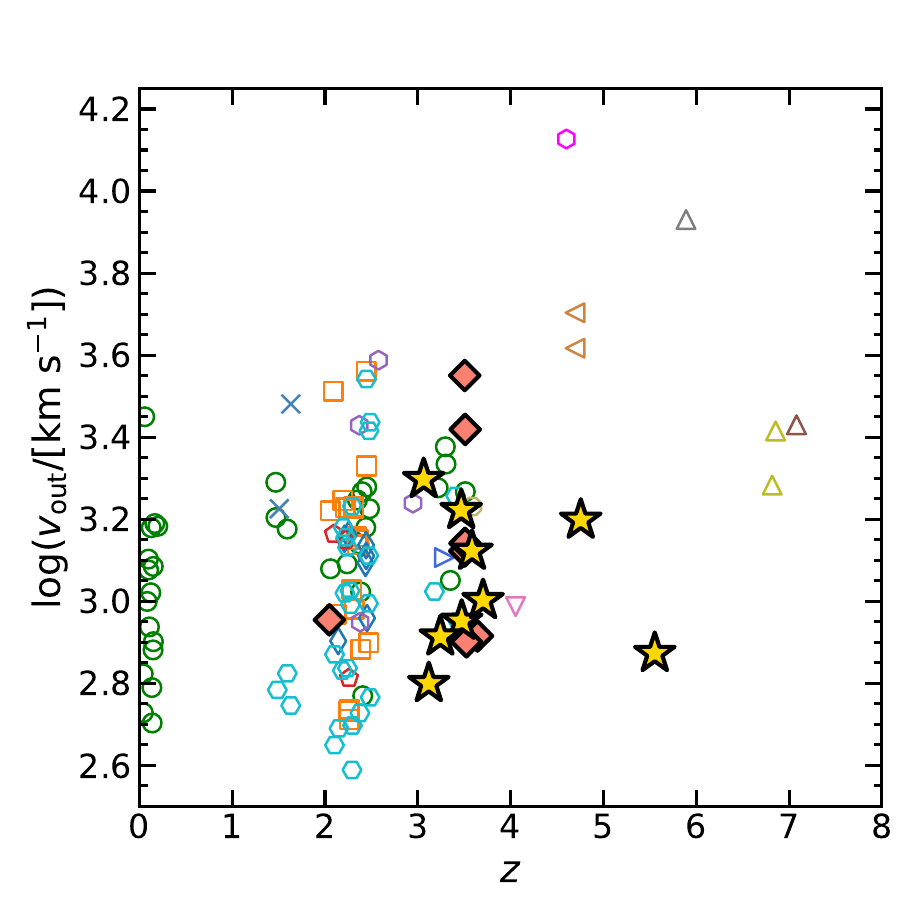}
    \includegraphics[width=0.33\linewidth]{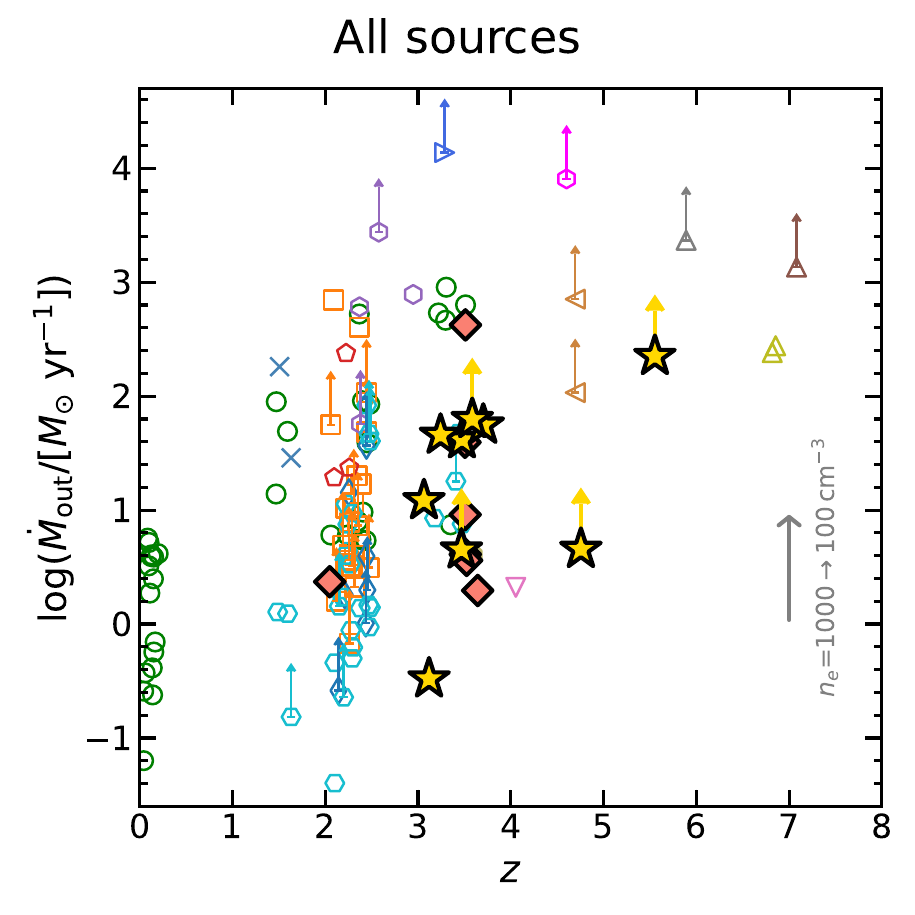}
    \includegraphics[width=0.33\linewidth]{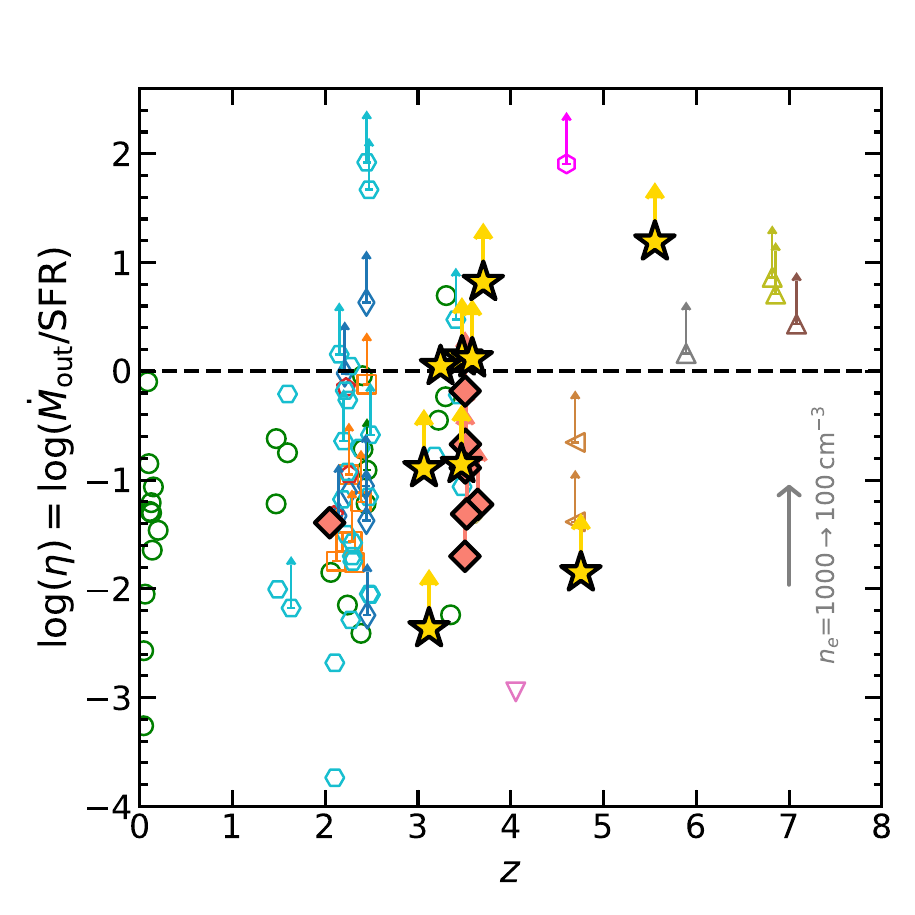}\\
    \includegraphics[width=0.33\linewidth]{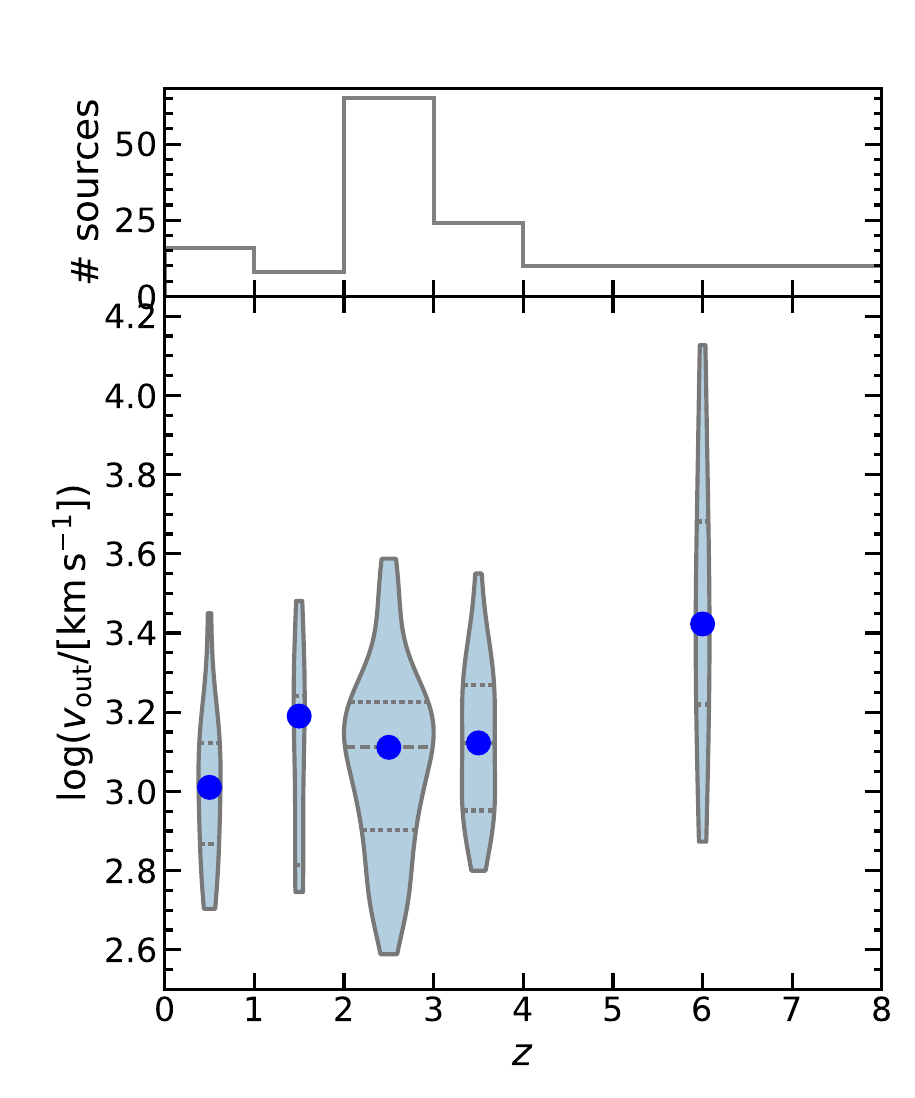}
    \includegraphics[width=0.33\linewidth]{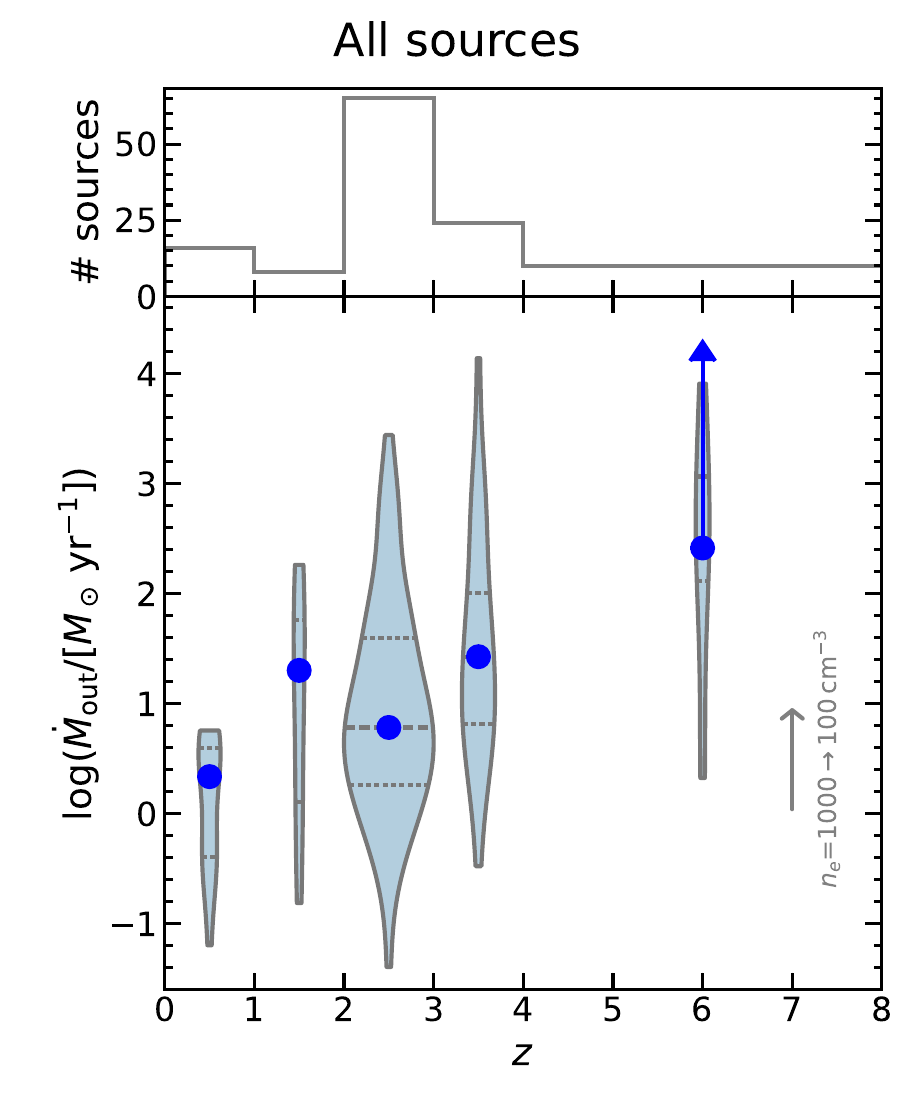}
    \includegraphics[width=0.33\linewidth]{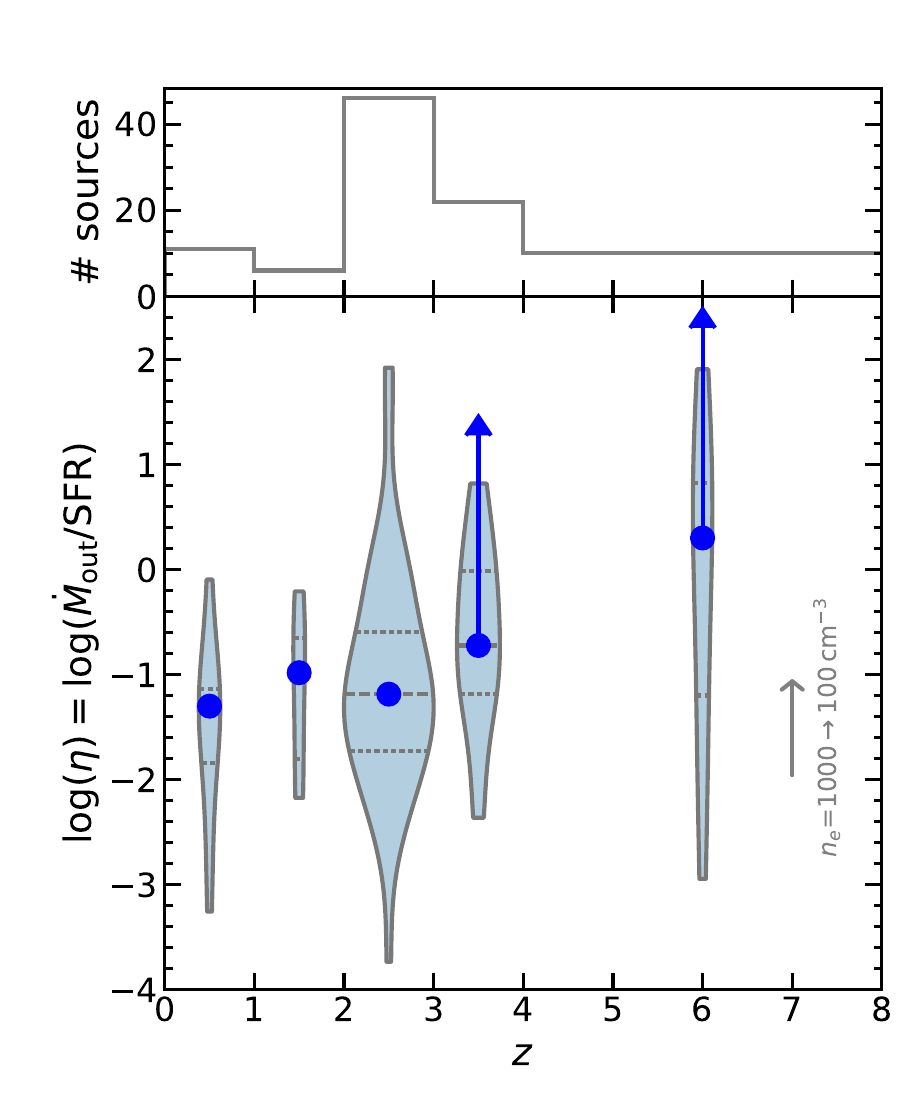}\\
    \includegraphics[width=0.33\linewidth]{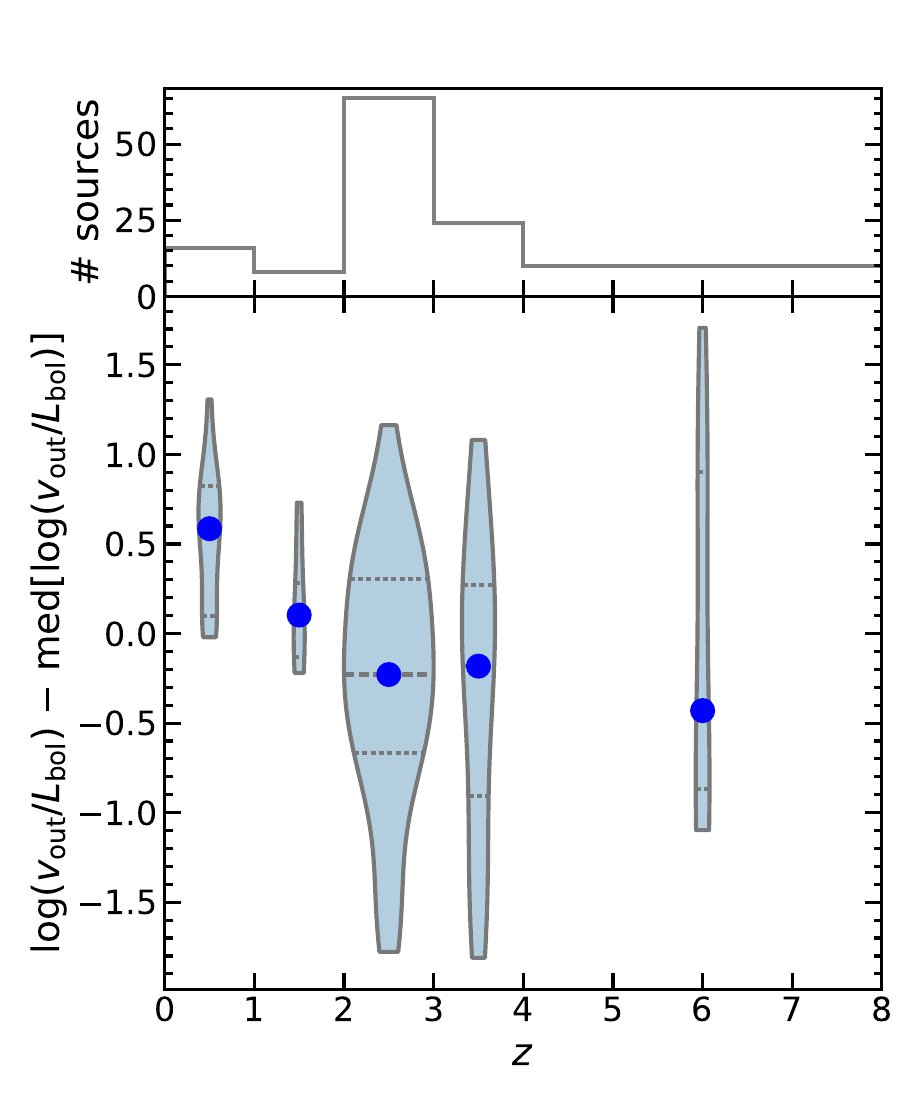}
    \includegraphics[width=0.33\linewidth]{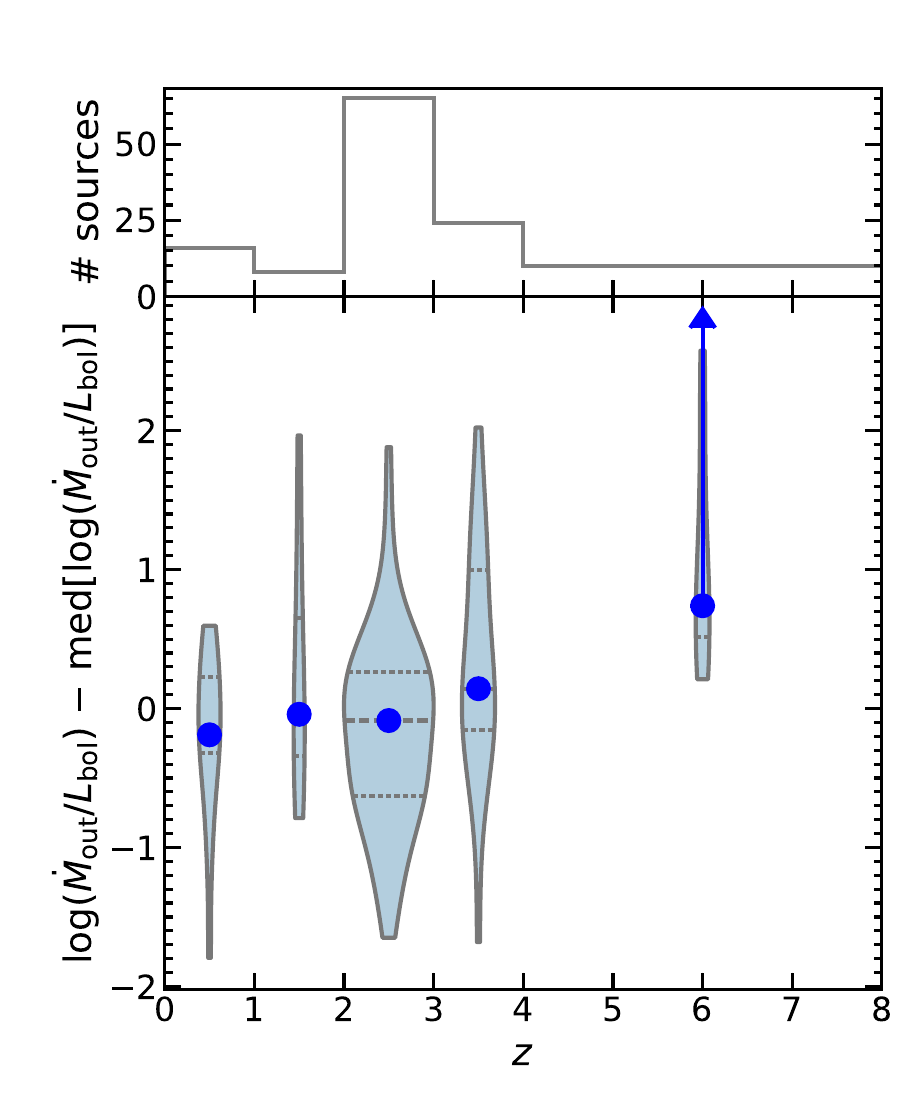}
    \includegraphics[width=0.33\linewidth]{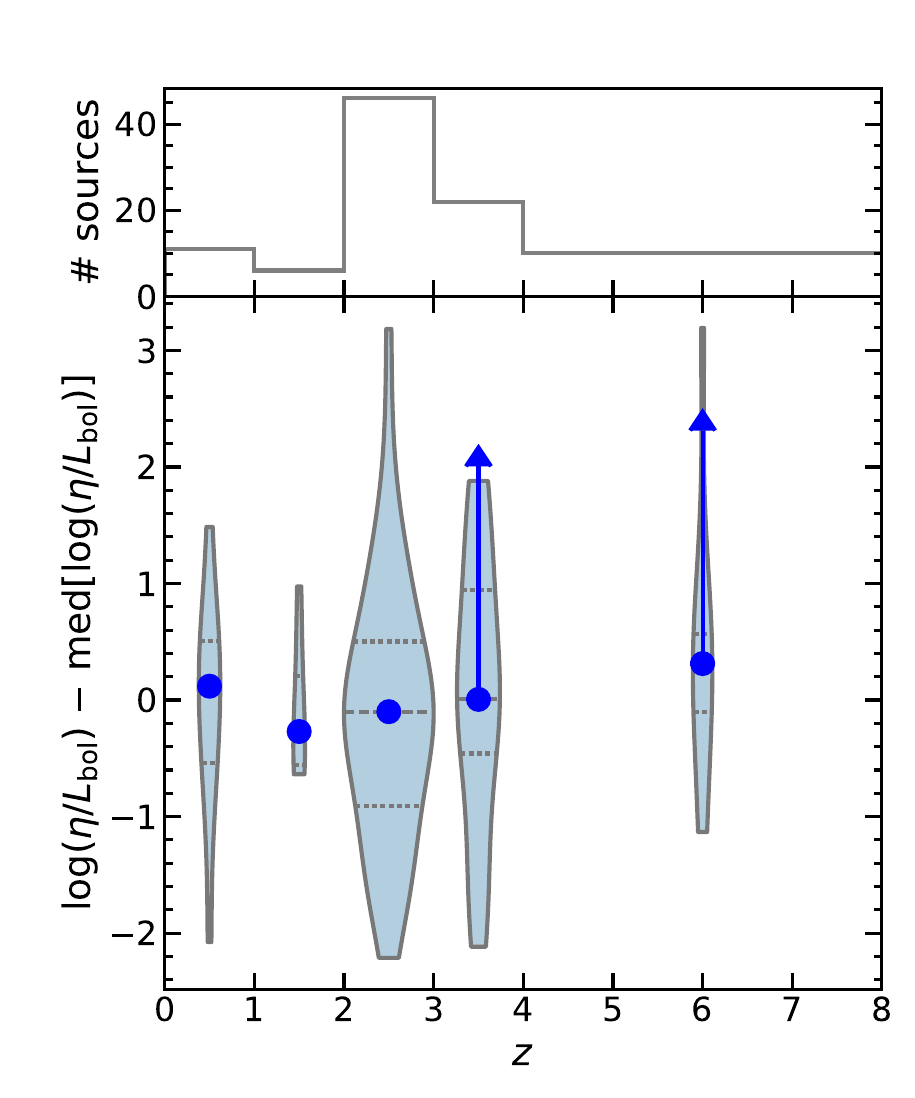}\\
    \caption{Redshift evolution of the outflow properties for all sources. Top and middle panels are as in Fig.~\ref{fig:zevol_lbolcut}. Bottom panels report the median of each quantity normalised by AGN bolometric luminosity (and subtracted by the median among all targets) in each redshift bin.}
    \label{fig:zevol_all}
\end{figure*}

In Fig.~\ref{fig:zevol_all} we show the outflow properties (\vout, \mdot, and $\eta$) as a function of redshift for all sources (top row for the single sources, middle row when binned in redshift), without applying any cut in \lbol as instead done in Fig.~\ref{fig:zevol_lbolcut}.
The same plots for all sources, but with the outflow quantities normalised by \lbol, are shown in the bottom row, in an attempt to remove the luminosity bias (as done in \citealt{Bertola2025}). These latter plots have the advantage of maximising the statistics, since all sources are considered, as opposed to those in Fig.~\ref{fig:zevol_lbolcut} where a cut at \lbol < $10^{46}$~\ergs is operated in order to get a luminosity-matched subsample.
However, we note that normalising the outflow properties by the AGN bolometric luminosity would not fully remove their dependence on that, unless the two quantities are related by a linear dependency of the form \mdot $\propto$ \lbol. 
If for example there is a log-linear dependency of the form log\,\mdot = $\alpha$ + $\beta$ log\,\lbol, as it is the case of the relation fitted by \citet[see our Fig.~\ref{fig:Lbolvoutprops}]{Fiore2017}, dividing by bolometric luminosity would give log\,(\mdot/\lbol) = log\,\mdot -- log\,\lbol = $\alpha$ + $\beta$ log\,\lbol -- log\,\lbol = $\alpha$ + ($\beta$ -- 1) log\,\lbol. 
There would thus be a remaining dependence as ($\beta$ -- 1) log\,\lbol, which in the case of the \cite{Fiore2017} relation ($\beta\simeq1.3$) would be $\simeq0.3 \log$\,\lbol. This is just an example, which can be generalised to other functional forms relating \mdot to \lbol, such as the second-order polynomial in logarithmic space fitted by \cite{Bischetti2019}. However, the intrinsic functional form relating \mdot (or any of the other outflow quantities) and \lbol is unknown, which would prevent us from ensuring the full removal of any luminosity bias by adopting any of them.

\end{appendix}

\end{document}